\documentclass{thesisclass}	

\hypersetup{
 pdfauthor={Dominik Kara},
 pdftitle={PhD Thesis},
 pdfsubject={Not set},
 pdfkeywords={Not set}
}

\newcommand{\mytitle}{Higher-Order Corrections to Higgs Boson\\Amplitudes with Full Quark Mass Dependence\\in Quantum Chromodynamics}

\def\e{\epsilon}
\def\d{\mathrm{d}}
\def\p{\partial}
\def\MS{\overline{\mathrm{MS}}}
\def\permille{\ensuremath{{}^\text{o}\mkern-5mu/\mkern-3mu_\text{oo}}}
\allowdisplaybreaks

\begin{document}

\selectlanguage{english}

\frontmatter
\pagenumbering{roman}
\newcommand{\diameter}{20}
\newcommand{\xone}{-15}
\newcommand{\xtwo}{160}
\newcommand{\yone}{15}
\newcommand{\ytwo}{-253}

\begin{titlepage}
	\vspace*{1.0cm}
	\begin{center}
		\huge{\mytitle}
		\noindent\rule{16cm}{0.4pt}\\~\\~\\
		{\Large
		\textbf{Dissertation\\~\\
		\vspace{-0.3cm}
		\textbf{zur}\\~\\
		\vspace{-0.3cm}
		Erlangung der naturwissenschaftlichen Doktorw\"urde\\
		(Dr. sc. nat.)\\~\\
		\vspace{-0.2cm}
		vorgelegt der\\~\\
		\vspace{-0.2cm}
		Mathematisch-naturwissenschaftlichen Fakult\"at\\~\\
		\vspace{-0.2cm}
		der\\~\\
		\vspace{-0.2cm}
		Universit\"at Z\"urich\\~\\
		\vspace{-0.2cm}
		von}\\~\\
		\vspace{-0.2cm}
		Dominik Kara\\~\\
		\vspace{-0.2cm}
		\textbf{aus} \\~\\
		\vspace{-0.2cm}
		Deutschland\\~\\
		\textbf{Promotionskommission}\\~\\
		\vspace{-0.2cm}
		Prof. Dr. Thomas Gehrmann (Vorsitz und Leitung der Dissertation)\\
		Prof. Dr. Massimiliano Grazzini\\
		Prof. Dr. Stefano Pozzorini\\
		Prof. Dr. Florencia Canelli
		\\~\\~\\
		\textbf{Z\"urich, 2018}\\}
	\end{center}
\end{titlepage}

\blankpage

\chapter*{Zusammenfassung in deutscher Sprache \\ \textbf{\Large{German Summary}}}

Die Entdeckung des Higgs-Bosons am Large Hadron Collider im Jahr 2012 als letztes fehlendes Teilchen im Standardmodell der Teilchenphysik setzte einen Meilenstein in der Geschichte der Teilchenphysik-Phänomenologie. Seitdem wurden die Eigenschaften und Kopplungen des Higgs-Bosons mit zunehmender Präzision gemessen, wodurch immer ge\-nauere theoretische Vorhersagen nötig werden.\\
Im Folgenden möchten wir versuchen, diesem Bedürfnis nachzukommen, indem wir Korrekturen höherer Ordnungen zur Störungsreihe von Streuamplituden, die das Higgs-Boson enthalten, im Rahmen der Quantenchromodynamik berechnen, wobei wir uns auf Prozesse mit schweren Quark-Loops konzentrieren.\\
Zunächst leiten wir die Korrekturen dritter Ordnung zum Form-Faktor her, der die Yukawa-Kopplung des Higgs-Bosons zu einem Paar von Bottom-Quarks beschreibt. Das kann erreicht werden, indem man dem herkömmlichen Ablauf von Multi-Loop-Rechnungen folgt, welcher auf der Verwendung von Feynman-Regeln sowie der Tensorzerlegung von Streuamplituden basiert.\\
Darüber hinaus berechnen wir die Zwei-Loop-Korrekturen zur $H\to Z\,\gamma$-Zerfallsrate, wobei wir die volle Abhängigkeit der internen Quarkmasse beibehalten. Das erreichen wir durch die Verwendung von Integration-by-Parts-Relationen, sodass sich die Streuamplitude als Funktion von sogenannten Masterintegralen ausdrücken lässt, welche wir mit Hilfe von Differentialgleichungen ermitteln.\\
Zuletzt beschreiben wir die Berechnung der planaren Masterintegrale, die für die Zwei-Loop-Amplitude der Higgs-plus-Jet-Produktion mit voller Quarkmassenabhängigkeit benö\-tigt werden. Im Gegensatz zur $H\to Z\,\gamma$-Zerfallsrate ist eine exakte Lösung der Differentialgleichungen zum jetzigen Zeitpunkt nicht möglich, da die Masterintegrale elliptische Strukturen beinhalten, für die eine standardisierte Lösungsmethode nach wie vor fehlt.

\chapter*{English Summary}

The discovery of the Higgs boson at the Large Hadron Collider in 2012 as the last missing particle from within the Standard Model of particle physics set a milestone in the history of particle physics phenomenology. Since then, the properties and couplings of the Higgs boson have been measured to higher and higher precision, thereby requiring increasingly accurate predictions on the theory side.\\
In the following, we attempt to satisfy this demand by evaluating higher-order corrections to the perturbative expansions of scattering amplitudes involving the Higgs boson in the framework of Quantum Chromodynamics, where we focus on processes that are mediated through heavy-quark loops.\\
First, we derive the third-order corrections to the form factor describing the Yukawa coupling of a Higgs boson to a pair of bottom quarks. This can be done by following the conventional workflow of multi-loop calculations, which is based on the application of Feynman rules and the tensor decomposition of scattering amplitudes.\\
Furthermore, we compute the two-loop corrections to the $H\to Z\,\gamma$ decay width by retaining the full dependence on the internal quark mass. We achieve this by applying Integration-by-Parts relations to express the scattering amplitude in terms of so-called Master Integrals, whose computation is carried out by means of the method of differential equations.\\
Finally, we describe the calculation of the planar Master Integrals relevant to the two-loop amplitude for Higgs-plus-jet production with full quark mass dependence by establishing a method to derive series expansions from differential equations. Unlike in the $H\to Z\,\gamma$ case, an exact evaluation of the differential equations is not feasible to date, since the set of Master Integrals involves elliptic structures, for the solution of which a standard procedure is still missing.
\blankpage

\tableofcontents
\blankpage

\mainmatter
\pagenumbering{arabic}

\chapter{Introduction}
\label{chap:introduction}

When Dirac was asked ``How did you find the Dirac equation?" he is said to have replied:
\begin{center}
\textit{``I found it beautiful."} \cite{Berry:1998}
\end{center}

This quotation from one of the greatest physicists of all time stands for the community's resolute faith in the fundamental feature of nature: Physical laws are described in terms of mathematical theories of great beauty and power. This spirit smoothed the way for a collaborative effort in the largest sense, spanning continents as well as decades, and culminating in the Standard Model of particle physics.\\

\textbf{\large{The Standard Model of Particle Physics as the Framework of This Thesis}}

\begin{figure}[tb]
	\begin{center}
	\includegraphics[scale=0.5]{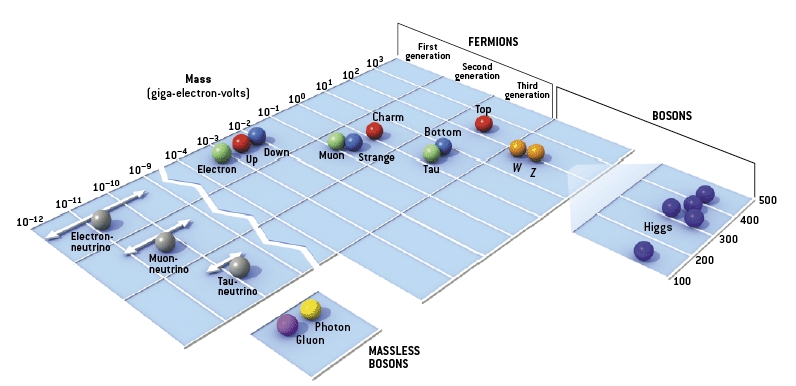}
	\caption[The Standard Model of particle physics]{\textbf{The Standard Model of particle physics} \cite{Kane:2006}}
	\label{fig:SM}
	\end{center}
\end{figure}

The Standard Model (Fig.~\ref{fig:SM}) is a relativistic quantum field theory and consists of two types of particles, which are separated according to the spin-statistics theorem: Half-integer spin particles are known as fermions whereas integer spin particles are referred to as bosons. Except for the Higgs boson, these bosons are called gauge bosons and carry the fundamental forces of nature:
\begin{itemize}	
\item \textbf{Gluons} $\boldsymbol{g}$ mediate the strong interaction associated with a non-abelian theory known as quantum chromodynamics (QCD). It describes the symmetry under $\mathrm{SU(3)_C}$ transformations, which preserves the color charge.
\item \textbf{Photons} $\boldsymbol{\gamma}$ mediate the electromagnetic force and are described by the theory of quantum electrodynamics (QED). QED is invariant under $\mathrm{U(1)_{em}}$ transformations and ensures the conservation of the electric charge $e$.
\item The $\boldsymbol{W^+}$, $\boldsymbol{W^-}$ \textbf{and} $\boldsymbol{Z}$ \textbf{bosons} mediate the weak interaction. The corresponding theory respects the symmetry under $\mathrm{SU(2)_L}$ transformations and preserves the weak isospin. Unlike gluons and photons, the weak gauge bosons are massive.
\item These weak bosons obtain their masses through the Higgs mechanism associated with the \textbf{Higgs boson}: The electroweak unification of the electromagnetic and the weak interaction is canceled by spontaneous symmetry breaking of the electroweak symmetry\footnote{\, Within this unification, $\mathrm{U(1)_Y}$ is the symmetry which preserves the hypercharge $Y$.} $\mathrm{SU(2)_L} \times \mathrm{U(1)_Y}$ to $\mathrm{U(1)_{em}}$.
\end{itemize}
Hence, the Standard Model is defined as the unification of the strong and the electroweak interactions, i.e. as the local $\mathrm{SU(3)_C} \times \mathrm{SU(2)_L} \times \mathrm{U(1)_Y}$ gauge symmetry. \\
Fermions are classified according to how they interact:

\begin{itemize}	
\item The defining property of the \textbf{quarks} $\boldsymbol{q}$ is their color charge. Consequently, their behavior is not only influenced by the electroweak interaction but particularly by the strong one. The third component of the weak isospin is given by $I_q^3 = \nicefrac{1}{2}$ for the up-type quarks $q=\{u,c,t\}$ and by $I_{q'}^3 = \nicefrac{-1}{2}$ for the down-type quarks $q'=\{d,s,b\}$. Similarly, they carry electric charge $Q_q = \nicefrac{2}{3}$ and $Q_{q'} = \nicefrac{-1}{3}$, respectively. The QCD particles known as quarks and gluons are often referred to as \textbf{partons}.
\item The remaining fermions are called \textbf{leptons}. Both the electromagnetic and the weak force act only on the charged leptons; the neutrinos do not carry electric charge, so they interact only through the weak force.
\end{itemize}

The Standard Model is widely accepted due to its outstanding experimental success: Today's formulation was finalized in the mid-1970s upon confirmation of the existence of quarks \cite{Bloom:1969, Breidenbach:1969}. Since then, discoveries of the bottom quark in 1977 \cite{Herb:1977}, the top quark in 1995 \cite{Abe:1995, Abachi:1995} and the tau neutrino in 2000~\cite{Kodama:2000} have given it further credence. The detection of the Higgs boson in 2012 completes the set of particles predicted by the SM~\cite{Aad:2012, Chatrchyan:2012}.\\
Unfortunately, the model does not incorporate the remaining fundamental interaction, namely gravitation as described by general relativity. However, gravitation is not relevant within the scales on which particle physics operates so that the Standard Model and in particular QCD is well suited to serving as the framework of this thesis.\\

\textbf{\large{Describing Collisions at the Large Hadron Collider in Perturbative QCD}}

With a circumference of 27 km and a center-of-mass energy of 7-8 TeV in its first research run dubbed \textit{Run~1}, the Large Hadron Collider (LHC) at CERN has become the world's largest and most powerful particle collider in 2010. After a two-year upgrade, the second run of the proton-proton collider, referred to as \textit{Run~2}, was launched at a combined energy level of 13 TeV. Last year, Run~2 reached a luminosity twice as high as the LHC's design value.\\
The discovery of the long-sought Higgs boson in Run~1, crowned by the award of the 2013 Nobel Prize in Physics, was one of the key goals of the LHC. Since then, the properties and couplings of the Higgs boson have been measured in detail, illustrating why highly precise Standard Model calculations are required on the theory side. This particularly applies to the mass of the Higgs boson, which has been determined as~\cite{Patrignani:2016}
\begin{equation}
m_H = 125.09 \pm 0.24\,\mathrm{GeV} \,.
\end{equation}
Another important goal of the LHC is the discovery of physics beyond the Standard Model, which has not yet been successful, although it is expected to emerge at the TeV energy level. In this context, Standard Model predictions of high precision are essential to distinguish effects that go beyond it, possibly opening the gate to new physics phenomena.\\
The crucial quantity measured at a particle collider is the so-called \textit{cross section}\footnote{The cross section $\sigma$ is replaced by the decay rate $\Gamma$ for decay kinematics.}. It describes the probability that two colliding kinds of particles yield a certain number of events of previously specified resulting particles. The \textit{inclusive} or \textit{total} cross section $\sigma$ is defined as the integral over the \textit{exclusive} or \textit{differential} cross section $\d\sigma$:
\begin{equation}
\sigma = \int \d\sigma \,.
\end{equation}
Since experiments measure final-state particles in a specific region of the phase space, only fully differential theoretical calculations can be compared to experimental data.\\
QCD with its strong couplings is the key theory for making precise predictions at a high-energy proton-proton collider like the LHC. At low energies, the strong coupling constant~$\alpha_s$ has a large value and color-charged particles cannot be isolated, but clump together to form hadrons instead. This phenomenon is referred to as confinement. In the high-energy regime of the LHC, however, the value of $\alpha_s$ decreases, implying that quarks and gluons are asymptotically free and can be treated with the method of perturbation theory. Accordingly, the cross section can be formulated as a power series in the strong coupling constant $\alpha_s$:
\begin{equation}
\d\sigma = \underbrace{\d\sigma^{(0)}}_\mathrm{LO} + \underbrace{\d\sigma^{(1)} \alpha_s}_\mathrm{NLO} + \underbrace{\d\sigma^{(2)} \alpha_s^2}_\mathrm{NNLO} + \underbrace{\d\sigma^{(3)} \alpha_s^3}_\mathrm{N^3LO} + \, \mathcal{O}(\alpha_s^4) \,.
\label{cs}
\end{equation}
Provided that the coupling constant is sufficiently small, this power series can be truncated at a certain order to approximate the full result.\\
The individual orders in Eq.~\eqref{cs} are labeled leading order (LO), next-to-leading order (NLO), next-to-next-to-leading order (NNLO), and so on. These so-called fixed-order computations are finally augmented by logarithmic resummation procedures in the form of a parton shower. This accounts for contributions from regions, where the naive perturbative expansion in $\alpha_s$ is no longer valid. More precisely, these contributions emerge at every order in $\alpha_s$ from large logarithms, that arise from an incomplete cancellation of soft and collinear divergences. These logarithms can be resummed to all orders due to the factorization of soft and collinear radiation from the hard process.\\
Beyond leading order, a fixed-order calculation is composed of real and virtual contributions. They are of the same order in the coupling constant and can be obtained from the previous order in perturbation theory by adding one real particle in the final state or one virtual loop, respectively:
\begin{align}
\begin{matrix} \includegraphics[scale=0.4]{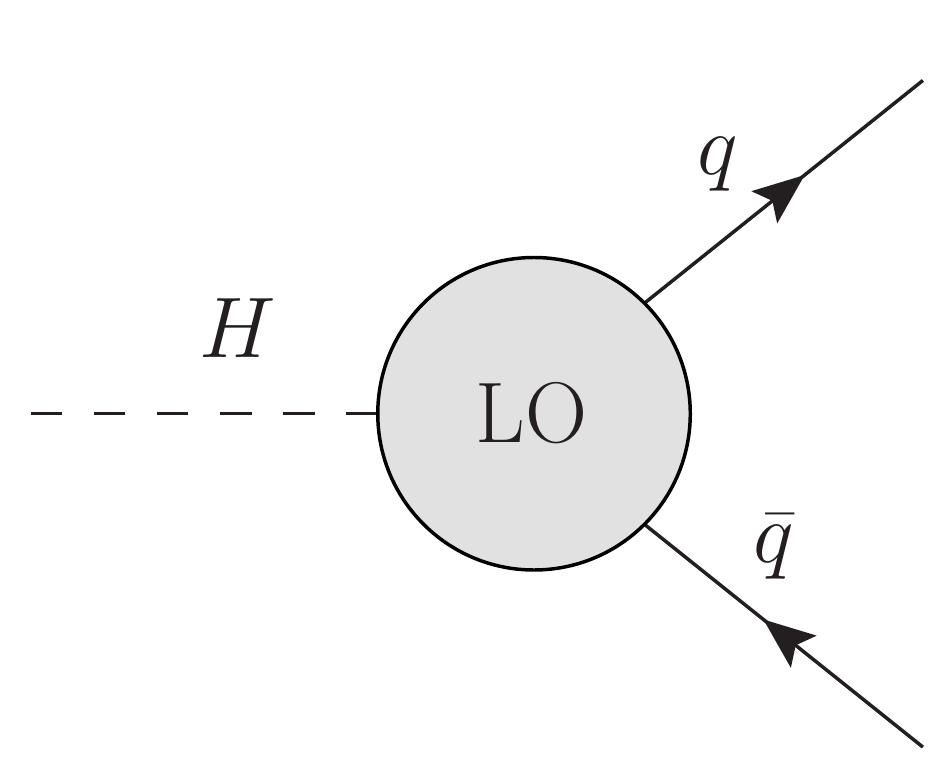} \end{matrix} &=
\begin{matrix} \includegraphics[scale=0.4]{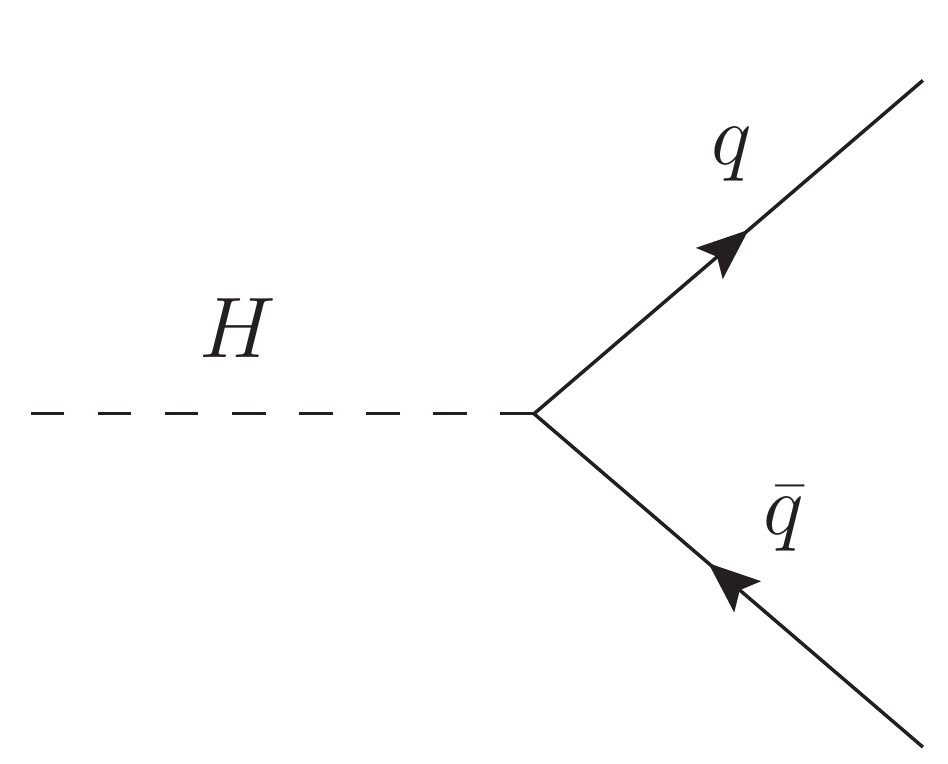} \end{matrix} \nonumber \\
\begin{matrix} \includegraphics[scale=0.4]{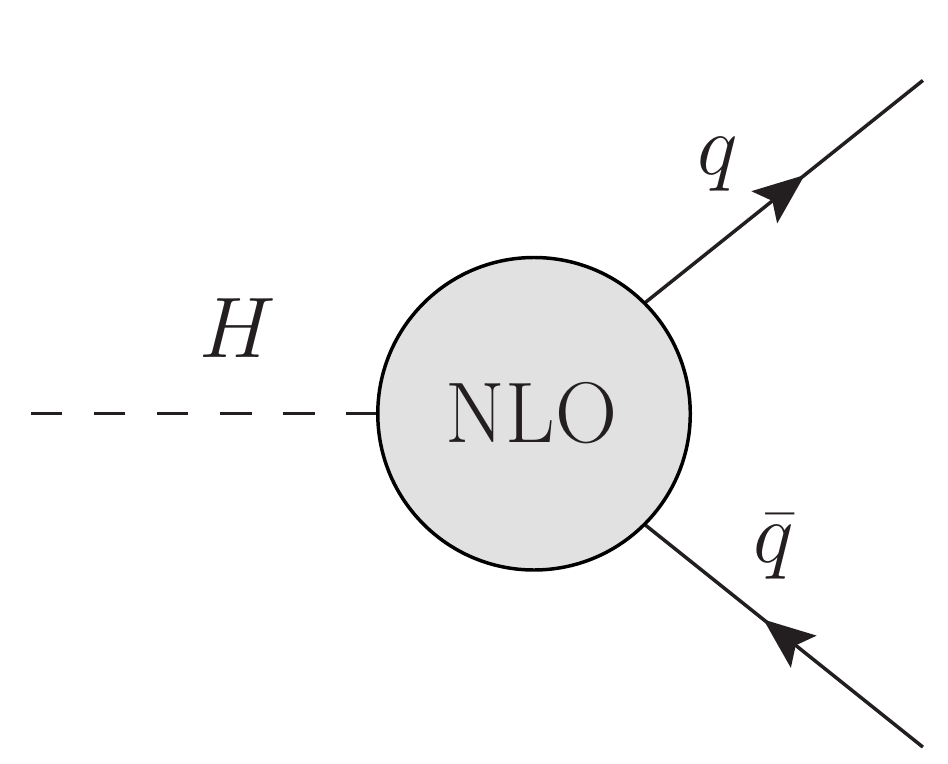} \end{matrix} &=
\begin{matrix} \includegraphics[scale=0.4]{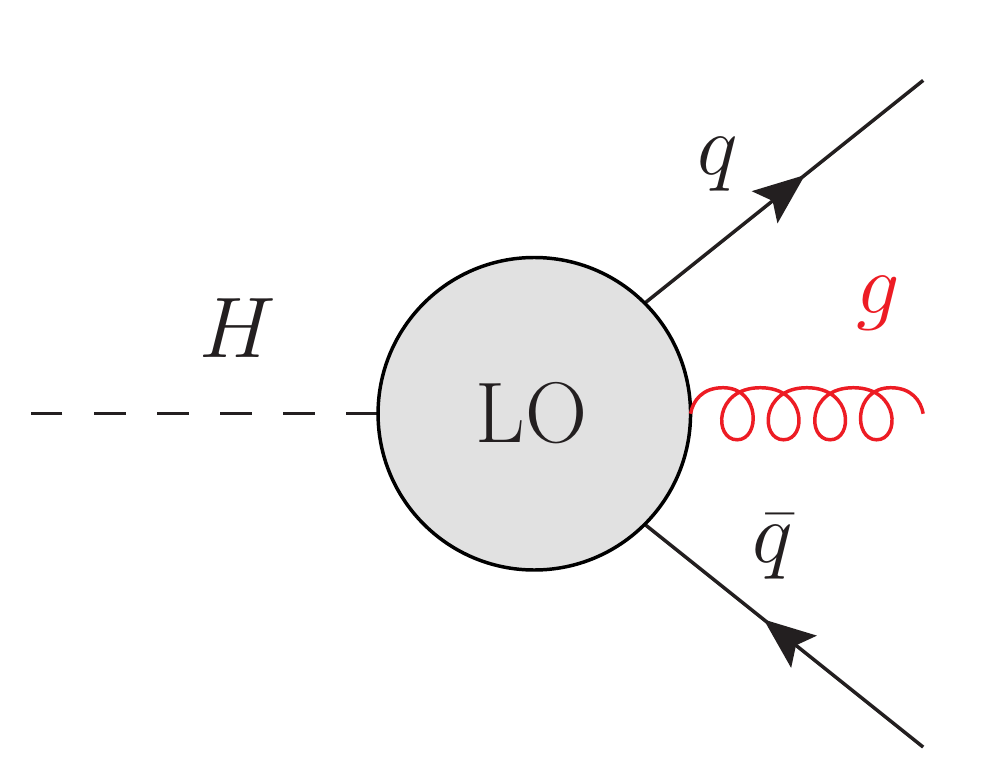} \end{matrix} + \cdots + \begin{matrix} \includegraphics[scale=0.4]{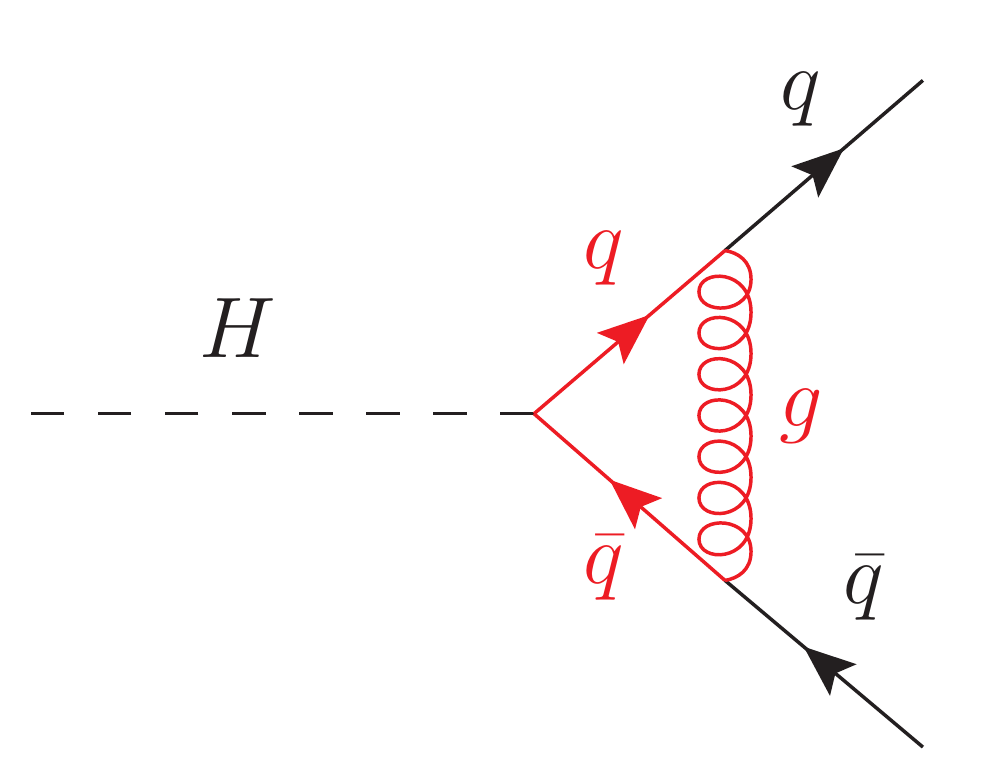} \end{matrix} + \cdots \nonumber \\
&\nonumber \\
\begin{matrix} \includegraphics[scale=0.4]{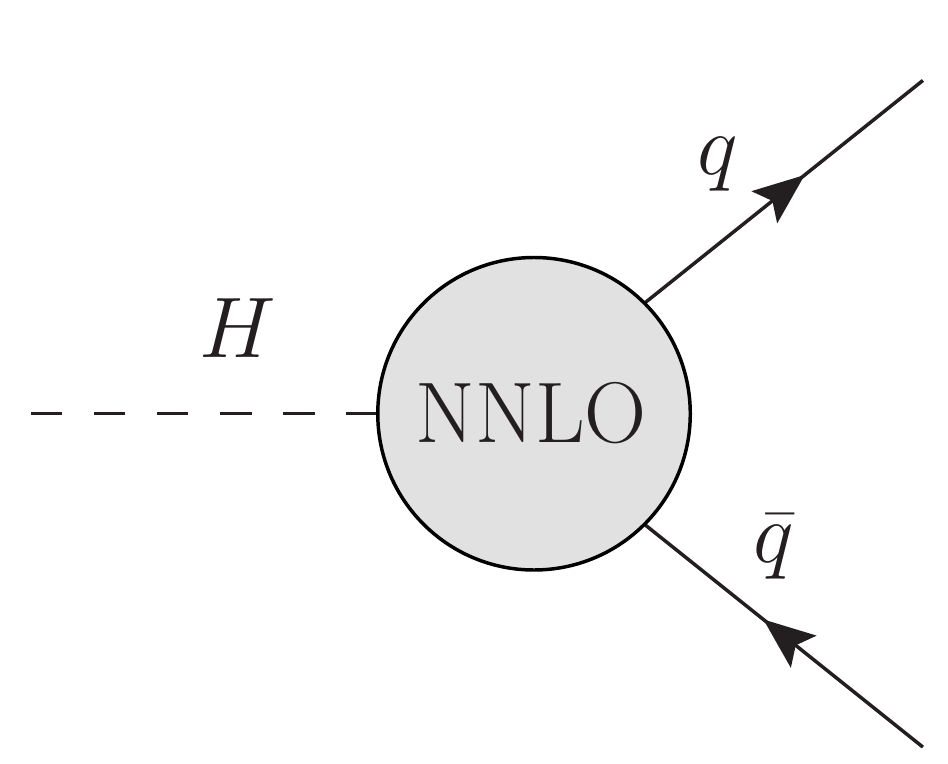} \end{matrix} &=
\begin{matrix} \includegraphics[scale=0.4]{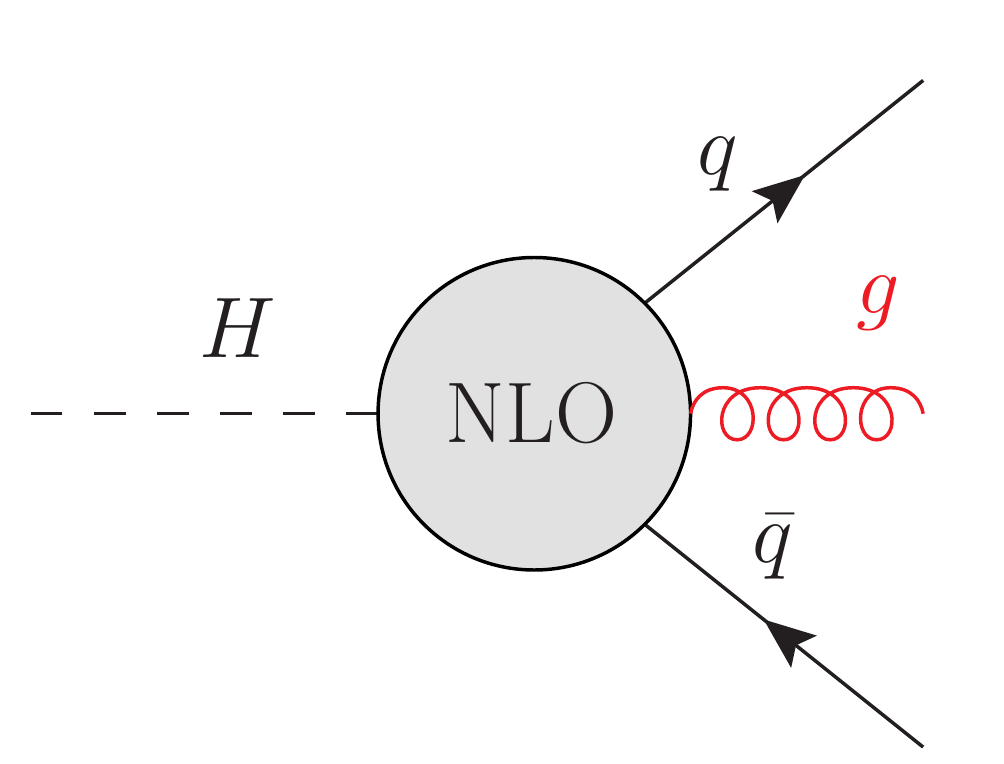} \end{matrix} + \cdots + \begin{matrix} \includegraphics[scale=0.4]{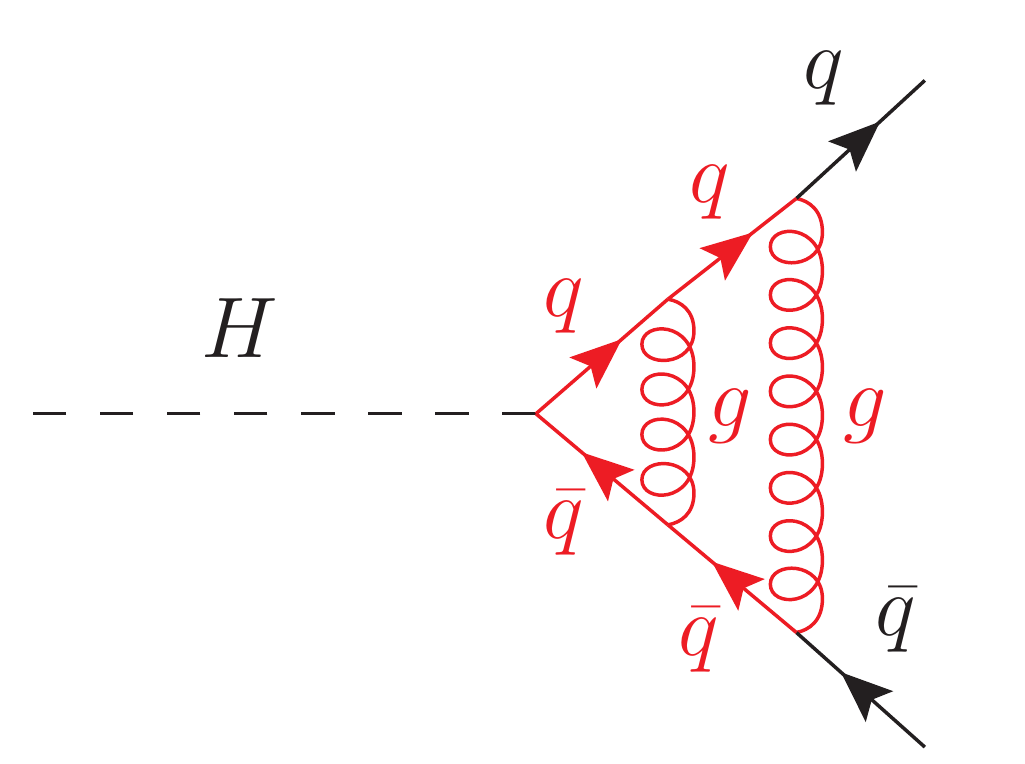} \end{matrix} + \cdots \nonumber
\end{align}
For a final state of $m$ particles generating the phase space $\Phi_m$, the cross section at NLO can then be written as
\begin{equation}
\d\sigma^{(1)} = \int_{\d\Phi_{m+1}} \left(\d\sigma^{(1)}_R - \d\sigma^{(1)}_S \right) + \left[ \int_{\d\Phi_{m+1}} \d\sigma^{(1)}_S + \int_{\d\Phi_m} \d\sigma^{(1)}_V \right] \,,
\label{subtraction}
\end{equation}
where the subscript $S$ denotes the subtraction terms for the real and virtual contributions, which are in turn denoted by the subscripts $R$ and $V$, respectively. Knowing that the sum of real and virtual contributions is infrared finite due to the KLN theorem~\cite{Kinoshita:1962,Lee:1964}, the subtraction terms are designed such that they cancel the infrared divergences of the corresponding contribution. Since the integrals in Eq.~\eqref{subtraction} and thus their singularities live in different phase spaces, the task of deriving the corresponding subtraction terms is highly non-trivial and several approaches can be found in the literature~\cite{Anastasiou:2003,Binoth:2004,Catani:2007,GehrmannDeRidder:2005a,GehrmannDeRidder:2005b,GehrmannDeRidder:2005c,Currie:2013,Czakon:2010,Boughezal:2011,Boughezal:2015a,Boughezal:2015b,Gaunt:2015,Somogyi:2008,DelDuca:2016}. This is especially true at NNLO, where Eq.~\eqref{subtraction} is extended to the double-real, real-virtual and double-virtual contributions.\\
Eq.~\eqref{subtraction} tells us that a fully differential distribution requires the computation of purely virtual contributions to the cross section, which are trivially linked to the \textit{Feynman amplitude} $\mathcal{M}$:
\begin{equation}
\d\sigma_V \propto \left|\mathcal{M}\right|^2 \,.
\label{sigmaV}
\end{equation}
This Feynman amplitude $\mathcal{M}$ represents the main building block for the three processes considered in this thesis. Its generic Lorentz-invariant structure is known and can be rephrased as the sum of all distinct tensor structures times gauge-independent coefficient functions, the latter of which are referred to as \textit{form factors}. These form factors can be expressed in terms of an irreducible set of so-called \textit{Master Integrals}, which have to be computed independently.\\

\textbf{\large{What This Thesis Is About}}

\begin{figure}[tb]
	\begin{center}
	\includegraphics[width=\textwidth]{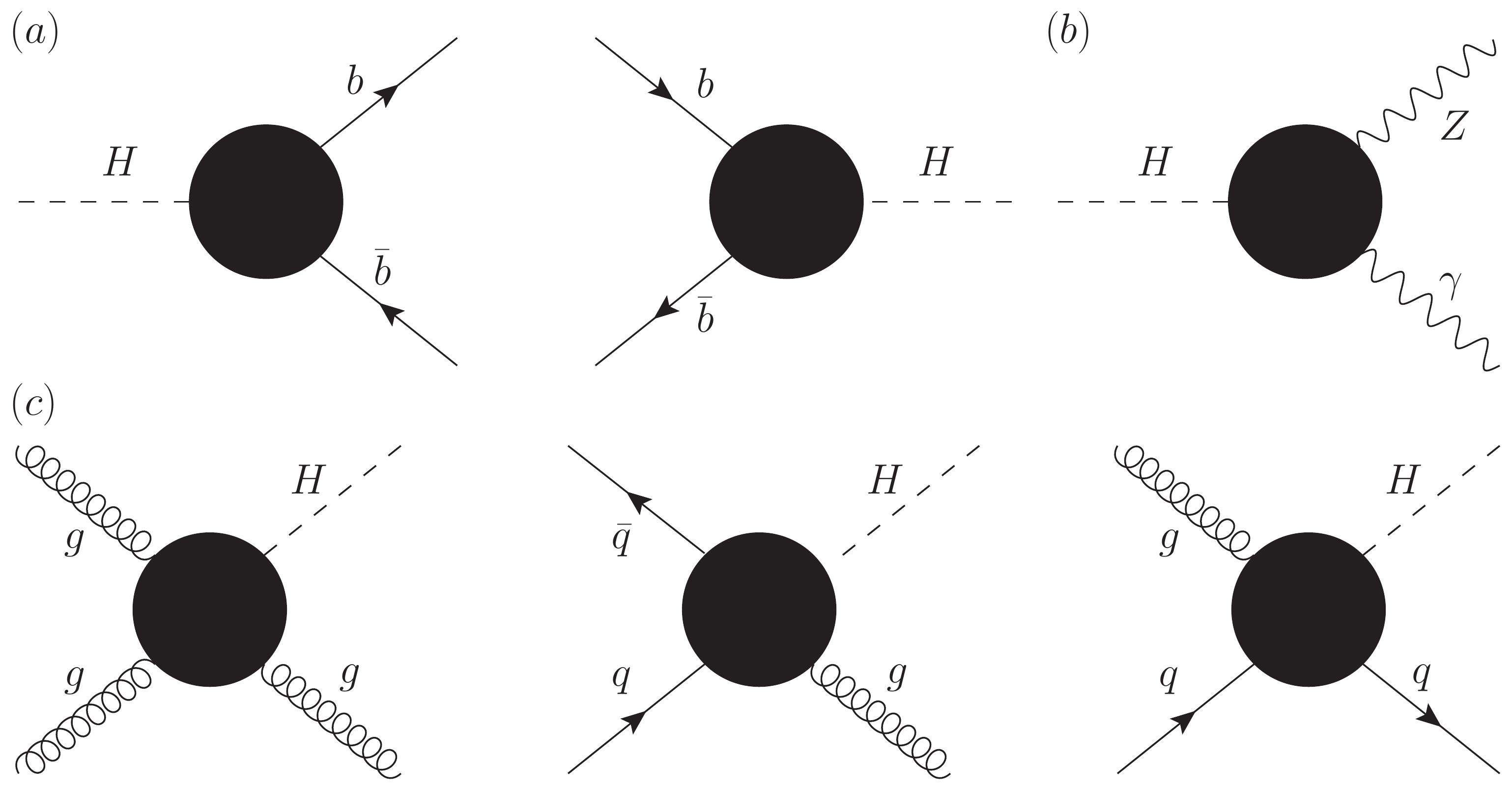}
	\caption[Shape of diagrams for the processes considered in this thesis]{\textbf{Shape of diagrams for the processes considered in this thesis.}\\
	$(a)$ Vertex diagrams for the $\mathcal{O}(\alpha_s^3)$ corrections to $H \to b \, \bar{b}$ and $b \, \bar{b} \to H$, where the black blob stands for the three-loop QCD corrections and involves bottom quarks plus gluons.\\
	$(b)$ Vertex diagrams for the $\mathcal{O}(\alpha_s)$ corrections to $H \to Z \, \gamma$, where the black blob stands for the two-loop QCD corrections and involves quarks plus gluons.\\
	$(c)$ Box diagrams for the $\mathcal{O}(\alpha_s)$ corrections to $g\,g\to H\,g$, $q\,\bar{q}\to H\,g$ and $q\,g\to H\,q$, where the black blob stands for the two-loop QCD corrections and involves quarks plus gluons.}
	\label{fig:blob}
	\end{center}
\end{figure}

The LHC as a high-energy hadron collider combined with the mass distribution of the Standard Model suggests that the Higgs boson as well as the bottom and top quarks $b$ and $t$ of the heaviest generation must play a vital role in today's particle physics phenomenology. Consequently, exploring processes that involve the couplings $Hb\bar{b}$ and $Ht\bar{t}$ of these particles in the framework of the strong interaction is obviously well suited for further understanding precision measurements at the LHC in the Standard Model and beyond. These couplings are common to all processes that are considered in this thesis:

\newpage

\begin{itemize}
\item[\textbf{(a)}] \textbf{Three-loop corrections to the $\boldsymbol{Hb\bar{b}}$ form factor in the limit of vanishing quark masses at $\boldsymbol{\mathrm{N^3LO}}$} \cite{Gehrmann:2014a}.\\
We calculate the three-loop QCD corrections to the vertex function for the Yukawa coupling of a Higgs boson to a pair of bottom quarks. This QCD form factor is a crucial ingredient of third-order QCD corrections for the fully differential decay rate of Higgs bosons to bottom quarks,
\begin{equation}
H \to b \, \bar{b}\,,
\end{equation}
and for the production of a Higgs boson in bottom quark fusion:
\begin{equation}
b \, \bar{b} \to H\,.
\end{equation}
The computation can be carried out for vanishing bottom quark mass, since it is mass suppressed by three orders of magnitude due the factor $m_b^2/m_H^2\approx 10^{-3}$. Fig.~\ref{fig:blob}$(a)$ shows the shape of the vertex diagrams for these $\mathcal{O}(\alpha_s^3)$ corrections with both bottom quarks and gluons running inside the loops.

\item[\textbf{(b)}] \textbf{Two-loop corrections to the $\boldsymbol{H\to Z\,\gamma}$ decay rate with full quark mass dependence at NLO} \cite{Gehrmann:2015a,Bonciani:2015}.\\
In the Standard Model, the decay
\begin{equation}
H \to Z \, \gamma
\end{equation}
is forbidden at tree-level and loop-mediated through a $W$ boson or a heavy quark. We analytically compute the exact QCD corrections of $\mathcal{O}(\alpha_s)$ to the heavy-quark loop, which is depicted in Fig.~\ref{fig:blob}$(b)$.

\item[\textbf{(c)}] \textbf{Two-loop corrections to the amplitude of Higgs-plus-jet production with full quark mass dependence at NLO.}\\
Fig.~\ref{fig:blob}$(c)$ illustrates that there are three possible channels for this process:
\begin{align}
g\,g&\to H\,g\,, \\
q\,\bar{q}&\to H\,g\,, \\
q\,g&\to H\,q\,.
\label{hjchannels}
\end{align}
Similarly to $(b)$, the amplitude does not exist at tree-level and is loop-mediated through a heavy quark. We calculate the QCD corrections of $\mathcal{O}(\alpha_s^2)$ to the amplitude in terms of Master Integrals. Furthermore, we describe the calculation of the planar Master Integrals as a series expansion, whose coefficients depend on the quark mass. By retaining the dependence on the quark mass, we are able to make reliable predictions at high transverse momenta of the Higgs boson, where the commonly used effective field theory description in the limit of infinite top quark mass is inappropriate, since the top quark loop is resolved by the recoiling jet.
\end{itemize}

Although it is striking that process $(a)$ is computed two orders higher in the perturbative expansion than processes $(b)$ and $(c)$, it is important to note that the three processes are specified in ascending order with respect to the complexity of the calculation. A measure for this could be given by
\begin{equation}
\mathcal{C} = \text{\# loops} + \text{\# legs} + \kappa \cdot \left( \text{\# scales} \right) \,.
\label{complexity}
\end{equation}
One may argue that summing up both the number of legs and the number of scales is not appropriate because they are not independent, however doing so accounts for massive internal propagators. Recent applications to processes with masses running in the loop have shown that they are more difficult to deal with compared to massless processes with an equivalent number of scales. The reasons for this are twofold: First, calculations including masses lead to much more cumbersome expressions, often exceeding the capacity of currently available computational resources. Second, massive integrals might produce classes of functions that are beyond the well-established framework of multi-loop calculations, so that new approaches have to be found. These effects may be underestimated in Eq.~\eqref{complexity}, however this could be compensated by the choice $\kappa>1$. The following chapters, in which we deal with corrections to heavy-quark loops, will clarify these statements.
\begin{SCtable}[3][tb]
\caption[Complexity of the processes considered in this thesis]{\textbf{Overview of the complexity of the processes $\boldsymbol{(a)}$, $\boldsymbol{(b)}$ and $\boldsymbol{(c)}$ from Fig.~\ref{fig:blob}}, expressed through the number of loops, the number of legs and the number of scales as defined in Eq.~\eqref{complexity} with $\kappa=1$.}
\begin{tabular}{lccc}
\toprule
Process &  $(a)$ & $(b)$ & $(c)$ \\
\midrule
\# loops & 3 & 2 & 2 \\
\# legs & 3 & 3 & 4 \\
\# scales & 1 & 3 & 4 \\
\midrule
$\qquad \mathcal{C}$ & 7 & 8 & 10 \\
\bottomrule
\label{tab:complexity}
\end{tabular}
\end{SCtable}
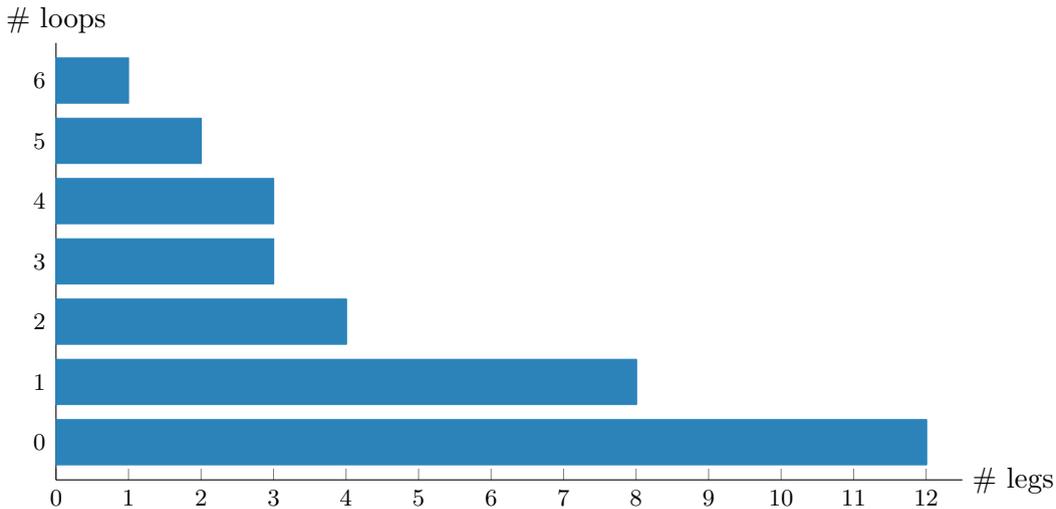
\begin{figure}
\definecolor{constructCluster}{HTML}{2B83BA}
\begin{center}
\begin{tikzpicture}
\begin{axis}[
    xbar stacked,
    legend style={
    legend columns=4,
        at={(xticklabel cs:0.5)},
        anchor=north,
        draw=none
    },
    ytick=data,
    axis y line*=left,
    axis x line*=bottom,
    tick label style={font=\footnotesize},
    legend style={font=\footnotesize},
    label style={font=\footnotesize},
    xtick={0,1,...,12},
    ytick={0,1,...,12},
    xlabel={\# legs},
    ylabel={\# loops},
    every axis x label/.style={at={(ticklabel* cs:1)},anchor=west},
    every axis y label/.style={at={(ticklabel* cs:1)},anchor=south},
    width=.9\textwidth,
    bar width=6mm,
    xmin=0,
    xmax=12.5,
    ymin=0,
    ymax=6,
    area legend,
    y=8mm,
    enlarge y limits={abs=0.625},
    visualization depends on=y \as \pgfplotspointy,
    every axis plot/.append style={fill}
]
\addplot[constructCluster] coordinates
  {(12,0) (0,1) (0,2) (0,3) (0,4) (0,5) (0,6)};
\addplot[constructCluster] coordinates
  {(0,0) (8,1) (0,2) (0,3) (0,4) (0,5) (0,6)};
\addplot[constructCluster] coordinates
  {(0,0) (0,1) (4,2) (0,3) (0,4) (0,5) (0,6)};
\addplot[constructCluster] coordinates
  {(0,0) (0,1) (0,2) (3,3) (0,4) (0,5) (0,6)};
\addplot[constructCluster] coordinates
  {(0,0) (0,1) (0,2) (0,3) (3,4) (0,5) (0,6)};
\addplot[constructCluster] coordinates
  {(0,0) (0,1) (0,2) (0,3) (0,4) (2,5) (0,6)};
\addplot[constructCluster] coordinates
  {(0,0) (0,1) (0,2) (0,3) (0,4) (0,5) (1,6)};
\end{axis}  
\end{tikzpicture}
\caption[Highest available numbers of loops and legs for processes in the literature]{\textbf{Highest available numbers of loops and legs for processes in the literature} at the level of physical results instead of individual integrals. As for process~$(c)$ in Fig.~\ref{fig:blob}, the highest available number of scales in the literature is four in case of $2\to 2$~scattering at two loops.}
\label{fig:complexity}
\end{center}
\end{figure}\noindent
For the processes considered in this thesis, the values appearing in Eq.~\eqref{complexity} are indicated in Table~\ref{tab:complexity}. Referring to physical results instead of individual integrals, Fig.~\ref{fig:complexity} reveals that process $(c)$ is at the border of feasibility in today's particle physics phenomenology. At present, only two processes with the same number of loops, legs and scales are available in the literature:
\newpage
\begin{itemize}
\item Two-loop QCD corrections to \textbf{off-shell vector boson pair production} in gluon fusion and quark-antiquark annihilation~\cite{Gehrmann:2015b,vonManteuffel:2015a,Grazzini:2015,Grazzini:2016a,Grazzini:2016b,Grazzini:2017,Caola:2014,Caola:2015a,Caola:2015b,Caola:2015c}.\\
On the one hand, the kinematics are more complicated than in the case of Higgs-plus-jet production, which becomes manifest in the non-factorizing Jacobi matrix associated with the kinematic invariants. On the other hand, massive propagators do not occur in the calculation, so that the results could be derived in fully analytic form.
\item Two-loop QCD corrections to \textbf{double-Higgs production} in gluon fusion with full quark mass dependence~\cite{Borowka:2016a, Borowka:2016b}\footnote{Very recently, the same method has been used to calculate the two-loop QCD corrections to Higgs-plus-jet production with full top quark mass dependence in a purely numerical form~\cite{Jones:2018}. However, the ratio~$m_H^2/m_t^2$ is fixed within this computation, thereby reducing the number of scales by one.}.\\
Compared to Higgs-plus-jet production, this process comes with a simpler tensor structure and more symmetries due to the additional scalar boson in the final state. This means that a lower number of Master Integrals has to be computed, however the results are only available in numerical form.
\end{itemize}
For processes $(b)$ and $(c)$, we are confronted with a combination of the main challenges of these two calculations, i.e. with analytic computation of the Master Integrals while retaining the dependence on the internal quark mass.\\

\textbf{\large{The Outline of This Thesis}}

This thesis is structured in a pedagogical way in the sense that every even-numbered chapter is designed to introduce the methods and tools required for the calculation of the phenomenological process presented in the subsequent odd-numbered chapter.\\
We start by elaborating on the foundations of perturbation theory within Quantum Chromodynamics in Chapter~\ref{chap:workflow1}, which is followed by establishing the link between cross sections and amplitudes at any loop order. In the same chapter, we describe how multi-loop scattering amplitudes can be decomposed in terms of Lorentz-invariant tensor structures, thereby enabling the computation of their scalar coefficients known as form factors. These form factors depend on a huge number of scalar integrals, which can be reduced to a minimal set of so-called Master Integrals through the application of Integration-by-Parts relations. Given that these Master Integrals are known, the standard workflow of any multi-loop calculation described in Chapter~\ref{chap:workflow1} can be used to make phenomenological predictions for particle physics processes, as presented in the case of three-loop corrections to the~$Hb\bar{b}$ form factor in Chapter~\ref{chap:hbb}.\\
For many applications, the Master Integrals are not available in the literature and thus have to be computed. For this purpose, we introduce the method of evaluating Master Integrals by solving differential equations with respect to the kinematic invariants of the problem under consideration, which has proven to be a powerful tool for many multi-scale processes in the past years. In this approach, the Master Integrals can be expressed in terms of iterated integrals referred to as Multiple Polylogarithms, whose numerical evaluation is straightforward. In Chapter~\ref{chap:hza}, we make use of these findings and elaborate on the exact computation of the two-loop corrections to the $H\to Z\,\gamma$ decay rate in terms of Multiple Polylogarithms by retaining the full dependence on the internal quark mass.\\
The Master Integrals required for the $H\to Z\,\gamma$ decay width consist of up to three-point functions, which form a subset of the planar Master Integrals necessary for the computation of the two-loop amplitudes of Higgs-plus-jet production with full quark mass dependence. However, the computation of the planar four-point functions occuring therein cannot be carried out in the same way as for the three-point functions mentioned above. This is due to the fact that the evaluation of elliptic integrals with multiple scales in the physical region is required, for which the literature currently lacks a well-tested standard procedure. In Chapter~\ref{chap:workflow3}, we therefore introduce a method designed to compute series expansions from differential equations in a parameter~$\lambda$, thereby effectively reducing multi-scale problems to the single-variable case. Multiple series expansions can then be matched in order to produce results over the whole physical region. In Chapter~\ref{chap:hj}, we apply this approach to all planar Master Integrals relevant to the two-loop amplitudes of Higgs-plus-jet production with full quark mass dependence, which includes the first analytical computation of elliptic multi-scale integrals in the physical region. We show that the results can be evaluated numerically in a fast and reliable way and compute the two-loop corrections to the scattering amplitudes for Higgs-plus-jet production with full quark mass dependence in terms of the planar and non-planar MIs. Finally, we conclude in Chapter~\ref{chap:summary}.\\

\textbf{\large{Assumptions and Notation}}

Throughout this thesis the following conventions are used unless stated otherwise:
\begin{itemize}
\item $\hbar = c = \varepsilon_0 = 1$ applies, i.e. we use \textbf{natural units}.
\item The \textbf{Einstein summation convention} is valid, i.e. we sum over all possible values of an index variable that appears twice in a single term.
\item All momenta are given in the four-dimensional \textbf{Minkowski space} equipped with a nondegenerate, symmetric bilinear form $g^{\mu\nu}$ with signature $\mathrm{(+,-,-,-)}$. External momenta are always denoted by the letter $q$ whereas $k$ and $l$ stand for loop momenta.
\item We use \textbf{dimensional regularization}, a method to isolate divergences within Feynman integrals preserving Lorentz invariance and gauge symmetry. The idea is to solve integrals in $D = 4-2 \e$ dimensions so that they can be analytically continued for all complex $D$. Singularities in the result then manifest themselves as poles in~$\e$ and those of ultraviolet character can be eliminated via renormalization (see Section \ref{sec:renormalization}). Whenever such poles appear within this thesis, they are accompanied by the additional terms $\log 4\pi - \gamma_E$ (with the Euler-Mascheroni constant $\gamma_E$), which will be omitted in the following for the sake of clarity.
\item Within dimensional regularization, the \textbf{renormalization scale} $\mu$ is introduced to ensure that physical quantities have the correct dimensions. In this thesis, we suppose $\mu = m_H$, which is evidently characteristic of processes involving the Higgs boson.
\item The integration measure of all integrals in this thesis is given by $\int \d^D k/(2\pi)^D$.
\item We work in the \textbf{Feynman-'t Hooft gauge}, i.e. $\xi=1$.
\item We use the Feynman slash notation $\slashed{p}=\gamma^\mu\,p_\mu$. The Dirac matrices $\gamma^\mu$ fulfill the Clifford algebra,
\begin{equation}
\{\gamma^\mu,\gamma^\nu\} = 2\,g^{\mu\nu} \,,
\end{equation}
and the matrix $\gamma^5$ is defined as the product of the four Dirac matrices:
\begin{equation}
\gamma^5 \equiv i\,\gamma^0\,\gamma^1\,\gamma^2\,\gamma^3\,.
\end{equation}
\end{itemize}

\chapter{The Workflow of Multi-Loop Calculations, Part I:\\From Feynman Diagrams to Amplitudes}
\chaptermark{Multi-Loop Calculations, Part I: From Feynman Diagrams to Amplitudes}
\label{chap:workflow1}

In this chapter, we introduce the theoretical foundations required to understand the calculation of the $Hb\bar{b}$ form factor in Chapter~\ref{chap:hbb}. We begin by outlining the basics of perturbation theory within QCD in Section~\ref{sec:QCD}, including the derivation of the QCD Feynman rules from the Lagrangian and the computation of well-defined finite quantities from ultraviolet-divergent parameters through the concept of \textit{renormalization}. As a next step, Section~\ref{sec:amplitudes} is intended to describe the application of these findings to scattering amplitudes, which are expressed in terms of scalar multi-loop integrals. Finally, we explain in Section~\ref{sec:reduction} how this huge number of loop integrals can be reduced to much fewer so-called \textit{Master Integrals}, before we complete this chapter with Section~\ref{sec:programs} by specifying the program packages we have used for each outlined step.

\section{The Perturbative Nature of QCD}
\label{sec:QCD}
As mentioned previously, all computations carried out in this thesis are embedded in the Standard Model of particle physics and, more precisely, in the framework of QCD. In this section, we therefore introduce the foundations of perturbative QCD, which are essential to understand the calculations described in the following chapters. The considerations made here closely follow Refs.~\cite{Peskin:1995, Ellis:1991, Collins:2011, Dissertori:2003}.

\subsection{From the Lagrangian to Feynman Rules}
\label{sec:rules}

The starting point in any quantum field theory is given by a fundamental quantity referred to as the \textit{action} $\mathcal{S}_A$:
\begin{equation}
\mathcal{S}_A = \int \mathcal{L}(\phi_i(x),\p_\mu \phi_i(x))\,\d^4x\,.
\label{action}
\end{equation}
This functional is obtained by integrating the so-called \textit{Lagrangian density} $\mathcal{L}$ over the four-dimensional space-time $\d^4x$, with $\mathcal{L}$ depending on the fundamental fields $\phi_i(x)$ and their derivatives $\p_\mu \phi_i(x)$. According to the principle of least action, minimizing the action by means of
\begin{equation}
\delta \mathcal{S}_A = 0
\end{equation}
determines the equations of motion for a given Lagrangian density, which in turn yield the values of the fields as their quantized solutions.\\
The Lagrangian density of QCD is invariant under $\mathrm{SU(3)}$ transformations and reads
\begin{equation}
\mathcal{L}_\mathrm{GI} = \bar{\psi}_{q,i} \, \left(i \, \slashed{D}_{ij} - m_q \, \delta_{ij}\right) \, \psi_{q,j} - \frac{1}{4} F^a_{\mu\nu} \, F^{\mu\nu,a} \,.
\label{LGI}
\end{equation}
The QCD Lagrangian describes the dynamics of the quark and antiquark field spinors $\psi_{q,j}$ and $\bar{\psi}_{q,i}$ of flavor $q$ and mass $m_q$, where the color indices $\{i,j,\dots\}$ run over $(1,\dots,N_c)$. With $N_c=3$, quarks come in three colors and are said to be in the fundamental representation of the $\mathrm{SU(3)}$ color group. The color indices~$\{a,b,\dots\}$ of the eight gluon fields~$A^a_\mu$, which are said to be in the adjoint representation of $\mathrm{SU(3)}$, can take the values $(1,\dots,N_c^2-1)$. They appear in the definitions of the gluonic field strength tensor $F^a_{\mu\nu}$ and the covariant derivative $D^\mu_{ij}$,
\begin{align}
F^a_{\mu\nu} &= \p_\mu \, A^a_\nu - \p_\nu \, A^a_\mu - g_s \, f^{abc} \, A^a_\mu \, A^b_\nu \,, \label{fieldstrength} \\
D^\mu_{ij} &= \p^\mu \, \delta_{ij} - i \, g_s \, T^a_{ij} \, A^{\mu,a} \,,
\label{covdev}
\end{align}
where the coupling strength $g_s$ of quarks to gluons is linked with the strong coupling constant according to $\alpha_s=\nicefrac{g_s^2}{4\pi}$. The structure constants $f^{abc}$ occur in the defining Lie algebra of the eight generators $T^a$ of the $\mathrm{SU(3)}$ gauge group,
\begin{equation}
[T^a,T^b] = i\,f^{abc}\,T^c\,,
\end{equation}
which can be expressed through the Gell-Mann matrices in the fundamental representation~\cite{Gell-Mann:1962}. The color algebra relations
\begin{align}
T^a_{ij} \, T^a_{jk} &= C_F \, \delta_{jk} \,,\\
f^{acd} \, f^{bcd} &= C_A \, \delta^{ab} \,,\\
T^a_{ij} \, T^b_{ij} &= T_F \, \delta^{ab}
\end{align}
explain the origin of the Casimir operators:
\begin{align}
C_F &\equiv \frac{N_c^2-1}{2\,N_c} = \frac{4}{3} \,, \nonumber \\
C_A &\equiv N_c = 3 \,, \nonumber \\
T_F &=\frac{1}{2} \,.
\label{casimirs}
\end{align}
They are associated with gluon emissions from quarks, gluon emissions from gluons and gluon-splittings to $q\bar{q}$ pairs, respectively.\\
The Lagrangian density defined in Eq.~\eqref{LGI} is locally gauge-invariant in the sense that the physical content does not change when the fields and the covariant derivative transform according to
\begin{align}
\psi_q &\to U(x) \, \psi_q \,, \\
\bar{\psi}_q &\to \bar{\psi}_q \, U^\dagger(x) \,, \\
A_\mu &\to U(x) \, A_\mu \, U^\dagger(x) + \frac{i}{g_s} \, \left(\p_\mu \, U(x) \right) \,  U^\dagger(x) \,, \\
D_\mu &\to U(x) \, D_\mu \, U^\dagger(x)
\end{align}
under $\mathrm{SU(3)}$ transformations of the kind
\begin{align}
U(x) = e^{ig_s \theta^a(x) T^a} \,.
\end{align}
It follows that there is a degeneracy between gluon field configurations, which can be transformed into each other through local gauge transformations and thus appear to be physically equivalent. However, quantum field theories are required to yield unambiguous solutions. This problem can be circumvented by adding a gauge-fixing term $\mathcal{L}_\mathrm{GF}$ to the Lagrangian density that forces the gluons into a specific gauge and allows defining a gluon propagator:
\begin{equation}
\mathcal{L}_\mathrm{GF} =  -\frac{1}{2\,\xi} \, \left(\p^\mu A_\mu^a \right)^2 \,.
\label{LGF}
\end{equation}
This group of so-called $R_\xi$~gauges is a generalization of the covariant Landau gauge \mbox{$\p^\mu A_\mu^a = 0$}. Evidently, physical quantities cannot depend on the actual choice of $\xi$. Therefore we make use of the fact that our calculations are simplest in the Feynman-'t~Hooft gauge, as in the case of most quantum field theory computations. Hence, we assume $\xi = 1$ whenever calculating physical quantities in this thesis.\\
The introduction of a gauge-fixing expression of covariant nature into the Lagrangian leads to non-transverse polarizations within the gluon propagator, ultimately causing a violation of unitarity. According to Faddeev and Popov, these unphysical degrees of freedom are canceled by supplementing the gauge-fixing with a gauge-compensating term,
\begin{equation}
\mathcal{L}_\mathrm{GC} = \p^\mu \, \eta^{a\dagger} \, \left(\p_\mu \, \delta^{ab} + g_s \, f^{abc} \, A^c_\mu \right) \eta^b \,,
\label{LGC}
\end{equation}
giving rise to unphysical states $\eta^a$ that appear as virtual particles only, called \textit{ghosts}. As anti-commuting scalar fields with spin 0, ghosts violate the spin-statistics theorem.\\
The complete QCD Lagrangian emerges from the sum of the gauge-invariant, gauge-fixing and gauge-compensating contributions from Eqs.~\eqref{LGI}, \eqref{LGF} and \eqref{LGC},
\begin{align}
\mathcal{L}_\mathrm{QCD} = \, &\mathcal{L}_\mathrm{GI} + \mathcal{L}_\mathrm{GF} + \mathcal{L}_\mathrm{GC} \nonumber \\
= \, &\bar{\psi}_{q,i} \, \left(i \, \slashed{\p} - m_q \right) \, \psi_{q,i} - \frac{1}{4} \left( \p_\mu \, A^a_\nu - \p_\nu \, A^a_\mu \right)^2 + \p^2 \eta^{a\dagger} \, \eta^a \nonumber \\
&-\frac{1}{2\,\xi} \left(\p^\mu A_\mu^a \right)^2 - g_s \, \, \bar{\psi}_{q,i} \, \slashed{A}^a \, \psi_{q,j} + g_s \, f^{abc} \left( \p_{\mu} \, A^a_\nu \right) A^{b,\mu} A^{c,\nu} \nonumber \\
&-\frac{g_s^2}{4} \left(f^{eab} \, A^a_\mu \, A^b_\nu \right) \left(f^{ecd} \, A^{c,\mu} \, A^{d,\nu} \right) - g_s \eta^{a\dagger} \, f^{abc} \, \p^\mu \, A^b_\mu \, \eta^c \,,
\label{LQCD}
\end{align}
where we fully expanded the relations from Eqs.~\eqref{fieldstrength} and \eqref{covdev}.
By construction, $\mathcal{L}_\mathrm{QCD}$ is no longer gauge-invariant, but invariant under the BRST~symmetry~\cite{Becchi:1974a,Becchi:1974b,Becchi:1975,Tyutin:1975}. This new symmetry allows deriving \textit{Slavnov-Taylor identities}~\cite{Taylor:1971,Slavnov:1972}, that correspond to the appropriate non-abelian version of the Ward-Takahashi identities and provide the necessary tool to prove renormalizability of $\mathcal{L}_\mathrm{QCD}$ to all orders in perturbation theory.\\
From the Lagrangian in Eq.~\eqref{LQCD}, one can immediately deduce all \textit{QCD Feynman rules}, that can be understood as diagrammatic representation of the couplings and fields occuring in the Lagrangian density. Using straight lines for quarks, curly lines for gluons and dotted lines for ghosts, they are given as follows in the $R_\xi$~gauge:
\newpage
\textbf{QCD Feynman Rules for the Propagators}\\\\
\includegraphics[scale=0.625]{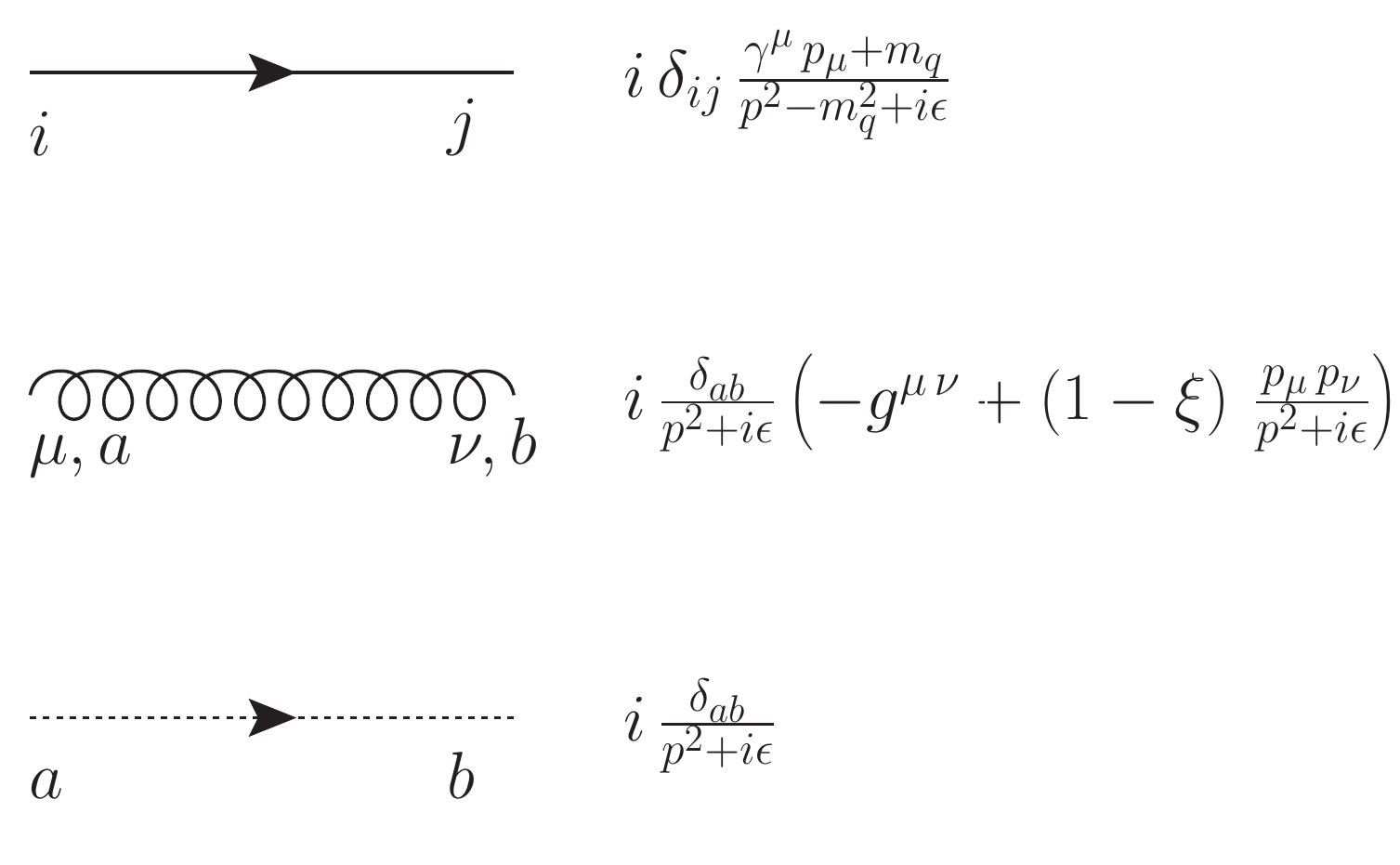}\\
The sign of the infinitesimal imaginary part $\e$ is chosen such that causality is ensured.\\\\
\textbf{QCD Feynman Rules for the Vertices}\\\\
\includegraphics[scale=0.625]{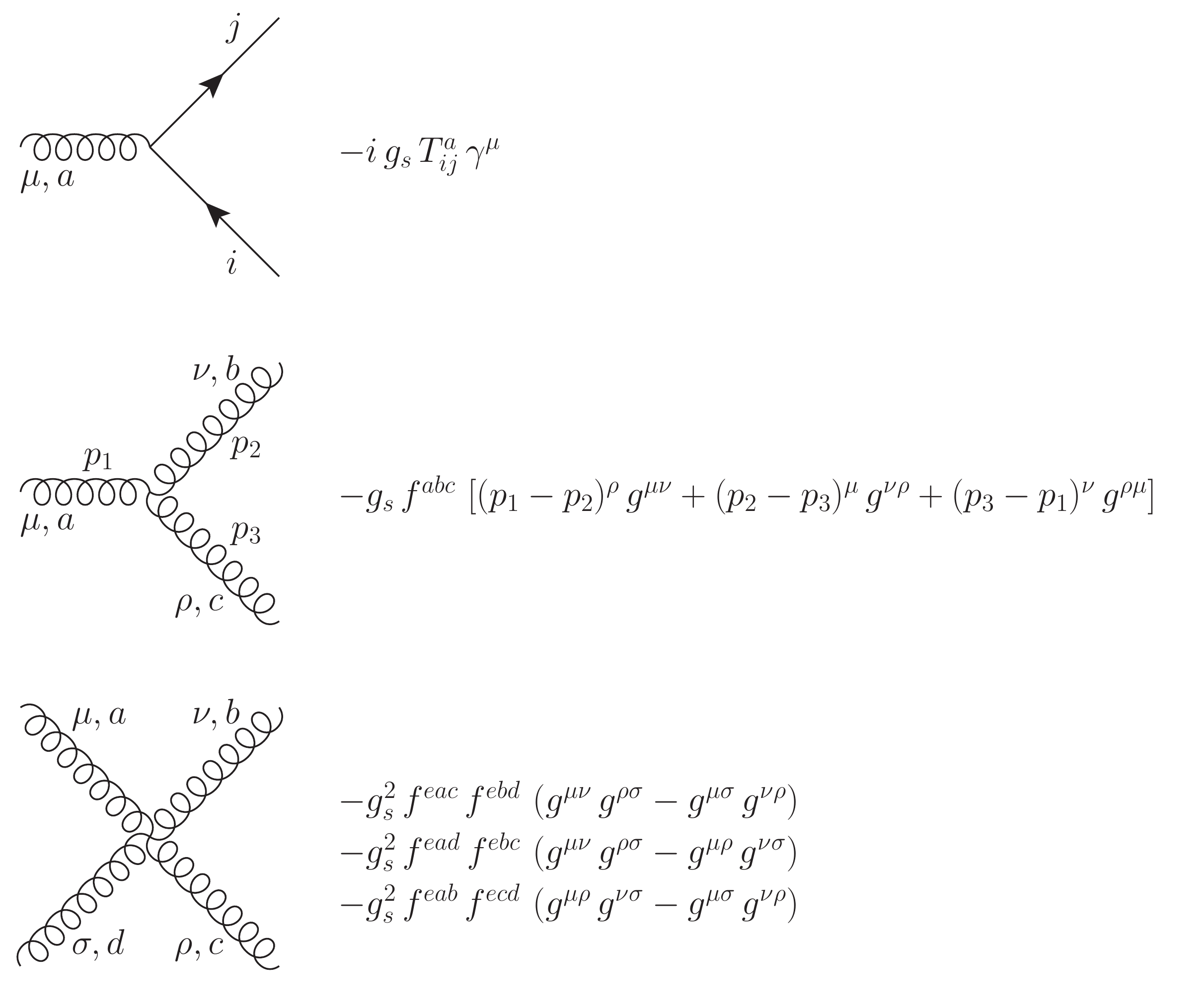}\\
\includegraphics[scale=0.625]{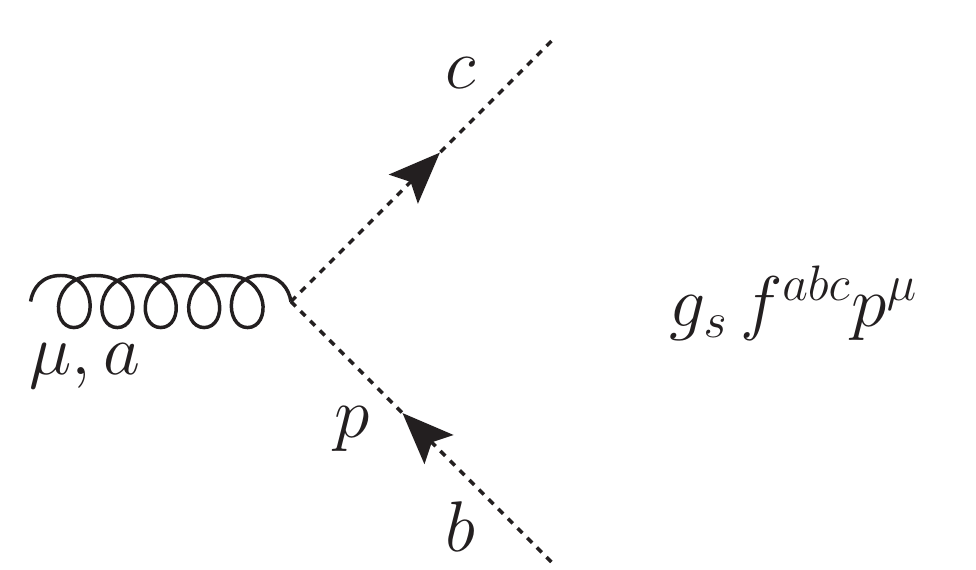}\\\\
Note that all momenta have to be taken as ingoing. Except for the ghost propagator and the ghost-gluon vertex, all these Feynman rules are needed for computing the amplitudes of the processes $(a)$, $(b)$ and $(c)$ described in Chapter~\ref{chap:introduction}. Process~$(b)$ is the only case where ghosts do not occur, since the calculated corrections correspond to the leading order in $\alpha_s$. Instead, the amplitude of process~$(b)$ requires knowledge of the neutral vector boson couplings to the quarks, and on top of that all processes have the Higgs boson coupling to the quarks in common, as mentioned previously. These additional Feynman rules can be obtained from the electroweak sector of the Standard Model Lagrangian in the same way as for the QCD Lagrangian. For the sake for completeness, we specify them without quoting the explicit electroweak Lagrangian in the Glashow-Weinberg-Salam theory~\cite{Glashow:1959,Salam:1959,Weinberg:1967}. In the following, dashed lines denote the Higgs boson and wavy lines correspond to the $Z$~boson as well as to the photon~$\gamma$:\\\\
\textbf{Additional Feynman Rules}\\\\
\includegraphics[scale=0.625]{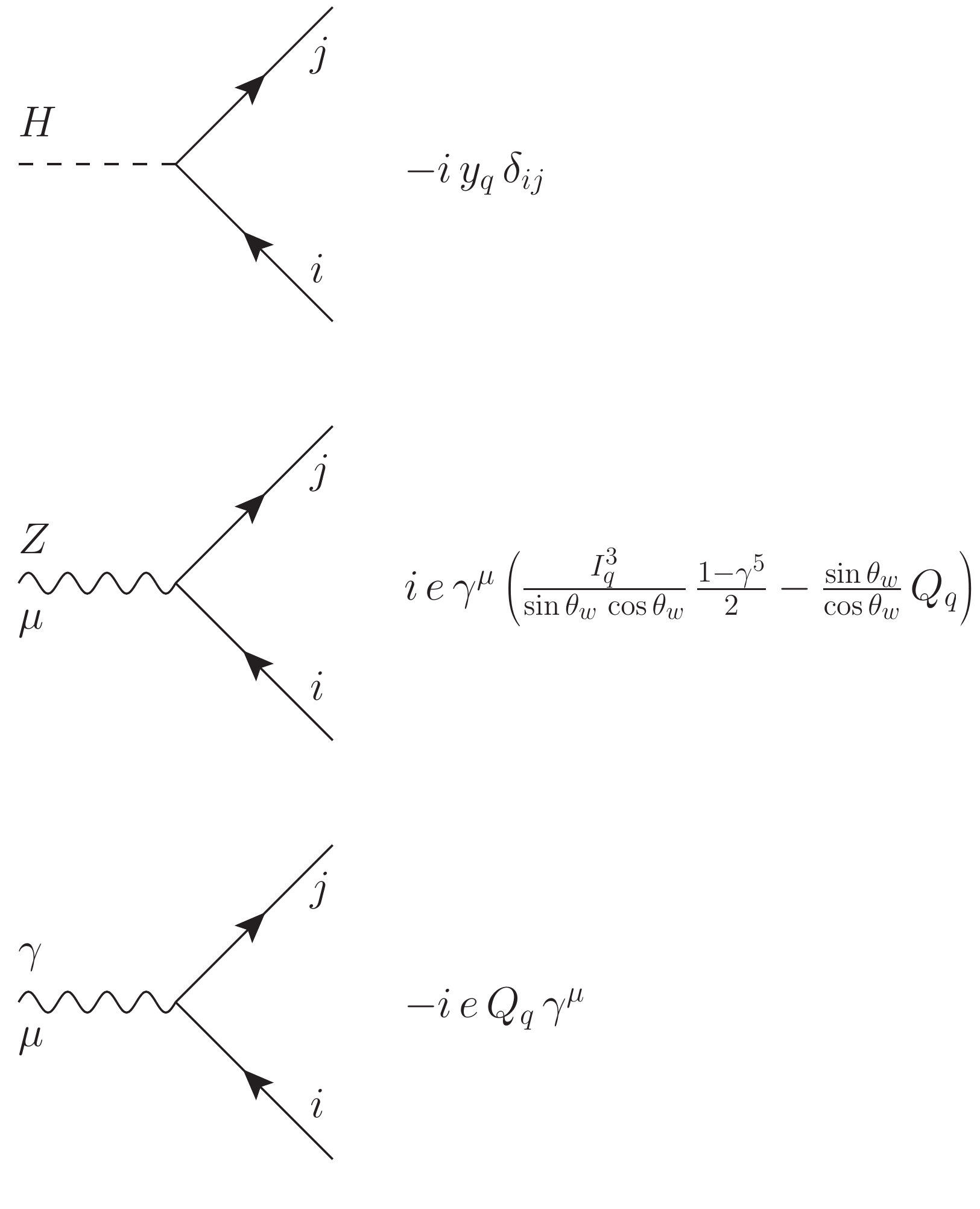}\\
The Yukawa coupling $y_q$ appearing in the Higgs-quark vertex is defined as
\begin{equation}
y_q = \frac{m_q}{v}\,,\qquad v=(\sqrt{2} \, G_F)^{-\frac{1}{2}} = \frac{2\,m_W\sin\theta_w}{e}
\label{yukawa}
\end{equation}
with the vacuum expectation value $v$ of the Higgs boson, Fermi's constant $G_F$, the $W$ boson mass $m_W$ and the weak mixing angle $\theta_w$. The matrix $\gamma^5$ is required to project the fields onto their left- and right-handed chirality components, however it will not play a role when the amplitude of process $(b)$ is derived. This is due to the the Lorentz structure of the external momenta being such that it does not saturate~$\gamma^5$.\\
For a given order in the coupling constant or equivalently for a given loop-order $l$, we can compute the scattering amplitude if we write down all possible combinations of the specified propagators and vertices that connect the initial- and final-state particles of the considered process. Subsequently we proceed as follows for each diagram $\mathcal{D}_j^{(l)}$:
\begin{enumerate}
\item For each internal loop $i$ up to $l$ loops, integrate over all loop momenta using the integration measure
\begin{equation}
\prod_{i=1}^l \int \frac{\d^D k_i}{(2\pi)^D} \,.
\end{equation}
\item Supply every fermionic loop, i.e. all quark and ghost loops, with a factor $(-1)$.
\item Multiply by an overall symmetry factor to account for equivalent permutations of internal propagators or external legs.
\end{enumerate}
The sum of all contributing Feynman diagrams at loop order $l$ then yields the operator~$\mathcal{S}^{(l)}$ at that loop order,
\begin{equation}
\mathcal{S}^{(l)} = \sum_{j=1}^{N_D} \mathcal{D}_j^{(l)} \,,
\label{scatteringamp2}
\end{equation}
and one is left with the evaluation of the remaining loop integrals. The operator $\mathcal{S}$, not to be confused with the action~$\mathcal{S}_A$ in Eq.~\eqref{action}, can be expanded as a power series in the same way as the cross section:
\begin{equation}
\mathcal{S} = \mathcal{S}^{(0)} + \mathcal{S}^{(1)} \, \alpha_s + \mathcal{S}^{(2)} \, \alpha_s^2 + \mathcal{S}^{(3)} \, \alpha_s^3 + \, \mathcal{O}(\alpha_s^4) \,.
\label{scatteringamp1}
\end{equation}
The scalar Feynman amplitude $\mathcal{M}$ is obtained from the tensorial operator $\mathcal{S}$ by contracting with fixed external states, and their precise connection will be specified in Section~\ref{sec:amplitudes}.\\
This diagrammatic approach has proven to be a successful tool for an endless number of computations within the Standard Model. Based on the key ideas presented in Ref.~\cite{Bern:1994} at the one-loop level, unitarity-based methods are currently being developed at two loops, in which the unitarity properties of the Feynman amplitude are explored. The advantage of this method is that the cumbersome computation of possibly high numbers of Feynman diagrams is circumvented. However this unquestionably elegant approach has only been applied to specific multi-loop processes of limited complexity and not been generalized beyond one loop. For this reason, we use to the well-established method of evaluation via Feynman diagrams.

\subsection{Renormalization}
\label{sec:renormalization}

The computation of perturbative corrections to the Feynman amplitude requires the evaluation of loop integrals, which are in general divergent, ill-defined quantities. In Eq.~\eqref{subtraction} we have seen how to take care of infrared divergences, i.e. of singularities that arise when loop momenta tend to zero or become collinear to another particle momentum. What about the case where loop momenta tend to infinity, corresponding to divergences in the ultraviolet regime? In fact, these singularities can already be removed at the level of the Lagrangian and thus of the Feynman amplitude, so that the remaining \textit{renormalized} Feynman amplitude contains only infrared divergences. In order to accomplish this, we use the powerful method of dimensional regularization, which isolates divergences within Feynman integrals. The idea is to solve integrals in $D=4-2\e$ dimensions so that they can be analytically continued for all complex $D$. Both ultraviolet and infrared singularities then manifest themselves as poles in $\e$ \cite{tHooft:1972}. On top of that, the coupling strength
\begin{equation}
g_s \to \mu_0^\e \, g_s
\label{mu0}
\end{equation}
has to be rescaled using the \textit{mass parameter} $\mu_0$ \textit{of dimensional regularization} in order to maintain a dimensionless coupling in the so-called \textit{bare} Lagrangian of Eq.~\eqref{LQCD},
\begin{align}
\mathcal{L}^B_\mathrm{QCD} \equiv \, &\mathcal{L}_\mathrm{GI} + \mathcal{L}_\mathrm{GF} + \mathcal{L}_\mathrm{GC} \nonumber \\
= \, &\bar{\psi}_{q,i}^B \, \left[i \, \slashed{\p} - m_q^B \right] \, \psi_{q,i}^B - \frac{1}{4} \left[ \p_\mu \, \left(A^a_\nu\right)^B - \p_\nu \,  \left(A^a_\mu \right)^B \right]^2 + \p^2 \left(\eta^{a\dagger}\right)^B \, \left(\eta^a\right)^B \nonumber \\
&-\frac{1}{2\,\xi^B} \left[\p^\mu \left(A_\mu^a\right)^B \right]^2 - \mu_0^\e \, g_s^B \, \bar{\psi}_{q,i}^B \,\left(\slashed{A}^a \right)^B \, \psi_{q,j}^B  \nonumber \\
&+ \mu_0^\e \, g_s^B \, f^{abc} \left[ \p_{\mu} \, \left(A^a_\nu\right)^B \right] \left(A^{b,\mu}\right)^B \left(A^{c,\nu}\right)^B \nonumber \\
&-\frac{\left(\mu_0^\e \, g_s^B\right)^2}{4} \left[f^{eab} \, \left(A^a_\mu\right)^B \, \left(A^b_\nu\right)^B \right] \left[f^{ecd} \, \left(A^{c,\mu}\right)^B \, \left(A^{d,\nu}\right)^B \right] \nonumber \\
&- \mu_0^\e \, g_s^B \, \left(\eta^{a\dagger}\right)^B \, f^{abc} \, \p^\mu \, \left(A^b_\mu\right)^B \, \left(\eta^c\right)^B \,,
\label{LQCDbare}
\end{align}
where we added the superscript `$B$' to all bare parameters\footnote{In $\left(A^b_\mu\right)^B$, do not confuse the inner superscript `$b$', which stands for a color index in the adjoint representation of $\mathrm{SU(3)}$, with the outer superscript `$B$' denoting a \textit{bare} field.}.\\
Let us start by rewriting this bare Lagrangian density through a simple change of variables from bare to renormalized, ultraviolet finite couplings and fields,
\begin{align}
\mu_0^\e \, g_s^B &= Z_g \, \mu^\e \, g_s \,, \label{rescaling1} \\
\psi_{q,i}^B &= Z_q^\frac{1}{2} \, \psi_{q,i} \,,\\
\left(A^a_\mu\right)^B &= Z_A^\frac{1}{2} \, A^a_\mu \,,\\
\left(\eta^a\right)^B &= Z_\eta^\frac{1}{2} \, \eta^a \,.
\label{rescaling2}
\end{align}
This is done with the help of \textit{renormalization constants} $Z_i$, which are designed to absorb the divergences of the bare quantities. For the sake of clarity, we assume vanishing quark masses and move the discussion of their peculiarities to Section~\ref{sec:schemes}. Substituting Eqs.~\eqref{rescaling1}--\eqref{rescaling2} into Eq.~\eqref{LQCD} yields the bare Lagrangian in terms of renormalized parameters, thus retaining physical predictions:
\begin{align}
\mathcal{L}^B_\mathrm{QCD} = \, &i\,Z_q \, \bar{\psi}_{q,i} \, \slashed{\p} \, \psi_{q,i} - \frac{Z_A}{4} \left( \p_\mu \, A^a_\nu - \p_\nu \, A^a_\mu \right)^2 + Z_\eta \, \p^2 \, \eta^{a\dagger} \, \eta^a \nonumber \\
&-\frac{Z_A}{2\,\xi^B} \left(\p^\mu A_\mu^a \right)^2 - \underbrace{Z_g \, Z_A^\frac{1}{2} \, Z_q}_{\equiv \, Z_{Aqq}} \, g_s \, \bar{\psi}_{q,i} \, \slashed{A}^a \, \psi_{q,j} + \underbrace{Z_g \, Z_A^\frac{3}{2}}_{\equiv \, Z_{A^3}} \, g_s \, f^{abc} \left( \p_{\mu} \, A^a_\nu \right) A^{b,\mu} A^{c,\nu} \nonumber \\
&-\underbrace{Z_g^2 \, Z_A^2}_{\equiv \, Z_{A^4}} \, \frac{g_s^2}{4} \left(f^{eab} \, A^a_\mu \, A^b_\nu \right) \left(f^{ecd} \, A^{c,\mu} \, A^{d,\nu} \right) - \underbrace{Z_g \, Z_A^\frac{1}{2} \, Z_\eta}_{\equiv \, Z_{A\eta\eta}} \, g_s \eta^{a\dagger} \, f^{abc} \, \p^\mu \, A^b_\mu \, \eta^c \,.
\label{LQCDren}
\end{align}
Therein, the overall coefficients of the quark-gluon, three-gluon, four-gluon and ghost-gluon vertices are denoted by $Z_{Aqq}$, $Z_{A^3}$, $Z_{A^4}$ and $Z_{A\eta\eta}$, respectively. In the framework of the BRST symmetry and the Slavnov-Taylor identities, it can be shown that these coefficients are not independent:	
\begin{equation}
\frac{Z_{Aqq}}{Z_A^\frac{1}{2} \, Z_q} = \frac{Z_{A^3}}{Z_A^\frac{3}{2}} = \sqrt{\frac{Z_{A^4}}{Z_A^2}} = \frac{Z_{A^3}}{Z_A^\frac{3}{2}} = Z_g \,.
\label{BRST}
\end{equation}
This equation simply states that the renormalized coupling constant $g_s$ is universal, i.e. that the $Z_g$'s coincide independently of the vertex they are associated with, which makes sense given the gauge invariance of the original Lagrangian. As a next step, we rewrite
\begin{equation}
Z_i = 1-\delta_i \quad (i=g,A,q,\eta) \,,
\label{Zdelta}
\end{equation}
which allows us to split the bare Lagrangian $\mathcal{L}^B_\mathrm{QCD}$ into a Lagrangian $\mathcal{L}_\mathrm{QCD}$, depending exclusively on renormalized quantities, and into a Lagrangian $\mathcal{L}_\mathrm{CT}$. Let us rephrase this as
\begin{equation}
\mathcal{L}_\mathrm{QCD} = \mathcal{L}^B_\mathrm{QCD} + \mathcal{L}_\mathrm{CT} \,.
\end{equation}
This simple yet elegant equation is remarkable in the sense that we manage to obtain well-defined, ultraviolet-finite fields and couplings by adding divergent counter terms of opposite sign to divergent bare parameters. The statement holds at every order in perturbation theory, therefore the counterterms can be expressed in the pictorial language of the Feynman diagrams as well, an example of which is provided by Fig.~\ref{fig:hzact} in Section~\ref{sec:hza2l}. We can see that this is usually indicated by putting a cross on the propagators and vertices from Section~\ref{sec:rules} and by multiplying the corresponding Feynman rule by the renormalization constant defined above. Since the number of counterterms required to cancel the ultraviolet singularities is finite, QCD belongs to the group of \textit{renormalizable} quantum field theories.\\
So far, we have ignored the discussion of the gauge-fixing term $-\frac{Z_A}{2\,\xi^B} \left(\p^\mu A_\mu^a \right)^2$ in Eq.~\eqref{LQCDren}, which we could have rescaled in the same way as the other parameters,
\begin{equation}
\xi^B = Z_\xi \, \xi \,,
\end{equation}
obtaining a renormalized gauge parameter $\xi$. Based on the fact that the longitudinal part of the gluon propagator is finite and does not need to be renormalized, the relation $Z_\xi=Z_A$ can be proven~\cite{Collins:1984}, so that the gauge-fixing term becomes independent of any renormalization constant. Equally, we could have used the freedom to set $\xi^B=\infty$~\cite{Weinberg:1996}, thus omitting the gauge-fixing term prior to the discussion without any loss of physical predictions, which is what we will do in the following for the sake of simplicity.\\
As can be seen from Eq.~\eqref{Zdelta}, all renormalization constants in QCD are fixed by renormalizing only four Green's functions, which are $n$-point correlation functions $G_{N_A,N_q}$ of a product of $n$ field operators with $N_A$ external gluon and $N_q$ external quark fields. Characterizing any external four-momenta present in the problem by $\{q\}$, the bare \textit{non-amputated} Green's function $G^B_{N_A,N_q}$ is obtained by appending propagators to the bare \textit{amputated} Green's function $\Gamma^B$, one propagator for each of the external legs, or vice versa
\begin{equation}
\Gamma^B = \frac{G^B_{N_A,N_q}}{\prod_{i=1}^{N_A} G^B_{2,0}(\{q\}) \, \prod_{j=1}^{N_q} G^B_{0,2}(\{q\})} \,.
\end{equation}
The scattering amplitude is then described by the amputated Green's function, which can be written in terms of either bare or renormalized quantities. A multiplicative renormalization procedure would lead to
\begin{equation}
\Gamma^B(\alpha_s^B,\{q\}) = Z_A^{-\frac{N_A}{2}} \, Z_q^{-\frac{N_q}{2}} \, \Gamma(\mu,\alpha_s,\{q\}) \,.
\end{equation}
Clearly, the left-hand side of this equation is independent of the renormalization scale $\mu$ occuring in Eq.~\eqref{rescaling1} and therefore this applies to the right-hand side as well, as it must be for unique physical predictions. Differentiating with respect to $\mu$ yields the renormalization group equation:
\begin{align}
\mu \, \frac{\d}{\d\mu} \, \Gamma^B &= \left[ \mu \, \frac{\p}{\p\mu} + \beta \, \frac{\p}{\p\alpha_s} - N_A \,  \gamma^A - N_q \, \gamma^q \right. \nonumber \\
&\left. \qquad\qquad+ \sum_{i,j} \frac{\tilde{T}_i \, \tilde{T}_j}{2} \, \log \left(\frac{\mu^2}{-2\,q_i\,q_j}\right) \, \gamma^\mathrm{cusp} \right] \Gamma = 0 \,.
\label{rge2}
\end{align}
Therein, the gluon and quark collinear anomalous dimensions $\gamma^A$ and $\gamma^q$ are given by
\begin{equation}
\gamma^i = \frac{\p\log Z_i}{\p\log \mu^2} \equiv -\gamma^i_0 \left(\frac{\alpha_s}{\pi}\right) - \gamma^i_1 \left(\frac{\alpha_s}{\pi}\right)^2 - \gamma^i_2 \left(\frac{\alpha_s}{\pi}\right)^3 +{\cal O}(\alpha_s^4) \qquad (i=A,q)
\end{equation}
and can be computed from the gluon and quark self energies at a given loop-order. It remains to comment on the last term within Eq.~\eqref{rge2} involving the cusp anomalous dimension $\gamma^\mathrm{cusp}$, which is related to Wilson loops with light-like segments\footnote{In simple terms, Wilson loops are phase factors represented by path-ordered exponentials of gauge fields.}~\cite{Korchemsky:1987,Korchemsky:1988,Korchemskaya:1992}. It can be decomposed in the same way as their gluon and quark collinear equivalents:
\begin{equation}
\gamma^\mathrm{cusp} \equiv -\gamma^\mathrm{cusp}_0 \left(\frac{\alpha_s}{\pi}\right) - \gamma^\mathrm{cusp}_1 \left(\frac{\alpha_s}{\pi}\right)^2 - \gamma^\mathrm{cusp}_2 \left(\frac{\alpha_s}{\pi}\right)^3 +{\cal O}(\alpha_s^4) \,.
\end{equation}
The sum within Eq.~\eqref{rge2} runs over the $n$ external partons of the considered process and their external momenta $q_i$ and $q_j$ are understood to be incoming. As described in Ref.~\cite{Catani:1996}, the color operators $\tilde{T}_i$ and $\tilde{T}_j$ associated with the partons $i$ and $j$ are linked with the color matrices in the respective fundamental or adjoint representation of $\mathrm{SU(3)}$ and their squares can be identified with the Casimirs in Eqs.~\eqref{casimirs}. The possibility of representing the last term within Eq.~\eqref{rge2} in this universal form, in the sense that all color dependence can be factorized, is a remarkable result. Recently it has been shown, however, that this conjecture breaks at four loops, as expected, due to the appearance of new color structures in terms of quartic Casimirs~\cite{Boels:2017}. Furthermore, process-dependent kinematic corrections have to be considered starting from three loops~\cite{Almelid:2015}, so that the process-independent behavior quoted in Eq.~\eqref{rge2} is only valid up to two loops. More details about the universal nature of the cusp anomalous dimension can be found in Refs.~\cite{Becher:2009a,Almelid:2017}.\\
Finally, the ordinary differential equation in Eq.~\eqref{rge2} is solved using an integrating factor along the parametrization $t=\log\frac{Q}{\mu}$:
\begin{equation}
\Gamma(\mu,\alpha_s(\mu),\{q\}) = \exp \left[-\int_0^t \d t' \left(N_A \, \gamma^A(t') + N_q \, \gamma^q(t')\right)\right] \Gamma(Q,\alpha_s(Q),\{q\}) \,.
\label{rgeint}
\end{equation}
This equation states that the theory defined at $\mu$ and $\alpha_s$ is equivalent to the one defined at $Q$, provided that the coupling is changed to the effective value $\alpha_s(Q)$. Hence, the strong coupling constant depends on the renormalization scale $\mu$ and is said to be \textit{running}. This behavior is governed by the function $\beta$ appearing in Eq.~\eqref{rge2} and will be explained in the following.

\subsection{Running Coupling}
\label{sec:running}

\begin{figure}[tb]
	\begin{center}
	\includegraphics[width=0.9\textwidth]{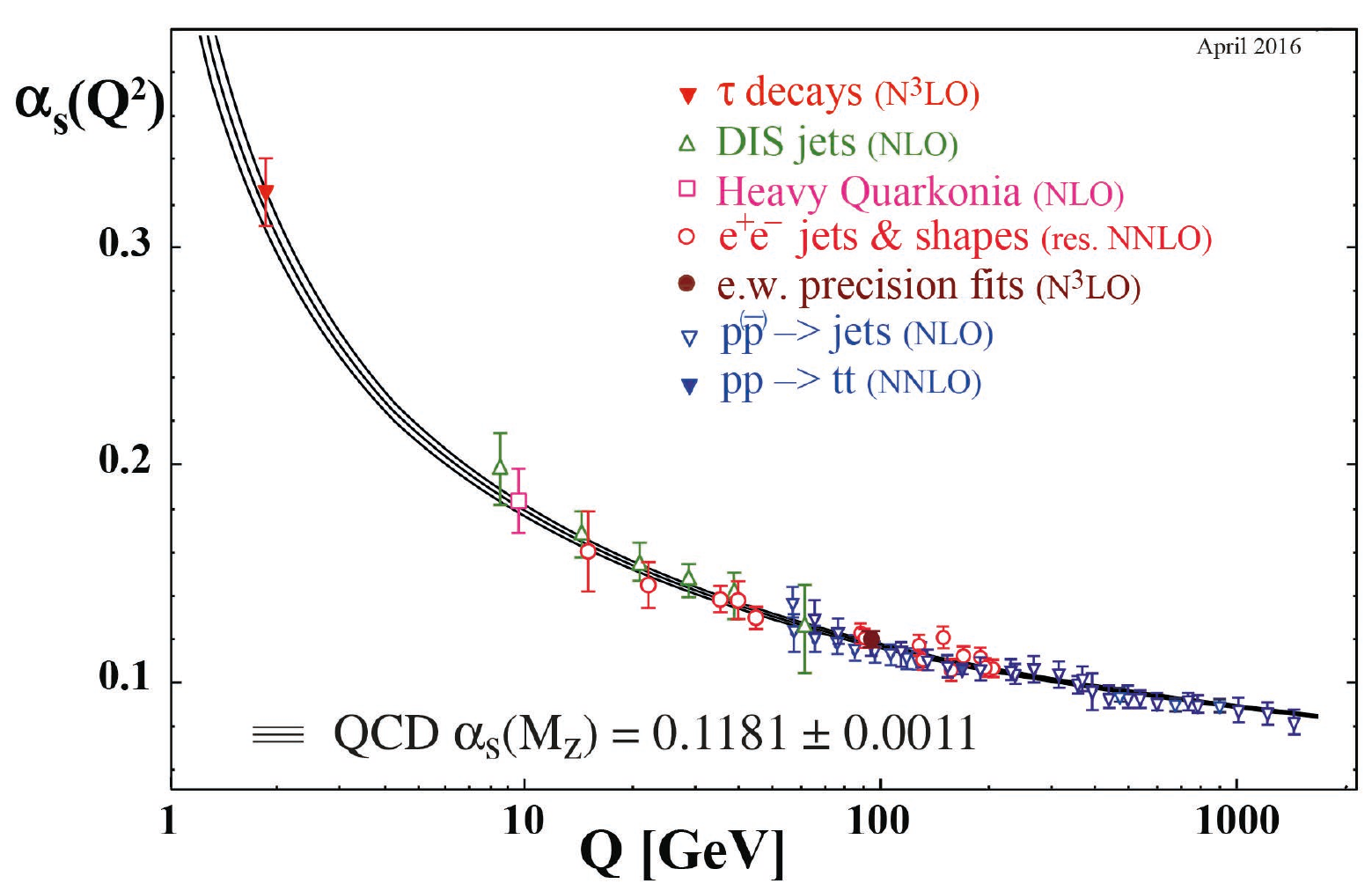}
	\caption[Summary of measurements of $\alpha_s$ as a function of the energy scale $Q$]{\textbf{Summary of measurements of~$\boldsymbol{\alpha_s}$ as a function of the energy scale~$\boldsymbol{Q}$.} The respective degree of QCD perturbation theory used in the extraction of $\alpha_s$ is indicated in parentheses~\cite{Patrignani:2016}.}
	\label{fig:alphas}
	\end{center}
\end{figure}

As pointed out in Chapter~\ref{chap:introduction}, the series expansion in Eq.~\eqref{cs} only converges if the coupling constant is sufficiently small. In any quantum field theory, the evolution of the coupling constant as a function of the energy is given by the renormalization group equation. For the strong coupling, it reads
\begin{equation}
\frac{\p}{\p\log\mu^2} \left(\frac{\alpha_s}{\pi} \right) \equiv \beta \left(\alpha_s\right) = -\frac{\alpha_s}{\pi} \, \left(\beta_0 \, \frac{\alpha_s}{\pi} + \beta_1 \, \left(\frac{\alpha_s}{\pi} \right)^2 + \beta_2 \, \left(\frac{\alpha_s}{\pi} \right)^3 +{\cal O}(\alpha_s^4) \right) \,,
\label{rge}
\end{equation}
where $\alpha_s=\alpha_s(\mu^2)$ is a function of the renormalization scale $\mu$. The coefficients $\beta_i$ in Eq.~\eqref{rge} are referred to as the $(i+1)$-loop beta function for the coupling of an effective theory. Therein, $N_F$ quark flavors are considered light, i.e. $m_q \ll \mu$, whereas the remaining heavier quark flavors decouple from the theory. We will make use of up to the first three beta function coefficients in the calculations of this thesis. They are given by~\cite{Gross:1973,Politzer:1973,Caswell:1974,Jones:1974,Egorian:1978,Tarasov:1980,Larin:1993}
\begin{align}
\beta_0 &= \frac{11}{12} C_A - \frac{1}{3} T_F N_F \,, \nonumber \\
\beta_1 &= \frac{17}{24} C_A^2 - \left(\frac{5}{12} C_A + \frac{1}{4} C_F\right) T_F N_F \,, \nonumber \\
\beta_2 &= \frac{2857 C_A^3}{3456}+\frac{1}{64} C_F^2 N_F-\frac{205 C_F C_A N_F}{1152}-\frac{1415 C_A^2 N_F}{3456}+\frac{11 C_F N_F^2}{576}+\frac{79 C_A N_F^2}{3456} \label{beta1} \\
\intertext{for general gauge group and simplify to}
\beta_0 &= \frac{11}{4} - \frac{1}{6} N_F \,, \nonumber \\
\beta_1 &= \frac{51}{8} - \frac{19}{24} N_F \nonumber \\
\beta_2 &= \frac{2857}{128} - \frac{5033}{1152} N_F + \frac{325}{3456} N_F^2 \label{beta2}
\end{align}
in case of the $\mathrm{SU(3)}$ symmetry of QCD. Beyond one loop, Eq.~\eqref{rge} only produces implicit solutions, but to leading order in $\alpha_s$ it can be solved exactly:
\begin{equation}
\alpha_s(Q) = \frac{\alpha_s(\mu)}{1 + \alpha_s(\mu) \, \frac{\beta_0}{\pi} \, \log\frac{Q^2}{\mu^2}} \overset{\alpha_s(\mu)\to\infty}{=} \frac{12\,\pi}{\left(33-2\,N_F\right)\log\frac{Q^2}{\Lambda^2}} \,.
\label{alphas}
\end{equation}
For QCD, $\Lambda$ lies between 100 and 300 MeV and characterizes the scale at which $\alpha_s$ diverges. This corresponds to the scale where confinement plays a non-negligible role and truncation of the power series is no longer justified.\\
Equation~\eqref{beta2} implies that $\beta_0$ is positive provided that there are at most 16 quark flavors, which is true in currently known strong interactions. Combined with the global minus sign on the right-hand side of Eq.~\eqref{rge}, we can deduce that the strength of the strong coupling constant $\alpha_s$ increases with distance and decreases with energy, eventually vanishing for $Q^2\to\infty$. The experimental results shown in Fig.~\ref{fig:alphas} are in agreement with these statements, enabling us to apply perturbation theory in the high-energy regime of asymptotic freedom as mentioned in Chapter~\ref{chap:introduction}. Experimental measurements are indeed required to provide a boundary value for $\alpha_s(\mu)$ within Eq.~\eqref{alphas}, so that fixed-order results can be produced by resumming logarithmic expressions of the form $\alpha_s(\mu) \, \log\frac{Q^2}{\mu^2}$. In general, one obtains partial results including logarithmic expressions of $\mu^2$, but measurable quantities as a whole, such as cross sections and decay rates, should not depend on this unphysical parameter. Hence, this scale can be chosen arbitrarily within fixed-order results. This is commonly done by setting~$\mu$ close to the scale of the momentum transfer~$Q$ in a given process so that it describes the characteristic energy scale of that process and thus large logarithms are avoided. The effective strength of the strong interaction is then
\begin{equation}
\alpha_s=\alpha_s(\mu^2\simeq Q^2)\,.
\end{equation}
An estimation of the uncertainty on the prediction from missing higher orders can be obtained by varying $\mu$ around that value.\\
In contrast to its non-abelian counterpart QCD, the theory of QED is an abelian quantum field theory that behaves differently: The coupling constant $\alpha$ of QED satisfies a renormalization group equation, where the one-loop beta function has the opposite sign compared to QCD if defined equivalently to Eq.~\eqref{rge}:
\begin{equation}
\beta^\mathrm{QED}_0  = -\frac{\alpha^2}{3\pi} \,.
\end{equation}
Therein, the coupling $\alpha$ is equal to the fine structure constant and increases with energy. However its value
\begin{equation}
\alpha(Q^2) = \frac{e^2}{4\pi} \simeq
\begin{cases}
\nicefrac{1}{137} \quad (Q^2=0)\,,\\
\nicefrac{1}{128} \quad (Q^2=m_W^2)
\end{cases}
\end{equation}
is sufficiently small and widely stable over the energy range on which particle physics colliders operate so that perturbation theory can be applied in QED as well.

\subsection{Renormalization Schemes and Quark Masses}
\label{sec:schemes}

Due to the confining property of QCD, free quarks have never been observed and are said to \textit{hadronize}\footnote{The top quark is an exception: It decays before it can hadronize because of its heavy mass.} on a timescale $1/\Lambda$. Hence, defining the quark mass is a complicated task and requires a specific prescription or $\textit{scheme}$. As mentioned in Section~\ref{sec:renormalization}, the divergences of the counterterms are designed to eliminate those of the bare Lagrangian. However, we skipped the discussion on the finite parts of the counterterms, which is exactly where different renormalization schemes come into play. Moreover, we have completely neglected the quark mass in Sections~\ref{sec:renormalization} and \ref{sec:running}. In fact, Eqs.~\eqref{rgeint}, \eqref{beta1} and \eqref{beta2} are only valid in mass-independent renormalization schemes, where the term \textit{mass-independent} refers to the finite parts of the counterterms. In order to account for that, let us add another relation to Eqs.~\eqref{rescaling1}--\eqref{rescaling2}:
\begin{equation}
m_q^B = Z_m \, m_q \,.
\end{equation}
In the renormalization group equation~\eqref{rge2}, this eventually produces an additional term
\begin{equation}
\left[ \mu \, \frac{\p}{\p\mu} + \beta \, \frac{\p}{\p \alpha_s} + m_q \, \gamma^m \, \frac{\p}{\p m_q} - N_A \,  \gamma^A - N_q \, \gamma^q + \sum_{i,j} \frac{\tilde{T}_i \, \tilde{T}_j}{2} \, \log \frac{\mu^2}{-2\,q_i\,q_j} \, \gamma^\mathrm{cusp} \right] \Gamma = 0
\end{equation}
involving the quark mass anomalous dimension $\gamma^m$:
\begin{equation}
\gamma^m = \frac{\p\log m_q}{\p\log\mu^2} \equiv -\gamma^m_0 \left(\frac{\alpha_s}{\pi}\right) - \gamma^m_1 \left(\frac{\alpha_s}{\pi}\right)^2 - \gamma^m_2 \left(\frac{\alpha_s}{\pi}\right)^3 +{\cal O}(\alpha_s^4)  \,.
\end{equation}
In order to indicate results for a parameter like this, we need to specify a renormalization scheme. The simplest scheme one can imagine is the minimal subtraction (MS) scheme \cite{tHooft:1973}, in which the counterterms cancel the divergences but nothing else, i.e. the finite part of the Lagrangian remains unchanged. The observation that there is an $\e$-dependent factor in one-loop calculations, leading to universally occuring terms in renormalized Green's functions, gave rise to a variant of the MS scheme. This is the modified minimal subtraction ($\MS$) scheme \cite{Bardeen:1978}, in which poles are accompanied by additional terms according to
\begin{equation}
\frac{S_\gamma}{\e} \equiv \frac{(4\pi)^\e}{\e\,e^{\gamma_E\,\e}} = \frac{1}{\e} + \log 4\pi - \gamma_E + \mathcal{O}(\e)
\label{Sgamma}
\end{equation}
with the Euler-Mascheroni constant $\gamma_E\approx 0.57721$. From the MS scheme, the $\MS$ scheme emerges through the replacement $\mu^2\to 4\pi\mu^2/e^{\gamma_E}$, removing potentially large contributions to radiative corrections, so that the convergence behavior of the perturbative series is improved. Furthermore, both the MS and $\MS$ schemes have counterterms with minimal mass dependence, i.e. they are chosen to be independent of the quark mass with the exception of the mass parameter. In the following, we will quote the results of renormalization constants and anomalous dimensions in the $\MS$ scheme, that will be needed in one of our calculations, up to the required order in the coupling constant.\\
The result of the quark mass anomalous dimension to three loop-order reads \cite{Vermaseren:1997, Chetyrkin:1997a,Czakon:2005}
\begin{align}
\gamma_m^0 &= \frac{3}{4} C_F \,, \nonumber \\
\gamma_m^1 &= \frac{1}{16} \left[ \frac{3}{2} C_F^2 + \frac{97}{6} C_F C_A - \frac{10}{3} C_F T_F N_F \right] \,, \nonumber \\
\gamma_m^2 &= \frac{1}{64} \left[ \frac{129}{2} C_F^3 - \frac{129}{4} C_F^2 C_A + \frac{11413}{108} C_F C_A^2 + C_F^2 T_F N_F \left(48\zeta_3 - 46\right) \right. \nonumber \\
& \qquad\quad \left. - C_F C_A T_F N_F \left( 48\zeta_3 + \frac{556}{27} \right) - \frac{140}{27} C_F T_F^2 N_F^2 \right] \,.
\end{align}
We will make use of this result to determine the quark mass renormalization constant $Z_m$ through the relation
\begin{equation}
\gamma_m = -\frac{\d\log \, Z_m}{\d\log \, \mu^2} = -\frac{\p\log \, Z_m}{\p\alpha_s} \frac{\d\alpha_s}{\d\log \, \mu^2} = -\pi \, \frac{\p\log \, Z_m}{\p\alpha_s} \left[-\e \, \left(\frac{\alpha_s}{\pi}\right) + \beta\left(\alpha_s\right) \right] \,,
\end{equation}
leading to
\begin{align}
Z_m &= Z_y \nonumber \\
&= 1 - \frac{3 \, C_F}{4 \, \e}\left(\frac{\alpha_s}{\pi}\right) 
\nonumber \\
&\quad+\frac{1}{16} \Bigg[C_F^2 \left(\frac{9}{2 \e^2}-\frac{3}{4 \e} \right) + C_F C_A \left(\frac{11}{2 \e^2}-\frac{97}{12 \e}\right) + C_F N_F \left(-\frac{1}{\e^2}+\frac{5}{6 \e}\right)\Bigg]
\left(\frac{\alpha_s}{\pi}\right)^2
\nonumber \\
&\quad +\frac{1}{64} \Bigg[
C_F^3 \left(-\frac{9}{2 \e^3}+\frac{9}{4 \e^2}-\frac{43}{2 \e} \right) + C_F^2 C_A \left(-\frac{33}{2 \e^3}+\frac{313}{12 \e^2}+\frac{43}{4 \e} \right)
\nonumber \\
& \qquad\quad + C_F C_A^2 \left(-\frac{121}{9 \e^3}+\frac{1679}{54 \e^2}-\frac{11413}{324 \e} \right) + C_F^2 N_F \left(\frac{3}{\e^3}-\frac{29}{6 \e^2}+\frac{1}{\e} \left(\frac{23}{3}-8\zeta_3\right) \right)
\nonumber \\
& \qquad\quad + C_F C_A N_F \left(\frac{44}{9 \e^3}-\frac{242}{27 \e^2}+\frac{1}{\e} \left(\frac{278}{81}+8\zeta_3\right) \right)
\nonumber \\
& \qquad\quad + C_F N_F^2 \left(-\frac{4}{9 \e^3}+\frac{10}{27 \e^2}+\frac{35}{81 \e} \right)
\Bigg]
\left(\frac{\alpha_s}{\pi}\right)^3
+{\cal O}(\alpha_s^4) \; .
\label{ZyMS}
\end{align}
From the definition of the Yukawa coupling in Eq.~\eqref{yukawa} it becomes clear that the same expression holds for the renormalization constant $Z_y$ \cite{Harlander:2003}, which is defined through
\begin{equation}
y_q^B = Z_y \, y_q \,.
\label{yukawarenorm}
\end{equation}
On top of that, the evaluation of the quark collinear anomalous dimension up to three loops in the $\MS$ scheme yields~\cite{Becher:2006,Becher:2009a}
{\allowdisplaybreaks
\begin{eqnarray}
\gamma^q_0 &=& -\frac{3}{4} C_F\,, \nonumber \\
\gamma^q_1 &=& 
\frac{1}{16} \left[ C_F^2\biggl(-\frac{3}{2}+2\pi^2-24\zeta_3\bigg) 
+C_FC_A\biggl(-\frac{961}{54}-\frac{11\pi^2}{6}+26\zeta_3\bigg) \right. \nonumber \\ & &
\qquad +\left. C_FN_F\biggl(\frac{65}{27}+\frac{\pi^2}{3}\bigg) \right] \,, \nonumber \\
\gamma^q_2 &=& 
\frac{1}{64} \left[ C_F^2N_F\biggl(\frac{2953}{54}-\frac{13\pi^2}{9}-\frac{14\pi^4}{27}+\frac{256\zeta_3}{9}\bigg) 
+C_FN_F^2\biggl(\frac{2417}{729}-\frac{10\pi^2}{27}-\frac{8\zeta_3}{27}\bigg) \right. \nonumber \\ & &
\qquad +C_FC_AN_F\biggl(-\frac{8659}{729}+\frac{1297\pi^2}{243}+\frac{11\pi^4}{45}-\frac{964\zeta_3}{27}\bigg) \nonumber \\ & &
\qquad +C_F^3\biggl(-\frac{29}{2}-3\pi^2-\frac{8\pi^4}{5}-68\zeta_3+\frac{16\pi^2\zeta_3}{3}+240\zeta_5\bigg)\nonumber \\ & &
\qquad +C_AC_F^2\biggl(-\frac{151}{4}+\frac{205\pi^2}{9}+\frac{247\pi^4}{135}-\frac{844\zeta_3}{3}-\frac{8\pi^2\zeta_3}{3}-120\zeta_5\bigg)\nonumber \\ & &
\qquad \left. +C_A^2C_F\biggl(-\frac{139345}{2916}-\frac{7163\pi^2}{486}-\frac{83\pi^4}{90}+\frac{3526\zeta_3}{9}-\frac{44\pi^2\zeta_3}{9}-136\zeta_5\bigg) \right]
\label{gammaq}
\end{eqnarray}
and the coefficients of the cusp anomalous dimension are given by~\cite{Moch:2005}:
\begin{eqnarray}
\gamma^\mathrm{cusp}_0 &=& 1\,, \nonumber \\
\gamma^\mathrm{cusp}_1 &=& 
\frac{C_A}{16} \bigg(\frac{268}{9}-\frac{4\pi^2}{3}\bigg) 
-\frac{5N_F}{18}\,, \nonumber \\
\gamma^\mathrm{cusp}_2 &=& 
\frac{1}{64} \left[ C_A^2\bigg(\frac{490}{3}-\frac{536\pi^2}{27}+\frac{44\pi^4}{45}+\frac{88\zeta_3}{3}\bigg) 
+C_AN_F\bigg(-\frac{836}{27}+\frac{80\pi^2}{27}-\frac{112\zeta_3}{3}\bigg) \right. \nonumber \\ & &
\qquad \left. +C_FN_F\bigg(-\frac{110}{3}+32\zeta_3\bigg) 
-\frac{16N_F^2}{27} \right] \,.
\label{gammacusp}
\end{eqnarray}
The renormalization of the strong coupling constant can be achieved by switching from Eq.~\eqref{rescaling1} to
\begin{equation}
\alpha_s^B \, \mu_0^{2\e} = Z_{\alpha_s} \, \mu^{2\e} \, \alpha_s(\mu^2) \,,
\label{alphasrenorm}
\end{equation}
i.e. the bare coupling $\alpha_s^B$ is replaced with the renormalized coupling $\alpha_s\equiv \alpha_s(\mu^2)$, which is evaluated at the renormalization scale $\mu^2$. Then
\begin{equation}
Z_{\alpha_s} = Z_g^2
\end{equation}
can be computed indirectly through Eq.~\eqref{BRST}. In doing this we obtain
\begin{align}
Z_{\alpha_s} = 1 &- \frac{\beta_0}{4 \, \e}\left(\frac{\alpha_s}{\pi}\right) 
+\frac{1}{16} \left(\frac{\beta_0^2}{\e^2}-\frac{\beta_1}{2\e}\right)
\left(\frac{\alpha_s}{\pi}\right)^2\nonumber \\
&-\frac{1}{64} \left(\frac{\beta_0^3}{\e^3}-\frac{7}{6}\frac{\beta_1\beta_0}{\e^2}+\frac{1}{3}\frac{\beta_2}{\e}\right)\left(\frac{\alpha_s}{\pi}\right)^3+{\cal O}(\alpha_s^4) \,,
\label{Zalphas}
\end{align}
where the coefficients $\beta_i$ of the beta function up to three loops have been indicated in Eqs.~\eqref{beta1} and \eqref{beta2}.\\
The use of a mass-independent scheme like MS or $\MS$ becomes unsuitable whenever the energy scale is much less than one of the quark masses. One may therefore choose `more physical' renormalization schemes when heavy quarks are involved, like the on-shell (OS) scheme. In this scheme, the finite parts of the quark wave function and mass renormalization constants $Z_q$ and $Z_m$ are adjusted so that the real part of the pole in the quark propagator is equal to the quark mass ($p^2=m_q^2)$ and has unit residue. Thereby, the finite parts of the counterterms may involve mass-dependent terms of the form $m_q/\mu$ in contrast to the minimal subtraction schemes. The calculations for process $(b)$ involve the renormalization of the quark mass and the Yukawa coupling in the OS scheme, both of which can be expressed through the renormalization constant $Z_\mathrm{OS}$ of the Yukawa coupling in this scheme. Evaluating the coefficient of $Z_\mathrm{OS}$, which is the OS equivalent of the $\MS$ renormalization constant $Z_y$ given in Eq.~\eqref{ZyMS}, yields at one-loop order~\cite{Bernreuther:2004}
\begin{equation}
Z_{\mathrm{OS}} = -\frac{\alpha_s}{\pi} \, \frac{C_F}{4} \, \frac{3-2\e}{\e\left(1-2\e\right)} \,.
\label{ZOS}
\end{equation}
The additional terms of $Z_{\mathrm{OS}}$ to finite order in $\e$ compared to $Z_y$ in Eq.~\eqref{ZyMS} are clearly visible and vanish in the limit $\e\to 0$. Bearing in mind that renormalization schemes only differ through this kind of shifts in the finite parts of the renormalization constants, it becomes obvious that the OS mass $M_q$ and the OS Yukawa coupling $Y_q$ are related to the $\MS$ quantities $\overline{m}_q$ and $\overline{y}_q$ by finite scheme transformations. To one loop, these slowly converging series read~\cite{Melnikov:2000,Chetyrkin:1999,Jamin:1997}
\begin{align}
M_q &= \overline{m}_q(\mu) \, \left( 1 + \Delta \right) \,, \nonumber \\
Y_q &= \overline{y}_q(\mu) \, \left( 1 + \Delta \right) \,, \nonumber \\
\Delta &= \frac{\alpha_s(\mu)}{\pi} \, C_F \left(1 + \frac{3}{4} \, \log \frac{\mu^2}{\overline{m}_q^2(\mu)} \right)
\label{OStoMS}
\end{align}
and are evaluated at a particular matching scale $\mu_m$.\\
A composite scheme may be most appropriate whenever both light and heavy quarks are involved~\cite{Collins:1978}. The number of \textit{active} quarks $N_F$ appearing in Eq.~\eqref{beta1} then parametrizes the subscheme of light quarks that come with normal $\MS$ counterterms. In contrast, the remaining heavier quarks, referred to as $\textit{inactive}$, decouple from this theory and are given in a more physical environment like the OS scheme.\\
For the reasons mentioned before, masses of heavier quarks are often quoted either as the OS mass $M_q$ or as the $\MS$ mass $\overline{m}_q(\overline{m}_q)$ evaluated at a scale equal to the mass itself, whereas light quark masses are mostly indicated in the $\MS$ scheme at a very low scale of around 2~$\mathrm{GeV}$.

\newpage

\section{Scattering Amplitudes}
\label{sec:amplitudes}

\subsection{From Amplitudes to Cross Sections}

In fact, Eq.~\eqref{sigmaV} has to be rewritten in the form
\begin{equation}
\d\sigma_V = \mathcal{F} \left|\mathcal{M}_{fi}\right|^2 \d\Phi_m \,, 
\label{cs2}
\end{equation}
where the overall constant $\mathcal{F}$ is called \textit{flux factor} and depends only on the kinematics of the process under consideration. 
The Feynman amplitude $\mathcal{M}_{fi}$ is defined as the part of the scattering matrix $S$ which is due to interactions,
\begin{equation}
\langle f|S|i \rangle = \delta_{fi} + i \, (2 \pi)^4 \, \delta(p_f - p_i) \, \mathcal{M}_{fi} \,,
\label{feynmanamp}
\end{equation}
with $p_i$ and $p_f$ denoting the sum of the four-momenta of the initial and final state $|i\rangle$ and $|f\rangle$, respectively. The $S$ matrix, in turn, describes the probability for a particle to pass from a normalized initial state into a final state via the transition matrix element
\begin{equation}
\left|\langle f|S|i \rangle\right|^2 \,.
\end{equation}
Consequently, the computation of the Feynman amplitude $\mathcal{M}_{fi} \equiv \mathcal{M}$ is required in order to obtain a result for the cross section, which ultimately can be compared to measurements at particle colliders. The full calculation of the Feynman amplitude is however not feasible in the framework of the Standard Model, thus we can only rely on the method of perturbation theory in high-energy collisions. As shown for the cross section in Eq.~\eqref{cs}, the Feynman amplitude can then be decomposed as a power series in the strong coupling constant:
\begin{equation}
\mathcal{M} = \mathcal{M}^{(0)} + \mathcal{M}^{(1)} \, \alpha_s + \mathcal{M}^{(2)} \, \alpha_s^2 + \mathcal{M}^{(3)} \, \alpha_s^3 + \, \mathcal{O}(\alpha_s^4) \,.
\label{feynmanampexp}
\end{equation}
Substituting Eq.~\eqref{feynmanampexp} into Eq.~\eqref{cs2} yields
\begin{align}
\d\sigma_V = \mathcal{F} &\left(\underbrace{\left|\mathcal{M}^{(0)}\right|^2}_\mathrm{LO} + \underbrace{2 \, \mathrm{Re} \left[\mathcal{M}^{(0)*} \mathcal{M}^{(1)}\right] \alpha_s}_\mathrm{NLO} \right. \nonumber \\
&\left.+\underbrace{\left(\left|\mathcal{M}^{(1)}\right|^2 + 2 \, \mathrm{Re} \left[\mathcal{M}^{(0)*} \mathcal{M}^{(2)}\right]\right)\alpha_s^2}_\mathrm{NNLO} + \, \mathcal{O}(\alpha_s^3) \right) \d\Phi_m \,,
\label{sigmaVpert}
\end{align}
i.e. the perturbative orders in the power series of the cross section are obtained by mixing the coefficients of the Feynman amplitude's power series according to their power in $\alpha_s$.

\subsection{Tensor Decomposition of Scattering Amplitudes}
\label{sec:tensordecomp}

In Section~\ref{sec:rules} we have seen how to derive Feynman rules from the Lagrangian density, and how these Feynman rules can be employed to compute amplitudes in terms of loop integrals. In that discussion, we have omitted the fact that the Feynman amplitude~$\mathcal{M}$ in Eq.~\eqref{cs2} is in general a scalar quantity that contains non-trivial Lorentz structures. As stated in Eq.~\eqref{scatteringamp2}, however, the application of the Feynman rules presented in Section~\ref{sec:rules} leads to the tensorial object $\mathcal{S}$, which involves Lorentz indices that have not been contracted. If the external particles are bosons or fermions, their state vectors contain polarization vectors or spinors, respectively, which can be amputated from $\mathcal{M}$. Consequently, the remainder becomes a tensorial operator in Lorentz or Dirac space, and it is this remainder~$\mathcal{S}$, which we have derived from the Feynman rules. In the following, we will complete this picture by explaining the method of \textit{tensor decomposition}, which enables us to factorize the full dependence on the Lorentz structure, and thus establishes a connection between the quantities~$\mathcal{S}$ and $\mathcal{M}$. With the help of the Feynman rules, we are then left with the computation of scalar quantities only, referred to as \textit{form factors}.\\
Let us consider a generic process in any renormalizable quantum field theory with $n$ external particles carrying momenta $q_1,\dots,q_n$. If $b$ denotes the number of external vector bosons, we can decompose the Feynman amplitude defined in Eq.~\eqref{feynmanamp} according to
\begin{equation}
\mathcal{M} = \mathcal{S}^{\mu_1\dots\mu_b}(q_1,\dots,q_n) \prod_{i=1}^b \e_{\mu_i}(q_i) \,.
\label{tensor1}
\end{equation}
Therein, we separated the dependence on the wave functions of the external bosonic states from the remaining tensorial object $\mathcal{S}$, whose Lorentz indices will eventually be contracted to produce a scalar quantity. Note that $\mathcal{S}$ still contains the full dependence on external fermionic states, and its perturbative coefficients~$\mathcal{S}^{(i)}$ defined in Eq.~\eqref{scatteringamp1} are the $D$-dimensional tensorial quantities obtained with the Feynman rules from Section~\ref{sec:rules} at a given loop order $i$. The most general tensor decomposition of $\mathcal{S}$ can be predicted using Lorentz and gauge invariance,
\begin{equation}
\mathcal{S}^{\mu_1\dots\mu_b}(q_1,\dots,q_n) = \sum_{i=1}^N \mathcal{A}_i(q_1,\dots,q_n) \, T_i^{\mu_1\dots\mu_b}(q_1,\dots,q_n) \,,
\label{decomposition}
\end{equation}
and involves $N$ linearly independent tensor structures $T_i^{\mu_1\dots\mu_b}(q_1,\dots,q_n)$ as well as their scalar coefficients $\mathcal{A}_i(q_1,\dots,q_n)$ within this decomposition. It is important to note that the complete dependence on the loop momenta $k_i$ is encoded in the scalar coefficients referred to as form factors. Due to Eq.~\eqref{scatteringamp2}, the decomposition must also hold at the level of each Feynman diagram:
\begin{equation}
\mathcal{D}^{\mu_1\dots\mu_b}(q_1,\dots,q_n) = \sum_{i=1}^N d_i(q_1,\dots,q_n) \, T_i^{\mu_1\dots\mu_b}(q_1,\dots,q_n) \,.
\end{equation}
All considerations made so far are independent of the perturbative order and thus valid to all orders in perturbation theory. In fact, the combination of Eqs.~\eqref{scatteringamp1} and \eqref{decomposition} tells us that the form factors must obey the same kind of power series as the Feynman amplitude $\mathcal{M}$ and the tensorial operator~$\mathcal{S}$:
\begin{equation}
\mathcal{A} = \mathcal{A}^{(0)} + \mathcal{A}^{(1)} \, \alpha_s + \mathcal{A}^{(2)} \, \alpha_s^2 + \mathcal{A}^{(3)} \, \alpha_s^3 + \mathcal{O}(\alpha_s^4) \,.
\label{seriesff}
\end{equation}
This statement is very powerful in the sense that it allows using the Feynman rules exclusively for the determination of the form factors, with the Lorentz structure being known a priori. In practice, this is achieved through the application of so-called \textit{projectors} $P_i$, which have the property of returning the corresponding form factor $\mathcal{A}_i$,
\begin{equation}
\sum_\mathrm{spins} P_i(q_1,\dots,q_n) \, \mathcal{S}^{\mu_1\dots\mu_b}(q_1,\dots,q_n) = \mathcal{A}_i(q_1,\dots,q_n) \,,
\label{proj1}
\end{equation}
where the sum runs over all spins and polarizations of the external particles. The explicit form of the projectors is obtained by expressing them in the basis of the gauge-invariant tensor structures, too:
\begin{equation}
P_i(q_1,\dots,q_n) = \sum_{i=1}^N \alpha_i(q_1,\dots,q_n) \left[T_i^{\mu_1\dots\mu_b}(q_1,\dots,q_n)\right]^\dagger \,.
\label{proj2}
\end{equation}
The scalar coefficients $\alpha_i$ are determined by substituting Eq.~\eqref{proj2} into Eq.~\eqref{proj1} and requiring that Eq.~\eqref{decomposition} is satisfied.\\
The helicity amplitudes
\begin{equation}
\mathcal{M}^{\lambda_1\dots\lambda_n} = \mathcal{S}^{\mu_1\dots\mu_b}(q_1,\dots,q_n,\lambda_{b+1},\dots\lambda_n) \prod_{i=1}^b \e_{\mu_i,\lambda_i}(q_i)
\end{equation}
are then obtained from the $D$-dimensional tensors in Eq.~\eqref{decomposition} by constraining the dimensionality of the Lorentz matrices to four and applying the usual four-dimensional helicity techniques \cite{Xu:1986,Berends:1982,Berends:1983,Dixon:1996}. Applications of the methods described in this section to Higgs-plus-jet production, i.e. to process $(c)$ of this thesis, can be found in Ref.~\cite{Gehrmann:2011}.

\section{Reduction to Master Integrals}
\label{sec:reduction}

With the help of the Feynman rules and the method of tensor decomposition described in Sections~\ref{sec:QCD} and \ref{sec:amplitudes}, the Feynman amplitude for any multi-loop scattering process can be cast into a product of tensor structures times scalar coefficients, where the latter include the full dependence on the remaining loop integrals. The determination of these loop integrals is the last missing step on the way to a well-defined amplitude, which can be numerically evaluated and used for phenomenological applications. Especially for multi-loop processes, however, one is faced with a huge number of such loop integrals, whose independent evaluation would be a formidable task and is often not feasible in practice. In fact, it was found out that these loop integrals are not independent, and that numerous relations exist which reduce the number of loop integrals to a smaller number of so-called \textit{Master Integrals} (MIs). The MIs can be chosen freely in the sense that these identities do not fix the MIs themselves, but only their number within a predefined set referred to as \textit{topology}, \textit{integral family} or \textit{sector}. Recently, it has been shown that this number is always finite~\cite{Smirnov:2010}.\\
Before we can classify the mentioned identities into groups, we have to elaborate on the concept of topologies. Any dimensionally regularized scalar $l$-loop integral, that remains after the application of tensor decomposition, can be cast into the form
\begin{equation}
I(q_1,\dots,q_n) =  \int \prod_{i=1}^l \frac{\d^Dk_i}{(2\pi)^D} \, \frac{S_1^{a_1} \dots S_\rho^{a_\rho}}{D_1^{b_1} \dots D_\sigma^{	b_\sigma}} \,,
\label{scalarint1}
\end{equation}
where $S_j=k_i\cdot p_k$ denotes a scalar product raised to integer power $a_j\in\mathbb{Z}$, and $D_j = p_j^2 + m_j^2$ stands for a propagator in the Euclidean space raised to integer power $b_j\in\mathbb{Z}$. In addition, $p$~represents any combination of internal and external momenta and the masses $m_j$ can either vanish or be different from zero. Given a set of $\sigma$ denominators $D_j$, the number $\rho$ of independent scalar products $S_j$ in the numerator cannot be infinitely large, but is constrained. Simple combinatorics leads to
\begin{equation}
\rho = l \left[n+\frac{1}{2}\left(l-1\right)\right]
\label{irreducible}
\end{equation}
independent scalar products, and consequently
\begin{equation}
\tau=\rho-\sigma
\label{irreducible2}
\end{equation}
of these scalar products can be expressed through the set of denominators $D_j$ without loss of generality. The $\tau$ scalar products that remain are called \textit{irreducible}. In case of a two-loop four point function, for example, Eq.~\eqref{irreducible} yields $\rho=9$ and $\tau=2$ with the input values $l=2$, $n=4$ and $\sigma=7$. In this irreducible representation, Eq.~\eqref{scalarint1} transforms into
\begin{equation}
I(q_1,\dots,q_n) =  \int \prod_{i=1}^l \frac{\d^Dk_i}{(2\pi)^D} \, \frac{S_1^{a_1} \dots S_\tau^{a_\tau}}{D_1^{b_1} \dots D_\sigma^{	b_\sigma}}
\label{scalarint2}
\end{equation}
with $a_j,b_j\in\mathbb{N}_0$. The topology of an integral as an interconnection of propagators and external momenta is uniquely characterized by specifying the set $D_1,\dots,D_t$ of $t$ distinct propagators, regardless of the powers they are raised to\footnote{In the following, we will use the expressions \textit{topology} and \textit{sector} interchangeably, whereas \textit{integral family} mostly refers to the complete tree of topologies.}. Any integral within this topology is then indicated through the set of scalar products $S_1,\dots,S_\tau$, plus the actual values of $a_1,\dots,a_\tau$ and $b_1,\dots,b_\sigma$.\\
With this information, we denote the class $I_{t,r,s}$ of integrals with $t$ distinct denominators raised to any positive power, where
\begin{equation}
r=\sum_j b_j \,, \quad s=\sum_j a_j
\label{rs}
\end{equation}
are the sum of powers of all propagators and the sum of powers of all scalar products, respectively. It is evident that only a variation of $t$ can change the topology, although different topologies with the same number of $t$ can occur, whereas a variation of $r$ or $s$ does not affect the definition of the topology itself. The total number of different integrals belonging to one class $I_{t,r,s}$ is given by the following product of binomial coefficients~\cite{Argeri:2007}:
\begin{equation}
N(I_{t,r,s}) = \binom{r-1}{t-1} \, \binom{s+\tau-1}{\tau-1} \,.
\label{Nints}
\end{equation}
By exploring all possible ways of removing denominators and thus reducing $t$ consecutively up to its minimal value, one can create a so-called \textit{subtopology tree}. In the pictorial language of Feynman integrals, this is often referred to as \textit{pinching} a line that stands for a propagator.\\
Finally, we would like to point out that Eq.~\eqref{scalarint2} can be reformulated as
\begin{equation}
I(q_1,\dots,q_n) =  \int \prod_{i=1}^l \frac{\d^Dk_i}{(2\pi)^D} \, \frac{1}{D_1^{b_1} \dots D_\rho^{b_\rho}}
\label{scalarint3}
\end{equation}
if one allows negative denominator powers $b_j\in\mathbb{Z}$, since all irreducible scalar products can be described as inverse propagators or combinations thereof. In this case, $r$ and $s$ are defined by
\begin{equation}
r=\sum_{j\in b_j>0} b_j \,, \quad s=\sum_{j\in b_j<0} b_j \,.
\label{rs2}
\end{equation}
In the following subsections, we will start from this representation in order to illustrate the reduction to MIs through various classes of relations, which are all valid in generic space-time dimension $D$.

\subsection{Integration-by-Parts Relations}
\label{sec:ibp}
The so-called \textit{Integration-by-Parts} (IBP) relations are by far the largest and most important class of identities and were first derived in Ref.~\cite{Chetyrkin:1981}\footnote{In fact, there is a footnote in Ref.~\cite{tHooft:1972} stating the existence of IBP relations. However, they have not been pursued further in that reference.}. As the name suggests, the idea is to relate various Feynman integrals based on the possibility of integrating by parts and always neglecting surface terms, which are integrals over the total derivative with respect to any loop-momentum.\footnote{Conventionally, scaleless integrals, i.e. massless tadpoles, vanish within dimensional regularization.} Applied to an integral of the form~\eqref{scalarint3} after differentiation with respect to one of the loop momenta $k_j$, this corresponds to
\begin{equation}
\frac{\p}{\p k_j^\mu} \, p^\mu \, I(q_1,\dots,q_n) =  \int \prod_{i=1}^l \frac{\d^Dk_i}{(2\pi)^D} \, \frac{\p}{\p k_j^\mu} \frac{p^\mu}{D_1^{b_1} \dots D_\rho^{	b_\rho}} = 0 \,,
\label{IBP1}
\end{equation}
where $p^\mu=\{k_1^\mu,\dots,k_l^\mu,q_1^\mu,\dots,q_n^\mu\}$ can be any of the internal or external momenta. In this way
\begin{equation}
N_\mathrm{IBP}=l \, (l+n-1)
\label{NIBP}
\end{equation}
identities are created from a single integrand. Explicitly, they are obtained by executing the derivative and contracting the Lorentz indices, ultimately leading to a scalar quantity. The resulting expression relates integrals belonging to the same topology tree, with the powers of the scalar products varying by $s-1,s,s+1$ and those of the propagators by $r,r+1$. It should be clear that the contraction of the Lorentz indices potentially produces reducible scalar products, i.e. scalar products that cancel denominators of the given topology, thus leading to a subtopology.\\
Let us illustrate the IBP method with the help of Fig.~\ref{fig:IBP1}, which for vanishing quark mass corresponds to the sectors $B_{5,103}$ and $B_{5,391}$ within process $(c)$ as defined in Appendix~\ref{sec:hjlaporta}. Assuming positive integer powers $b_j$ of the propagators, the Feynman integral reads
\begin{align}
I(q) &= \iint \frac{\d^Dk \, \d^Dl}{(2\pi)^{2D} \, (k^2)^{b_1} \, \left[(q-k)^2 \right]^{b_2} \, (l^2)^{b_3} \, \left[(q-l)^2 \right]^{b_4} \, \left[(k-l)^2 \right]^{b_5}} \nonumber \\
& \equiv \mathcal{I}(b_1,b_2,b_3,b_4,b_5) \,,
\label{IBP2}
\end{align}
where the loop momenta are labeled $k$ and $l$. From Eqs.~\eqref{IBP1} and \eqref{IBP2}, the following IBP identity can be derived:
\begin{equation}
	\iint \frac{\d^Dk \, \d^Dl}{(2\pi)^{2D} \, \left[(l^2)^{b_3} \, \left[(q-l)^2 \right]^{b_4} \right]} \, \frac{\p}{\p k_\mu} \left(\frac{k_\mu}{(k^2)^{b_1} \, \left[(q-k)^2 \right]^{b_2} \, \left[(k-l)^2 \right]^{b_5}} \right) = 0 \, .
\end{equation}
By differentiating and recovering Eq.~\eqref{IBP2} within the remaining expression, we end up with the relation
\begin{equation}
	(b_1 + b_2 + 2 \, b_5 -D) \, \mathcal{I} = \left[b_1 \, \mathbf{1}^+ \left(\mathbf{3}^- - \mathbf{5}^- \right) + b_2 \, \mathbf{2}^+ \left(\mathbf{4}^- - \mathbf{5}^- \right) \right] \, \mathcal{I} \, ,
	\label{IBP3}
\end{equation}
where the operator $\mathbf{j}^\pm$ changes the power of the $j$-th propagator by $\pm 1$, e.g.
\begin{equation}
	\mathbf{2}^+ \, \mathbf{5}^- \mathcal{I}(b_1,b_2,b_3,b_4,b_5) = \mathcal{I}(b_1,b_2+1,b_3,b_4,b_5-1) \, .
\end{equation}
Actually, Eq. \eqref{IBP3} is a consequence of the `triangle rule' originating from Fig. \ref{fig:IBP2}:
\begin{equation}
	\frac{1}{D-b_1-b_2-2 \, b_3} \left[b_1 \, \mathbf{1}^+ \left(\mathbf{3}^- -q_1^2 \right) + b_2 \, \mathbf{2}^+ \left(\mathbf{3}^- - q_2^2 \right) \right] = 1 \, .
\end{equation}
The IBP relation in Eq. \eqref{IBP3} follows from the application of this rule to the left triangle in Fig. \ref{fig:IBP1}.\\
Equation~\eqref{IBP3} reveals a typical feature of IBP relations: The sum $b_3+b_4+b_5$ of the integrals on the right-hand side is one less than that on the left-hand side. Hence, this equation can be used as a recurrence relation for the given family of integrals. By successively applying Eq. \eqref{IBP3}, any integral within this family can be reduced to integrals with at least one vanishing index. This reduction procedure is the IBP method's main field of applicability.\\
Beyond that, IBPs can be used to evaluate integrals in particularly simple cases such as this one. Choosing the integral with all $b_j=1$ unknown, we obtain by means of Eq.~\eqref{IBP3}:
\begin{equation}
	\mathcal{I}(1,1,1,1,1) = \frac{1}{\e} \left[\mathcal{I}(2,1,0,1,1) - \mathcal{I}(2,1,1,1,0) \right] \,.
\end{equation}
The integrations over $k$ and $l$ within $\mathcal{I}(2,1,1,1,0)$ decouple, leading to a product of two one-loop integrals, which can be carried out with the help of Refs.~\cite{Smirnov:2002, Smirnov:2006}:
\begin{equation}
	\int \frac{\d^Dk}{(k^2)^{\lambda_1} \, \left[(q-k)^2 \right]^{\lambda_2}} = \frac{i}{(4\pi)^{D/2}} \, \frac{\Gamma(\lambda_1 + \lambda_2 + \e - 2) \, \Gamma(2-\e-\lambda_1) \, \Gamma(2-\e-\lambda_2)}{(q^2)^{\lambda_1 + \lambda_2 + \e - 2} \, \Gamma(\lambda_1) \, \Gamma(\lambda_2) \, \Gamma(4-\lambda_1-\lambda_2-2 \, \e)} \, .
\end{equation}
In addition, the successive application of this one-loop formula serves to recursively determine $\mathcal{I}(2,1,0,1,1)$. Adding up both integrals and expanding the $\Gamma$ functions as a Laurent series in $\e$ yields the well-known result
\begin{equation}
	\mathcal{I}(1,1,1,1,1) = -\frac{3 \, \zeta_3}{128 \, \pi^4 \, q^2} + \mathcal{O}(\e) \, .
\end{equation}
In this simple example, the use of one IBP identity has fulfilled our needs. In general, the topology is more complex and requires including all possible IBP relations, which can amount to several thousands for complicated processes.
\begin{figure}[tb]
\begin{minipage}[t]{.5\textwidth}
	\begin{center}
	\includegraphics[scale=0.6]{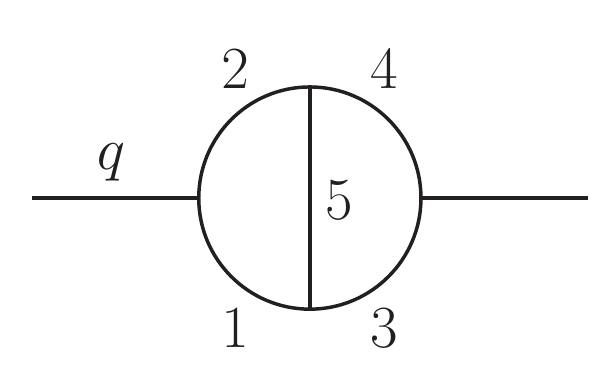}
	\caption[Example for the IBP method]{\textbf{Example for the IBP method} in case of a massless two-loop two-point diagram with external momentum~$q$}
	\label{fig:IBP1}
	\end{center}
\end{minipage}
\hfill
\begin{minipage}[t]{.4\textwidth}
	\begin{center}
	\includegraphics[scale=0.6]{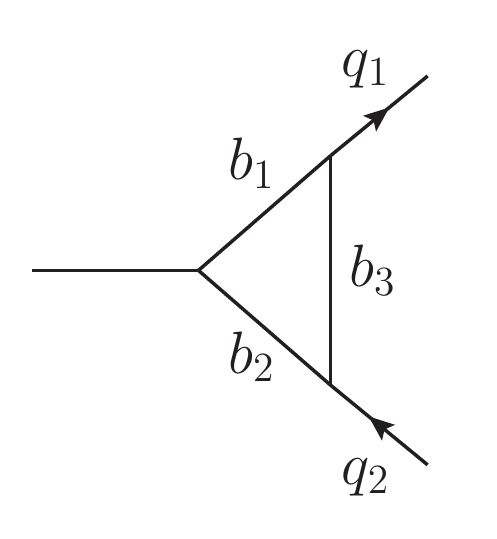}
	\caption[Triangle diagram designed to create IBP relation]{\textbf{Triangle diagram designed to create IBP relation} with external momenta $q_1$ and $q_2$}
	\label{fig:IBP2}
	\end{center}
\end{minipage}
\end{figure}

\subsection{Lorentz Invariance Relations}
\label{sec:LI}

By construction, integrals of the types~\eqref{scalarint2} and \eqref{scalarint3} are Lorentz scalars and as such, they are invariant under Lorentz transformations of the form
\begin{equation}
q^\mu \to q^\mu + \delta q^\mu = q^\mu + \delta\omega^\mu_\nu \, q^\nu
\label{infinitesimal}
\end{equation}
with a totally antisymmetric tensor
\begin{equation}
\delta\omega^\mu_\nu = - \delta\omega^\nu_\mu \,.
\label{antisymmetric}
\end{equation}
One can then exploit the fact that this should leave the scalar Feynman integral unchanged upon expanding in infinitesimal quantities $\delta$:
\begin{align}
I(q_1+\delta q_1,\dots,q_n+\delta q_n) &= I(q_1,\dots,q_n) + \sum_{i=1}^n \delta q_i^\mu \frac{\p}{\p q_i^\mu} I(q_1,\dots,q_n) \nonumber \\
&\overset{!}{=} I(q_1,\dots,q_n) \,.
\end{align}
Using the notation of Eq.~\eqref{infinitesimal}, we arrive at the equations
\begin{equation}
\delta \omega^\mu_\nu \sum_{i=1}^n q_i^\nu \frac{\p}{\p q_i^\mu} I(q_1,\dots,q_n) = 0 \,,
\end{equation}
which are however not linearly independent due to the antisymmetry of $\omega^\mu_\nu$. The linearly independent set can be obtained by taking Eq.~\eqref{antisymmetric} into account, leading to
\begin{equation}
\sum_{i=1}^n \left(q_i^\nu \frac{\p}{\p q_{i,\mu}} - q_i^\mu \frac{\p}{\p q_{i,\nu}} \right) I(q_1,\dots,q_n) = 0 \,,
\end{equation}
which can be finally used to produce scalar Lorentz invariance (LI) identities by contracting the equation with all possible antisymmetric combinations of $q_{j,\mu} \, q_{k,\nu}$.\\
Clearly, the number of LI identities for a given topology depends on the number of independent external momenta and therefore on the multiplicity. In case of a three-point function, for example, there are two linearly independent momenta, from which only one antisymmetric combination can be constructed, resulting in one LI identity ($N_\mathrm{LI}=1$). Similar considerations for a four-point function leads to three LI identities ($N_\mathrm{LI}=3$), showing us that the full potential of this class of relations can only be exploited for integrals with more than four legs. For multiplicities of five or higher, six linearly independent antisymmetric combinations of external momenta can be created, fully examining the potential of maximally available linear independent LI identities ($N_\mathrm{LI}=6$). This can be summarized as
\begin{equation}
N_\mathrm{LI} =
\begin{cases}
\frac{(n-1)(n-2)}{2} \quad &(n\leq 5)\,,\\
6 \quad &(n\geq 5) \,.
\end{cases}
\label{NLI}
\end{equation}
A few years ago, it was proven that LI relations are actually a subset of IBP identities, provided that a sufficiently large system of IBPs is generated \cite{Lee:2008}. In this sense, they do not add new information to the much wider class of IBP identities presented in the previous section, however they are still very valuable for practical purposes. This is because the generation and solution of the corresponding additional set of IBP relations is very expensive in terms of computational resources, thus replacing them by LI identities leads to a substantial speed-up when implemented in computer codes for automated reduction to MIs.

\subsection{Symmetry Relations}
\label{sec:symmetryrel}

We would like to point out that another class of so-called \textit{symmetry relations} exists, which can further reduce the number of MIs per topology, independently of the IBP and LI identities. An exceptional feature of these identities is that, on top of relating integrals within the same topology, they can be established for integrals belonging to different topologies. This is achieved by shifting loop momenta according to
\begin{equation}
k_i \to k_i + p \,,
\end{equation}
where $p$ denotes any combination of internal and external momenta apart from $k_i$, provided that this shift has a unit Jacobian. The fully automatic implementation of this kind of identities is highly non-trivial and only a limited number of public codes have approached this challenge to date. \textsc{Reduze}~\cite{vonManteuffel:2012}, for example, provides two methods to derive symmetry relations: Starting from one graph per topology, the first method relies on a matroid-based algorithm, that performs all relevant twists resulting in a unique equivalent of its matroid class. The second method, in contrast, generates all possible graphs through combinatorics and chooses the minimal graph to represent the topology.\\
However, there is an exception of a small class of symmetries that is not recognized by \textsc{Reduze} with the help of these methods and has to be supplied manually: Imagine that two integrals result in the same value, but this equivalence cannot be identified at the level of the integrand and thus is not detected by a graph-based approach. For example, this can occur when a triangle integral depends on two scales and hence on one ratio only, like $I_{24}$ in Fig.~\ref{fig:hjmaster}. After interchanging the legs with momenta $q_1$ and $q_2$ of that integral, the integral still depends on the same single ratio and the value of the integral must remain unchanged after permutation of the massless legs, although both integrals are topologically different. Note that in the following, such a permutation of external momenta will be referred to as a \textit{crossing} leading to a \textit{crossed} integral, or in this case more precisely to a $q_1\leftrightarrow q_2$ or x12~\textit{crossing}.

\subsection{Laporta's Algorithm}
\label{sec:laporta}

In the previous sections, we have elaborated on the existence of identities that relate a large number of integrals, eventually leading to a much smaller number of MIs, and in particular on how to generate these identities. As a next step, this system needs to be solved by inversion, which is more of a practical difficulty than a conceptual one. With such an enormous number of relations, the question of how to solve a given system in the most efficient way is of utmost importance, and an algorithmic procedure is desirable.\\
Before answering the question of $\textit{how}$ to solve the system, we should tackle the issue of $\textit{whether}$ or to which extent the system can be solved. Neglecting the possibility of deriving symmetry relations, the sum of IBP and LI identities that can be derived amounts to $(N_\mathrm{IBP}+N_\mathrm{LI}) \, N(I_{t,r,s})$ for a given topology of $N(I_{t,r,s})$ integrals. These identities will involve $N(I_{t,r+1,s})+N(I_{t,r+1,s+1})$ so far unconsidered integrals of more complicated structure, which do not belong to the topology tree of $N(I_{t,r,s})$. Recalling Eqs.~\eqref{Nints}, \eqref{NIBP} and \eqref{NLI}, it becomes clear that the number of these new integrals is overcompensated by the number of new IBP and LI identities, and hence the system as a whole is overconstrained so that the equations cannot all be independent. This observation considerably complicates the task of formulating an efficient way of solving the system, since it is not known a priori how many and particularly which equations are in fact linearly independent.\\
Decisive steps in this direction were taken in Refs.~\cite{Laporta:1996,Laporta:2001}, providing a set of criteria such that the intermediate expression swell can be minimized at each step of inverting the system of equations. These criteria, in a nutshell referred to as the \textit{Laporta algorithm}, especially concern the ordering in which the equations are solved, since it is well-known that the size of intermediate expressions strongly depends on this choice, although every possible choice must lead to the same final result. The Laporta algorithm has been implemented in numerous public and private codes, which will be presented in the next section, and has enabled the computation of countless applications within the Standard Model. As a result, the reduction to MIs has been considered conceptually solved for a long time, provided that sufficient computational resources are available.\\
However, the required amount of these resources, as a consequence of the desire to calculate processes with increasing number of scales, grows faster than their availability. Although there have been attempts to optimize the procedure by finding linear dependencies before solving the system of equations~\cite{Kant:2013}, observations suggest that a completely new approach may be necessary. First promising steps into this direction have been initiated by Ref.~\cite{vonManteuffel:2014a}, whose key idea is to construct algebraic identities from numerical samples obtained from reductions over finite fields. Certainly, these ideas are still being developed and have exclusively been implemented in private computer codes. For this reason, they have only been applied to a very limited number of computations, so that we retain the conservative approach of the Laporta algorithm in the following.\\
Finally, we would like to point out that, apart from the ordering that is established by the algorithm, there is another possibility of minimizing intermediate expressions: We have learned that, although the number of MIs for a given topology is fixed by reduction identities, the actual choice of MIs is not. A good choice can substantially simplify the solution of the system, and in Chapter~\ref{chap:workflow2} we will clarify the meaning of the term \textit{good} in this respect.

\section{Program Packages}
\label{sec:programs}

In this section, we briefly summarize the program packages, which we use for the computational steps described so far.\\
We generate all relevant Feynman diagrams contributing to the processes $(a)$, $(b)$ and $(c)$ as described in Section~\ref{sec:QCD} with \textsc{Qgraf}~\cite{Nogueira:1991}, which is written in \textsc{Fortran 77}. The user has to supply the vertices and propagators of the underlying theory in a so-called \texttt{model} file. Subsequently, the incoming and outgoing fields from within this \texttt{model} file as well as the number of loops have to be defined. In addition, topological restrictions can be imposed, e.g. by allowing only one-particle-irreducible diagrams, and the number of internal propagators of a certain kind may be limited. The actual output of the diagrams is provided via text files, which are passed to \textsc{Form}~\cite{Ruijl:2017}. With the help of \textsc{Form}, Feynman rules are inserted and the resulting expression is projected onto the tensor structure from Section~\ref{sec:amplitudes} by contracting the corresponding Lorentz indices. Finally, the remaining loop integrals are classified through a mapping onto previously defined topologies.\\
As a next step, these loop integrals are fed into reduction programs, whereof several public implementations exist, written in either \textsc{Mathematica} or \textsc{C++}. All these programs work through the steps described in Section~\ref{sec:reduction} and are provided with the following information:
\begin{itemize}
\item The external momenta and loop momenta have to be defined.
\item The topology has to be specified by indicating the occuring propagators including masses and momenta.
\end{itemize}
The aforementioned \textsc{Reduze} code is our main tool for the purpose of reduction to MIs for multiple reasons: First, it can import \textsc{Qgraf} output in a fully automatic way, which can be used in order to check if the generated Feynman diagrams belong to the set of predefined topologies. Second, it provides the largest number of symmetry relations, including the possibility of finding symmetries between crossed sectors and across different topologies, therefore producing the lowest number of MIs. Finally, \textsc{Reduze} offers many more tools that we have not tested, such as the automatic generation of amplitudes up to a certain level of complexity, and other ones that will be useful in Chapter~\ref{chap:workflow2}, like the automatic generation of differential equations for MIs. Although \textsc{Reduze} is able to parallelize its workflow to a large extent, its usage is quite resource intensive. For this reason, we have tried several other implementations of the Laporta algorithm, however with limited success compared to \textsc{Reduze}, namely the combination \textsc{LiteRed}+\textsc{Fire}~\cite{Lee:2013a,Smirnov:2014} as well as the only recently published version of \textsc{Kira}~\cite{Maierhoefer:2017}. On top of that, the private implementation \textsc{Crusher}~\cite{crusher} was employed. The only publicly available implementation that was not tested is \textsc{Air}~\cite{Anastasiou:2004}, which is written in \textsc{Maple}.\\
Manipulations of intermediate expressions were carried out by means of either \textsc{Form}, \textsc{Mathematica} or \textsc{Fermat}~\cite{fermat}, and all Feynman diagrams in this thesis were drawn using \textsc{Jaxodraw}~\cite{Binosi:2003,Binosi:2008}.

\chapter{Three-Loop Corrections to the $\boldsymbol{Hb\bar{b}}$ Form Factor}
\chaptermark{Three-Loop Corrections to the $Hb\bar{b}$ Form Factor}
\label{chap:hbb}

In this chapter, we compute the three-loop QCD corrections to the vertex function for the Yukawa coupling of a Higgs boson to a pair of bottom quarks in the limit of vanishing quark masses. After motivating the importance of this calculation and recapitulating related results in Section~\ref{sec:hbbintro}, we define the $Hb\bar b$ form factor and discuss its renormalization in Section~\ref{sec:hbbrenorm}. In Section~\ref{sec:ff}, we summarize results at one and two loops, before we proceed with the calculation of the three-loop form factor along the lines of the calculations of the three-loop QCD corrections to the vector and scalar form factors~\cite{Baikov:2009,Gehrmann:2010a,Gehrmann:2010b}. The infrared pole structure is analyzed in Section~\ref{sec:ir}, and we conclude with an outlook in Section~\ref{sec:conc1}.

\section{Introduction}
\label{sec:hbbintro}

It is imperative to study the production mechanisms and decay channels of the Higgs boson to high precision in order to fully validate the mechanism of electroweak symmetry breaking and to uncover potential deviations from its Standard Model realization. As indicated by Fig.~\ref{fig:run2}, new experimental data is expected from Run~2 of the LHC for the decay of the Higgs boson into bottom quarks. The interpretation of this increasingly accurate data demands equally precise theoretical predictions, requiring the inclusion of higher orders in the perturbative expansion for decay channels and production processes.\\
\begin{figure}[tb]
	\begin{center}
	\includegraphics[width=\textwidth]{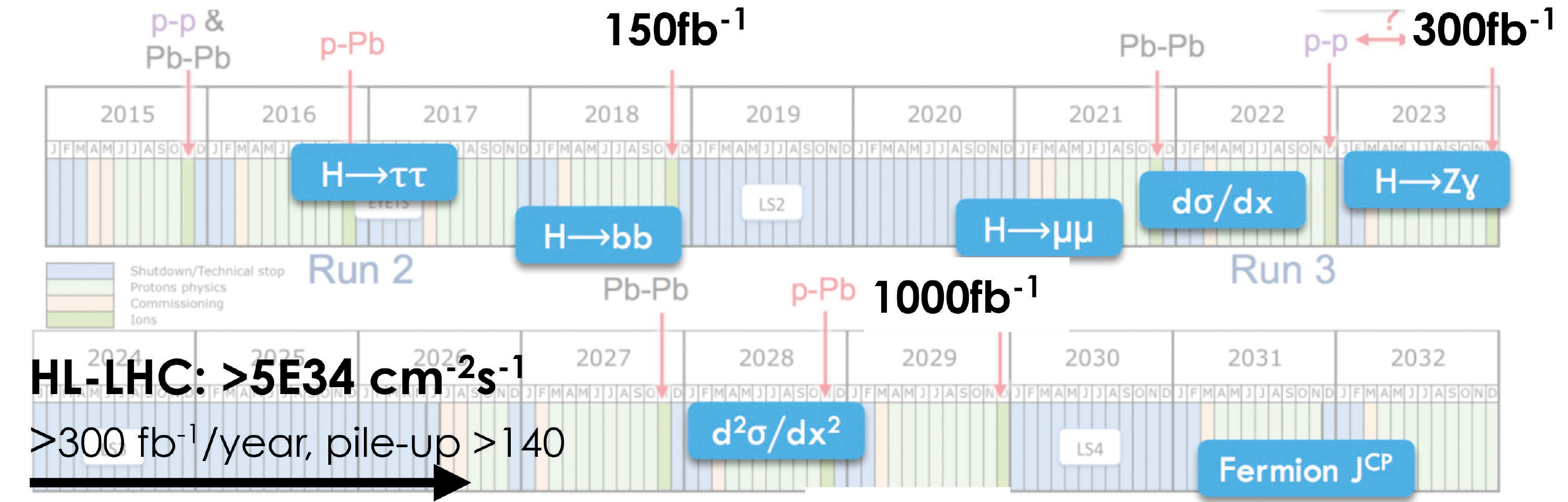}
	\caption[Timeline of the LHC beyond Run2]{\textbf{Timeline of the LHC}~\cite{Meridiani:2017}. Further events for the Higgs boson decay into bottom quarks are expected to be measured in 2018.}
	\label{fig:run2}
	\end{center}
\end{figure}\noindent
As mentioned, the three-loop $Hb\bar{b}$ form factor is a crucial ingredient of third-order QCD corrections for the production of the Higgs boson in bottom quark fusion, and for the fully differential decay rate of Higgs bosons to bottom quarks. Currently, fully differential results are known for Higgs boson production in gluon fusion~\cite{Anastasiou:2005,Grazzini:2008}, bottom quark annihilation~\cite{Buehler:2012} and associated production with vector bosons~\cite{Ferrera:2011,Ferrera:2013} to NNLO in QCD. Fully differential Higgs boson production through vector boson fusion is known to NNLO~\cite{Cacciari:2015,Cruz-Martinez:2018}, whereas associated production with top quarks was computed to NLO accuracy~\cite{Beenakker:2002,Dawson:2003a,Frederix:2011}. The inclusive decay rates of the Higgs boson have been derived to fourth order in QCD for the decay mode to hadrons~\cite{Chetyrkin:1997b} and to bottom quarks~\cite{Baikov:2005}. To study the dominant decay mode to bottom quarks, especially the associated production with vector bosons is of relevance, and a fully differential description of production and decay is demanded. The decay distributions to NNLO have been derived in Ref.~\cite{Anastasiou:2011}, and a combined description with the associated production at NNLO was obtained in Refs.~\cite{Ferrera:2014,Ferrera:2017,Caola:2017}.\\
In the Standard Model, the dominant Higgs boson production process is gluon fusion, while bottom quark annihilation contributes to the total production only at the per-cent level. In extensions of the Standard Model with an enlarged Higgs sector, the coupling of Higgs bosons to bottom quarks can be enhanced, such that bottom quark annihilation could become their dominant production process. Bottom quark annihilation is moreover of conceptual interest, since it allows the study of different prescriptions for the treatment of bottom quark-induced processes at hadron colliders. In the fixed flavor number scheme (FFNS), bottom quarks are produced only from gluon splitting, introducing potentially large logarithmic corrections at each order. These initial-state splittings are resummed into bottom quark parton distributions in the variable flavor number scheme (VFNS). To the same order in the strong coupling constant, the leading-order process in the FFNS corresponds to NNLO in the VFNS. Higgs production from bottom quark annihilation is known to NLO in the FFNS~\cite{Dittmaier:2003,Dawson:2003b,Wiesemann:2014} and to NNLO in the VFNS~\cite{Buehler:2012,Harlander:2003,Harlander:2014}. Using the NLO calculation of Higgs-plus-jet production in bottom quark annihilation~\cite{Harlander:2010}, the Higgs production with a jet veto was also derived to NNLO~\cite{Harlander:2011}. Most calculations are carried out in the limit of vanishing bottom quark mass, which is justified by the large mass hierarchy between the bottom quark and the Higgs boson.\\
As explained in Eq.~\eqref{cs2}, the calculation of perturbative higher-order QCD corrections requires the derivation of virtual loop corrections to the relevant matrix elements of the Feynman amplitude. In case of Higgs production and decay involving bottom quarks, the form factor describing the Yukawa coupling of the Higgs boson to bottom quarks is the crucial ingredient. Corrections up to two loops were derived for this form factor with massless bottom quarks~\cite{Buehler:2012,Harlander:2003,Ravindran:2006} and also including the full mass dependence~\cite{Bernreuther:2005}. Two-loop corrections to the Higgs decay amplitude describing the decay to a pair of bottom quarks and a gluon were also derived~\cite{Ahmed:2014a} in massless QCD. The three-loop QCD corrections to the $Hb\bar b$ form factor, whose pole structure can be predicted~\cite{Ravindran:2006} from factorization properties of QCD amplitudes~\cite{Catani:1998,Sterman:2002,Moch:2005,Becher:2009b,Gardi:2009}, enters the N$^3$LO corrections to the Higgs production cross section in bottom quark annihilation and the differential description of Higgs decays to bottom quarks at this order. Both types of applications require a substantial extension of current technical methods in order to perform calculations of collider observables to N$^3$LO. First steps in this direction have been taken in Refs.~\cite{Anastasiou:2013a,Duhr:2013,Li:2013,Anastasiou:2013b,Li:2014,Hoschele:2014}, cumulating in the calculation of the N$^3$LO threshold contribution to Higgs production in gluon fusion~\cite{Anastasiou:2014}\footnote{Recently, the exact result has been computed in Ref.~\cite{Mistlberger:2018}.}. Exploiting universal QCD factorization properties at threshold \cite{Ahmed:2014b,Catani:2014}, the result of Ref.~\cite{Anastasiou:2014} could be combined with the form factor derived here to obtain the N$^3$LO threshold contribution to Higgs production in bottom quark annihilation~\cite{Ahmed:2014c}.

\section{Kinematics and Renormalization}
\label{sec:hbbrenorm}

In QCD, form factors are scalar functions which couple an external off-shell current with four-momentum $q^2 = s_{12}$ to a pair of partons with on-shell momenta $p_1$ and $p_2$. They are computed by contracting the respective basic vertex functions with projectors.
In the $Hb\bar{b}$ case, the unrenormalized form factor $\mathcal{A}^B$ is obtained from a scalar vertex function $\Gamma$ according to
\begin{equation}
{\cal A}^B  = -\frac{1}{2 q^2}\, \mathrm{Tr} \left( p_1 \!\!\!\! / \, p_2 \!\!\!\! / \, \Gamma \right)\,.
\label{projq}
\end{equation}
It is described by a single form factor only in the case of massless partons. In fact, we consider a Higgs boson coupling to the bottom quarks as defined in Eq.~\eqref{yukawa} via an unrenormalized Yukawa coupling $y^B_b \equiv y^B$,
\begin{equation}
y^B = \frac{m^B}{v} \,,
\end{equation}
where $m^B_b \equiv m^B$ is the bare mass of the bottom quark. However, we treat the bottom mass as independent of the Yukawa coupling and suppose it to be massless in the calculation of matrix elements. This is justified by the fact that the Higgs boson is much heavier than the bottom quark. Here and in what follows, the bottom quark could be equally replaced by any other lighter quark, i.e. we consider a process with $N_F=5$ active flavors, where the top quark decouples from the theory. In fact, the contribution from top quarks running in the loops is suppressed by their heavy mass so that it does not need to be taken into account.\\
At a given loop order, the unrenormalized form factor is obtained as an expansion in powers of the coupling constant by evaluating the Feynman diagrams that contribute to the vertex function in perturbative QCD, examples of which are given in Fig.~\ref{fig:hbbdiagrams}. With the mass parameter $\mu_0^2$ introduced in Eq.~\eqref{mu0} and the definition of $S_\gamma$ in Eq.~\eqref{Sgamma} this expansion can be written as
\begin{eqnarray}
{\cal A}^B (\alpha_s^B, s_{12}) &=& y^B\left(1 + \sum_{n=1}^{\infty} \left( \frac{\alpha_s^B}{\pi}\right)^n \left(\frac{-s_{12}}{\mu_0^2}\right)^{-n\e}
S_\gamma^n \,{\cal A}^B_n\right) \, .
\end{eqnarray}
Each power of the coupling constant corresponds to a virtual loop, i.e. Eq.~\eqref{projq} is normalized in such a way that the tree-level form factor is equal to unity.\\
\begin{figure}[tb]
\begin{center}
\begin{minipage}[t]{0.24\textwidth}
\vspace{-\ht\strutbox}\includegraphics[width=\textwidth]{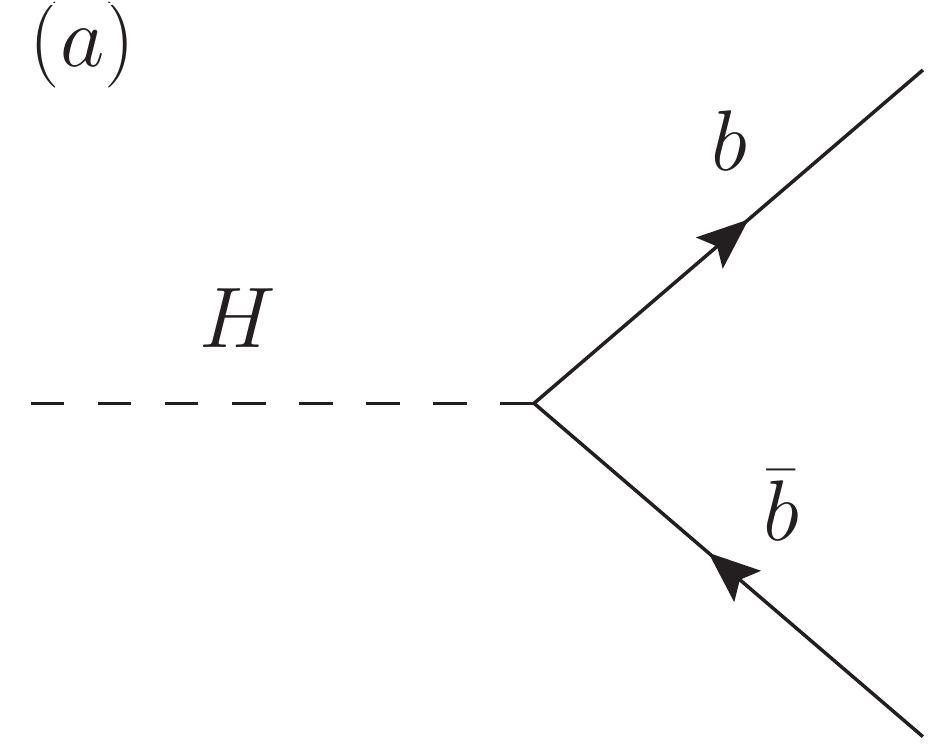}
\end{minipage}
\hfill
\begin{minipage}[t]{0.24\textwidth}
\vspace{-\ht\strutbox}\includegraphics[width=\textwidth]{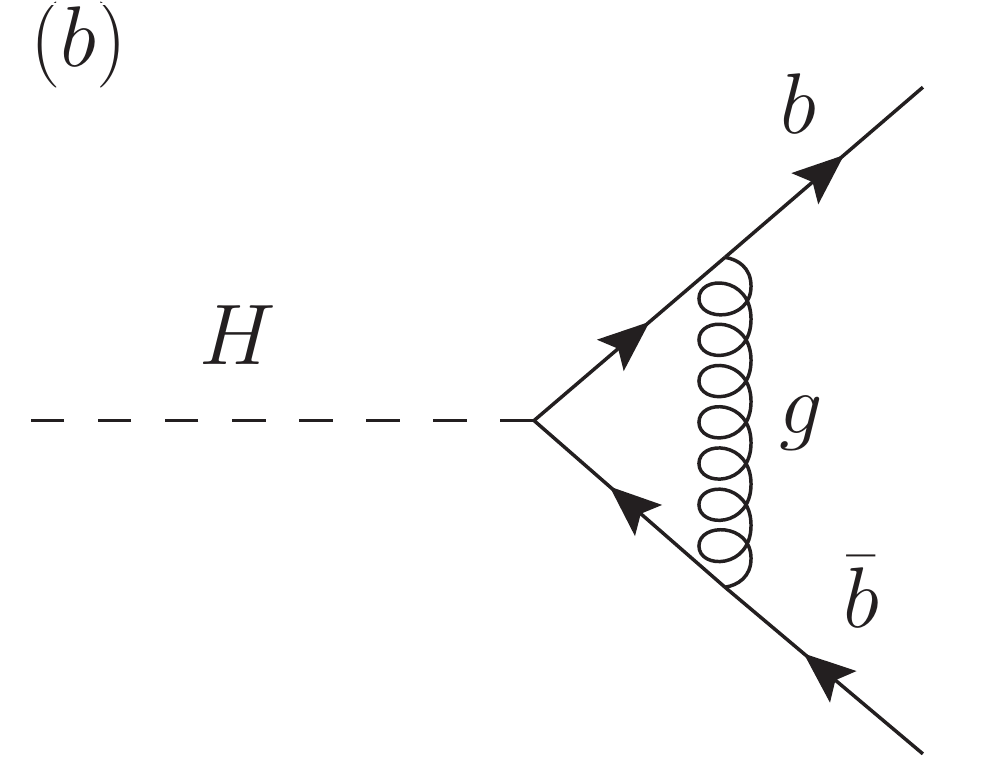}
\end{minipage}
\begin{minipage}[t]{0.24\textwidth}
\vspace{-\ht\strutbox}\includegraphics[width=\textwidth]{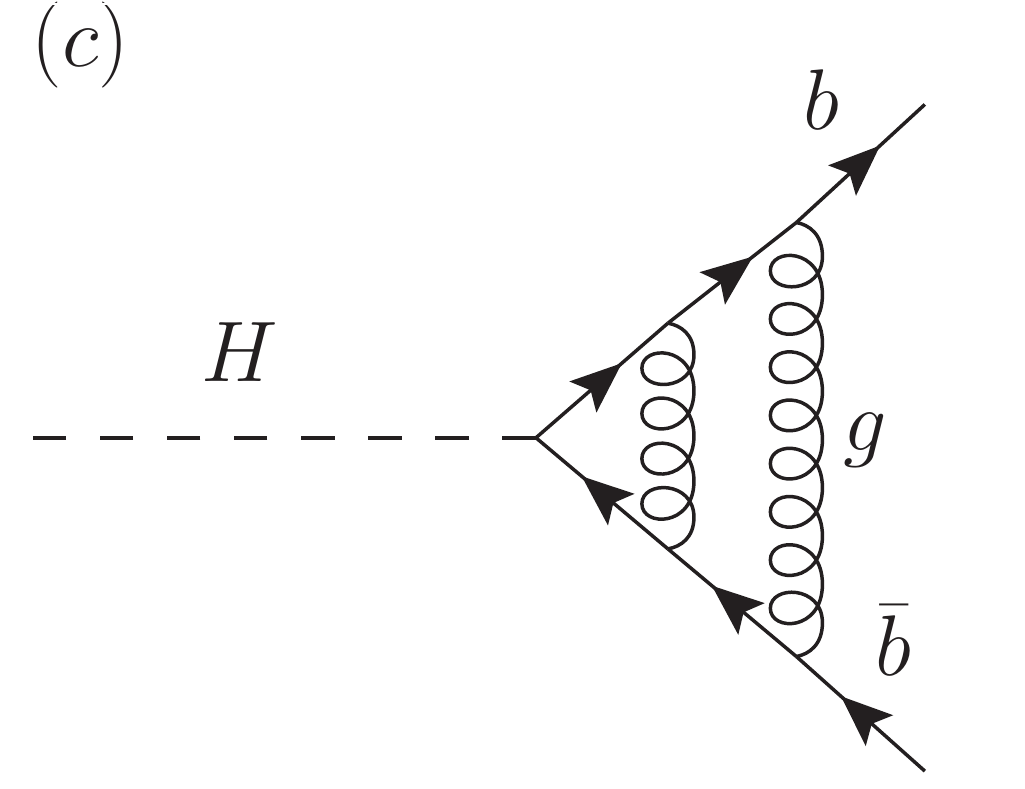}
\end{minipage}
\begin{minipage}[t]{0.24\textwidth}
\vspace{-\ht\strutbox}\includegraphics[width=\textwidth]{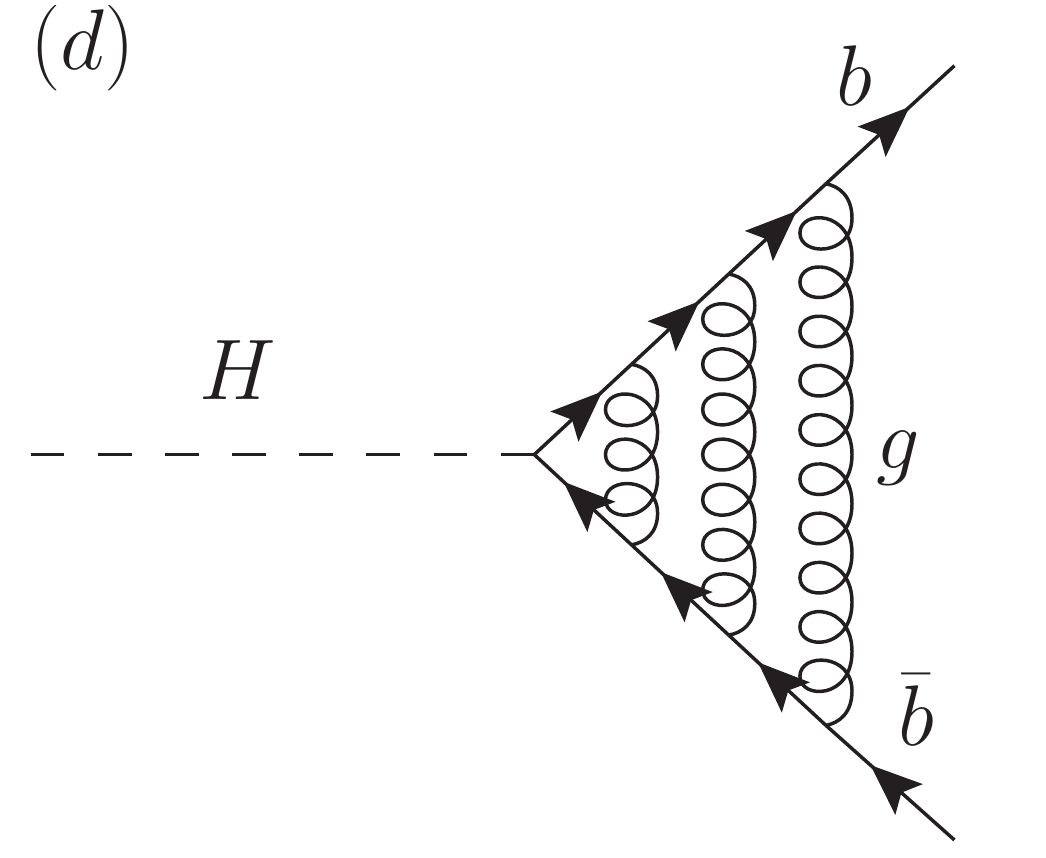}
\end{minipage}
\caption[Example diagrams for the calculation of the $Hb\bar{b}$ form factor]{\textbf{Example diagrams for the calculation of the $\boldsymbol{Hb\bar{b}}$ form factor}\\up to three loops in decay kinematics with $(a)$ the LO diagram, $(b)$ a NLO diagram, $(c)$ a NNLO diagram and $(d)$ a N$\mathrm{^3}$LO diagram. In this case, the loop diagrams are obtained from the tree-level diagram $(a)$ by attaching one gluon per loop order to the bottom-quark final states.}
\label{fig:hbbdiagrams}
\end{center}
\end{figure}\noindent
The ultraviolet renormalization of the form factor requires two ingredients: First, the bare coupling $\alpha_s^B$ is replaced with the renormalized coupling $\alpha_s\equiv \alpha_s(\mu^2)$, which is evaluated at the renormalization scale $\mu^2$, as described in Eq.~\eqref{alphasrenorm}, by using the result of the renormalization constant $Z_{\alpha_s}$ in Eq.~\eqref{Zalphas}. Second, the renormalization of the Yukawa coupling is carried out by replacing the bare coupling $y^B$ with the renormalized coupling $y\equiv y(\mu^2)$ according to Eq.~\eqref{yukawarenorm} with the help of the result for the renormalization constant of the Yukawa coupling in Eq.~\eqref{ZyMS}. The renormalized form factor $\mathcal{A}$ is then defined as follows:
\begin{eqnarray}
\mathcal{A} (\alpha_s(\mu^2), s_{12},\mu^2= |s_{12}|) &=& y \left(1 + \sum_{n=1}^{\infty} \left( \frac{\alpha_s(\mu^2)}{\pi}\right)^n   
  \,\mathcal{A}_n\right).
\end{eqnarray}
In order to derive the $i$-loop contributions $\mathcal{A}_i$ to the renormalized form factor from the unrenormalized coefficients ${\cal{A}}^B_i$, two possible configurations have to be distinguished: The partons can be both either in the initial or in the final state ($s_{12}>0$, time-like) or one parton can be in the initial and one in the final state ($s_{12}<0$, space-like). We will indicate the results for the renormalized form factors in the time-like case, which corresponds to the Higgs decay into bottom quarks or to the $b\bar{b}$ annihilation process into a Higgs boson. In this case, the renormalized form factor acquires imaginary parts from the $\e$-expansion of
\begin{equation}
\Delta(s_{12}) = (-\mathrm{sgn}(s_{12})-i0)^{-\e} \, .
\end{equation}
Up to three loops, the renormalized coefficients for the $Hb\bar{b}$ form factor are then obtained as
\begin{eqnarray}
\mathcal{A}_1   &=& 
 {\cal A}^B_1 \Delta(s_{12})
-\frac{3 C_F}{4 \e}   ,  \nonumber \\
\mathcal{A}_2  &=& 
 {\cal A}^B_2 \left(\Delta(s_{12})\right)^2
+ \Bigg[ -\frac{3 C_F}{4 \e}-\frac{11 C_A}{12 \e}+\frac{N_F}{6 \e} \Bigg]
{\cal A}^B_1  \Delta(s_{12}) \nonumber \\
&& + \frac{1}{16} \Bigg[ C_F^2 \left(\frac{9}{2 \e^2}-\frac{3}{4 \e}\right) + C_F C_A \left(\frac{11}{2 \e^2}-\frac{97}{12 \e}\right) + C_F N_F \left(-\frac{1}{\e^2}+\frac{5}{6 \e}\right) \Bigg] , \nonumber \\
\mathcal{A}_3  &=& 
 {\cal A}^B_3 \left(\Delta(s_{12})\right)^3
+ \Bigg[ -\frac{3 C_F}{4 \e}-\frac{11 C_A}{6 \e}+\frac{N_F}{3 \e} \Bigg] {\cal A}^B_2  \left(\Delta(s_{12})\right)^2 \nonumber \\
&& + \frac{1}{16} \Bigg[ C_F^2 \left(\frac{9}{2 \e^2}-\frac{3}{4 \e}\right) + C_F C_A \left(\frac{33}{2 \e^2}-\frac{97}{12 \e}\right) + C_A^2 \left(\frac{121}{9 \e^2}-\frac{17}{3 \e}\right) \nonumber \\
&& \quad + C_F N_F \left(-\frac{3}{\e^2}+\frac{11}{6 \e}\right) + C_A N_F \left(-\frac{44}{9 \e^2}+\frac{5}{3 \e}\right) + \frac{4 N_F^2}{9 \e^2} \Bigg] {\cal A}^B_1\Delta(s_{12}) \nonumber \\
&& + \frac{1}{64} \Bigg[ C_F^3 \left(-\frac{9}{2 \e^3}+\frac{9}{4 \e^2}-\frac{43}{2 \e}\right) + C_F^2 C_A \left(-\frac{33}{2 \e^3}+\frac{313}{12 \e^2}+\frac{43}{4 \e}\right) \nonumber \\
&& \quad + C_F C_A^2 \left(-\frac{121}{9 \e^3}+\frac{1679}{54 \e^2}-\frac{11413}{324 \e}\right) + C_F^2 N_F \left(\frac{3}{\e^3}-\frac{29}{6 \e^2}+\frac{1}{\e} \left(\frac{23}{3}-8\zeta_3\right) \right) \nonumber \\
&& \quad + C_F C_A N_F \left(\frac{44}{9 \e^3}-\frac{242}{27 \e^2}+\frac{1}{\e} \left(\frac{278}{81}+8\zeta_3\right) \right) \nonumber \\
&& \quad + C_F N_F^2 \left(-\frac{4}{9 \e^3}+\frac{10}{27 \e^2}+\frac{35}{81 \e}\right) \Bigg] .
\label{reng}
\end{eqnarray}
The one- and two-loop relations agree with those in Ref.~\cite{Ahmed:2014a}.

\section{Calculation of the Form Factor}
\label{sec:ff}
\subsection{Results at One Loop}

We define
\begin{equation}
S_{R} = \frac{16\pi^2 \, S_{\Gamma}}{S_\gamma} = \frac{e^{\gamma_E\,\e}}{\Gamma(1-\e)} \, ,
\end{equation}
where
\begin{equation}
S_{\Gamma} = \frac{(4\pi)^\e}{16\pi^2 \, \Gamma(1-\e)}
\label{SGamma}
\end{equation}
corresponds to the normalization of the one-loop bubble integral $B_{2,1}$. With this, the unrenormalized one-loop form factor can be written as
\begin{equation}
{\cal A}^B_1/S_R = C_F\,B_{2,1} \left(\frac{1}{(D-4)}+\frac{D}{4}\right) \,.
\label{f1g}
\end{equation}
The exact result for the one-loop bubble integral is indicated in Ref.~\cite{Gehrmann:2005} under the name $A_{2,\mathrm{LO}}$, i.e. Eq.~\eqref{f1g} can be understood as an expression valid to all orders in~$\e$. The $\e$-expansion of $B_{2,1}$ can be found in Appendix~A of Ref.~\cite{Gehrmann:2010a}. Inserting this expansion and keeping terms through to ${\cal O}(\e^4)$, we obtain
\begin{align}
{\cal A}^B_1 = \frac{C_F}{4}\Biggl[
&-\frac{2}{\e^2}
+\left(\zeta_2-2\right)
+\e\left(\frac{14\zeta_3}{3}-4\right)
+\e^2\left(\frac{47\zeta_2^2}{20}+\zeta_2-8\right)\nonumber \\
&-\e^3\left(\frac{7\zeta_2\zeta_3}{3}-2\zeta_2-\frac{14\zeta_3}{3}-\frac{62\zeta_5}{5}+16\right)\nonumber \\
&+\e^4\left(\frac{949\zeta_2^3}{280}+\frac{47\zeta_2^2}{20}-\frac{49\zeta_3^2}{9}+4\zeta_2+\frac{28\zeta_3}{3}-32\right)
\Biggr] \, .
\end{align}
By renormalizing this result as described in Eq.~\eqref{reng}, we find that the one-loop form factor agrees with the $\e$-expansion of Eq.~(3.2) in Ref.~\cite{Anastasiou:2011}.

\subsection{Results at Two Loops}

Written in terms of the two-loop MIs specified in Appendix~A of Ref.~\cite{Gehrmann:2010a}, the unrenormalized two-loop form factor is given by
\begin{align}
{\cal A}^B_2/S_R^2 = \frac{C_F^2}{16} \Biggl[
&B_{4,2}
\left(D^2+\frac{32}{(D-4)}+\frac{16}{(D-4)^2}+8\right)\nonumber \\
&-C_{4,1}
\left(\frac{7D^2}{8}-\frac{137D}{16}-\frac{265}{32(2D-7)}-\frac{58}{(D-4)}-\frac{40}{(D-4)^2}-\frac{239}{32}\right)\nonumber \\
&+ B_{3,1} 
\left(\frac{27D^2}{8}-\frac{969D}{16}+\frac{1855}{32(2D-7)}-\frac{3}{2(D-3)}\right.\nonumber \\
&\left.\qquad \qquad-\frac{730}{(D-4)}-\frac{720}{(D-4)^2}-\frac{288}{(D-4)^3}-\frac{3079}{32}\right)\nonumber \\
&- C_{6,2}
\left(\frac{D^2}{16}-\frac{21D}{32}-\frac{53}{64(2D-7)}+\frac{29}{64}\right)
\Biggr]\nonumber \\
+\frac{C_F C_A}{16} \Biggl[
&-C_{4,1}
\left(\frac{D^2}{16}-\frac{7D}{32}+\frac{265}{64(2D-7)}+\frac{1}{3(D-1)}\right.\nonumber \\
&\left.\qquad \qquad+\frac{53}{3(D-4)}+\frac{16}{(D-4)^2}+\frac{367}{64}\right)\nonumber \\
&- B_{3,1} 
\left(\frac{75D^2}{16}-\frac{1129D}{32}+\frac{1855}{64(2D-7)}+\frac{1}{4(D-3)}\right.\nonumber \\
&\left.\qquad \qquad-\frac{241}{(D-4)}-\frac{228}{(D-4)^2}-\frac{96}{(D-4)^3}-\frac{903}{64}\right)\nonumber \\
&+C_{6,2}
\left(\frac{D^2}{32}-\frac{21D}{64}-\frac{53}{128(2D-7)}+\frac{29}{128}\right)
\Biggr]\nonumber \\
+\frac{C_F N_F}{16} \Biggl[
&-C_{4,1}
\left(D+\frac{2}{3(D-1)}+\frac{4}{3(D-4)}-2\right)
\Biggr] \,.
\end{align}
As in the one-loop case, the all-order result is obtained by replacing $B_{4,2}$, $B_{3,1}$, $C_{4,1}$ and $C_{6,2}$ with $A_{2,\mathrm{LO}}^2$, $A_3$, $A_4$ and $A_6$ from Ref.~\cite{Gehrmann:2005}, respectively.\\
Inserting the expansion of the two-loop MIs and keeping terms through to ${\cal O}(\e^2)$ yields
\begin{align}
{\cal A}^B_2 = \frac{C_F^2}{16} \Biggl[
&\frac{2}{\e^4}
-\frac{1}{\e^2}\left(2\zeta_2-4\right)
-\frac{1}{\e}\left(\frac{64\zeta_3}{3}-6\zeta_2-8\right)-\left(13\zeta_2^2-12\zeta_2+30\zeta_3-22\right)\nonumber \\
&-\e\left(\frac{96\zeta_2^2}{5}-\frac{112\zeta_2\zeta_3}{3}-36\zeta_2+\frac{404\zeta_3}{3}+\frac{184\zeta_5}{5}-64\right)\nonumber \\
&+\e^2\left(\frac{223\zeta_2^3}{5}-\frac{426\zeta_2^2}{5}+2\zeta_2\zeta_3+\frac{2608\zeta_3^2}{9}+106\zeta_2-\frac{1744\zeta_3}{3}-126\zeta_5+192\right) 
\Biggr]\nonumber \\
+\frac{C_F C_A}{16} \Biggl[
&-\frac{11}{6\e^3}
+\frac{1}{\e^2}\left(\zeta_2-\frac{67}{18}\right)
-\frac{1}{\e}\left(\frac{11\zeta_2}{6}-13\zeta_3+\frac{220}{27}\right)\nonumber \\
&+\left(\frac{44\zeta_2^2}{5}-\frac{103\zeta_2}{18}+\frac{305\zeta_3}{9}-\frac{1655}{81}\right)\nonumber \\
&+\e\left(\frac{1171\zeta_2^2}{60}-\frac{89\zeta_2\zeta_3}{3}-\frac{490\zeta_2}{27}+\frac{2923\zeta_3}{27}+51\zeta_5-\frac{12706}{243}\right)\nonumber \\
&-\e^2\left(\frac{809\zeta_2^3}{70}-\frac{11819\zeta_2^2}{180}+\frac{127\zeta_2\zeta_3}{9}+\frac{569\zeta_3^2}{3}+\frac{4733\zeta_2}{81}\right.\nonumber \\
&\left.\qquad \quad-\frac{30668\zeta_3}{81}-\frac{2411\zeta_5}{15}+\frac{99770}{729}\right)
\Biggr]\nonumber \\
+\frac{C_F N_F}{16} \Biggl[
&\frac{1}{3\e^3}
+\frac{5}{9\e^2}
+\frac{1}{\e}\left(\frac{\zeta_2}{3}+\frac{46}{27}\right)
+\left(\frac{5\zeta_2}{9}-\frac{26\zeta_3}{9}+\frac{416}{81}\right)\nonumber \\
&-\e\left(\frac{41\zeta_2^2}{30}-\frac{46\zeta_2}{27}+\frac{130\zeta_3}{27}-\frac{3748}{243}\right)\nonumber \\
&-\e^2\left(\frac{41\zeta_2^2}{18}+\frac{26\zeta_2\zeta_3}{9}-\frac{416\zeta_2}{81}+\frac{1196\zeta_3}{81}+\frac{242\zeta_5}{15}-\frac{33740}{729}\right) 
\Biggr] \, .
\end{align}
After renormalization, we find full agreement with Eq.~(3.6) of Ref.~\cite{Anastasiou:2011} through to ${\cal O}(\e^0)$ and provide the next two terms in the expansion.

\subsection{Outline of the Calculation at Three Loops}

As any multi-loop computation, the calculation of the $Hb\bar{b}$ three-loop form factor can be separated into the steps described in Chapter~\ref{chap:workflow1}: Initially, one calculates the matrix elements in terms of three-loop integrals. Next, the algebraic reduction of all three-loop integrals appearing in the relevant Feynman diagrams is performed. Eventually, the remaining MIs are computed. Let us elaborate on these three steps.\\
In order to determine the three-loop vertex function, we generated 244 Feynman diagrams contributing to the $Hb\bar{b}$ form factor at three loops with the help of \textsc{Qgraf}. Every diagram is then contracted with the projector in Eq.~\eqref{projq} and can be expressed as a linear combination of many scalar three-loop Feynman integrals with up to nine different propagators. The integrands depend on the three loop momenta and on the two on-shell external momenta, leading to twelve different combinations of scalar products involving loop momenta. Hence, we are left with three irreducible scalar products in the numerator since they do not cancel against all linearly independent combinations of denominators, corresponding to the integral representation in Eq.~\eqref{scalarint2}.\\
The reduction to MIs was carried out using \textsc{Reduze}. For this purpose, we switch to the representation in Eq.~\eqref{scalarint3} and define a minimal set of integral families, onto which all MIs can be mapped. According to the considerations made above each family clearly consists of a set of twelve linearly independent propagators, our choice of which is given in Table~\ref{tab:hbbtopo}. After the reduction, we are left with 13 MIs for family I, 8 MIs for family II and 1 MI for family III, leading to a total of 22 MIs. All of them were computed analytically in the past~\cite{Gehrmann:2006,Heinrich:2007,Heinrich:2009,Lee:2010} and are summarized in detail in Ref.~\cite{Gehrmann:2010a} so that they will not be reproduced here.
\begin{table}[tb]
\caption[Integral families for the reduction to Master Integrals of the three-loop $Hb\bar{b}$ form factor]{\textbf{Integral families for the reduction to Master Integrals of the three-loop $\boldsymbol{Hb\bar{b}}$ form factor.} Each of the three families I, II and III are given by a set of twelve propagators, where the loop momenta are labeled $k_1$, $k_2$, $k_3$ and $p_1$, $p_2$ denote the external momenta of the bottom quarks occuring in Eq.~\eqref{projq}.}
\setlength{\tabcolsep}{1.25cm}
\begin{tabularx}{\textwidth}{lll}
\toprule
\multicolumn{1}{c}{Family I} & \multicolumn{1}{c}{Family II} & \multicolumn{1}{c}{Family III} \\
\midrule
$k_1^2$ & $k_1^2$ & $k_1^2$ \\
$k_2^2$ & $k_2^2$ & $k_2^2$ \\
$k_3^2$ & $k_3^2$ & $k_3^2$ \\
$(k_1-k_2)^2$ & $(k_1-k_2)^2$ & $(k_1-k_2)^2$ \\
$(k_1-k_3)^2$ & $(k_1-k_3)^2$ & $(k_1-k_3)^2$ \\
$(k_2-k_3)^2$ & $(k_2-k_3)^2$ & $(k_1-k_2-k_3)^2$ \\
$(k_1-p_1)^2$ & $(k_1-k_3-p_2)^2$ & $(k_1-p_1)^2$ \\
$(k_1-p_1-p_2)^2$ & $(k_1-p_1-p_2)^2$ & $(k_1-p_1-p_2)^2$ \\
$(k_2-p_1)^2$ & $(k_2-p_1)^2$ & $(k_2-p_1)^2$ \\
$(k_2-p_1-p_2)^2$ & $(k_1-k_2-p_2)^2$ & $(k_2-p_1-p_2)^2$ \\
$(k_3-p_1)^2$ & $(k_3-p_1)^2$ & $(k_3-p_1)^2$ \\
$(k_3-p_1-p_2)^2$ & $(k_3-p_1-p_2)^2$ & $(k_3-p_1-p_2)^2$ \\
\bottomrule
\label{tab:hbbtopo}
\end{tabularx}
\end{table}

\subsection{Results at Three Loops}

The unrenormalized three-loop form factor can be decomposed into the following color structures:
\begin{align}
{\cal A}^B_3/S_R^3 &=
C_F^3 ~X_{C_F^3}
+C_F^2C_A  ~X_{C_F^2C_A} 
+ C_FC_A^2  ~X_{C_FC_A^2} 
+ C_F^2 N_F  ~X_{C_F^2 N_F}
\nonumber \\
&
+C_FC_A N_F ~X_{C_FC_A N_F}
+C_FN_F^2  ~X_{C_FN_F^2} \,.
\label{fq3lbare}
\end{align}
It should be noted that, in contrast to the quark form factor for a photonic coupling~\cite{Baikov:2009,Gehrmann:2010a}, no contribution from the Higgs boson coupling to closed quark loops appears at three loops in massless QCD. This contribution requires a helicity flip on both the internal and external quark lines, and is consequently mass-suppressed. After the reduction of the integrals appearing in the Feynman diagrams, the coefficients $X_i$ of the color structures include linear combinations of MIs. These coefficients are somewhat lengthy and will not be presented here. Inserting the expansion of the three-loop MIs and keeping terms through to ${\cal O}(\e^0)$, we find that the unrenormalized three-loop coefficient is given by
\begin{align}
{\cal A}^B_3 = 
 \frac{C_F^3}{64} \Biggl[
&-\frac{4}{3\e^6}
+\frac{1}{\e^4}\left(2\zeta_2-4\right)
-\frac{1}{\e^3}\left(12\zeta_2-\frac{100\zeta_3}{3}+8\right)\nonumber \\
&+\frac{1}{\e^2}\left(\frac{213\zeta_2^2}{10}-26\zeta_2+60\zeta_3-28\right)\nonumber \\
&+\frac{1}{\e}\left(\frac{126\zeta_2^2}{5}-\frac{214\zeta_2\zeta_3}{3}-94\zeta_2+\frac{784\zeta_3}{3}+\frac{644\zeta_5}{5}-\frac{238}{3}\right)\nonumber \\
&-\left(
          \frac{9095\zeta_2^3}{252}
          - \frac{887\zeta_2^2}{10}
          - 202\zeta_2\zeta_3
          + \frac{1826\zeta_3^2}{3}
\right. \nonumber \\
&    \left.    \qquad  
          + \frac{1085\zeta_2}{3}
          - 538\zeta_3
          - 676\zeta_5
          + \frac{385}{3}
\right)
\Biggr]\nonumber \\
+\frac{C_F^2 C_A}{64} \Biggl[
&\frac{11}{3\e^5}
-\frac{1}{\e^4}\left(2\zeta_2-\frac{67}{9}\right)
+\frac{1}{\e^3}\left(\frac{11\zeta_2}{6}-26\zeta_3+\frac{539}{27}\right)\nonumber \\
&-\frac{1}{\e^2}\left(\frac{83\zeta_2^2}{5}-\frac{631\zeta_2}{18}+135\zeta_3-\frac{4507}{81}\right)\nonumber \\
&-\frac{1}{\e}\left(\frac{31591\zeta_2^2}{360}-\frac{215\zeta_2\zeta_3}{3}-\frac{10199\zeta_2}{54}+\frac{1721\zeta_3}{3}+142\zeta_5-\frac{38012}{243}\right)\nonumber \\
&-\left(
          \frac{18619\zeta_2^3}{1260}
          + \frac{305831\zeta_2^2}{1080}
          + \frac{1663\zeta_2\zeta_3}{18}
          - \frac{1616\zeta_3^2}{3}
 \right. \nonumber \\
& \left.          \qquad
          - \frac{131161\zeta_2}{162}
          + \frac{17273\zeta_3}{9}
          + \frac{27829\zeta_5}{45}
          - \frac{332065}{729}
\right) \Biggr]\nonumber \\
+\frac{C_F C_A^2}{64} \Biggl[
&-\frac{242}{81\e^4}
+\frac{1}{\e^3}\left(\frac{88\zeta_2}{27}-\frac{3254}{243}\right)
-\frac{1}{\e^2}\left(\frac{88\zeta_2^2}{45}+\frac{553\zeta_2}{81}-\frac{1672\zeta_3}{27}+\frac{9707}{243}\right)\nonumber \\
&+\frac{1}{\e}\left(\frac{802\zeta_2^2}{15}-\frac{88\zeta_2\zeta_3}{9}-\frac{15983\zeta_2}{243}+\frac{8542\zeta_3}{27}-\frac{136\zeta_5}{3}
-\frac{385325}{4374}\right)\nonumber \\
& -\left(
          \frac{6152\zeta_2^3}{189}
          - \frac{100597\zeta_2^2}{540}
          + \frac{980\zeta_2\zeta_3}{9}
          + \frac{1136\zeta_3^2}{9}
 \right. \nonumber \\
& \left.        \qquad
          + \frac{478157\zeta_2}{1458}
          - \frac{306992\zeta_3}{243}
          - \frac{3472\zeta_5}{9}
          + \frac{1870897}{26244}
\right) 
\Biggr]\nonumber \\
+\frac{C_F^2 N_F}{64} \Biggl[
&-\frac{2}{3\e^5}-\frac{10}{9\e^4}-\frac{1}{\e^3}\left(\frac{\zeta_2}{3}+\frac{104}{27}\right)-\frac{1}{\e^2}\left(\frac{53\zeta_2}{9}-\frac{146\zeta_3}{9}+\frac{865}{81}\right)\nonumber \\
&+\frac{1}{\e}\left(\frac{337\zeta_2^2}{36}-\frac{736\zeta_2}{27}+\frac{1882\zeta_3}{27}-\frac{15511}{486}\right)\nonumber \\
&+\left(\frac{15769\zeta_2^2}{540}-\frac{343\zeta_2\zeta_3}{9}-\frac{16885\zeta_2}{162}+\frac{27812\zeta_3}{81}+\frac{278\zeta_5}{45}-\frac{307879}{2916}\right) 
\Biggr]\nonumber \\
+\frac{C_F C_A N_F}{64} \Biggl[
&\frac{88}{81\e^4}-\frac{1}{\e^3}\left(\frac{16\zeta_2}{27}-\frac{1066}{243}\right)+\frac{1}{\e^2}\left(\frac{316\zeta_2}{81}-\frac{256\zeta_3}{27}+\frac{3410}{243}\right)\nonumber \\
&-\frac{1}{\e}\left(\frac{44\zeta_2^2}{5}-\frac{5033\zeta_2}{243}+\frac{5140\zeta_3}{81}-\frac{90305}{2187}\right)\nonumber \\
&-\left(\frac{3791\zeta_2^2}{135}-\frac{368\zeta_2\zeta_3}{9}-\frac{63571\zeta_2}{729}+\frac{23762\zeta_3}{81}+\frac{208\zeta_5}{3}-\frac{1451329}{13122}\right) 
\Biggr]\nonumber \\
+\frac{C_F N_F^2}{64} \Biggl[
&-\frac{8}{81\e^4}-\frac{80}{243\e^3}-\frac{1}{\e^2}\left(\frac{4\zeta_2}{9}+\frac{32}{27}\right)-\frac{1}{\e}\left(\frac{40\zeta_2}{27}-\frac{136\zeta_3}{81}+\frac{9616}{2187}\right)\nonumber \\
&-\left(\frac{83\zeta_2^2}{135}+\frac{16\zeta_2}{3}-\frac{1360\zeta_3}{243}+\frac{109528}{6561}\right) 
\Biggr] \;.
\end{align}
The ultraviolet renormalization of the $Hb\bar{b}$ form factor has been derived in Section~\ref{sec:hbbrenorm}. Applying Eq.~\eqref{reng} yields the expansion coefficients of the renormalized form factors. In the time-like kinematics, the real part reads
\begin{align}
\operatorname{Re} {\cal A}_3 = 
 \frac{C_F^3}{64} \Biggl[
&-\frac{4}{3\e^6}
-\frac{6}{\e^5}
+\frac{1}{\e^4}\left(38\zeta_2-13\right)
+\frac{1}{\e^3}\left(66\zeta_2+\frac{100\zeta_3}{3}-23\right)\nonumber \\
&-\frac{1}{\e^2}\left(\frac{1947\zeta_2^2}{10}-\frac{191\zeta_2}{2}-124\zeta_3+\frac{235}{4}\right)\nonumber \\
&+\frac{1}{\e}\left(\frac{861\zeta_2^2}{5}-\frac{2914\zeta_2\zeta_3}{3}+\frac{899\zeta_2}{4}+\frac{1117\zeta_3}{3}+\frac{644\zeta_5}{5}-\frac{550}{3}\right)\nonumber \\
&-\left(
            \frac{19301\zeta_2^3}{252}
          - \frac{4495\zeta_2^2}{8}
          + 2298\zeta_2\zeta_3
          + \frac{1826\zeta_3^2}{3}
 \right. \nonumber \\
&    \left.    \qquad  
          - \frac{3635\zeta_2}{6}
          - \frac{1877\zeta_3}{2}
          - \frac{3932\zeta_5}{5}
          + \frac{1060}{3}
\right)
\Biggr]\nonumber \\
+\frac{C_F^2 C_A}{64} \Biggl[
& - \frac{11}{\e^5}
-\frac{1}{\e^4}\left(
            2 \zeta_2
          + \frac{361}{18}
\right)
+ \frac{1}{\e^3}\left(
            \frac{181\zeta_2}{2}
          - 26 \zeta_3
          + \frac{79}{54}
\right)\nonumber \\
&+\frac{1}{\e^2}\left(
            \frac{187\zeta_2^2}{5}
          - \frac{2789\zeta_2}{18}
          - \frac{158\zeta_3}{9}
          + \frac{4699}{324}
\right)\nonumber \\
&-\frac{1}{\e}\left(
            \frac{8267\zeta_2^2}{72}
          - \frac{2321\zeta_2\zeta_3}{3}
          + \frac{28031\zeta_2}{108}
          + \frac{1135\zeta_3}{3}
          + 142\zeta_5
          - \frac{16823}{972}
\right)\nonumber \\
&+\left(
            \frac{239933\zeta_2^3}{1260}
          + \frac{78529\zeta_2^2}{270}
          + \frac{3917\zeta_2\zeta_3}{2}
          + \frac{1616\zeta_3^2}{3}
 \right. \nonumber \\
& \left.          \qquad
          - \frac{30463\zeta_2}{81}
          - \frac{7765\zeta_3}{6}
          - \frac{4514\zeta_5}{9}
          + \frac{31618}{729}
\right) \Biggr]\nonumber \\
+\frac{C_F C_A^2}{64} \Biggl[
&-\frac{1331}{81\e^4}
-\frac{1}{\e^3}\left(
            \frac{110\zeta_2}{27}
          - \frac{2866}{243}
\right)
-\frac{1}{\e^2}\left(
            \frac{88\zeta_2^2}{45}
          - \frac{1625\zeta_2}{81}
          + \frac{902\zeta_3}{27}
          - \frac{11669}{486}
\right)\nonumber \\
&-\frac{1}{\e}\left(
            \frac{166\zeta_2^2}{15}
          + \frac{88\zeta_2\zeta_3}{9}
          + \frac{7163\zeta_2}{243}
          - \frac{3526\zeta_3}{27}
          + \frac{136\zeta_5}{3}
          + \frac{139345}{8748}
\right)\nonumber \\
& +\left(
            \frac{19136\zeta_2^3}{945}
          - \frac{3137\zeta_2^2}{135}
          - \frac{1258\zeta_2\zeta_3}{3}
          - \frac{1136\zeta_3^2}{9}
\right. \nonumber \\
& \left.        \qquad  
          + \frac{380191\zeta_2}{1458}
          + \frac{107648\zeta_3}{243}
          + \frac{106\zeta_5}{9}
          + \frac{5964431}{26244}
\right) 
\Biggr]\nonumber \\
+\frac{C_F^2 N_F}{64} \Biggl[
& \frac{2}{\e^5}+\frac{35}{9\e^4}
-\frac{1}{\e^3}\left(
          17\zeta_2
          + \frac{23}{27}
\right)+\frac{1}{\e^2}\left(
            \frac{199\zeta_2}{9}
          - \frac{110\zeta_3}{9}
          - \frac{641}{162}
\right)\nonumber \\
&+\frac{1}{\e}\left(
            \frac{577\zeta_2^2}{36}
          + \frac{3235\zeta_2}{54}
          + \frac{442\zeta_3}{27}
          - \frac{967}{486}
\right)\nonumber \\
&-\left(
            \frac{8822\zeta_2^2}{135}
          + 85\zeta_2\zeta_3
          - \frac{22571\zeta_2}{162}
          - \frac{15131\zeta_3}{81}
          + \frac{386\zeta_5}{9}
          + \frac{145375}{2916}
\right) 
\Biggr]\nonumber \\
+\frac{C_F C_A N_F}{64} \Biggl[
& \frac{484}{81\e^4}+\frac{1}{\e^3}\left(
            \frac{20\zeta_2}{27}
          - \frac{752}{243}
\right)-\frac{1}{\e^2}\left(
            \frac{476\zeta_2}{81}
          - \frac{212\zeta_3}{27}
          + \frac{2068}{243}
\right)\nonumber \\
&+\frac{1}{\e}\left(
            \frac{44\zeta_2^2}{15}
          + \frac{2594\zeta_2}{243}
          - \frac{964\zeta_3}{81}
          - \frac{8659}{2187}
\right)\nonumber \\
&-\left(
            \frac{836\zeta_2^2}{135}
          - \frac{148\zeta_2\zeta_3}{3}
          + \frac{59999\zeta_2}{729}
          + \frac{2860\zeta_3}{27}
          + \frac{4\zeta_5}{3}
          + \frac{521975}{13122}
\right) 
\Biggr]\nonumber \\
+\frac{C_F N_F^2}{64} \Biggl[
&-\frac{44}{81\e^4}-\frac{8}{243\e^3}+\frac{1}{\e^2}\left(
            \frac{4}{9}\zeta_2
          + \frac{46}{81}
\right)-\frac{1}{\e}\left(
            \frac{20}{27}\zeta_2
          + \frac{8}{81}\zeta_3
          - \frac{2417}{2187}
\right)\nonumber \\
&+\left(
            \frac{172}{135}\zeta_2^2
          + \frac{388}{81}\zeta_2
          - \frac{200}{243}\zeta_3
          + \frac{2072}{6561}
\right) 
\Biggr]\;. \label{f3qr}
\end{align}
For the sake of completeness, we also provide the imaginary part of the ultraviolet renormalized three-loop form factor:
\begin{align}
\frac{\operatorname{Im} {\cal A}_3}{\pi} = 
 \frac{C_F^3}{64} \Biggl[
&-\frac{4}{\e^5}
-\frac{12}{\e^4}
+\frac{1}{\e^3}\left(42\zeta_2-21\right)+\frac{1}{\e^2}\left(24\zeta_2+100\zeta_3-\frac{93}{2}\right)\nonumber \\
&-\frac{1}{\e}\left(\frac{873\zeta_2^2}{10}-\frac{15\zeta_2}{2}-308\zeta_3+141\right)\nonumber \\
&+\left(
            372\zeta_2^2
          - 1114\zeta_2\zeta_3
          - \frac{177\zeta_2}{4}
          + 985\zeta_3
          + \frac{1932\zeta_5}{5}
          - \frac{773}{2}
\right)
\Biggr]\nonumber \\
+\frac{C_F^2 C_A}{64} \Biggl[
& -\frac{55}{3\e^4}
- \frac{1}{\e^3}\left(
            6\zeta_2
          - \frac{1}{3}
\right)
+\frac{1}{\e^2}\left(
            \frac{283\zeta_2}{6}
          - 78\zeta_3
          + \frac{715}{18}
\right)\nonumber \\
&+\frac{1}{\e}\left(
            \frac{21\zeta_2^2}{5}
          - \frac{502\zeta_2}{3}
          - \frac{1531\zeta_3}{9}
          + \frac{1768}{27}
\right)\nonumber \\
&-\left(
            \frac{5669\zeta_2^2}{40}
          - 917\zeta_2\zeta_3
          - \frac{253\zeta_2}{36}
          + \frac{4222\zeta_3}{3}
          + 426\zeta_5
          - \frac{35539}{162}
\right) \Biggr]\nonumber \\
+\frac{C_F C_A^2}{64} \Biggl[
&-\frac{242}{27\e^3}
-\frac{1}{\e^2}\left(
            \frac{44\zeta_2}{9}
          - \frac{2086}{81}
\right)
-\frac{1}{\e}\left(
            \frac{88\zeta_2^2}{15}
          - \frac{536\zeta_2}{27}
          + \frac{44\zeta_3}{9}
          + \frac{245}{9}
\right)\nonumber \\
& +\left(
            2\zeta_2^2
          - \frac{88\zeta_2\zeta_3}{3} 
          + \frac{1036\zeta_2}{81}
          + \frac{13900\zeta_3}{27}
          - 136\zeta_5
          - \frac{10289}{1458}
\right) 
\Biggr]\nonumber \\
+\frac{C_F^2 N_F}{64} \Biggl[
& \frac{10}{3\e^4}+\frac{2}{3\e^3}
-\frac{1}{\e^2}\left(
            \frac{29\zeta_2}{3}
          + \frac{71}{9}
\right)
+\frac{1}{\e}\left(
            \frac{76\zeta_2}{3}
          - \frac{74\zeta_3}{9}
          - \frac{403}{27}
\right)\nonumber \\
&+\left(
            \frac{3\zeta_2^2}{4}
          + \frac{487\zeta_2}{18}
          + \frac{1192\zeta_3}{9}
          - \frac{9649}{162}
\right) 
\Biggr]\nonumber \\
+\frac{C_F C_A N_F}{64} \Biggl[
& \frac{88}{27\e^3}+\frac{1}{\e^2}\left(
            \frac{8\zeta_2}{9}
          - \frac{668}{81}
\right)-\frac{1}{\e}\left(
            \frac{80\zeta_2}{27}
          - \frac{56\zeta_3}{9}
          - \frac{418}{81}
\right)\nonumber \\
&+\left(
            \frac{12\zeta_2^2}{5}
          - \frac{196\zeta_2}{81}
          - \frac{724\zeta_3}{9}
          + \frac{7499}{729}
\right) 
\Biggr]\nonumber \\
+\frac{C_F N_F^2}{64} \Biggl[
&-\frac{8}{27\e^3}+\frac{40}{81\e^2}+\frac{8}{81\e}
-\left(
            \frac{16}{27}\zeta_3
          + \frac{928}{729}
\right) 
\Biggr] \, .
\end{align}
It should be stressed that for every color structure of Eq.~\eqref{fq3lbare}, the coefficients of the leading poles in $\e$ agree with the ones of the $\gamma^*q\bar q$ form factor of Ref.~\cite{Gehrmann:2010a}, as expected.

\section{Infrared Pole Structure}
\label{sec:ir}
A more powerful check consists in analyzing the complete infrared pole structure of our three-loop results. \\
As outlined in Refs.~\cite{Catani:1998,Sterman:2002,Moch:2005,Becher:2009b,Gardi:2009}, it can be predicted from infrared factorization properties of QCD. Accordingly, the infrared pole structure of the renormalized $Hb\bar{b}$ form factors $\mathcal{A}_1$, $\mathcal{A}_2$ and $\mathcal{A}_3$ can be derived from the same formulae as for the $\gamma^*q\bar q$ form factor in Ref.~\cite{Gehrmann:2010a}. They read
\begin{eqnarray}
\label{polef1q}
{\cal P}oles{({\cal A}_1)} &=& 
-\frac{C_F \gamma^\mathrm{cusp}_0}{2 \e^2}+\frac{\gamma^q_0}{\e}\,,\\
\label{polef2q}
{\cal P}oles{({\cal A}_2)} &=& 
\frac{3 C_F \gamma^\mathrm{cusp}_0 \beta_0}{8 \e^3}+\frac{1}{\e^{2}}\biggl(-\frac{\beta_0 \gamma^q_0}{2}-\frac{C_F 
\gamma^\mathrm{cusp}_1}{8}\biggr)
+\frac{\gamma^q_1}{2 \e} +\frac{\left({\cal A}_1\right)^2}{2}\,,\\
\label{polef3q}
{\cal P}oles{({\cal A}_3)} &=& 
-\frac{11 \beta_0^2 C_F \gamma^\mathrm{cusp}_0}{36 \e^4}+\frac{1}{\e^{3}}
\biggl( \frac{5 \beta_0 C_F \gamma^\mathrm{cusp}_1}{36}+\frac{\beta_0^2 \gamma^q_0}{3}+\frac{2 C_F \gamma^\mathrm{cusp}_0 \beta_1}{9}\biggr)\nonumber \\ & &
+\frac{1}{\e^{2}}\biggl(-\frac{\beta_0 \gamma^q_1}{3}-\frac{C_F \gamma^\mathrm{cusp}_2}{18}-\frac{\beta_1 \gamma^q_0}{3}\biggr)
+\frac{\gamma^q_2}{3 \e}
-\frac{\left({\cal A}_1\right)^3}{3}+{\cal A}_2 {\cal A}_1\,.
\end{eqnarray}
These equations require knowing the coefficients $\gamma^\mathrm{cusp}_i$ of the cusp soft anomalous dimension up to three loops, which were presented in Eq.~\eqref{gammacusp}. Moreover, $\gamma^{q}_i$ denotes the coefficients of the quark collinear anomalous dimension. To three-loop order, they are given in Eq.~\eqref{gammaq}.\\
The deepest infrared pole for the $i$-loop form factor ${\cal A}_i$ is proportional to $\e^{-2i}$. Due to the last term of Eq.~\eqref{polef3q}, we thus need to include the renormalized form factors ${\cal A}_1$ through to ${\cal O}(\e^3)$ and ${\cal A}_2$ through to ${\cal O}(\e)$, both stated in Section \ref{sec:hbbrenorm} above. In doing so, we succeed in reproducing the infrared poles of the renormalized form factor up to three loops. 

\section{Conclusions}
\label{sec:conc1}
In this chapter, we have derived the three-loop QCD corrections to the form factor describing the Yukawa coupling of a Higgs boson to a pair of bottom quarks. We have neglected the bottom quark mass in internal propagators and external states, which is justified by the large mass hierarchy between the Higgs boson  and the bottom quark. The pole structure of our result is in agreement with the prediction of 
infrared factorization formulae~\cite{Catani:1998,Sterman:2002,Moch:2005,Becher:2009b,Gardi:2009}.\\
Our results can be applied to derive the third-order QCD corrections to Higgs boson production from bottom quark fusion~\cite{Ahmed:2014c} and the fully differential description of Higgs boson decays into bottom quarks. Besides the three-loop corrections derived here, these processes also require two-loop corrections to the matrix element for $Hb\bar bg$, derived in Ref.~\cite{Ahmed:2014a}, and higher multiplicity tree-level and one-loop matrix elements that can by now be derived using standard methods. The integration of all subprocess contributions over the relevant phase spaces is far from trivial, and methods are currently under intensive development~\cite{Anastasiou:2013a,Duhr:2013,Li:2013,Anastasiou:2013b,Li:2014,Hoschele:2014}.\\
A more imminent application was the N$^3$LO soft-virtual threshold approximation to Higgs boson production in bottom quark fusion~\cite{Ahmed:2014c}, using the derived result for Higgs boson production in gluon fusion to this order~\cite{Anastasiou:2014}, combined with universal factorization properties~\cite{Ahmed:2014b,Catani:2014}.

\chapter{The Workflow of Multi-Loop Calculations, Part II:\\Exact Calculation of Master Integrals from Differential Equations}
\chaptermark{Multi-Loop Calculations, Part II: Exact Calculation of Master Integrals}
\label{chap:workflow2}

In Chapter~\ref{chap:hbb} we have seen that a physical quantity is obtained by applying the mechanical procedures of Chapter~\ref{chap:workflow1} step-by-step, where we ultimately inserted the results for the MIs available in the literature. In complex cases, however, these MIs may be still unknown\footnote{The question whether the result for a MI already exists in the literature is more difficult to answer than expected, so that there have recently been efforts to create a database\cite{Bogner:2017a}.} and their evaluation is in this case often the bottleneck for multi-scale multi-loop applications, especially when massive propagators are involved, as in processes $(b)$ and $(c)$. In this chapter, we would like to elaborate on the exact analytical computation of MIs, which we aim for with the help of the method of \textit{differential equations}. After giving a short overview of the spectrum of available approaches in Section~\ref{sec:overview}, the differential equation method is presented in Section~\ref{sec:deq1}. In the same section, we clarify what is meant by a differential equation in \textit{canonical form}, and why this form serves to considerably simplify the calculation of MIs. Section~\ref{sec:mpl} is designed to introduce the class of functions known as \textit{Multiple Polylogarithms}, which naturally arise from the canonical form. Finally, we can proceed with the integration of differential equations in Section~\ref{sec:deq2} in terms of these functions, which have to be supplied with appropriate boundary conditions.

\section{Overview of Various Approaches}
\label{sec:overview}

Analytical and numerical approaches for the integration of Feynman integrals have competed for a long time, but should be rather understood as complementary. Given the complexity for these integrals, it seems like an obvious solution to perform their highly non-trivial integration numerically. This is pursued by the method of sector decomposition~\cite{Hepp:1966,Roth:1996,Binoth:2000,Heinrich:2008}, for example, whereof several public implementations exist~\cite{Borowka:2015,Borowka:2017,Smirnov:2015} and which has led to a few impressive results (see e.g. Refs.~\cite{Borowka:2016a,Borowka:2016b,Czakon:2013}). However, these calculations have also shown that it requires great effort to produce sufficiently precise results in a reasonable amount of time over the whole phase space, in particular when it comes to the computation of differential quantities. This is partially due to the fact that loop integrals are in general ill-defined, divergent objects in the limit $D\to 4$, and in terms of numerical stability it is far from trivial to deal with the regularization and integration of these singularities in a fully automatic way. First steps in a promising direction were made in Ref.~\cite{vonManteuffel:2014b}, where IBP identities in arbitrary dimensions are used in order to rotate to a basis of finite Feynman integrals. The latter are defined in shifted dimensions with higher powers of the propagators, thus making both infrared and ultraviolet divergences explicit, which has been explored in Refs.~\cite{Borowka:2016a,Borowka:2016b,vonManteuffel:2015b,vonManteuffel:2017a}. Nevertheless, the issue of numerical stability as well as a number of other open questions remain, thus we retain the standard procedure of analytical integration in this thesis.\\
The analytical approach has led to a countless number of results, which cannot be all quoted at this point. It is reliable in terms of speed and precision, provided that it is possible to perform the actual integration. On top of that, a result in analytical form may reveal significant properties of the considered amplitude and lead to a deeper insight into the theoretical surroundings, once the characteristics of the underlying  class of functions is investigated.\\
Prior to the development of the differential equation approach, there have been attempts to derive analytical results for integrals based on the use of dispersion relations, which are derived by exploring the unitarity properties of the $S$ matrix~\cite{Remiddi:1981,Laporta:1995}. Having a dispersion relation at one's disposal greatly simplifies the calculation of a Feynman integral, in the sense that only its imaginary part remains to be computed, which can in turn be achieved through the application of the \textit{Cutkosky rules}~\cite{Cutkosky:1960,Veltman:1963}. Although restricting the calculation to the imaginary part is without any doubt advantageous, its computation with the help of the Cutkosky rules or equivalently the \textit{optical theorem} requires the evaluation of more complicated topologies than the original one and thus can still be a challenging task. Nevertheless, this approach led to the groundbreaking result of Ref.~\cite{Laporta:1996}.\\
A completely independent way of solving Feynman integrals analytically consists in the use of the \textit{Feynman parametrization}, the application of \textit{Mellin-Barnes transformations} to the propagators~\cite{Smirnov:1999,Tausk:1999} or the method of \textit{negative dimensions}~\cite{Anastasiou:1999}. All these techniques rely on the explicit integration over the loop momenta and have proven to be useful for processes with a restricted number of legs or massless internal propagators, thus limiting the number of scales. Recent developments in the same direction, where the representation in terms of Feynman parameters is combined with the criterion of \textit{linear reducibility}~\cite{Brown:2008}, look more promising (see for example Ref.~\cite{Chavez:2012}) and have been implemented in a public computer code~\cite{Panzer:2014}. A systematic generalization to arbitrary number of scales, in particular to four-point functions and to diagrams with massive internal propagators, was however still missing.\\
This gap was closed by the method of differential equations~\cite{Kotikov:1990,Remiddi:1997,Caffo:1998a,Caffo:1998b}, which avoids the explicit integration over the loop momenta by differentiating with respect to kinematic invariants. On top of that, in contrast to other techniques it is completely straightforward to derive the Laurent expansion in the dimensional regulator $\e$, since the expansion can already be performed at the level of the differential equation. It should be noted, however, that differential equations generally have to be supplied with appropriate boundary conditions. The procedure was finally made state-of-the-art by Refs.~\cite{Gehrmann:1999,Gehrmann:2000,Gehrmann:2001a} through systematic application to a non-trivial class of two-loop four-point functions. The method of \textit{difference equations}~\cite{Laporta:2001} can be viewed as the discrete equivalent of the differential equation approach and is obtained by deriving the same set of identities with respect to one of the integer denominator powers. By taking the dimensional parameter $D$ as integer variable for difference equations, one arrives at the well-known Tarasov-Lee shifts, that relate integrals in an integer number of space-time dimensions~\cite{Tarasov:1996,Lee:2009}.

\section[Differential Equations for Master Integrals, Part I:\\The Path to Canonical Form]{Differential Equations for Master Integrals, Part I:\\The Path to Canonical Form%
\sectionmark{Differential Equations for Master Integrals: The Path to Canonical Form}}
\sectionmark{Differential Equations for Master Integrals: The Path to Canonical Form}
\label{sec:deq1}
\subsection{Derivation of Differential Equations}

With the knowledge that IBP relations emerge from the differentiation with respect to loop momenta as described in Eq.~\eqref{IBP1}, it seems natural to repeat this procedure by differentiating with respect to external momenta, or more generally with respect to any kinematic invariant of a given process.\\
Let us consider a process with $n$ external legs of momenta $q_1,\dots,q_n$, $n-1$ of which are independent due to momentum conservation. Let us further choose them to be $q_1,\dots,q_{n-1}$ so that we can produce $n \, (n-1)/2$ distinct external invariants from this independent set, which we denote the \textit{Mandelstam variables}~$s_{ij}$:
\begin{equation}
s_{ij} \equiv q_i \cdot q_j \,.
\end{equation}
Next, let us express the derivative with respect to any external momentum $q_i$ as a linear combination of the derivatives with respect to the invariants $s_{ij}$ with the help of the chain rule,
\begin{equation}
\frac{\p}{\p q_m^\mu} = \sum_{k=1}^{n(n-1)/2} \frac{\p s_k}{\p q_m^\mu} \frac{\p}{\p s_k} \,,
\end{equation}
where we labeled $s_k \equiv (s_{ij} )_k$ for the sake of readability. In order to produce scalar quantities, this differential operator can in principle be contracted with any of the $n-1$ independent external momenta, producing $(n-1)^2$ relations:
\begin{equation}
q_i^\mu \frac{\p}{\p q_m^\mu} = q_i^\mu \sum_{k=1}^{n(n-1)/2} \frac{\p s_k}{\p q_m^\mu} \frac{\p}{\p s_k} \,.
\label{chainrule1}
\end{equation}
Assuming a process with at most five external legs and recalling Eq.~\eqref{NLI}, it can be shown that only
\begin{equation}
(n-1)^2 - N_\mathrm{LI} = (n-1)^2 - \frac{(n-1)(n-2)}{2} = \frac{n(n-1)}{2}
\end{equation}
of them are independent by using the Lorentz Invariance identities from Section~\ref{sec:LI}, which corresponds exactly to the number of kinematic invariants. As a consequence, the set of relations~\eqref{chainrule1} turns into a square system, which can be inverted, leading to
\begin{equation}
\frac{\p}{\p s_k} = \sum_m a^{(k)}_m \, q_i^\mu \, \frac{\p}{\p q_m^\mu} \,,
\label{chainrule2}
\end{equation}
where the coefficients $a^{(k)}_m=a^{(k)}_m(s_1,\dots,s_{n(n-1)/2})$ of the inverted system may depend themselves on the external invariants in form of rational functions.\\
So far, we have completely neglected the discussion of massive internal propagators. For every internal mass $m_i$ within $m_1,\dots,m_{N_m}$ running in the loops, we obtain another differential operator of the kind
\begin{equation}
m_i^2 \, \frac{\p}{\p m_i^2}
\label{massdeq}
\end{equation}
on top of the ones in Eq.~\eqref{chainrule2}. For the sake of clarity, let us combine the set of $N_m$ internal masses squared $m_i^2$ and the set of $n(n-1)/2$ external invariants $s_k$ to a set of
\begin{equation}
N_x \equiv N_m+n(n-1)/2
\end{equation}
kinematic invariants $\vec{x}=(x_1,\dots,x_{N_x})$ with derivatives
\begin{align}
\frac{\p}{\p x_1} &= \frac{\p}{\p m_1^2} \,, \dots \,, \frac{\p}{\p x_{N_m}} = \frac{\p}{\p m_{N_m}} \,, \nonumber \\
\frac{\p}{\p x_{N_m+1}} &= \frac{\p}{\p s_1} \,, \dots \,, \frac{\p}{\p x_{N_x}} = \frac{\p}{\p s_{n(n-1)/2}} \,.
\label{derset}
\end{align}
In fact, this system of differential operators is still not linearly independent, but contains one redundant operator due to the properties of the integral under rescaling of all kinematic invariants:
\begin{equation}
I_{t,r,s}(D;\vec{x}) = \lambda^{-\alpha(D,r,s)} \, I_{t,r,s}(\lambda^2 \, \vec{x},D) \,.
\end{equation}
Therein, the mass dimension
\begin{equation}
\alpha(D,r,s) = l \cdot D + 2\left(s-r\right)
\label{massdim}
\end{equation}
of the integral depends on the space-time dimension $D$, the number of loops $l$ and the values of $r$ and $s$ defined in Eq.~\eqref{rs}. The \textit{rescaling equation} or equivalently the \textit{scaling relation} can then be formulated as
\begin{equation}
\left(\sum_{i=1}^{N_x} x_i \, \frac{\p}{\p x_i} - \frac{\alpha}{2} \right) I_{t,r,s}(D;\vec{x}) = 0 \,,
\label{scaling}
\end{equation}
serving as a check for the validity of the derived differential equations. Hence, the system of $N_x$ operators can be reduced by one, or, in other words, it is useful to form $N_x-1$ independent ratios out of the $N_x$ kinematic invariants.\\
We point out that the current version of \textsc{Reduze} comes with the useful feature of producing equations of the type~\eqref{derset} and \eqref{scaling} in a fully automatic way, given that the integral families are provided. However, the further processing of these relations has to be performed manually. Within the \textsc{Reduze} code, the fact that the independent number of scales is less by one compared to the overall number of scales is reflected by the possibility of setting one of the scales to unity, which is ultimately reconstructed from dimensional considerations. A deliberate choice of this scale can considerably minimize the complexity of the reduction relations.\\
As an example for the relations~\eqref{derset}, let us consider a process with $n=4$ external legs of momenta $q_1,\dots,q_4$ in the Euclidean region. Let us further suppose that all external particles are massless except for one, i.e. $q_1^2=q_2^2=q_3^2=0$ and $q_4^2=M^2$, and that there is one massive propagator of mass~$m$. We start by choosing the Mandelstam invariants
\begin{equation}
s\equiv (q_1+q_2)^2 \,, \quad t\equiv (q_1+q_3)^2 \,, \quad u\equiv (q_2+q_3)^2
\label{mandelstam}
\end{equation}
as independent variables, so that the external mass $M^2$ within the Mandelstam relation
\begin{equation}
M^2 = s+t+u
\label{convervation}
\end{equation}
depends on these Mandelstam invariants. As a next step, we express their derivatives in terms of $q_1,\dots,q_3$, yielding the equivalent of the identities~\eqref{derset} in this special case:
\begin{align}
s \, \frac{\p}{\p s} &= \frac{1}{2} \left(q_1^\mu \, \frac{\p}{\p q_1^\mu} + q_2^\mu \, \frac{\p}{\p q_2^\mu} - q_3^\mu \, \frac{\p}{\p q_3^\mu} \right) \,, \nonumber \\
t \, \frac{\p}{\p t} &= \frac{1}{2} \left(q_1^\mu \, \frac{\p}{\p q_1^\mu} - q_2^\mu \, \frac{\p}{\p q_2^\mu} + q_3^\mu \, \frac{\p}{\p q_3^\mu} \right) \,, \nonumber \\
u \, \frac{\p}{\p u} &= \frac{1}{2} \left(-q_1^\mu \, \frac{\p}{\p q_1^\mu} + q_2^\mu \, \frac{\p}{\p q_2^\mu} + q_3^\mu \, \frac{\p}{\p q_3^\mu} \right) \,, \nonumber \\
m^2 \, \frac{\p}{\p m^2} & \,.
\label{deqmandelstam}
\end{align}
Constructing three ratios out of this according to
\begin{equation}
x = \frac{s}{m^2} \,, \quad z = \frac{u}{m^2} \,, \quad h = \frac{s+t+u}{m^2} \,, \quad m^2=\tilde{m}^2
\label{ratios}
\end{equation}
corresponds to a change of variables from the set $s,t,u,m^2$ to the set $x,z,h,\tilde{m}^2$. The derivatives with respect to the new set can be computed in terms of the derivatives with respect to the old set by applying the chain rule in the same way as for the transition $\p/\p p_i\to \p/\p s_k$:
\begin{align}
\frac{\p}{\p x} &= m^2 \, \left(\frac{\p}{\p s} - \frac{\p}{\p t}\right) \,, \nonumber \\
\frac{\p}{\p z} &= m^2 \, \left(\frac{\p}{\p u} - \frac{\p}{\p t}\right) \,, \nonumber \\
\frac{\p}{\p h} &= m^2 \, \frac{\p}{\p t} \,, \nonumber \\
m^2 \, \frac{\p}{\p m^2} &= \frac{\alpha}{2} \,.
\label{deqratios}
\end{align}
This is exactly the set of derivatives that will be used in the following, particularly in the context of Higgs-plus-jet production in Chapter~\ref{chap:hj}. Note that, although we could have kept the Mandelstam invariant $t$, we replaced it by the mass $M$ of the massive external leg with the help of momentum conservation as stated in Eq.~\eqref{convervation} for reasons which will be outlined at a later point. Evidently, the differential equation with respect to the mass $m$ turns into the rescaling relation, confirming the previously made statement that one of the equations becomes redundant under this kind of variable transformation.\\
We emphasize that all statements made so far are independent of the loop order. Let us now consider a sector at a certain loop order with at least one MI (without counting the MIs belonging to its subtopology tree), which is given in terms of the integral representation~\eqref{scalarint3}. After making a choice for the MIs, the differential operators in Eq.~\eqref{derset} can be directly applied to this integral representation and leave us with a linear combination of subsectors plus integrals of the sector under consideration, which do not necessarily correspond to our choice of MIs. This can be corrected, however, by applying IBP identities as described in Section~\ref{sec:ibp} in order to rotate to the basis of our MIs. Let us point out that the IBP relations needed in this context do not necessarily cover the same range of the parameters $r$ and $s$ compared to the identities required for reducing the amplitude as described in Section~\ref{sec:reduction}. Commonly, the maximum values of $r$ are higher and those of $s$ are lower than in case of the reduction for the amplitude, which is due to the fact that the differentiation leads to higher powers of the denominators of the integral representation, whereas the Lorentz structure of the amplitude introduces more involved dependencies on the momenta in the numerator. After rotating to the chosen basis of MIs with the help of reduction identities, the resulting system of first-order differential equations reads
\begin{equation}
\frac{\p}{\p x_i} I_j(D;\vec{x}) = C^{(i)}_{jk}(D;\vec{x}) \, I_k(D;\vec{x}) + G^{(i)}_{jl} \, m_l(D;\vec{x}) \,,
\label{genericdeq}
\end{equation}
where $C^{(i)}_{jk}$ and $G^{(i)}_{jl}$ are rational coefficient functions of the integrals $I_k$ belonging to the considered sector and of the subsector integrals $m_l$, respectively.

\subsection{The Case of One Master Integral}
\label{sec:oneMI}

Before discussing the general case, we consider a sector with only one MI, i.e. $j=1$ within Eq.~\eqref{genericdeq}, which turns into 
\begin{equation}
\frac{\p}{\p x_i} I(D;\vec{x}) = C^{(i)}(D;\vec{x}) \, I(D;\vec{x}) + G^{(i)}_l(D;\vec{x}) \, m_l(D;\vec{x}) \,.
\end{equation}
The solution is then straightforward:
\begin{enumerate}
\item Choose one of the $N_x-1$ independent derivatives of the kinematic invariants, for example $\p/\p x_1$.
\item Compute the \textit{integrating factor} or \textit{homogeneous solution} $h_1(D;\vec{x})$ of the MI $I(D;\vec{x})$ by integrating the corresponding coefficient over $\d x_1$:
\begin{equation}
h_1(D;\vec{x}) = \exp\left(\int C^{(1)}(D;\vec{x}) \, \d x_1 \right) \,.
\label{homsol}
\end{equation}
\item The full solution can then be phrased through the so-called \textit{variation of constants} by including the subsector coefficients $G^{(1)}_l(D;\vec{x})$ and the known solutions $m_l(D;\vec{x})$ of the subsectors:
\begin{equation}
I(D;\vec{x}) = h_1(D;\vec{x}) \, \int \frac{G^{(1)}_l(D;\vec{x}) \, m_l(D;\vec{x})}{h_1(D;\vec{x})} \d x_1 + c_1(D;x_2,\dots,x_{N_x-1}) \,.
\end{equation}
\item Now the solution is known up to a constant $c_1$, which depends on all independent kinematic invariants except for $x_1$. Plug this solution into the differential equation with respect to another kinematic invariant, say $x_2$. This leads to a first-order differential equation for the constant $c_1$ with respect to $x_2$, which must be satisfied. Solving it in the exact same manner as the differential equation with respect to $x_1$ leaves us with a solution up to a constant $c_2$, which depends on all kinematic invariants except for $x_1$ and $x_2$.
\item By repeating this procedure, one obtains the full solution up to a constant $c_{N_x-1}$, which only depends on the dimensional parameter $D$ and has to be determined by appropriate boundary conditions. We will come back to the issue of finding boundary conditions in Section~\ref{sec:boundary}.
\end{enumerate}
This procedure illustrates that the system must be solved bottom-up, i.e. starting from the topologies with the lowest number $t$ of different propagators and increasing $t$ gradually. Remarkably, we have turned the problem of directly integrating over $l$ loop momenta in $D$~dimensions into a single one-dimensional integration per kinematic invariant plus the determination of an appropriate boundary condition. The general case of two or more MIs is more complicated and will be approached in the next section.

\subsection{The Case of Two or More Master Integrals:\\
Coupled Differential Equations and Triangular Form}
\label{sec:moreMIs}

A great feature of this method is that it allows performing operations at the level of the differential equations before the actual integration, so that the integration can be substantially simplified or carried out at all. This is particularly useful in the case of more than one MI per sector, where in general all MIs of a given sector appear on the right-hand side of the differential equations in Eq.~\eqref{genericdeq}, which is called \textit{coupled} in the following. Let us discuss one such feature before answering the question of how to \textit{decouple} such a system in order to enable the actual integration: In the case of one MI in the last section, we have performed all operations on the exact $D=4-2\e$-dimensional integral $I$. For practical purposes though, this is not required and one is only interested in the coefficients of the Laurent expansion in $\e$:
\begin{equation}
I(D;\vec{x}) = \sum_{k=a}^\infty \e^k \, I^{(k)}(\vec{x}) \,.
\label{laurent}
\end{equation}
The negative integer~$a$ indicates the power of the deepest pole and its lowest possible value can be obtained from simple power counting, leading to $a\geq -4$ for the at most two-loop four-point functions considered in the remainder of this thesis. The deepest pole for a given MI can for instance be determined from a sector decomposition approach. Since the system of differential equations is solved bottom-up, however, the deepest pole of the integral under consideration emerges automatically from the knowledge of the lowest-order integral in the method of differential equations, where `lowest order' refers to the number $t$ of distinct propagators. At the other end, the sum in Eq.~\eqref{laurent} runs in principle to $\infty$, but in practice only the first few orders are of interest. Substituting this ansatz into Eq.~\eqref{genericdeq} leads to a set of coupled first-order differential equations for the coefficients $I^{(k)}(\vec{x})$ of the Laurent series. This new set of differential equations is simpler in the sense that the dependence on the dimensional parameter $\e$ is completely factorized from the solution, so that the solution can be determined order by order in $\e$. Furthermore, one arrives in a direct way at the physically more relevant result in terms of a Laurent series, without taking a detour via the exact $D$-dimensional solution, whose subsequent Laurent expansion might not even be feasible in complicated cases.\\
Let us start from a system of $N$ coupled differential equations for the MIs $\vec{I}=(I_1,\dots,I_N)$ of a given sector. The equivalent of Eq.~\eqref{genericdeq} in matrix form can then be written as
\begin{equation}
\frac{\p}{\p x_i} \vec{I}(D;\vec{x}) = C^{(i)}(D;\vec{x}) \, \vec{I}(D;\vec{x}) + G^{(i)}(D;\vec{x}) \, \vec{m}(D;\vec{x}) \,.
\label{matrixdeq}
\end{equation}
Assuming the solutions of the subsectors or the \textit{inhomogeneous} part of the equations to be known, the task of $\textit{decoupling}$ the differential equations of a sector comes down to the question of whether it is possible to cast the matrix $C^{(i)}(D;\vec{x})$ of the \textit{homogeneous} part into triangular form. Moreover, taking into account that we are only interested in the Laurent series around the point $\e=0$, the requirement can be weakened to whether the matrix $C^{(i)}(4;\vec{x})$ can be triangularized. Given that a triangular form of $C^{(i)}$ is reached in $D=4$ dimensions, the Laurent coefficients of the MIs can be determined one after another at every order in $\e$.\\
Although there is no general principle or algorithm explaining how to get to a triangular form, it could be achieved for a lot of processes in the literature and specifying all of them would go beyond the scope of this thesis. However we would like to point out that, although it is a challenging task, this is in agreement with our observation that a triangular form can always be accomplished in practice, provided that it exists, by following a set of guidelines. In the following, we will elaborate on these guidelines.

\subsection{Basis Choice}
\label{sec:basischoice}

There are two ways to enter the path of finding a triangular basis of MIs: First, one can simply vary the choice of MIs within the Laporta algorithm or the IBP identities, derive their differential equations and hope to find a triangular form. Second, one may define a set of MIs, which does not triangularize the matrix of the homogeneous part of the differential equation, and look for linear combinations thereof leading to a triangular form. Due to the structure of the IBP relations, these linear combinations can be as well understood as a new choice of MIs. Starting from a linearly independent set of MIs, which can be easily verified by checking the validity of their scaling relations, we present a set of rules for the first option:
\begin{itemize}
\item[\textbf{a)}] \textbf{Coefficient Size}\\
A reasonable first indicator is given by the mere size of the coefficients in the differential equations. Inappropriate basis choices can actually be recognized already at the level of the IBP identities: Naturally, there is a correlation between the size of the coefficients of the IBP identities and the size of the coefficients of the differential equations with respect to a certain choice of MIs. More precisely, we have observed that MIs, which are part of IBP relations whose coefficients are considerably larger than others, are much more likely to lead to unsuitable basis integrals.

\item[\textbf{b)}] \textbf{Factorizable Denominators}\\
Our aim is to obtain a form in which the dimensional regulator $\e$ is factorized from the remaining part of the coefficients to the largest extent possible. Since the numerators can be partial fractioned with respect to the denominators, this factorization should be present at the level of the denominators of the differential equations and concerns particularly the space-time dimension $D$. To sum up, differential equations, whose denominators contain non-factorizable terms of both $D$ and the Mandelstam invariants, should be avoided.

\item[\textbf{c)}] \textbf{Mass Dimension}\\
Starting from the Feynman parameter representation\footnote{For the definitions of the $U$ and $W$ polynomials occuring in the Feynman parameter representation, we refer the reader to the standard literature, see e.g. Ref.~\cite{Smirnov:2012}.}
\begin{equation}
I(D;\vec{x}) \propto \int \prod_{i,j} \d \alpha_i \, \alpha_j^{\kappa_j-1} \, \delta\left(\sum_k \alpha_k-1\right) \, \frac{U(\vec{\alpha})^{e_U}}{W(\vec{x},\vec{\alpha})^{e_W}}
\label{feynmanpar1}
\end{equation}
with the vector~$\vec{\alpha}$ of Feynman parameters $\alpha_i$, it can be argued that integrals of the form
\begin{equation}
I(D;\vec{x}) \propto \int \prod_i \frac{\d \alpha_i}{g(\vec{x},\vec{\alpha})^k} \,, \qquad k\in\mathbb{N} \,,
\label{feynmanpar2}
\end{equation}
where $g(\vec{x},\vec{\alpha})$ is an irreducible polynomial, are more likely to lead to a suitable basis than integrals of different form~\cite{Hoschele:2014,Henn:2013a}. Comparing Eq.~\eqref{feynmanpar1} to Eq.~\eqref{feynmanpar2} puts constraints on the powers of the $U$ and $W$ polynomials within the Feynman parametrization, which are defined as
\begin{align}
e_W &= \kappa - \frac{l\,D}{2} = -\frac{\alpha(D,r,s)}{2} \geq 1 \,, \nonumber \\
e_U &= e_W - \frac{D}{2} = -\frac{D+\alpha(D,r,s)}{2} \leq 0 \,.
\label{UWpowers}
\end{align}
In the last step, we applied Eq.~\eqref{massdim} to obtain relations in terms of the mass dimension $\alpha(D,r,s)$ and the cumulated power
\begin{equation}
\kappa = \sum_j \kappa_j = r-s
\label{alphapower}
\end{equation}
of the Feynman parameters can be identified with the difference of the indices $r$ and $s$ introduced in Eq.~\eqref{rs}. Rewriting Eq.~\eqref{UWpowers} reveals that MIs with certain mass dimension should be preferred, namely the ones fulfilling
\begin{equation}
\frac{l\,D}{2} + 1 \leq \kappa \leq \frac{l\,D}{2} + \frac{D}{2} \,.
\end{equation}
For processes $(b)$ and $(c)$, we have $D=4$ and $l=2$ so that this equation turns into $5 \leq \kappa \leq 6$, in other words into the recommendations of either $r-s=5$ or $r-s=6$. This guideline can be understood as the quantitative version of the qualitative statement that one should raise $r$ for integrals with $t<5$, referred to as \textit{putting dots} on the propagators, whereas $s$ should be increased for integrals with $t>6$, corresponding to more scalar products in the numerator of the integral representation. From Appendix~\ref{sec:hjlaporta}, we can deduce that this rule applies, with very few exceptions, to every MI within the two-loop corrections of $H\to Z\gamma$ and Higgs-plus-jet production, and thus has proven to be very useful in our calculations.

\item[\textbf{d)}] \textbf{Symmetry Considerations}\\
According to guideline c), one has to put either dots or scalar products to achieve the suggested mass dimension. We have observed that this should be done such that the entity of the graphs of all MIs of a given sector is as much symmetric as possible with respect to the graph of the corner integral\footnote{The corner integral is defined as the one with $r=t$ and $s=0$}. Let us illustrate this with two examples:
\begin{itemize}
\item \textit{Example for increasing r}\\
Sector $A_{5,182}$ is defined in Appendix~\ref{sec:hjlaporta} and includes four MIs. We obtain a triangular form by choosing the basis integrals $I_{39}\text{--}I_{42}$ in Fig.~\ref{fig:hjmaster}, where the graph of the corner integral $I_{39}$ is by definition symmetric to itself. Besides, placing a dot on the central diagonal line, like in case of $I_{40}$, obviously leads to a symmetric graph with respect to this corner integral as well. Finally, putting another dot on the upper line, as for $I_{41}$, requires squaring the lower propagator line of $I_{42}$, too, in order to retain the graph symmetry of the entire sector with respect to the corner integral.
\item \textit{Example for increasing s}\\
The definition of sector $A_{7,247}$ is given in Appendix~\ref{sec:hjlaporta} and contains four MIs. We achieve a triangular form by selecting $I_{67}\text{--}I_{70}$ in Fig.~\ref{fig:hjmaster} as MIs, where the graph of the corner integral $I_{67}$ is again by definition symmetric to itself. With a sector of seven and an integral family of nine propagators, there are only two possible ways of adding one scalar product by choosing negative powers of the remaining two denominators, leading to the graphs of $I_{68}$ and $I_{69}$. The remaining basis integral $I_{70}$ is then determined in such a way that both of these scalar products occur, which is still in agreement with the requirement of the mass dimension described in paragraph c).
\end{itemize}

\item[\textbf{e)}] \textbf{Infrared-Divergent Behavior}\\
In the literature, it is often stated that MIs with dots on massive lines should be preferred over MIs with dots on massless lines, because the mass acts as a regulator and cures the infrared-divergent behavior in the limit $k\to 0$. However, our observations suggest that this statement is superseded by all other considerations made so far, especially when conflicting with the ones of c) and d). For example, this means that to first approximation all statements made in d) are valid without distinguishing between massive and massless propagators. Let us illustrate this with the help of sector $A_{5,174}$, which contains three MIs. The corresponding integrals $I_{33}\text{--}I_{35}$ are provided in Appendix~\ref{sec:hjlaporta} and are depicted in Fig.~\ref{fig:hjmaster}. According to c) and d), it makes sense to select the corner integral $I_{33}$ as one MI. As soon as the second MI is identified with $I_{34}$ by squaring the propagator of the leftmost line, the most suitable choice that completes the set of MIs is $I_{35}$. This selection corresponds to the only one, where a dot is put on a massless propagator, and at the same time to the only one, which leads to conservation of the graph symmetry with respect to the corner integral.

\item[\textbf{f)}] \textbf{Known Results}\\
It can be useful to study integrals available in the literature that are similar to the ones under consideration in terms of kinematics and mass configuration. In the spirit of guideline e), we particularly recommend reviewing basis choices of previous computations, that are up to the mass distribution topologically identical to integrals whose differential equations ought to be decoupled. Even if only massless equivalents, which come in general with a lower number of MIs, are provided in the literature, they may propose candidate integrals suitable for a triangular basis choice. In this sense, for instance, Ref.~\cite{Caron-Huot:2014} has proven to be beneficial in the context of Higgs-plus-jet production, i.e. of process~(c).
\end{itemize}
We would like to emphasize that this approach is far from being algorithmic or emerging from general principles, however this set of rules directed us to the desired goal of finding a triangular basis in every case we tried. Needless to say, applying such loose guidelines requires a certain amount of experience and intuition.\\
Let us return to the second option, which consists in finding linear combinations of MIs that decouple the differential equations:
\begin{itemize}
\item[\textbf{g)}] \textbf{Linear Combinations}\\
Let us start from a system of coupled differential equations of two MIs $I_1(\vec{x}),I_2(\vec{x})$, i.e. the matrix $C(4;\vec{x})$ in Eq.~\eqref{matrixdeq} corresponds to a $2\times 2$ matrix with non-zero entries. One may then attempt to find a solution in terms of a linear combination of the given MIs by making an ansatz for one of the two MIs through the following replacement:
\begin{equation}
\begin{pmatrix}I_1(\vec{x})\\I_2(\vec{x})\end{pmatrix} \to \begin{pmatrix}J_1(\vec{x})\\J_2(\vec{x})\end{pmatrix} = \begin{pmatrix}I_1(\vec{x})\\c_1(\vec{x}) \, I_1(\vec{x}) + c_2(\vec{x}) \, I_2(\vec{x})\end{pmatrix} \,.
\end{equation}
Next, one could derive the differential equations of the integrals $J_1(\vec{x}),J_2(\vec{x})$ and impose that the matrix element $\mathcal{C}^{(i)}_{21}(4;\vec{x})$ of the new basis is zero, corresponding to a vanishing coefficient of $J_1(\vec{x})$ in the differential equation of $J_2(\vec{x})$. As a result, one obtains two linear first-order differential equations for the coefficients $c_1(\vec{x})$ and $c_2(\vec{x})$, which have to be solved themselves. Although solving them might be easier in practice than doing this for the original differential equations, this brings us formally back to the original problem. It would therefore be desirable to determine these coefficients in an algorithmic way, without the need of solving differential equations, especially if more than two MIs are involved, which is exactly what is provided by Ref.~\cite{Tancredi:2015}\footnote{Recently, similar ideas have been presented in Ref.~\cite{Adams:2017a} by using the language of Picard-Fuchs operators presented in Section~\ref{sec:identell}.}. Let us state the key idea by starting from a set of $N$ MIs $I_1(\vec{x}),\dots,I_N(\vec{x})$, whose differential equations are coupled in four dimensions:
\begin{enumerate}
\item Derive IBP identities for this set of MIs in a fixed even number of dimensions, i.e. $D=2\,k \,, k\in\mathbb{N}$, and set the subsectors to zero within these relations.
\item Scan the derived IBP identities for degeneracies in the sense that the MIs become linearly dependent, e.g.
\begin{equation}
\sum_{k=1}^N b_{i,k}(\vec{x}) \, I_k(\vec{x}) = 0 \,.
\end{equation}
If all $N$ differential equations should be decoupled, $N-1$ such relations are required, which means that $i=1,\dots,N-1$.
\item Supposing these relations to be found in $D=d$ dimensions with fixed, even~$d$, define a new set of basis integrals $J_1(d;\vec{x}),\dots,J_N(d;\vec{x})$, which corresponds exactly to the degeneracies found in the last step except for the $N$-th integral, i.e.
\begin{align}
J_i(d;\vec{x}) &\equiv \sum_k^N b_{i,k}(\vec{x}) \, I_k(\vec{x}) \,, \nonumber \\
J_N(d;\vec{x}) &\equiv I_N(\vec{x}) \,.
\end{align}
By construction, the differential equations of the basis $J_1(d;\vec{x}),\dots,J_N(d;\vec{x})$ decouple in $d$ dimensions.
\item If $d\neq 4$, apply Tarasov-Lee shifts as explained in Refs.~\cite{Tarasov:1996,Lee:2009}, which perform a change of basis from $J_1(d;\vec{x}),\dots,J_N(d;\vec{x})$ to 
\begin{equation}
J_i(d\pm 2;\vec{x}) = \sum_k^N c_{i,k}(d;\vec{x}) \, J_k(d;\vec{x}) \,.
\end{equation}
Repeating this procedure sufficiently many times leads to $J_1(4;\vec{x}),\dots,J_N(4;\vec{x})$, which is a basis that decouples in four dimensions. By construction, it fulfills the same differential equations as the basis $J_1(d;\vec{x}),\dots,J_N(d;\vec{x})$ in $d$ dimensions, but with the replacement $d\to 4$.
\end{enumerate}
In our calculations, we have successfully used this method several times, so let us illustrate it in the context of sector $A_{5,213}$, which is the planar sector with the highest number of MIs in the context of Higgs-plus-jet production. By applying the previous guidelines a) to e), we obtain the basis $I_{48}\text{--}I_{52}$ of five MIs, which is defined in Appendix~\ref{sec:hjlaporta} and depicted in Fig.~\ref{fig:hjmaster}. The system is \textit{almost} in triangular form at this point: The only integrals that remain to be decoupled are $I_{49}$ and $I_{50}$, which means that we have to find one linearly dependent IBP relation of these two in a fixed, even number of dimensions. This seems to work in $d=2$, where the identity reads
\begin{equation}
I_{50}(2;x,z,h) - \frac{x}{z} \, I_{49}(2;x,z,h) = 0 \,.
\end{equation}
Therein, the variables $x,z,h$ correspond to the definitions in Eq.~\eqref{ratios} and at this point, the definition of $m$ and $M$ does not play a role. Replacing $I_{50}(2;x,z,h)$ with the linear combination
\begin{equation}
J_{50}(2;x,z,h) \equiv I_{50}(2;x,z,h) - \frac{x}{z} \, I_{49}(2;x,z,h)
\end{equation}
leads to decoupled differential equations of the new basis
\begin{equation}
J_{49}(2;x,z,h)\equiv I_{49}(2;x,z,h) \,, \quad J_{50}(2;x,z,h)
\end{equation}
in $d=2$ dimensions, and by performing Tarasov-Lee shifts to $d+2$ dimensions we obtain the desired linear combination in $d=4$:
\begin{align}
J_{50}(4;x,z,h) = &\left(z\,(h-2)-2\,x\right) \, I_{50}(4;x,z,h) \nonumber \\
&+ \left(x\,(h-2)-2\,z\right) \, I_{49}(4;x,z,h) \,.
\end{align}
Finally, we would like to point out that the only case where this procedure has not proven to be successful is sector $A_{6,215}$, which is defined through the basis of four MIs denoted by $I_{59}\text{--}I_{62}$ in Appendix~\ref{sec:hjlaporta} and is depicted in Fig.~\ref{fig:hjmaster}. More precisely, we manage to find sufficiently many relations in $d=2$ dimensions designed to fully decouple the basis integrals in $d=4$, which however fails after performing Tarasov-Lee shifts. In fact, this sector eventually turns out to be beyond the class of functions of Multiple Polylogarithms with integrals belonging to a class of functions, whose first-order differential equations \textit{cannot} be decoupled. Consequently, they require a different approach, which we will come back to in Chapters~\ref{chap:workflow3} and \ref{chap:hj}, but for now let us capture the fact that this enables us to turn the argument of Ref.~\cite{Tancredi:2015} around: It might be the case that the decoupling cannot be performed at all, whenever the degenerate IBP identities in fixed, even number of dimensions do not lead to a decoupling in $d=4$ . This would serve as a way to recognize this kind of integrals a priori, especially since there is currently no straightforward way of doing this.
\end{itemize} 

\subsection{The Canonical Form}
\label{sec:canon}

Let us assume that we have managed to triangularize the matrix $C^{(i)}(4;\vec{x})$ within Eq.~\eqref{matrixdeq} with the methods learned in the last section. At this point, one could in principle already proceed with the integration by inserting the ansatz for the Laurent series from Eq.~\eqref{laurent} into the differential equations, subsequently expanding them in $\e$ and finally integrating the equations. At fixed $\e$-order, one would start integrating the differential equation with the lowest number of MIs belonging to a given sector appearing on the right-hand side for $\e=0$, and continue with the equation with the second-lowest number of MIs, ultimately exploring the full length of the triangle within $C^{(i)}(4;\vec{x})$ in the last integration. The integrations themselves would then proceed along the lines of the single-MI case described in Section~\ref{sec:oneMI}. In the beginning of Section~\ref{sec:moreMIs}, however, we pointed out that an exceptional feature of the differential equation approach consists in the possibility of performing operations on the level of the differential equations, prior to the actual integration, in order to simplify the latter or to enable it at all. In the following, we will exploit this feature in its full depth.\\
In fact, although the integration is in principle feasible starting from a triangular form of the differential equations in $D=4$, it often requires a huge effort to accomplish in practice due to large intermediate expressions and unnecessarily complex representations of the final result. A breakthrough in this direction was achieved by Refs.~\cite{Kotikov:2010, Henn:2013b}, which suggest casting the differential equations into the so-called \textit{canonical} form, where the complete dependence on the dimensional parameter $\e$ is factorized from the kinematics:
\begin{equation}
\frac{\p}{\p x_i} \vec{M}(D;\vec{x}) = \e \, A^{(i)}(\vec{x}) \, \vec{M}(D;\vec{x}) \,.
\label{canonicaldeq}
\end{equation}
In the following, we will refer in this context to \textit{canonical integrals}~$\vec{M}(D;\vec{x})$ as opposed to \textit{Laporta integrals}~$\vec{I}(D;\vec{x})$ appearing in Eq.~\eqref{matrixdeq}. In order to call a differential equation canonical, another constraint on the structure of the matrix $A^{(i)}(\vec{x})$ must be fulfilled on top of Eq.~\eqref{canonicaldeq}, namely there must exist a total differential
\begin{equation}
\d A(\vec{x}) = \sum_k A_k \, \d\log(l_k) \,,
\label{defdlog}
\end{equation}
so that the entries of the matrix $\d A(\vec{x})$ are said to be in $\d\log$-form. Therein, $A_k$ are rational numbers and the entity of the arguments $l_k$ is referred to as the \textit{alphabet} of the differential equations. The alphabet is composed of rational functions of the kinematic invariants and could equally be read off from the denominators of the homogeneous differential equation within Eq.~\eqref{matrixdeq}. Equation~\eqref{defdlog} allows rewriting Eq.~\eqref{canonicaldeq} as
\begin{equation}
\d\vec{M}(D;\vec{x}) = \e \, \d A(\vec{x}) \, \vec{M}(D;\vec{x}) \,,
\label{totdiff}
\end{equation}
where the differential `d' acts on all kinematic invariants.\\
Having a differential equation in canonical form translates into the fact that all equations are fully decoupled in the limit $\e\to 0$. This means, in turn, that the coefficient of a given order in the Laurent series can be determined through a one-dimensional integration over the coefficients of the previous order in $\e$. By definition, this induces results with iterative structure over analytic integration kernels, referred to as \textit{Chen iterated integrals}~\cite{Chen:1977}. In Section~\ref{sec:mpl}, we will present a special class of these iterated integrals called \textit{Multiple Polylogarithms}. For now, let us state that an outstanding advantage of integrating differential equations starting from the canonical form is that it leads to particularly compact representations of the result in terms of $\textit{pure functions of uniform transcendental weight}$~\cite{Henn:2013b}, which means that the structure of the result can be predicted prior to the integration. This serves as a useful check and will be helpful in the context of constructing an ansatz in a different approach, see Chapter~\ref{chap:workflow3}. Moreover, rotating to a canonical basis shifts complexity from the integration to the basis search. This powerful shift is advantageous in the sense that even with a canonical form at hand the integration is still the bottleneck, as we will see in Chapter~\ref{chap:hj}. On the other hand, it is not a problem to find a canonical basis in practice starting from a triangular form, which we will elaborate on in the following.

\subsection{From Triangular to Canonical Form}
\label{sec:triangulartocanonical}

Until recently, it was not known how to predict the existence of the canonical form for a given sector~\cite{Lee:2017a}. In order to recognize candidates suitable for canonical differential equations, Henn proposed to analyze the structure of the unitarity cut~\cite{Henn:2014} based on findings in $N=4$ supersymmetric Yang–Mills theory~\cite{Cachazo:2008,ArkaniHamed:2010}, which can be understood as a loose rule far from being algorithmic. A few years ago, Lee presented an algorithm which, starting from arbitrary basis integrals, is designed to find a change of basis so that the differential equations of the resulting integrals can be cast into canonical form~\cite{Lee:2014}. By now, several implementations are publicly available~\cite{Gituliar:2016,Gituliar:2017,Prausa:2017}, however they are confined to two-scale integrals, which can be expressed by a single dimensionless ratio. The same statement holds for the procedure described in Ref.~\cite{Ablinger:2015}, where coupled differential equations are solved via indefinite nested sum representations in Mellin space. Recently, an extended algorithm applicable to multiple scales was introduced in Refs.~\cite{Meyer:2016,Meyer:2017}, which comes with the restriction of allowing only rational dependence on both the dimensional parameter $\e$ and the kinematic invariants within the coefficients of the canonical differential equations. Unfortunately, both approaches turn out to be useless for the calculations of processes $(b)$ and $(c)$ presented in Chapters~\ref{chap:hza} and \ref{chap:hj}, respectively: According to Table~\ref{tab:complexity}, the two of them belong to the class of multi-scale problems so that they are beyond the scope of Lee's algorithm. In addition, the computation of process $(b)$ had been completed at the time when Refs.~\cite{Meyer:2016,Meyer:2017} became available. Finally, the differential equations of process $(c)$ involve square roots of the kinematic invariants which, to our best knowledge, cannot be converted to purely rational expressions. To sum up, although the mentioned algorithms are obviously useful for a number of calculations, they turn out to be ineffective in the context of our computations.\\
Nonetheless, we are in a lucky situation: Let us recall that, by using the guidelines outlined in Section~\ref{sec:basischoice}, we are able to attain a triangular form in $\e=0$ for the coefficient matrices in all eligible differential equations of processes $(b)$ and $(c)$. Beyond that, these differential equations involve at most linear dependence on $\e$, so that the missing piece is supplied by Ref.~\cite{Argeri:2014}, which allows rephrasing the rotation to a canonical basis in terms of the \textit{Magnus series}. We make use of a conceptually different, but equivalently algorithmic procedure presented in Ref.~\cite{Gehrmann:2014b}, which moreover admits polynomial dependence on $\e$ concerning the subsector coefficients. Let us elaborate on its idea by starting from a basis of the form~\eqref{matrixdeq},
\begin{equation}
\frac{\p}{\p x_i} \vec{I}(D;\vec{x}) = C^{(i)}(D;\vec{x}) \, \vec{I}(D;\vec{x}) + G^{(i)}(D;\vec{x}) \, \vec{m}(D;\vec{x}) \,,
\label{matrixdeq2}
\end{equation}
where the matrix
\begin{equation}
C^{(i)}(D;\vec{x}) = C^{(i)}(4;\vec{x}) + \e \, C^{(i)}_\e(\vec{x})
\end{equation}
depends only linearly on $\e$ and the slice $C^{(i)}(4;\vec{x})$ in $D=4$ is triangular. In case of a single MI, the triangularity is trivially satisfied since the matrix $C^{(i)}(D;\vec{x})$ reduces to a scalar. For a fixed number $t$ of different propagators and a sector of $N$ MIs $\vec{I}=(I_1,\dots,I_N)$ we proceed as follows:
\begin{enumerate}
\item Initially, we derive the solution of the homogeneous system in $D=4$,
\begin{equation}
\frac{\p}{\p x_i} \vec{I}(4;\vec{x}) = C^{(i)}(4;\vec{x}) \, \vec{I}(4;\vec{x}) \,,
\label{matrixdeq3}
\end{equation}
which can be separated into two steps: First, one has to compute the \textit{integrating factor} of every MI by means of Eq.~\eqref{homsol}, which sets the diagonal entries within $C^{(i)}(4;\vec{x})$ to zero and leads to a variation of the non-diagonal entries. Second, the non-diagonal entries within $C^{(i)}(4;\vec{x})$ are eliminated by integrating them over $\d x_i$, which is referred to as \textit{integrating out} the coefficients of the remaining MIs of the given sector. This procedure turns the differential equation~\eqref{matrixdeq2} into
\begin{equation}
\frac{\p}{\p x_i} \vec{J}(D;\vec{x}) = \e \, \tilde{C}^{(i)}(\vec{x}) \, \vec{J}(D;\vec{x}) + \tilde{G}^{(i)}(D;\vec{x}) \, \vec{m}(D;\vec{x})
\label{matrixdeq4}
\end{equation}
after performing the basis rotation $\vec{I}(D;\vec{x})\to \vec{J}(D;\vec{x})$, which is given in terms of algebraic functions. Consequently, we allow the introduction of roots of polynomials at this point, although the matrices $C^{(i)}(D;\vec{x})$ and $G^{(i)}(D;\vec{x})$ within Eq.~\eqref{matrixdeq2} have at most rational dependence on the kinematic invariants due to the rational nature of the IBP identities. 

\item Next, we have to take care of the subsector coefficients $\tilde{G}^{(i)}(D;\vec{x})$ in Eq.~\eqref{matrixdeq4}. Since we follow a bottom-up approach, by starting with the MIs with the lowest number of~$t$ and repeating this procedure for increasing $t$, we assume the differential equations of the subsectors $\vec{m}(D;\vec{x})$ to be in canonical form already. However, this does in general not apply to their coefficients $G^{(i)}(D;\vec{x})$ and $\tilde{G}^{(i)}(D;\vec{x})$ in the differential equations~\eqref{matrixdeq2} and \eqref{matrixdeq4} of the sector under consideration, with both admitting the decomposition
\begin{equation}
\tilde{G}^{(i)}(D;\vec{x}) = \tilde{G}_0^{(i)}(\vec{x}) + \e \, \tilde{G}_1^{(i)}(\vec{x}) + \frac{\tilde{G}_\e^{(i)}(\vec{x})}{u+v\,\e} \qquad (u,v\in\mathbb{Z}) \,.
\label{decompsubsector}
\end{equation}
In a first step, we have to eliminate the non-canonical structure $\tilde{G}_\e^{(i)}(\vec{x})$ since the corresponding basis rotation will affect the coefficients of the matrices $\tilde{G}_0^{(i)}(\vec{x})$ and $\tilde{G}_1^{(i)}(\vec{x})$ upon partial fractioning in $\e$. The rotation can be determined by making the ansatz
\begin{equation}
\vec{J}(D;\vec{x})\to \vec{K}(D;\vec{x}) = \vec{J}(D;\vec{x}) + \frac{\tilde{\mathcal{G}}_\e^{(i)}(\vec{x})}{u+v\,\e} \, \vec{m}(D;\vec{x})
\end{equation}
and deriving the differential equations for the new basis $\vec{K}(D;\vec{x})$, imposing that all coefficients proportional to $\tilde{\mathcal{G}}_\e^{(i)}(\vec{x})$ cancel. This leads to coupled linear first-order differential equations for the entries of the matrix $\tilde{\mathcal{G}}_\e^{(i)}(\vec{x})$, whose solution could however be deduced without any effort in the rare cases where we had to carry out a shift of this kind.

\item Through the steps described so far we obtain differential equations almost in canonical form,
\begin{align}
\frac{\p}{\p x_i} \vec{K}(D;\vec{x}) &= \e \, \left[\tilde{C}^{(i)}(\vec{x}) \, \vec{K}(D;\vec{x}) + E^{(i)}(D;\vec{x}) \, \vec{m}(D;\vec{x}) \right] \nonumber \\
&\quad+ F^{(i)}(D;\vec{x}) \, \vec{m}(D;\vec{x}) \,,
\end{align}
which require $F^{(i)}(D;\vec{x})$ to be removed in a final step. We therefore proceed with the rotation
\begin{equation}
\vec{K}(D;\vec{x})\to \vec{M}(D;\vec{x}) = \vec{K}(D;\vec{x}) + \mathcal{F}^{(i)}(\vec{x}) \, \vec{m}(D;\vec{x}) \,,
\label{canonicalshift}
\end{equation}
where $\mathcal{F}^{(i)}(\vec{x})$ is determined by integrating out the coefficients in the same way as when eliminating the non-diagonal entries of $C^{(i)}(4;\vec{x})$ in step 1. Eventually, we arrive at the canonical form of Eq.~\eqref{canonicaldeq}:
\begin{equation}
\frac{\p}{\p x_i} \vec{M}(D;\vec{x}) = \e \, A^{(i)}(\vec{x}) \, \vec{M}(D;\vec{x}) \,.
\label{matrixdeq5}
\end{equation}
\end{enumerate}
As mentioned, this procedure can be performed bottom-up, beginning with the lowest order in $t$ and gradually increasing its value. This ensures that at every point of casting a triangular form into a canonical differential equation, all subsectors appearing in the inhomogeneous part are already in canonical form. The method has proven to be successful in all cases we tried, which will be presented in Chapters~\ref{chap:hza} and \ref{chap:hj}.\\
It remains to comment on several issues:
\begin{itemize}
\item Note that all steps have to be performed simultaneously for every independent variable $x_i$. This means that for every coefficient function appearing in the basis rotations from above, e.g. for $\mathcal{F}^{(i)}(\vec{x})$ in Eq.~\eqref{canonicalshift}, a single function must exist which fulfills the requirements of the resulting differential equations in all $x_i$ at once.
\item In Ref.~\cite{Gehrmann:2014b}, Eq.~\eqref{decompsubsector} describing the subsector coefficients is extended to polynomial dependence on $\e$ as well as to higher powers of the denominators $u+v\,\e$. This extension can be achieved in a straightforward way, however it was not needed in the context of our calculations, which is why we skip the details.
\item We emphasize that Ref.~\cite{Gehrmann:2014b}, like any algorithm available in the literature, limits this procedure to rational dependence on the kinematic invariants throughout. Nevertheless, we have managed to extend it to square root expressions of the kinematic invariants. This is particularly important in the context of Higgs-plus-jet production, process~$(c)$, where a variable transformation to rational expressions does not exist to the best of our knowledge. Note, however, that this does not mean that the differential equations can be integrated in terms of the functions introduced in the next section, and we will return to this issue in Chapter~\ref{chap:workflow3}. In the differential equations of process~$(b)$, square roots are induced at the point of identifying the homogeneous solution. In terms of the variables introduced in Eq.~\eqref{ratios}, these square roots read
\begin{equation}
\sqrt{x\,(x-4)} \,, \qquad \sqrt{h\,(h-4)} \,,
\end{equation}
corresponding to two-particle cuts of massive propagators with mass $m$. They can be rationalized through transformations to Landau-type variables $\tilde{x}$, $\tilde{h}$:
\begin{equation}
x = -\frac{(1-\tilde{x})^2}{\tilde{x}} \,, \qquad h = -\frac{(1-\tilde{h})^2}{\tilde{h}} \,.
\label{landau1}
\end{equation}
\item Furthermore, we would like to stress that the algebraic dependence mentioned in the last bullet point is maximal in the sense that one cannot go beyond if the goal is to obtain canonical differential equations. This becomes clear when we recall the additional requirement of the canonical form, which presumes the existence of a total differential as in Eq.~\eqref{defdlog}\footnote{Very recently, an $\e$-factorized form was suggested for elliptic integrals in Ref.~\cite{Adams:2018} by allowing transcendental dependence on the kinematic invariants. However, the issue of finding a total differential as for the `conventional' canonical form was not addressed and does presumably not exist in this case.}. Before we elaborate on this, let us recall that the structure of the homogeneous solutions $\vec{I}(\vec{x})$ of the differential equations~\eqref{matrixdeq3} will reappear in the differential equation of every subsequent basis rotation, starting from Eq.~\eqref{matrixdeq4}, as a result of the first basis rotation $\vec{I}(D;\vec{x})\to \vec{J}(D;\vec{x})$. If these homogeneous solutions are beyond algebraic expressions, it will in general not be possible to cast the coefficient matrix $A^{(i)}(\vec{x})$ appearing in Eq.~\eqref{matrixdeq5} into $\d\log$ form.\\
In our calculations, we encountered several cases in which the homogeneous solutions were given in terms of logarithmic expressions so that the corresponding basis choice had to be discarded. One such case is sector $A_{5,213}$ with five MIs, whose basis choice in Appendix~\ref{sec:hjlaporta} and Fig.~\ref{fig:hjmaster} has already been analyzed as an example of guideline g) in Section~\ref{sec:basischoice}. According to guideline e) in that section, it would have been more natural to select the basis which includes the corner integral plus all possibilities of putting one dot on every massive propagator. This basis was discarded for the reasons outlined here.
\end{itemize}

\section[Excursus on a Special Class of Functions: Multiple Polylogarithms]{Excursus on a Special Class of Functions:\\Multiple Polylogarithms}
\label{sec:mpl}
Before proceeding with the actual integration, let us discuss the class of functions that we intend to express our results with. As mentioned in Section~\ref{sec:canon}, the canonical form leads to results which can be expressed by iterative integrals over analytic functions, known as Chen iterated integrals. A special class of Chen iterated integrals emerges from linear, rational integration kernels and is referred to as \textit{Multiple Polylogarithms} (MPLs). They were first mentioned by Kummer~\cite{Kummer:1840}, further studied by Poincaré~\cite{Poincare:1883} and recently attracted renewed attention in the mathematical literature thanks to Goncharov~\cite{Goncharov:1998}, which is why they are sometimes also called \textit{Goncharov Polylogarithms}.\\
From the particle physics point of view, it was revealed that the \textit{Dilogarithm} or \textit{Spence's function} and the \textit{Nielsen Polylogarithms}~\cite{Nielsen:1909} as its generalization are required to represent the results of one-loop Feynman integrals long ago. At higher orders in the perturbative expansion, it quickly became evident that these functions are not sufficient, which triggered the rediscovery of the single-variable \textit{Harmonic Polylogarithms} (HPLs)~\cite{Remiddi:1999} in the context of two-scale integrals (i.e. one independent ratio) and subsequently of its extension for one additional scale, referred to as the \textit{two-dimensional Harmonic Polylogarithms}~\cite{Gehrmann:2000}. Eventually, this culminated in the introduction of the generalization of HPLs to an arbitrary number of scales under the name of \textit{Generalized Harmonic Polylogarithms}, which are equivalent to the MPLs. They have proven to be beneficial in a countless number of multi-loop applications due to their fast and reliable numerical evaluation, whereof several implementations exist~\cite{Gehrmann:2001b,Gehrmann:2001c,Vollinga:2004,Maitre:2005,Maitre:2007,Buehler:2011}, and their useful algebraic properties, which have been analyzed in detail~\cite{Zagier:1991,Goncharov:1991,Goncharov:2001,Goncharov:2005,Goncharov:2010,Duhr:2011,Duhr:2012}.

\subsection{Definition}

When it comes to numerical evaluations, the preferred representation for the definition of the MPLs is given in terms of an infinite series. Nevertheless, we will define them through their integral representation, which naturally arises from the iterative structure induced by the differential equations. As a first step, let us recall the integral representation of the natural logarithm:
\begin{equation}
\log\left(x\right) = \int_1^x \frac{\d t}{t} \,, \qquad \log\left(1-\frac{x}{a}\right) = \int_0^x \frac{\d t}{t-a} \quad (a\neq 0) \,.
\label{deflog}
\end{equation}
Integrating over the product of a logarithm and a rational function yields the dilogarithm
\begin{equation}
\mathrm{Li}_2(x) = -\int_0^x \frac{\d t}{t} \, \log\left(1-t\right) = -\int_0^x \frac{\d t}{t} \int_0^t \frac{\d u}{u-1} \quad \left(x\in\mathbb{C}\backslash[1,\infty]\right) \,,
\end{equation}
where we inserted Eq.~\eqref{deflog} in the last step to make explicit the iterative two-dimensional integration over rational kernels. By extending this to arbitrary number of iterations, we obtain the definition of the \textit{classical polylogarithms}:
\begin{align}
\mathrm{Li}_n(x) &= \int_0^x \frac{\d t}{t} \, \mathrm{Li}_{n-1}(t) \quad \left(x\in\mathbb{C}\backslash[1,\infty]\right) \,, \nonumber \\
\mathrm{Li}_1(x) &= -\log\left(1-x\right) \,.
\label{defpolylog}
\end{align}
The iterative structure allows introducing the notion of the so-called \textit{transcendental weight}, which indicates the number $n$ of repeated integrations.\\
Let us remark that all integrations within Eq.~\eqref{defpolylog} are performed over the kernel $1/t$ except for the first, which is carried out over the kernel $1/(t-1)$ and breaks the symmetry with respect to the integration kernels. The natural procedure of restoring this symmetry consists in extending the allowed integration kernels to $1/t$, $1/(t-1)$ and $1/(t+1)$, leading to the definition of the weight-one HPLs:
\begin{align}
G(0;x) &= \log(x) \,, \nonumber \\
G(1;x) &= \int_0^x \frac{\d t}{t-1} = \log(1-x) \,, \nonumber \\
G(-1;x) &= \int_0^x \frac{\d t}{t+1} = \log(1+x) \,. 
\end{align}
These functions are single-valued and real in the domain $0\leq x\leq 1$. Note that upon their introduction, HPLs were denoted by $H$ instead of $G$ and the historical definition of $H(1;x)=-\log(1-x)$ came with a minus sign compared to $G(1;x)$.\\
The HPLs of higher weight are obtained by repeated integrations in the sense of
\begin{align}
G\left(w_1,\ldots,w_n;x\right) &= \int_0^x \d t \, \frac{1}{t-w_1} \, G\left(w_2,\ldots,w_n;t\right) \,, \nonumber \\
G\left(\vec{0}_n;x\right) &= \frac{\log^nx}{n!} \,, \quad \vec{0}_n = \{\underbrace{0,\dots,0}_{n \text{ times}}\}\,,
\label{defmpl}
\end{align}
where the length $n$ of the index vector $\vec{w}=(w_1,\dots,w_n)$ indicates the number of integrations and thus determines the weight. In fact, Eq.~\eqref{defmpl} already provides the full information on the definition of the MPLs by allowing any algebraic dependence of $\vec{w}$ on the kinematic invariants in the complex plane. From this general set, the HPLs emerge by confining the entries of $\vec{w}$ to the three values $\{0,1,-1\}$.\\
As a final remark, let us point out that all MPLs up to weight four can be expressed as linear combinations of the classical polylogarithms and their extension
\begin{equation}
\mathrm{Li}_{2,2}(x_1,x_2) = G\left(0,\frac{1}{x_1},0,\frac{1}{x_1\,x_2};1\right) \,,
\label{defli22}
\end{equation}
e.g. by using the program package \textsc{Gtolrules}~\cite{Frellesvig:2016}. For example, the generic weight-two relation is given by
\begin{equation}
G(w_1,w_2;x) = \mathrm{Li}_2\left(\frac{w_2-x}{w_2-w_1}\right) - \mathrm{Li}_2\left(\frac{w_2}{w_2-w_1}\right) + \log\left(1-\frac{x}{w_2}\right) \, \log\left(\frac{x-w_1}{w_2-w_1}\right)
\label{gtolrules}
\end{equation}
provided that $\left|\mathrm{Im}\left(\frac{w_1}{x}\right)\right|>\left|\mathrm{Im}\left(\frac{w_2}{x}\right)\right|$.

\subsection{Properties}
\label{sec:mplproperties}

Let us briefly summarize the main properties of the MPLs:
\begin{itemize}
\item[\textbf{a)}] \textbf{Shuffle Algebra}\\
Iterated integrals and in particular MPLs fulfill a Hopf algebra, the so-called shuffle algebra,
\begin{align}
G(w_1,\dots,w_{n_1};x) \, &G(w_{n_1+1},\dots,w_{n_1+n_2};x) \nonumber \\
&= \sum_{\sigma\in\Sigma(n_1,n_2)} G(w_{\sigma(1)},\dots,w_{\sigma(n_1+n_2)};x) \,,
\label{MPLshuffle}
\end{align}
where the sum runs over all possibilities $\Sigma(n_1,n_2)$ of riffle shuffling the two sets of $n_1$ and $n_2$ indices. This corresponds to shuffling two decks of cards with $n_1$ and $n_2$ cards, which means that the order of the cards within the original decks is preserved. As a consequence of this identity, any MPL of a given weight and certain indices can be expressed by other MPLs of the same structure modulo products of lower-weight MPLs.

\item[\textbf{b)}] \textbf{Scaling Relation}\\
Given that the rightmost index $w_n\neq 0$, referred to as \textit{trailing zero}, any MPL fulfills the scaling relation,
\begin{equation}
G(\vec{w};x) = G(\lambda\,\vec{w};\lambda \, x) \quad \left(\lambda\in \mathbb{C}^*\right) \,,
\label{MPLscaling}
\end{equation}
stating that one variable within the set of argument~$x$ and index vector~$\vec{w}$ is redundant. Note that this relation is still useful even for MPLs with trailing zeros, since they can always be converted to MPLs without trailing zeros through recursive application of the systematic inversion of Eq.~\eqref{MPLshuffle} referred to as \textit{co-algebra}. In fact, a unique characteristic of Hopf algebras is that such a co-algebra exists.

\item[\textbf{c)}] \textbf{H\"older Convolution}\\
Another identity is provided by the so-called H\"older Convolution, which allows expressing
\begin{align}
G(\vec{w};1) = \sum_{i=0}^n (-1)^i \, &G\left(1-w_i,1-w_{i-1},\dots,1-w_1;1-\frac{1}{p}\right) \nonumber \\
&\times G\left(z_{i+1},\dots,z_n;\frac{1}{p}\right)
\end{align}
in the case $w_1\neq 0$, $w_n\neq 0$ and $p\in \mathbb{C}^*$. This relation is used to improve the convergence of the series representation prior to numerical evaluation and transforms into
\begin{equation}
G(\vec{w};1) = (-1)^n \, G\left(1-w_n,\dots,1-w_1;1\right) \,.
\end{equation}
in the limit $p\to\infty$.

\item[\textbf{d)}] \textbf{Cut Structure}\\
In the context of analytic continuation, it may be worth mentioning that all information about the cut structure is contained in the index vector of the MPLs. In order to make the cut structure explicit, we remove trailing zeros by recursively applying the co-algebra as explained in~b). After this procedure, the only MPLs with trailing zeros correspond to the logarithms defined in Eq.~\eqref{defmpl}, which develop an imaginary part if $x<0$. Subsequently, we make use of the scaling relation~\eqref{MPLscaling} to map the remaining MPLs of the form $G(w_1,\dots,w_n;x)$ with $w_n=0$ to a representation with $x>0$. These MPLs only develop an imaginary part if the argument $x$ is larger than at least one of the indices $w_i$, i.e. if $\exists i\in\{1,\dots,n\}$ such that $x>w_i$. Note that this statement holds only if $\vec{w}\in\mathbb{R}$, which is true of most physical applications. In the case $\vec{w}\in\mathbb{C}$, the above statements can be generalized to whether the function $G(w_1,\dots,w_n;x)$ develops a branch cut\footnote{The definition of a branch cut will be explained in Section~\ref{sec:singularities}.}.
\end{itemize}

\subsection{Symbol and Coproduct Formalism}
\label{sec:symbol}

In the next section, we will see that integrating differential equations and imposing boundary conditions in a multi-scale problem requires deriving functional identities among MPLs. We distinguish two cases of such transformations:
\begin{itemize}
\item[\textbf{i)}] \textbf{Limiting Identities}\\
The first type of relations can be phrased as
\begin{equation}
G\left(w_1(y),\ldots,w_n(y);y\right) \to G\left(a_1,\ldots,a_n;y\right) \,,
\label{trafo1}
\end{equation}
where the transformed indices $a_i$ are nothing but complex numbers. In a three-scale problem with two independent ratios $y_1$ and $y_2$, for example, they arise when MPLs of the kind $G\left(w_1(y_2),\ldots,w_n(y_2);y_1\right)$ are subject to limits of the form $y_2\to f(y_1)$, which emerge from imposing boundary conditions.

\item[\textbf{ii)}] \textbf{Identities of Interchanging Arguments}\\
Further transformations of the type
\begin{equation}
G\left(w_1(y_2),\ldots,w_n(y_2);y_1\right) \to G\left(c_1(y_1),\ldots,c_n(y_1);y_2\right)
\label{trafo2}
\end{equation}
become necessary when the integration is performed in a different variable compared to the integration of a subtopology, which enters the differential equation under consideration.
\end{itemize}
One way to derive this kind of transformations is to differentiate MPLs with respect to their argument, thereby generating linear combinations of MPLs times rational functions. For a MPL of weight $k$, this procedure can be iterated $k$ times until only rational functions are left, whose trivial transformation properties are obtained straightforwardly. Finally, combining this outstanding yet simple property with the repeated use of integration-by-parts at the level of the integral representation and the method of partial fractioning allows the derivation of relations of the above type. Let us illustrate this with an example of case $ii)$:
\begin{align}
G(1,1-{y_2};{y_1}) &= G(1,1;{y_1}) + \int_0^{y_2} \d {y_2}' \, \frac{\d}{\d {y_2}'} \, G(1,1-{y_2}';{y_1}) \nonumber \\
&= G(1,1;{y_1}) + \int_0^{y_2} \d {y_2}' \, \frac{\d}{\d {y_2}'} \int_0^{y_1} \frac{\d t_1}{1-t_1} \int_0^{t_1} \frac{\d t_1}{1-t_2-{y_2}'} \nonumber \\
&= G(1,1;{y_1}) + \int_0^{y_2} \d {y_2}' \int_0^{y_1} \frac{\d t_1}{1-t_1} \int_0^{t_1} \frac{\d t_2}{(1-t_2-{y_2}')^2} \nonumber \\
&= G(1,1;{y_1}) + \int_0^{y_2} \d {y_2}' \int_0^{y_1} \d t_1 \left[-\left(\frac{1}{{y_2}}+\frac{1}{1-{y_2}}\right) \, \frac{1}{1-t_1} + \frac{1}{{y_2}\,(1-t_1-{y_2})} \right] \nonumber \\
&= G(1,1;{y_1}) + \int_0^{y_2} \d {y_2}' \left[\left(\frac{1}{{y_2}}+\frac{1}{1-{y_2}}\right) \, G(1;{y_1}) - \frac{G(1-{y_2};{y_1})}{{y_2}} \right] \nonumber \\
&= G(1,1;{y_1}) + G(1;{y_1}) \, \left[ G(0;{y_2}) - G(1;{y_2}) \right] \nonumber \\
&\quad - \int_0^{y_2} \frac{\d {y_2}'}{{y_2}'} \left[G(1-{y_1};{y_2}) + G(1;{y_1}) - G(1;{y_2}) \right] \nonumber \\
&= G(1,1;{y_1}) - G(0,1-{y_1};{y_2}) + G(0,1;{y_2}) - G(1;{y_1}) \, G(1;{y_2}) \,.
\label{symbolexp}
\end{align}
As we can see, the procedure requires detailed inspection of the integrands at every step of the derivation and thus a more systematic approach is desirable, which was achieved by introducing the so-called \textit{symbol map}~\cite{Duhr:2011}. Entering the details of this quite involved formalism is beyond the scope of this thesis, but let us state that the symbol map captures the decisive information on the properties of the MPLs so that identities among MPLs can be translated into identities among their symbols. More precisely, the symbol allows verifying equivalences, like for the left- and right-hand sides of Eq.~\eqref{symbolexp}, through multiple differentiations in a systematic and algebraic way. However, it is not constructive in the sense that equivalence relations cannot be derived from scratch, but possible equivalence relations have to be provided, which can ultimately be tested for their validity.\\
Vice versa, the task of finding the appropriate MPL, if the corresponding tensor in the vector space of one-forms is provided, turns out to be much more challenging and is often referred to as the \textit{inverse problem} in the literature or to \textit{integrating a symbol}. This is due to the fact that information on the appearance of transcendental functions like the $\zeta$ values is missing, which has been recently supplemented by extending the symbol to the coproduct formalism~\cite{Duhr:2012}. Opposed to the symbol map, the coproduct retains information about the $\zeta$ values, which renders the inverse problem feasible. The result is a coproduct-augmented symbol formalism, which allows deriving identities of types $i)$ and $ii)$ from general principles in a fully automatic way. Note that, although the definition in Eq.~\eqref{defmpl} in principal allows for algebraic dependence of the index vector $\vec{w}$ on the kinematic invariants, the methods described here are restricted to at most rational dependence. At present, it is not clear how to derive identities of the above kind for indices that go beyond, e.g. for square roots of kinematic invariants, and a multivariate extension of the concept of generalized weights as described in Ref.~\cite{vonManteuffel:2013} is desirable. Since this is not available to date, we have to follow other approaches in such a case, which will be elaborated on in Chapter~\ref{chap:workflow3}.\\
Let us close this section by pointing out that we apply the program package \textsc{CSimplify}~\cite{Gehrmann:2013,Weihs:2013} in the context of our calculations, which is an inhouse \textsc{Mathematica} implementation of the symbol and coproduct formalism. Moreover, we make use of the subroutine~\textsc{MPLEval} of this package for the numerical evaluation of MPLs, which includes a link to the \textsc{GiNaC} implementation presented in Refs.~\cite{Vollinga:2004,Bauer:2000}.

\section{Differential Equations for Master Integrals, Part II:\\From Canonical Form to Results}
\sectionmark{Differential Equations for Master Integrals: From Canonical Form to Results}
\label{sec:deq2}

\subsection{Integration}
\label{sec:int}

Let us assume that we are left with differential equations with respect to the variables $x_1,\dots,x_n$ in canonical form, whose alphabet has non-linear dependence on at most one variable denoted by $x_n$. The criterion of linearity of the variables determines the order of the integration in a multi-scale problem: The integration of the differential equation with respect to $x_n$ should be postponed until the end, which assures that the only roots occur in the index vector of the MPLs $G(\vec{w};x_n)$ and depend exclusively on complex numbers. The remaining dependence on $x_n$ is then completely encoded in MPLs, which emerge from previous integrations with respect to $x_1,\dots,x_{n-1}$ and are of the form $G(\vec{w}(x_n);x_{i\neq n})$. The key point of this procedure is that despite the non-linear dependence of the alphabet on $x_n$, the index vector $\vec{w}(x_n)$ involves by construction at most rational expressions of $x_n$ (and of all other variables), so that it is possible to derive the identities of Section~\ref{sec:symbol}. We can then proceed with the integration in terms of MPLs in a bottom-up approach with respect to the number $t$ of different propagators. Moreover, after inserting the Laurent expansion of the canonical integrals,
\begin{equation}
M(D;\vec{x}) = \sum_{k=a}^\infty \e^k \, M^{(k)}(\vec{x}) \,,
\label{laurent2}
\end{equation}
the bottom-up procedure also applies to the weight of the result or equivalently to the order of that expansion. We would like to stress at this point that the computation of the finite part of two-loop observables requires the evaluation of MPLs up to weight four. Starting from a tree-level diagram, this can be shown by generating loops through soft gluon insertions, which are attached to at least one massless leg, thereby generating a factor of $1/\e^2$ per loop. Finally, the actual integration can be performed in an either \textit{improper} or \textit{proper} manner, so let us elaborate on their differences.\\
For the improper integration, one integrates the differential equation with respect to any variable different from $x_n$, say $x_1$, so that the solution is known up to a constant $c_1$, which depends on all kinematic invariants except for $x_1$. This solution is then plugged into the differential equation with respect to another kinematic invariant different from $x_n$, say $x_2$, which leads to a first-order differential equation for the constant $c_1$. Solving this equation in the exact same way as the differential equation with respect to $x_1$ leaves us with a solution up to a constant $c_2$, which depends on all kinematic invariants except for $x_1$ and $x_2$. It is obvious that, by iterating this procedure sufficiently many times, we obtain the full solution up to a constant $c_n\in\mathbb{C}$, which remains to be determined by boundary conditions.\\
For the proper type of integration, we assume that the boundary conditions are imposed at the values $x^c_i$, so that the formal solution is given by a single equation,
\begin{equation}
M(x_1,\dots,x_n) = M(x^c_1,\dots,x^c_n) + \left. \sum_{j=1}^n \sum_{i=1}^j \int_{x^c_j}^{x_j} \frac{\p M(x_1,\dots,x_n)}{\p x_i}\right|_{x_{k<j}=x^c_k} \, \frac{\p x^c_i}{\p x_j} \, \d x_j \,,
\label{defint1}
\end{equation}
where we introduced the shorthand notation
\begin{align}
\frac{\p x^c_i}{\p x_i} &\equiv 1 \,, \\
\left. \frac{\p}{\p x_i} M(x_1,\dots,x_n)\right|_{x_{k<j}=x^c_k} &\equiv \lim_{x_{j-1}\to x^c_{j-1}} \dots \lim_{x_1\to x^c_1} \left[ \frac{\p}{\p x_i} M(x_1,\dots,x_n) \right] \,.
\label{shorthand}
\end{align}
Equation~\eqref{defint1} considerably simplifies in the special case that $x^c_i\in\mathbb{C}$ for all boundary values, i.e. that they do not depend on any other kinematic invariant $x^c_{j\neq i}$. The formula then turns into
\begin{align}
M(x_1,\dots,x_n) = M(x^c_1,\dots,x^c_n) + \sum_{i=1}^n \int_{x^c_i}^{x_i} \frac{\p M(x^c_1,\dots,x^c_{i-1},x_i,\dots,x_n)}{\p x_i} \, \d x_i  \,,
\label{defint2}
\end{align}
and the equivalent of Eq.~\eqref{shorthand} becomes
\begin{equation}
\frac{\p}{\p x_i} M(x^c_1,\dots,x^c_{i-1},x_i,\dots,x_n) \equiv \frac{\p}{\p x_i} \left[ \lim_{x_{i-1}\to x^c_{i-1}} \dots \lim_{x_1\to x^c_1} M(x_1,\dots,x_n) \right] \,.
\end{equation}
In Eqs.~\eqref{defint1} and \eqref{defint2}, $M(x^c_1,\dots,x^c_n)\in\mathbb{C}$ is the constant that remains to be fixed by imposing appropriate boundary conditions, equivalent to the constant $c_n$ in the improper approach.

\subsection{Boundary Conditions}
\label{sec:boundary}

Within the workflow of computing exact results for MIs through differential equations, the determination of boundary conditions represents the last missing ingredient. The need to impose boundary conditions can be understood as a price to pay for the reduced effort of computing only single-dimensional integrals compared to conventional approaches, like direct integration of the Feynman parameter representation. This task is non-trivial in the sense that independent information in the form of $n$ boundary conditions has to be supplied to a system of $n$ differential equations. One possibility is to calculate integrals at specific kinematic points, which comes down to computing them at a reduced number of scales. This can be attempted either by direct integration or by applying limiting procedures like asymptotic expansions~\cite{Beneke:1997,Harlander:1998,Harlander:1999,Smirnov:2002}. In a lucky situation, these lower-scale integrals can be recycled from results in the literature.\\
This is not required, however, for many practical applications. Thanks to another powerful property of the method of differential equations, information about the regularity of the MIs can be read off at the level of the differential equations. Let us recall the structure of the canonical differential equation~\eqref{canonicaldeq}, where the coefficient matrix $A^{(i)}(\vec{x})$ involves at most algebraic dependence on the kinematic invariants. In particular, this means that the roots of the denominator correspond to the singularities of the problem, referred to as \textit{thresholds}. From physical arguments, one can show that this produces singularities that cannot truly exist. These non-physical divergences occur spuriously and are called \textit{pseudo-thresholds}. By imposing regularity in such a pseudo-threshold, say in the kinematic point $\vec{x}^c_i$, one obtains algebraic identities of the form
\begin{equation}
\lim_{\vec{x}_i \to \vec{x}^c_i}\frac{\p}{\p x_i} \vec{M}(D;\vec{x}_i) \stackrel{!}{=} 0 = \e \, A^{(i)}(\vec{x}^c_i) \vec{M}(D;\vec{x}^c_i) \,.
\end{equation}
These relations are useful if they lead to non-trivial solutions for the MIs of the considered sector in terms of the subsector results in the given kinematic point, where the subsectors are already expected to be known due to the bottom-up approach. The operation can also be carried out at the level of the triangular differential equations~\eqref{matrixdeq}, whose denominator structure may contain additional information since the homogeneous solution has not been integrated out yet. Equivalently, it is much simpler to read off this extra piece of information directly from the homogeneous solution or integrating factor obtained from solving Eq.~\eqref{matrixdeq3}. If taking the limit in the chosen kinematic point leads both to regular Laporta MIs $\vec{I}(D;\vec{x})$ and to vanishing integrating factors, the boundary conditions of the canonical MIs have the very simple form of being zero to all orders in $\e$. We will show in Chapter~\ref{chap:hj} that this is the case for all planar MIs appearing in Higgs-plus-jet production, process~$(c)$. Although it is possible to proceed in the same way in the context of $H\to Z\gamma$, we will deviate from this approach in Chapter~\ref{chap:hza} for the sake of illustrating the flexibility of this method.

\subsection{Example: The One-loop Massive Bubble}
\label{sec:oneloopsunrise}

Let us close this chapter by giving an example of integrating a differential equation which appears in our calculations. Despite its simplicity, the chosen example displays the most important features of both the integration procedure and the application of boundary conditions. We consider the one-loop massive bubble~$I_2$ in the context of the two-loop amplitude of $H\to Z\,\gamma$, which is depicted in Fig.~\ref{fig:bubble}. Its integral representation in the Euclidean region can be inferred from the definition of integral family~$A$ in Table~\ref{tab:hjtopo},
\begin{equation}
I_2 = \int \frac{\d^D k}{(2\pi)^D} \, \frac{1}{(k^2+m_q^2)^2 \, ((k-q_1-q_2)^2+m_q^2)} \, \int \frac{\d^D l}{(2\pi)^D} \frac{1}{(l^2+m_q^2)^2} \,,
\end{equation}
where the second integral corresponds to a factorized tapole in order to remain in the two-loop notation. In this integral representation, we took one of the propagators of the massive bubble squared, because we prefer to solve the differential equations of its canonical equivalent $M_2$. It is given in Appendix~\ref{sec:hjcan} as a function of $I_2$ and in terms of the Landau variable $\tilde{x}$ introduced in Eq.~\eqref{landau1},
\begin{equation}
M_2 = -\e^2 \, m_q^2 \frac{(\tilde{x}+1)\,(\tilde{x}-1)}{\tilde{x}} \, I_2 \,,
\label{sunrisecan}
\end{equation}
where the $\e$-dependent prefactor is normalized such that the Laurent series
\begin{equation}
M_2(D;\tilde{x}) = \sum_{k=0}^4 \e^k \, M_2^{(k)}(\tilde{x})
\end{equation}
starts at least at order $\e^0$. As pointed out in Section~\ref{sec:int}, the finite part of two-loop observables requires the evaluation of MPLs up to weight four, according to which the upper bound of the Laurent expansion has been chosen. The differential equation is obtained through the procedures described in Section~\ref{sec:deq1} and evaluates to
\begin{align}
\frac{\p M_2}{\p\tilde{x}} &= \e \left[\left(\frac{1}{\tilde{x}} - \frac{2}{\tilde{x}+1} \right) \, M_2 + \frac{1}{\tilde{x}} \, M_1 \right] \,, \nonumber \\
\leftrightarrow \frac{\p M_2^{(k+1)}}{\p\tilde{\tilde{x}}} &= \left(\frac{1}{\tilde{x}} - \frac{2}{\tilde{x}+1} \right) \, M_2^{(k)} + \frac{1}{\tilde{x}} \, M_1^{(k)} \,.
\label{deqexample}
\end{align}
\begin{figure}[tb]
\begin{center}
\includegraphics[width=0.5\textwidth]{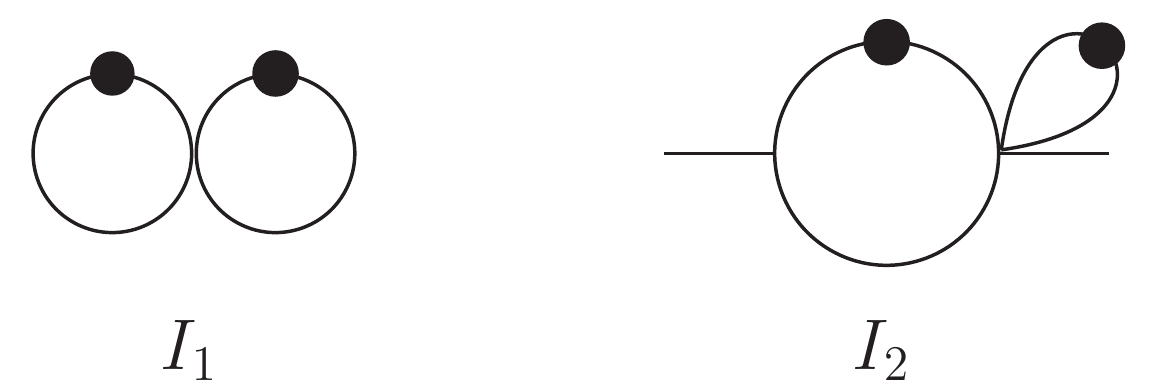}
\caption[One-loop massive bubble integral~$I_2$]{\textbf{One-loop massive bubble integral~$\boldsymbol{I_2}$}, which is multiplied with a one-loop tadpole in order to remain in the two-loop notation. The only subsector appearing in its differential equations is given by the double-tadpole~$I_1$. Dotted propagators are taken to be squared and internal solid lines denote propagators with mass $m_q$. These integrals are defined in Appendix~\ref{sec:hjlaporta} and appear both in the calculations of processes~$(b)$ and~$(c)$ presented in Chapter~\ref{chap:introduction}, where the external lines correspond to virtualities $m_Z^2$ and $s$, respectively.}
\label{fig:bubble}
\end{center}
\end{figure}\noindent
The double tadpole $M_1$ is the only appearing subsector, and its result is given in Eq.~\eqref{tadpole} by
\begin{equation}
M_1 = \e^2 \, I_1 = 1
\label{tadpolecan}
\end{equation}
if we neglect the normalization factor $S_\e$. The last missing ingredient required for integration consists in the boundary condition, which can in principle be determined implicitly from the differential equation~\eqref{deqexample}. For this purpose, we choose the limit $\tilde{x}\to 1$, which corresponds to $x=s/m_q^2\to 0$ or equivalently to the infinite quark mass limit, in which all integrals involving massive quark loops turn into tadpoles. However the differential equation~\eqref{deqexample} is trivially satisfied in this limit, so that we ought to return to the differential equation of the Laporta integral as stated in Section~\ref{sec:boundary}:
\begin{equation}
\frac{\p I_2}{\p\tilde{x}} = \left(\frac{1}{\tilde{x}} - \frac{1}{\tilde{x}+1} - \frac{1}{\tilde{x}-1}\right) + \e \left[\left(\frac{1}{\tilde{x}} - \frac{2}{\tilde{x}+1}\right) \, I_2 + \frac{1}{2} \left(\frac{1}{\tilde{x}+1} - \frac{1}{\tilde{x}-1} \right) \, I_1 \right] \,.
\end{equation}
By imposing regularity in this point, we obtain
\begin{align}
\lim_{\tilde{x}\to 1} \, (\tilde{x}-1) \frac{\p I_2}{\p\tilde{x}} &= \lim_{\tilde{x}\to 1} \left[-I_2 -\frac{1}{2} \, \e \, I_1 \right] \stackrel{!}{=} 0 \,, \nonumber \\
\leftrightarrow \lim_{\tilde{x}\to 1} I_2 &= \lim_{\tilde{x}\to 1} \left[-\frac{1}{2} \, \e \, I_1 \right] \,,
\end{align}
and combining this with Eqs.~\eqref{sunrisecan} and \eqref{tadpolecan} finally leads to
\begin{equation}
\lim_{\tilde{x}\to 1} M_2^{(k)} = 0 \qquad \forall \, i \,.
\label{exampleboundary}
\end{equation}
Bearing this in mind, we can start integrating the canonical differential equation~\eqref{deqexample} through the improper approach described in Section~\ref{sec:int}. Due to the fact that the tadpole result is of order $\e^0$, the first non-vanishing order of $M_2$ arises for $i=1$,
\begin{align}
M_2^{(1)} &= \int \d \tilde{x} \left[\left(\frac{1}{\tilde{x}} - \frac{2}{\tilde{x}+1} \right) \, M_2^{(0)} + \frac{1}{\tilde{x}} \, M_1^{(0)} \right] \nonumber \\
&= \int \frac{\d \tilde{x}}{\tilde{x}} \nonumber \\
&= G(0;\tilde{x}) + c_2^{(1)} \,,
\end{align}
where we applied the definition of the MPLs in Eq.\eqref{defmpl}. The boundary constant $c_2^{(1)}$ is then fixed by using Eq.~\eqref{exampleboundary}:
\begin{equation}
c_2^{(1)} = \lim_{\tilde{x}\to 1} \left[ M_2^{(1)} - G(0;\tilde{x}) \right] = -G(0;1) = -\log 1 = 0 \,.
\end{equation}
Since the boundary value $\tilde{x}^c$ within the limit $\tilde{x}\to\tilde{x}^c$ is independent of any other variable, limiting identities of the kind~\eqref{trafo1} are not necessary here. Since we are dealing with a one-variable problem, this is trivially the case here, but it may change for problems with more than one scale.\\
As a next step, we determine the coefficient of order $i=2$ in the Laurent expansion, which proceeds in an iterative manner by using the result of order $i=1$:
\begin{align}
M_2^{(2)} &= \int \d \tilde{x} \left[\left(\frac{1}{\tilde{x}} - \frac{2}{\tilde{x}+1} \right) \, M_2^{(1)} + \frac{1}{\tilde{x}} \, M_1^{(1)} \right] \nonumber \\
&= \int \d \tilde{x} \, G(0;\tilde{x}) \left(\frac{1}{\tilde{x}} - \frac{2}{\tilde{x}+1} \right) \nonumber \\
&= G(0,0;\tilde{x}) - 2 \, G(-1,0;\tilde{x}) + c_2^{(2)} \,.
\end{align}
The boundary constant is obtained through
\begin{align}
c_2^{(2)} &= \lim_{\tilde{x}\to 1} \left[ M_2^{(2)} - G(0,0;\tilde{x}) + 2 \, G(-1,0;\tilde{x})\right] \nonumber \\
&= - \frac{1}{2} \, \log^2 1 + 2 \, \left(-\frac{\pi^2}{12}\right) = -\frac{\pi^2}{6} \,.
\end{align}
The coefficient at order $\e^3$ is computed in the exact same way, so that the overall expression up to weight three can be written as
\begin{align}
M_2 &= \e \, G(0;\tilde{x}) \nonumber \\
&\quad + \e^2 \left[ G(0,0;\tilde{x}) - 2 \, G(-1,0;\tilde{x}) -\frac{\pi^2}{6} \right] \nonumber \\
&\quad + \e^3 \left[ G(0,0,0;\tilde{x}) - 2 \, G(0,-1,0;\tilde{x}) - 2 \, G(-1,0,0;\tilde{x}) + 4 \, G(-1,-1,0;\tilde{x}) \right. \nonumber \\
&\qquad\quad \left. - \frac{\pi^2}{6} \, G(0;\tilde{x}) + \frac{\pi^2}{3} \, G(-1;\tilde{x}) - 2\,\zeta_3 \right] + \mathcal{O}\left(\e^4\right) \,.
\label{sunriseexact}
\end{align}
Due to its length, we refrain from quoting the weight-four result here.

\chapter{Two-Loop Corrections to the Decay $\boldsymbol{H\to Z\gamma}$ with Full Quark Mass Dependence}
\chaptermark{Two-Loop Corrections to the Decay $H\to Z\gamma$}
\label{chap:hza}

In the Standard Model, the rare Higgs boson decay $H\to Z\gamma$ is forbidden at tree level and it is loop-mediated through a $W$ boson or a heavy quark. In this chapter, we analytically compute the QCD correction to the heavy quark loop, confirming earlier purely numerical results, that were obtained for on-shell renormalization.\\
The outline is as follows: After a short introduction, we establish the notation and discuss the different contributions to the $H\to Z\gamma$ decay as well as the kinematics in Section~\ref{sec:notation}. Section~\ref{sec:calc} describes the calculation of the amplitude, including a detailed discussion of the relevant two-loop three-point integrals and of the renormalization. On top of that, we elaborate on the small quark mass expansion of the decay matrix element, which turns out to contain only single-logarithmic contributions at each perturbative order in contrast to the double logarithms observed in $H\to \gamma\gamma$. In Section~\ref{sec:results}, we investigate the numerical interplay of bottom and top quark contributions and discuss the dependence of the result on the renormalization scheme. We close this chapter by concluding in Section~\ref{sec:conc2}.

\section{Introduction}

It is anticipated that the upcoming LHC data taking period at higher energy and luminosity will provide more precise measurements and open up new observables that were previously inaccessible. In this context, the different decay modes of the Higgs boson play a vital role. The decays to massive gauge bosons and to fermions are allowed at tree level and provide direct measurements of the Higgs boson couplings.  The rare decays $H\to \gamma\gamma$ and $H\to Z\gamma$ are forbidden at tree level, they are mediated through loops containing massive particles~\cite{Ellis:1975,Cahn:1978,Bergstrom:1985}. As such, they are more sensitive to new physics effects from high energy scales than the tree-level dominated decay modes.\\
The $H\to \gamma\gamma$ decay mode was among the most significant signatures in the Higgs boson discovery~\cite{Aad:2012,Chatrchyan:2012}, it has been measured in the meantime with a relative precision of below 20~percent~\cite{Aad:2014a,Khachatryan:2014}. The branching ratio for $H\to Z\gamma$, including the leptonic branching ratio of the $Z$ boson, is considerably lower and only upper bounds could be established on it up to now~\cite{Aad:2014b,Chatrchyan:2013}. From Fig.~\ref{fig:run2}, we can infer that new experimental data on $H\to Z\gamma$ is expected beyond Run~2 of the LHC, starting from 2023. Once analyzed in detail, this decay will provide access to a broader spectrum of observables than $H\to \gamma \gamma$, since the decay of the $Z$ boson to leptons will enable the study of spin-dependent particle correlations. It should be noted that the decay $H\to Z\gamma$ will be identified through a mass cut on the pair of decay leptons, and that it should be considered to be a pseudo-observable~\cite{Passarino:2013}.\\
Higher-order QCD corrections to $H\to Z\gamma$ from gluon exchange in the top quark loop were derived in Ref.~\cite{Spira:1991} by performing a purely numerical evaluation of the relevant two-loop integrals in terms of five-dimensional Feynman parameter representations. The results derived in Ref.~\cite{Spira:1991} use an on-shell renormalization for the top quark mass and the Yukawa coupling. Electroweak corrections to this decay are not known at present~\cite{Passarino:2013,Denner:2011}. It is our aim to rederive the QCD corrections to $H\to Z\gamma$ in an analytical form and to quantify uncertainties on them arising from scheme and scale dependence.\\
Besides its phenomenological implications for $H\to Z\gamma$, our calculation also provides an important subset of two-loop integrals relevant to the two-loop amplitudes of Higgs-plus-jet production with full top quark mass dependence. The calculation of NLO QCD corrections with exact top quark mass dependence is recognized as high-priority aim~\cite{Dittmaier:2012,Heinemeyer:2013} and will be discussed in Chapter~\ref{chap:hj}, with the integrals derived here being an important step towards it.

\section{Kinematics and Notation}
\label{sec:notation}

The Standard Model does not allow a tree-level coupling of the process
\begin{equation}
H(q_4) \to Z(q_{12}) \, \gamma(q_3) \qquad (q_{12} \equiv q_1+q_2) \,,
\label{process}
\end{equation}
where the momenta $q_i$ of the external particles are chosen such that they fit the notation established for Higgs-plus-jet production in Chapter~\ref{chap:hj}. It is instead mediated through a virtual particle loop, containing either a $W$ boson or a massive quark~\cite{Cahn:1978,Bergstrom:1985}. The Lorentz structure of its Feynman amplitude is constrained by gauge invariance to contain only a single scalar form factor $\mathcal{A}$. The generic equation emerging from inserting Eq.~\eqref{decomposition} into Eq.~\eqref{tensor1} simplifies in this case to
\begin{equation}
{\cal M} = \mathcal{A}\, \e_{1,\mu}(q_{12},\lambda_1) \, \e_{2,\nu}(q_3,\lambda_2) \, \frac{P^{\mu\nu}}{P^2} \,,
\label{tensamp}
\end{equation}
where the projector $P^{\mu\nu}$ defined in Eqs.~\eqref{proj1} and \eqref{proj2} and its absolute square are given by
\begin{align}
P^{\mu\nu} &= q_3^\mu \, q_{12}^\nu - (q_{12}\cdot q_3) \, g^{\mu \nu} \,, \\
P^2 &= \frac{1-\e}{2} \, \left( m_H^2-m_Z^2\right)^2 \,.
\end{align}
Due to the absence of external partons in the process, there is no need to consider real radiation contributions from Eq.~\eqref{subtraction} and Eq.~\eqref{cs2} is equally valid for the full decay width. By inserting the phase space factor for a two-particle decay and the flux factor into Eq.~\eqref{cs2},
\begin{align}
\Phi_m &= \frac{\int \d\Omega}{32 \, \pi^2 \, m_H^2} = \frac{1}{8 \, \pi \, m_H^2} \,, \\
\mathcal{F} &= \frac{m_H^2-m_Z^2}{2 \, m_H} \,,
\end{align}
the decay width can be written as
\begin{equation}
\Gamma = \frac{G_F^2 \, \alpha \, m_W^2}{4 \, m_H^3 \left(m_H^2-m_Z^2\right)} \, |A|^2 \,,
\label{gamma}
\end{equation}
where we substituted the Feynman amplitude from Eq.~\eqref{tensamp} as the last missing ingredient. This relation, which includes the Higgs and $Z$~boson masses $m_H$ and $m_Z$, respectively, is in agreement with the formulae available in the literature\footnote{We would like to point out that there is a misprint in Eq.~(7) of Ref.~\cite{Spira:1991}, which has to be multiplied by~$1/4$.}~\cite{Bergstrom:1985,Spira:1991,Djouadi:2005}. Note that in the transition from Eq.~\eqref{tensamp} to Eq.~\eqref{gamma}, we redefined the form factor $\mathcal{A} = c\,A$ by scaling out an overall normalization factor $c$ emerging from the Feynman rules from Section~\ref{sec:rules}. Depending on the particle coupled to the external Higgs boson, this amplitude can be further decomposed into contributions from the $W$~boson and the fermions~$q$:
\begin{equation}
A = c_W A_W + \sum_q c_q A_q\,.
\end{equation}
The coupling factors are
\begin{equation}
c_W = \mathrm{cos} \, \theta_w \,, \qquad c_q = N_c \, \frac{2 \, Q_q \left(I^3_q - 2 \, Q_q \, \mathrm{sin}^2 \, \theta_w \right)}{\mathrm{cos} \, \theta_w}  \;.
\end{equation}
Due to the mass hierarchy of the particles involved, we will only consider the dominant pieces coming from the $W$ boson, the top quark and the bottom quark, i.e. the theory of $N_F=6$ active quark flavors is split into $N_F=N_h+N_l$ with $N_h=2$ heavy and $N_l=4$ light quarks. $N_l$ and thus $N_F$ would only enter in a potential renormalization of the gauge coupling, which is however not required since the amplitudes considered here correspond at most to leading order in $\alpha_s$.\\
The Born-level contribution to the amplitude arises at one loop, it is written as:
\begin{equation} 
A^{(1)} = c_W A^{(1)}_W + c_t A^{(1)}_t + c_b A^{(1)}_b\,.
\end{equation}
Higher-order perturbative corrections are obtained by a loop expansion of the amplitude~$A$ as suggested by Eq.~\eqref{seriesff}. NLO QCD corrections affect only $A_t$ and $A_b$, they correspond to two-loop graphs with an internal mass:
\begin{equation}
A_q(m_H,m_Z,m_q,\alpha_s,\mu) = A^{(1)}_q (m_H,m_Z,m_q) + \frac{\alpha_s(\mu)}{\pi} A^{(2)}_q(m_H,m_Z,m_q,\mu) \,.
\label{amp}
\end{equation}
For the dominant top quark contribution, the two-loop correction $A^{(2)}_t$ was computed numerically (based on the Feynman parameter representation of the amplitude) using an on-shell renormalization for the quark mass and the Yukawa coupling in Ref.~\cite{Spira:1991}. Example diagrams for both the one- and the two-loop amplitudes are depicted in Fig.~\ref{fig:diagrams}$(a)$.
\begin{figure}[t]
\begin{center}
\includegraphics[width=0.92\textwidth]{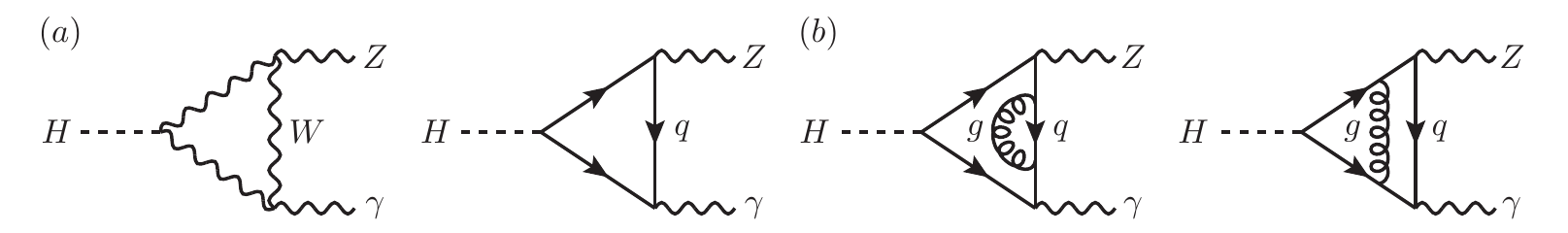}
\caption[Example diagrams for the computation of the $H\to Z\,\gamma$ amplitudes]{\textbf{$\boldsymbol{H\to Z\,\gamma}$ example diagrams}\\$(a)$ for the computation of the one-loop amplitudes $A^{(1)}_W$ and $A^{(1)}_q$ and\\$(b)$ for the computation of the two-loop amplitude $A^{(2)}_q$.}
\label{fig:diagrams}
\end{center}
\end{figure}

\section{Calculation of the Amplitude}
\label{sec:calc}

In order to compute the one-loop amplitudes $A^{(1)}_W$ and $A^{(1)}_q$ as well as the two-loop QCD contribution $A^{(2)}_q$ to the quark-mediated amplitude for $H\to Z\gamma$, we project all relevant Feynman diagrams generated by \textsc{Qgraf} onto the tensor structure~\eqref{tensamp} using \textsc{Form}. The resulting Feynman integrals have a remarkably simple numerator structure ($s\leq 1$) and are reduced to a set of MIs with the help of the \textsc{Reduze} code. After the reduction, the amplitude can be expressed in terms of a certain number of MIs depending on the loop~order.\\
As described in Section~\ref{sec:reduction}, the procedure outlined so far requires defining integral families, onto which the MIs can be mapped. In fact, the MIs required for the $H\to Z\,\gamma$ decay rate form a subset of those needed for the amplitude of Higgs-plus-jet production presented in Chapter~\ref{chap:hj}. Therefore, we can map all MIs introduced here onto the integral families specified in Table~\ref{tab:hjtopo} of Section~\ref{sec:hjdeq}, provided that the momentum assignment in Eq.~\eqref{process} is respected. More precisely, the MIs required for the computation of the $H\to Z\,\gamma$ amplitude contain at most one massless propagator and thus form an independent sub-tree within integral family~$A$ of Table~\ref{tab:hjtopo}. At the level of the two-loop Feynman diagrams, this connection can be identified by pinching the leftmost propagator of the diagram associated with the sector~$A_{7,247}$ in Fig.~\ref{fig:hjdiag}, which depicts a top-level topology of Higgs-plus-jet production within family~$A$. As a consequence, the massless legs of momenta $q_1$ and $q_2$ collapse to a massive one. This massive leg can be identified with the $Z$ boson, leading to the rightmost diagram in Fig.~\ref{fig:diagrams}.\\
Each of the MIs emerging from the reduction has a specific mass dimension, which can be scaled out by multiplying with the appropriate power of the mass $m_i$ running in the loop thanks to the scaling relation~\eqref{scaling}. The resulting dimensionless integrals are only functions of the mass ratios $x=m_Z^2/m_i^2$, $h=m_H^2/m_i^2$, which coincide with the definitions in Eqs.~\eqref{mandelstam} and \eqref{ratios} for $M\equiv m_H$ and $m\equiv m_i$. This establishes another connection to Higgs-plus-jet production, this time at the level of the kinematic configuration, since the set of variables presented in Eqs.~\eqref{mandelstam} and \eqref{ratios} are exactly the ones employed in Chapter~\ref{chap:hj}. We parametrize the dependence on the dimensionless ratios $x$ and $h$ by using the Landau-type variables introduced in Eq.~\eqref{landau1}:
\begin{equation}
h = -\frac{(1-\tilde{h}_i)^2}{\tilde{h}_i}\,, \qquad x = -\frac{(1-\tilde{x}_i)^2}{\tilde{x}_i} \qquad \left(i=W,q\right)\,.
\label{landau2}
\end{equation}
For the sake of readability, we drop the subscript $q$ of these variables in the following whenever we deal with quark-mediated amplitudes, i.e. $\tilde{h}_q \equiv \tilde{h}$, $\tilde{x}_q \equiv \tilde{x}$. 

\subsection{The One-Loop Amplitude}

The Born-level one-loop amplitudes for the contributions of $W$ bosons $A^{(1)}_W$ and heavy quarks $A^{(1)}_q$ to  $H\to Z\gamma$ were derived in Ref.~\cite{Cahn:1978,Bergstrom:1985}. In terms of the Landau variables introduced above, these results read
\begin{align}
A^{(1)}_W = \frac{4\,i\,S_\e\,m_W^4}{\tilde{h}_W\tilde{x}_W^2} &\left[\left\{(\tilde{h}_W^2+1) (\tilde{x}_W^2+1) - 4 \tilde{h}_W (\tilde{x}_W+1)^2 \right\} \, \left\{ \frac{(\tilde{h}_W-\tilde{x}_W) (\tilde{h}_W\tilde{x}_W-1)}{\tilde{h}_W} \right. \right. \nonumber \\
	&\left. \quad- \frac{(\tilde{h}_W+1) (\tilde{x}_W-1)^2}{\tilde{h}_W-1} \, \log(\tilde{h}_W) + (\tilde{x}_W^2-1) \, \log(\tilde{x}_W) \right\} \nonumber \\
	&\,+\left\{ (\tilde{x}_W + \tilde{x}_W^2 (\tilde{x}_W+4)) (\tilde{h}_W^2+1) - 2\tilde{h}_W (\tilde{x}_W^2-1)^2 \right\} \, \nonumber \\
	&\,\left. \quad \left\{ \log^2(\tilde{h}_W) - \log^2(\tilde{x}_W) \right\} \right] + {\cal O}(\e) \,, \\
A^{(1)}_q = i\,S_\e\,m_q^3\,y_q\,v&\left[8\,\left\{\tilde{h}+\frac{1}{\tilde{h}}-\left(\tilde{x}+\frac{1}{\tilde{x}}\right)\right.\right. \nonumber \\
	&\left. \qquad- \frac{(\tilde{h}+1) (\tilde{x}-1)^2}{\tilde{x}(\tilde{h}-1)} \, \log(\tilde{h}) + \left(\tilde{x}-\frac{1}{\tilde{x}}\right) \, \log(\tilde{x}) \right\} \nonumber \\
	&\,\left.-2\left\{\tilde{h}+\frac{1}{\tilde{h}}-\tilde{x}-\frac{1}{\tilde{x}}+4\right\} \left\{ \log^2(\tilde{h}) - \log^2(\tilde{x}) \right\} \right] + {\cal O}(\e)
\label{LO}
\end{align}
with both the Standard Model Higgs vacuum expectation value~$v$ and the Yukawa coupling~$y_q$ associated with the quark $q$ defined in Eq.~\eqref{yukawa}. The normalization factor
\begin{equation}
S_\e = i \, S_\Gamma \, \Gamma\left(1+\e\right) \, \Gamma\left(1-\e\right) \, \left(\frac{\mu_0^2}{m_i^2}\right)^\e = \frac{i \, \Gamma\left(1+\e\right)}{16\pi^2} \, \left(\frac{4\pi \mu_0^2}{m_i^2}\right)^\e
\label{norm}
\end{equation}
arises from the integration measure $\int \d^D k/(2\pi)^D$ of the MIs in $D=4-2\e$ dimensions, where $\mu_0$ is the mass parameter of dimensional regularization introduced in Eq.~\eqref{mu0} and $S_\Gamma$ is defined in Eq.~\eqref{SGamma}.

\subsection{Differential Equations and Master Integrals}
\label{sec:hzadeq}

The two-loop amplitude for the quark contribution $A^{(2)}_q$ has been computed purely numerically in terms of a five-dimensional Feynman parameter integral in Ref.~\cite{Spira:1991}. We derive an analytical expression for this amplitude, through a reduction of all two-loop integrals to a set of MIs.\\
To compute the two-loop MIs, we use the method of differential equations described in Chapter~\ref{chap:workflow2}. In this method, differential equations in internal masses and external invariants are derived for each integral by performing the differentiation on the integrand, which is then related to the original MI by the IBP identities. More precisely, the differential equations in the Landau variables can be derived from the IBP identities required for Higgs-plus-jet production by using the differential operators presented in Eq.~\eqref{deqmandelstam} together with the relations
\begin{align}
\frac{\p}{\p \tilde{x}} &= m_q^2 \, \frac{1-\tilde{x}^2}{\tilde{x}^2} \, \left(\frac{\p}{\p \bar{s}} - \frac{\p}{\p \bar{t}}\right) \,, \nonumber \\
\frac{\p}{\p \tilde{h}} &= m_q^2 \, \frac{1-\tilde{h}^2}{\tilde{h}^2} \, \frac{\p}{\p \bar{t}} \,, \nonumber \\
m_q^2 \, \frac{\p}{\p m_q^2} &= \frac{\alpha}{2}
\end{align}
following from the chain rule. With this, we obtain inhomogeneous differential equations in either Landau variable, plus a trivial homogeneous equation in $m_q$ for each integral. \\
The differential equations are solved in a bottom-up approach as explained in Section~\ref{sec:int}, i.e. starting from the MIs with the lowest number of different propagators because they will show up in differential equations of higher topologies.\\
Thanks to the non-linear transformation~\eqref{landau2}, the coefficients of the individual MIs in the homogeneous and inhomogeneous terms of the differential equations turn from square root expressions in the Higgs-plus-jet case into rational functions of $\tilde{x}$ and $\tilde{h}$. Upon partial fractioning, only a limited number of polynomials in $\tilde{x}$ and $\tilde{h}$ appear, which form the alphabet associated with this set of MIs:
\begin{equation}
\{ l_1,\ldots, l_{12} \} 
\label{denom}
\end{equation}
with
\begin{align*}
l_1 &= \tilde{x} \,, \\
l_2 &= \tilde{x}+1 \,, \\
l_3 &= \tilde{x}-1 \,, \\
l_4 &= \tilde{h} \,, \\
l_5 &= \tilde{h}+1 \,, \\
l_6 &= \tilde{h}-1 \,, \\
l_7 &= \tilde{h}-\tilde{x} \,, \\
l_8 &= \tilde{h} \, \tilde{x}-1 \,, \\
l_9 &= \tilde{h}^2-\tilde{h} \, \tilde{x}-\tilde{h}+1 \,, \\
l_{10} &= \tilde{h}^2 \, \tilde{x}-\tilde{h} \, \tilde{x}-\tilde{h}+\tilde{x} \,, \\
l_{11} &= \tilde{x}^2-\tilde{h} \, \tilde{x}-\tilde{x}+1 \,, \\
l_{12} &= \tilde{x}^2 \, \tilde{h}-\tilde{h} \, \tilde{x}-\tilde{x}+\tilde{h} \,.
\end{align*}
If the full system of differential equations for all 39 MIs (written as 39-component vector~$\vec{M}$) can be cast into the canonical form of Eq.~\eqref{canonicaldeq}, it can be written as
\begin{equation}
\d \vec{M}(D;\tilde{x},\tilde{h}) = \e \sum_{k=1}^{12} A_k \, \d\log (l_k) \, \vec{M} (D;\tilde{x},\tilde{h})\,,
\label{canon}
\end{equation}
where the matrices $A_k$ contain only rational numbers. In this case, two important features can be exploited. First, the differential equations can be integrated order by order in~$\e$ in terms of the MPLs introduced in Eq.~\eqref{defmpl}. Second, the results will be expressed as a linear combination of MPLs of homogeneous weight. With the methods described in Section~\ref{sec:basischoice}, we arrive at a total differential of the form \eqref{canon} starting from a triangular Laporta basis $\vec{I}(0;\tilde{x},\tilde{h})$ in $D=4$, which is depicted in Fig.~\ref{fig:hzamaster} and defined in Appendix~\ref{sec:hzalaporta}. Subsequently, we apply the algorithm described in Section~\ref{sec:triangulartocanonical}. We would like to point out that crossings of MIs of lower-level topologies appear as subsectors in the differential equations of higher topologies. These MIs can be either computed by explicitly applying the crossing to the result of the original MIs, or by considering the crossed MIs as independent. Since we are forced to treat them as independent in the context of Higgs-plus-jet production for reasons outlined in Chapter~\ref{chap:hj}, we will do the same here, i.e. these 11 crossed MIs are part of the 39-component vector $\vec{M}$ within Eq~\eqref{canon}. The definition of the Laporta basis~$\vec{I}$ and the canonical basis~$\vec{M}$ of both the original MIs and their crossings are given in Appendix~\ref{chap:hzaMIs}.\\
The alphabet~\eqref{denom} is not linear in the Landau variables, so that we further decompose it to enable the integration in either $\tilde{x}$ or $\tilde{h}$, which yields a solution up to an integration constant that only depends on the other variable. Let us stress that the non-linear letters always occur in only one of the two variables for a given equation, which is a necessary condition for the integration in terms of MPLs as explained in Section~\ref{sec:int}. The remaining boundary value is then determined by imposing regularity in special kinematic points, where the integrals are known to be regular from physical arguments. In our case, these points are given by
\begin{align}
\tilde{h} = 1 \qquad &\leftrightarrow \qquad m_H^2 = 0 \,, \nonumber \\
\tilde{x} = 1 \qquad &\leftrightarrow \qquad m_Z^2 = 0 \,, \nonumber \\
\tilde{h} = \tilde{x} \qquad &\leftrightarrow \qquad m_H^2 = m_Z^2 \,, \nonumber \\
\tilde{h} = \frac{1}{\tilde{x}} \qquad &\leftrightarrow \qquad m_H^2 = m_Z^2 \,,
\label{hzabc}
\end{align}
i.e. they correspond to the limit where the masses of the external particles either vanish or coincide. Choosing one of these points such that the rational prefactors in Eqs.~\eqref{appendix} are equal to zero considerably reduces the complexity of the integration constants. This is actually the case for the rational prefactors of all MIs in the first two boundary points specified in Eq.~\eqref{hzabc}, which correspond to the infinite quark mass limit $m_q\to\infty$, where all Laporta integrals turn into tadpoles and thus are regular. Nevertheless, we use the last two relations of Eq.~\eqref{hzabc} to fix the boundary constants of some of the MIs for $H\to Z\,\gamma$ in order to demonstrate the flexibility of the method: These MIs will reappear in the context of Higgs-plus-jet production in Chapter~\ref{chap:hj}, and applying different boundary conditions in both cases serves as a very strong check.\\
By taking limits in these kinematic points, we are left with MPLs that contain the same variable $y\in\{\tilde{x},\tilde{h}\}$ both in the argument and in the indices. In order to simplify the result and to obtain a unique representation, we rely on the symbol and coproduct formalism as explained in Section~\ref{sec:symbol} to perform transformations of the type~\eqref{trafo1}. In doing so, we end up with GHPLs up to weight four, which are given by
\begin{align}
G\left(a_1,\ldots,a_n;\tilde{h}\right) \quad &\text{with} \quad a_i \in \{ 0,\pm1,\tilde{x},\frac{1}{\tilde{x}},J_x,\frac{1}{J_x},K_x^\pm, L_x^\pm \} \,, \nonumber \\
G\left(b_1,\ldots,b_n;\tilde{x}\right) \quad &\text{with} \quad b_i \in \{ 0,\pm1,c,\bar{c} \} \,,
\label{alphabet}
\end{align}
where 
\begin{align}
c &= \frac{1}{2} \left(1 + i\sqrt{3}\right) \,, \nonumber \\
J_x &= \frac{\tilde{x}}{1 - \tilde{x} + \tilde{x}^2} \,, \nonumber \\
K_x^\pm &= \frac{1}{2} \left(1 + \tilde{x} \pm \sqrt{\text{--}3 + 2\,\tilde{x} + \tilde{x}^2}\right) \,, \nonumber \\
L_x^\pm &= \frac{1}{2\,\tilde{x}} \left(1 + \tilde{x} \pm \sqrt{1 + 2\,\tilde{x} - 3\,\tilde{x}^2}\right)
\label{hzamplindices}
\end{align}
for the underlying set of MIs. Further transformations of the kind~\eqref{trafo2} become necessary when the integration is performed in a different variable compared to the integration of a MI of a subtopology which enters the differential equation under consideration.\\
It remains to comment on one issue: Transformations of the types \eqref{trafo1} and \eqref{trafo2} are performed for all MIs except for $M_{43}\text{--}M_{46}$, where MPLs of the form
\begin{equation}
G\left(w_1(y),\ldots,w_n(y);y\right)
\label{nontrans}
\end{equation}
with $w_i=\{K_y^\pm,L_y^\pm\}$ occur. The transformations were not necessary in this case because the results of these integrals do not enter the differential equation of any MI of higher topology. However, transformations of these types could become desirable when the Higgs-plus-jet four-point functions are computed, where these four coupled MIs appear as subtopologies. As pointed out in Section~\ref{sec:symbol}, the symbol and coproduct formalism is restricted to at most rational dependence of the index vector $w_i$ on the kinematic invariants $y$. Therefore, a multivariate extension of the concept of generalized weights should be attempted~\cite{vonManteuffel:2013}, by working on non-linear indices of the form~\eqref{denom} rather than on linear ones as in Eq.~\eqref{alphabet}. As outlined in Chapter~\ref{chap:hj}, we will follow other methods to solve this issue, so that the unavailability of such an extension is not a problem.
\begin{figure}[H]
\begin{center}
\includegraphics[width=\textwidth]{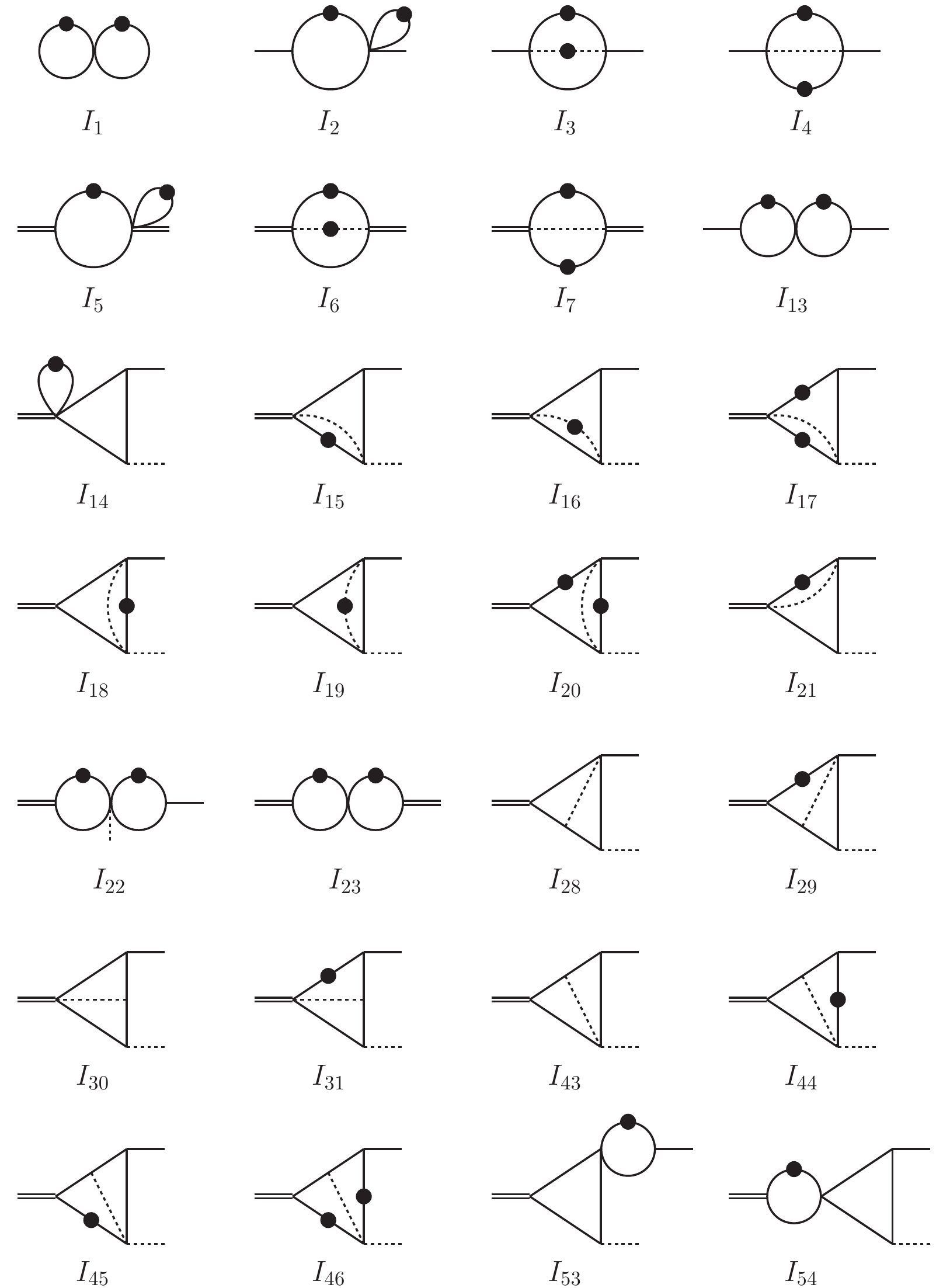}
\caption[Two-loop Laporta MIs for the calculation of $A^{(2)}_q$]{\textbf{Two-loop Laporta MIs for the calculation of $\boldsymbol{A^{(2)}_q}$},\\
which are defined in Appendix~\ref{sec:hzalaporta}. Dashed lines are massless, whereas internal solid lines denote propagators with mass $m_q$. Double and solid external lines correspond to virtualities $m_H^2$ and $m_Z^2$, respectively. Dotted propagators are taken to be squared. Note that the numbering is chosen such that it fits the corresponding definition of the MIs for Higgs-plus-jet production in Appendix~\ref{sec:hjlaporta}.}
\label{fig:hzamaster}
\end{center}
\end{figure}\noindent
We would like to state that the results of the MIs were checked in several ways. We performed transformations of the type \eqref{trafo2} and verified that the solution fulfills the differential equation in the other variable. This check works only up to a constant, which is why we compared each MI numerically against \textsc{SecDec}~\cite{Borowka:2015} and found agreement to high precision. Beyond that, we used different boundary conditions compared to the computation of these MIs in the context of Higgs-plus-jet production, where they appear as a subset, and verified that both approaches lead to the same numerical results. Finally, the results of our MIs were confirmed by an independent calculation of a collaborator~\cite{AvM1} in terms of classical polylogarithms as defined in Eqs.\eqref{defpolylog} and \eqref{defli22}, the procedure of which will be elaborated on in Chapter~\ref{chap:workflow3}. The analytic expressions of the MIs are rather lengthy and will not be reproduced here. They are available in \textsc{Mathematica} and \textsc{Form} format together with the arXiv submission of Ref.~\cite{Gehrmann:2015a}\footnote{We would like to point out that some definitions of the canonical MIs in Ref.~\cite{Gehrmann:2015a} deviate from the ones in Appendix~\ref{sec:hzacan} by a minus sign, which therefore holds for the corresponding analytic expressions as well.}.

\subsection{Calculation of the Two-Loop Amplitude}
\label{sec:hza2l}

For the two-loop amplitude, six generic Feynman diagrams and their permutations have to be evaluated. They emerge from the one-loop diagram in Fig.~\ref{fig:diagrams}$(a)$ by attaching one gluon propagator to the fermion lines in every possible way, which is depicted in Fig.~\ref{fig:diagrams}$(b)$. After the manipulations described in the beginning of Section~\ref{sec:calc} and after inserting the analytic results of the master integrals, we are left with the unrenormalized two-loop amplitude.\\
For its renormalization, we consider three different prescriptions:
\begin{itemize}
\item[(a)] quark mass and Yukawa coupling in the OS scheme.
\item[(b)] quark mass in the OS scheme, Yukawa coupling in the $\MS$ scheme.
\item[(c)] quark mass and Yukawa coupling in the $\MS$ scheme.
\end{itemize}
Since the two-loop amplitude is the leading order in both $\alpha_s$ and $\alpha$, no renormalization of the gauge couplings is required. Moreover, we do not need to perform wave function renormalization due to the absence of external partons. Although the numerical result of Ref.~\cite{Spira:1991} was only derived in scheme (a), repeating the calculation in the schemes~(b) and (c) is motivated by the observation of Ref.~\cite{Spira:1995}, which points out that it is more appropriate to renormalize the $Hb\bar{b}$ Yukawa coupling in the $\MS$ scheme for physical values of $m_H\approx 125~\mathrm{GeV}$.\\
The differences of the OS and the $\MS$ scheme have been explained in Section~\ref{sec:schemes} in great detail. From there it becomes clear that all three prescriptions yield the same pole parts of the renormalization counter terms and produce finite expressions for the renormalized amplitude. They are related by finite scheme transformations, which is why we choose to compute the renormalized amplitude in scheme (a) and use it to derive the results in schemes~(b) and (c). In the pure OS scheme, the quantity
\begin{equation}
- 16\,i\,\pi^2 \,S_\e \, \left(\frac{1}{m_q}\,\delta m_{\mathrm{OS}}\,C^{(1)}_q + Z_{\mathrm{OS}}\,A^{(1)}_q \right)
\label{renormOS}
\end{equation}
has to be added to the unrenormalized two-loop amplitude in order to remove its divergences, where the Yukawa and mass renormalization constants $Z_{\mathrm{OS}}$ and $\delta m_{\mathrm{OS}}=m_q \, Z_{\mathrm{OS}}$, respectively, result from Eq.~\eqref{ZOS}. Note that Eq.~\eqref{renormOS} requires the calculation of the one-loop amplitude $A^{(1)}_q$ and the mass counterterm $C^{(1)}_q$ up to $\mathcal{O}(\e)$. $C^{(1)}_q$ can be computed from the sum of the three diagrams depicted in Fig.~\ref{fig:hzact}.\\
\begin{figure}[tb]
\begin{center}
\vspace{-\ht\strutbox}\includegraphics[width=0.6\textwidth]{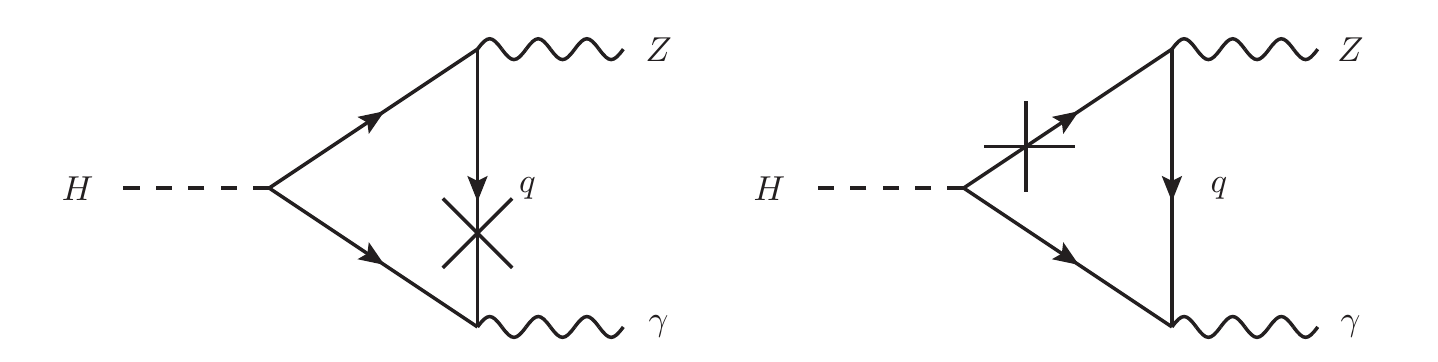}
\hspace{-0.5cm}
\vspace{-\ht\strutbox}\includegraphics[width=0.31\textwidth]{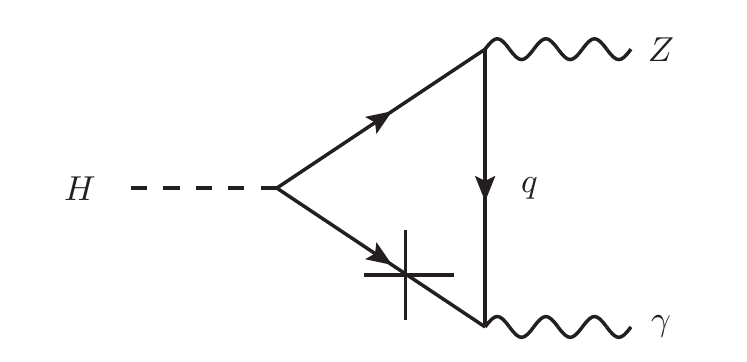}
\caption[Counterterm diagrams for the OS renormalization of the $H\to Z\,\gamma$ two-loop amplitude]{\textbf{Counterterm diagrams for the OS renormalization of the $\boldsymbol{H\to Z\,\gamma}$ two-loop amplitude.}\\These diagrams are obtained by putting a mass insertion into one of the fermion lines within Fig.~\ref{fig:diagrams}$(a)$ and their sum yields $C^{(1)}_q$ in Eq.~\eqref{renormOS}.}
\label{fig:hzact}
\end{center}
\end{figure}\noindent
Next, we express the OS quantities $M_q$ and $Y_q$ in terms of $\MS$ quantities 
$\overline{m}_q$ and $\overline{y}_q$ at a particular matching scale $\mu_m$ using the standard relations from Eq.~\eqref{OStoMS}. We perform this matching at the scale of the running $\MS$ quark mass $\overline{m}_q$. Starting from the OS result $A^{(2,a)}_q(m_H,m_Z,M_q)$, these scheme transformations 
induce finite shifts in the amplitudes of the prescriptions (b) and (c):
\begin{align} 
A^{(2,b)}_q(m_H,m_Z,\overline{m}_q,\mu) &= A^{(2,a)}_q(m_H,m_Z,\overline{m}_q(\mu)) +  \Delta \cdot \, A^{(1)}_q(m_H,m_Z,\overline{m}_q(\mu)) \nonumber \,,\\
A^{(2,c)}_q(m_H,m_Z,\overline{m}_q,\mu) &= A^{(2,b)}_q(m_H,m_Z,\overline{m}_q(\mu)) + \overline{\Delta} \cdot
\left. \frac{\p  A^{(1)}_q(m_H,m_Z,M_q)}{\p
M_q} \right|_{M_q=\overline{m}_q(\mu)} \,.
\end{align}
In practice, the coefficient $\overline{\Delta}$ emerges by making the following replacements in Eq.~\eqref{LO}, where the Landau variables $\bar{x}$ and $\bar{h}$ are defined according to Eq.~\eqref{landau2} with \mbox{$m_q = \overline{m}_q(\mu)$}:
\begin{align}
\tilde{x} &= \bar{x} - 2 \, \Delta \, \bar{x} \, \frac{\bar{x}-1}{\bar{x}+1} \,, \nonumber \\
\tilde{h} &= \bar{h} - 2 \, \Delta \, \bar{h} \, \frac{\bar{h}-1}{\bar{h}+1} \,.
\end{align}
The amplitudes in the schemes (a), (b) and (c) have a common polynomial structure, which contains only a limited number of combinations of the denominators~\eqref{denom}:
\begin{align}
A^{(2)}_q = 16\pi^2\,S_\e^2\,m_q^3\,y_q\,v\,C_F\,\cdot &\left[ \frac{c_1}{l_1} + \frac{c_2}{l_1 \, l_5} + \frac{c_3}{l_1 \, l_6} + \frac{c_4}{l_1 \, l_6^2} + \frac{c_5}{l_1 \, l_9 \, l_{10}} + \frac{c_6}{l_2 \, l_4} + \frac{c_7}{l_3 \, l_4} + \frac{c_8}{l_4} \right. \nonumber \\
	&\left.\; + \frac{c_9}{l_4 \, l_8} + \frac{c_{10}}{l_4 \, l_{11}} + \frac{c_{11}}{l_5 \, l_7} + \frac{c_{12}}{l_5 \, l_8} + \frac{c_{13}}{l_7} + \frac{c_{14}}{l_9} + \frac{c_{15}}{l_{10}}  + \frac{c_{16}}{l_{12}} \nonumber \right] \nonumber \\
	&+ {\cal O}(\e)\,.
\label{NLO}
\end{align}
The coefficients $c_i$ are linear combinations of MPLs multiplied by some power of $\tilde{x}$ or $\tilde{h}$ with positive exponent. The complete analytic expression of the two-loop amplitude exceeds the scope of this thesis and is attached to the arXiv submission of Ref.~\cite{Gehrmann:2015a}.

\subsection{Small Quark Mass Limit of the Two-Loop Amplitude}
\label{sec:hzaexp}

The analytic result of the full two-loop amplitude enables us to derive its limit for small quark masses and gain information about its logarithmic structure. We perform this expansion by removing trailing zeros in the indices of the MPLs using the shuffle relation~\eqref{MPLshuffle} so that the logarithmic singularities become explicit. Subsequently, we apply the scaling relation~\eqref{MPLscaling} to the remaining MPLs of the form~\eqref{alphabet}, which turn into
\begin{align}
G\left(a_1,\ldots,a_n;1\right) \quad &\text{with} \quad a_i \in \{ 0,\pm\frac{1}{\tilde{h}},\frac{\tilde{x}}{\tilde{h}},\frac{1}{\tilde{x}\,\tilde{h}},\frac{J_x}{\tilde{h}},\frac{1}{J_x\,\tilde{h}},\frac{K_x^\pm}{\tilde{h}}, \frac{L_x^\pm}{\tilde{h}} \} \,, \nonumber \\
G\left(b_1,\ldots,b_n;1\right) \quad &\text{with} \quad b_i \in \{ 0,\pm\frac{1}{\tilde{x}},\frac{c}{\tilde{x}},\frac{\bar{c}}{\tilde{x}} \} \,,
\end{align}
where $a_n \neq 0$ and $b_n \neq 0$. This procedure shifts the dependence on the quark mass from the argument to the indices of the MPLs and allows expanding the integrand of their integral representation without the need to take care of the integration boundaries. Consequently, we solve the definition of the Landau variables in Eq.~\eqref{landau2} for $\tilde{x}$ and $\tilde{h}$, replace them in the integral representation and expand the result in $m_q$. All limiting expressions of the MPLs required to derive this expansion can be found in Appendix~\ref{chap:hzaexpand}.\\
A subtlety occurs when MPLs of the form~\eqref{nontrans} are analyzed for small quark masses. The particular case $w_1=L_x^-$ gives rise to further singularities due to
\begin{equation}
\lim_{m_q \to 0} L_x^{-} = \tilde{x} +  \mathcal{O}(\tilde{x}^2) \,,
\end{equation}
which is why the corresponding MPLs are isolated with the help of the shuffle relation. Finally, we expand the remaining MPLs with $w_i=\{K_x^\pm,L_x^\pm\} \, (i \neq 1)$ in small quark masses, apply transformations of the type~\eqref{trafo1} and treat the resulting MPLs in the same way as the ones above. In doing so, we obtain the following expression for the amplitude from Eq.~\eqref{amp} in the limit of small quark masses, renormalized in the OS scheme:
\begin{align}
\lim_{M_q \to 0} A_q^{(a)}(m_H,m_Z,M_q) &= 4\,i\,S_\e\,M_q\,Y_q\,v\,\left(m_H^2-m_Z^2\right)\,\log\left(\frac{M_q^2}{m_Z^2}\right)\,\log\left(\frac{m_Z^2}{m_H^2}\right) \label{exp} \\
&\quad \left[1 + C_F\,\frac{\alpha_s(\mu)}{\pi} \, \log\left(\frac{M_q^2}{m_Z^2}\right) \right. \nonumber \\
&\quad\left.\left\{ \frac{3}{4} - \frac{1}{8} \, \log\left(\frac{m_Z^2}{m_H^2}\right) + \frac{1}{2} \, \log\left(1-\frac{m_Z^2}{m_H^2}\right) + \frac{\mathrm{Li_2}\left(\frac{m_Z^2}{m_H^2}\right) - \zeta_2}{2 \, \log\left(\frac{m_Z^2}{m_H^2}\right)} \right\} \right] \nonumber \,.
\end{align}
To check the expansions, we numerically validated (using the \textsc{GiNaC} implementation within \textsc{MPLEval}) that the individual MPLs converge towards their expansions in the limit of small quark masses. 
In addition, we rederived the corresponding limit of the $H\to\gamma\gamma$ amplitude by starting from the $H\to Z\gamma$ amplitude and setting the $Z$ boson mass to zero. With the manipulations described above, we agree with the previously available ratio of the two-to-one-loop amplitude for the process $H\to\gamma\gamma$ in small quark masses \cite{Spira:1995,Akhoury:2001}.\\
It is important to note that Eq.~\eqref{exp} contains only single-logarithmic terms, which is in contrast to the $H\to\gamma\gamma$ case. The non-trivial cancellation of the double-logarithmic terms, which originate from the Sudakov region, takes place both in the one-loop and in the two-loop amplitudes of Eqs.~\eqref{LO}~and~\eqref{NLO}, leaving only single-logarithmic terms at each order. A double-logarithmic Sudakov resummation, as performed for $H\to\gamma\gamma$ in Ref.~\cite{Akhoury:2001} is therefore not needed for $H\to Z\gamma$.  We observe from Eq.~\eqref{exp} that the introduction of a running Yukawa coupling, scheme (b), resums the single-logarithmic contribution that is independent of $m_Z^2/m_H^2$. 

\section{Numerical Results}
\label{sec:results}

The calculation of the MIs and the amplitude outlined in Section~\ref{sec:calc} is performed for the values
\begin{equation}
0 < \tilde{x} < 1 \,, \quad 0 < \tilde{h} < 1 \,.
\end{equation}
This corresponds to the Euclidean region, where $m_H^2$ and $m_Z^2$ in Eq.~\eqref{landau2} are negative and the MIs are real. In order to get a physical expression, the results have to be analytically continued to the physical Minkowski region, where we distinguish three kinematic regions:
\begin{itemize}
\item \makebox[2cm][l]{Region I:} $m_Z^2 < m_H^2 < 4 \, m_q^2$ \,,
\item \makebox[2cm][l]{Region II:} $m_Z^2 < 4 \, m_q^2 < m_H^2$ \,,
\item \makebox[2cm][l]{Region III:} $4 \, m_q^2 < m_Z^2 < m_H^2$ \,.
\end{itemize}
In Region~I, the virtualities of the external massive particles are below the threshold induced by the particle running in the loop and the amplitude is real~\cite{tHooft:1978}. The two solutions of each variable in Eq.~\eqref{landau2} become imaginary in this region with
\begin{align}
\tilde{x} &= \mathrm{e}^{\pm2i\phi_x}  \,, \\
\tilde{h} &= \mathrm{e}^{\pm2i\phi_h} \,.
\end{align}
$\phi_x$ and $\phi_h$ are phase factors given by
\begin{equation}
\phi_i = \mathrm{arctan} \sqrt{\frac{m_i^2}{4 m_q^2 - m_i^2}} \,,
\end{equation}
i.e. the variables lie on the unit circle in the complex plane and the second solution is the complex conjugated of the first one. In general, care has to be taken when choosing one of them such that it is in agreement with the common $+i0$ prescription for the Mandelstam variables $m_Z^2$ and $m_H^2$. Due to the real amplitude, however, there is no ambiguity in this case and the imaginary parts of $\tilde{x}$ and $\tilde{h}$ can be chosen freely.\\
In Region III, $m_Z^2$ and $m_H^2$ are beyond the threshold and the amplitude picks up an imaginary part. The two solutions of each variable in Eq.~\eqref{landau2} are real, with one fulfilling
\begin{equation}
-1 < \tilde{x} < 0 \,, \quad -1 < \tilde{h} < 0 \,,
\end{equation}
and the other one being its inverse. In contrast to Region I, one has to be careful when assigning a small imaginary part to $\tilde{x}$ and $\tilde{h}$ in order to fix branch cut ambiguities related to the Mandelstam variables. In this case, a positive imaginary part leads to the correct result if $\left|\tilde{x}\right| < 1$ or $\left|\tilde{h}\right| < 1$ and a negative imaginary part has to be applied when $\left|\tilde{x}\right| > 1$ or $\left|\tilde{h}\right| > 1$. Let us illustrate this through the analytic continuation of $h=m_H^2/m_q^2$ from negative to positive values of $\tilde{h}$ using the first expression of Eq.~\eqref{landau2}:
\begin{align}
h &= -\frac{(1-\tilde{h})^2}{\tilde{h}} \,, \qquad \tilde{h} \to -\tilde{h} \pm i\,\e \nonumber \\
&= \frac{(1+\tilde{h})^2}{\tilde{h}} \mp i\,\e \left(1-\frac{1}{\tilde{h}} \right) + \mathcal{O}\left(\e^2\right) \nonumber \\
&\approx \frac{(1+\tilde{h})^2}{\tilde{h}} + i\,\e \,.
\end{align}
Therein, the upper and lower sign of the prefactor of the infinitesimal imaginary part~$i\,\e$ correspond to the regions $\left|\tilde{h}\right| < 1$ and $\left|\tilde{h}\right| > 1$, respectively. The derivation of the appropriate imaginary part of $\tilde{x}$ proceeds along the same line of argument.\\
From the input parameters specified below, it is obvious that the top quark amplitude is calculated in Region I, while the bottom quark amplitude is computed in Region III. Region II is not needed for the physical values of the masses.\\
Since the analytical expression for $A^{(2)}_q$ is given in terms of MPLs, it can be evaluated using the \textsc{GiNaC} link within the \textsc{MPLEval} implementation. For masses and couplings, we use the input values
\begin{alignat}{4}
\alpha_s^{(5)} (m_H) &= 0.1130114 \, , &\quad \alpha &= 1/128 \, , &\quad G_F &= 1.1663787 \cdot 10^{-5} \, \mathrm{GeV}^{-2} \, , \notag \\
m_H &= 125.7 \, \mathrm{GeV} \, , &\quad m_Z &= 91.1876 \, \mathrm{GeV} \, , &\quad m_W &= 80.385 \, \mathrm{GeV} \, , \notag \\
M_t &= 173.21 \, \mathrm{GeV} \, , &\quad \overline{m}_t(m_H) &= 167.21 \, \mathrm{GeV} \, , &\quad M_b &= 4.7652 \, \mathrm{GeV} \, , \notag \\
\overline{m}_b(m_H) &= 2.7832 \, \mathrm{GeV} \, , &\quad \mathrm{sin}^2 \, \theta_w &= 0.23126 \, , &\quad \mathrm{cos}^2 \, \theta_w &= 0.76874 \, , \notag \\
Q_t &= 2/3 \, , &\quad Q_b &= -1/3 \, , &\quad I_t^3 &= 1/2 \, , \notag \\
I_b^3 &= -1/2 \, , &\quad C_F &= 4/3 \, , &\quad N_c &= 3 \, .
\end{alignat}
They were obtained by evolving the values of the Particle Data Group Collaboration~\cite{Agashe:2014} to the scale $\mu=m_H$ with the two-loop renormalization group equations implemented in \textsc{RunDec}~\cite{Chetyrkin:2000}.
To resum potentially large single logarithms in $\overline{m}_q/m_H$ to all orders in the perturbative expansion, we use the two-loop renormalization group equations~\cite{Chetyrkin:2000} to evolve the $\MS$ quark mass (and accordingly the Yukawa coupling) from the matching scale to $\mu=m_H$. This leads to the following NLO decay width $\Gamma^{(2)}$ in the renormalization schemes (a), (b) and (c):
\begin{align}
\Gamma^{(2,a)} &\overset{\textcolor{white}{\mu=m_H}}{=} \left[ 7.07533 + 0.42800 \, \frac{\alpha_s(\mu)}{\pi} \right] \mathrm{keV} \nonumber \\
	&\overset{\mu=m_H}{=} 7.09072 \, \mathrm{keV} \,, \\
\Gamma^{(2,b)} &\overset{\mu=m_H}{=} \left[ 7.09409 \right. \nonumber \\
&\qquad\quad \left.+ \frac{\alpha_s(m_H)}{\pi} \left( -0.53266 - 0.76661 \, \log \frac{m_H^2}{\overline{m}_t^2(m_H)} + 0.01229 \, \log \frac{m_H^2}{\overline{m}_b^2(m_H)} \right) \right] \mathrm{keV} \nonumber \\
	&\overset{\textcolor{white}{\mu=m_H}}{=} 7.09403 \, \mathrm{keV} \,, \label{widthhyb} \\
\Gamma^{(2,c)} &\overset{\mu=m_H}{=} \left[ 7.05934 \right. \nonumber \\
&\qquad\quad \left.+ \frac{\alpha_s(m_H)}{\pi} \left( 0.64587 + 0.10597 \, \log \frac{m_H^2}{\overline{m}_t^2(m_H)} + 0.01453 \, \log \frac{m_H^2}{\overline{m}_b^2(m_H)} \right) \right] \mathrm{keV} \nonumber \\
	&\overset{\textcolor{white}{\mu=m_H}}{=} 7.08438 \, \mathrm{keV} \,. \label{widthms}
\end{align}
The breakup of the terms in the first lines of Eqs.~\eqref{widthhyb} and \eqref{widthms} is to illustrate the relative numerical importance of the individual contributions.\\
An estimation of the uncertainty on the prediction from missing higher orders is provided by varying $m_H/2<\mu<2\,m_H$ in Fig.~\ref{fig:varscale}. For every data point $\mu=\bar{\mu}$, the $\MS$~quark mass $\overline{m}_q(\bar{\mu})$ and the strong coupling constant $\alpha_s(\bar{\mu})$ are evolved to the scale $\bar{\mu}$ using the two-loop renormalization group equations~\cite{Chetyrkin:2000}.\\
\begin{figure}[tb]
\begin{center}
\includegraphics[width=0.92\textwidth]{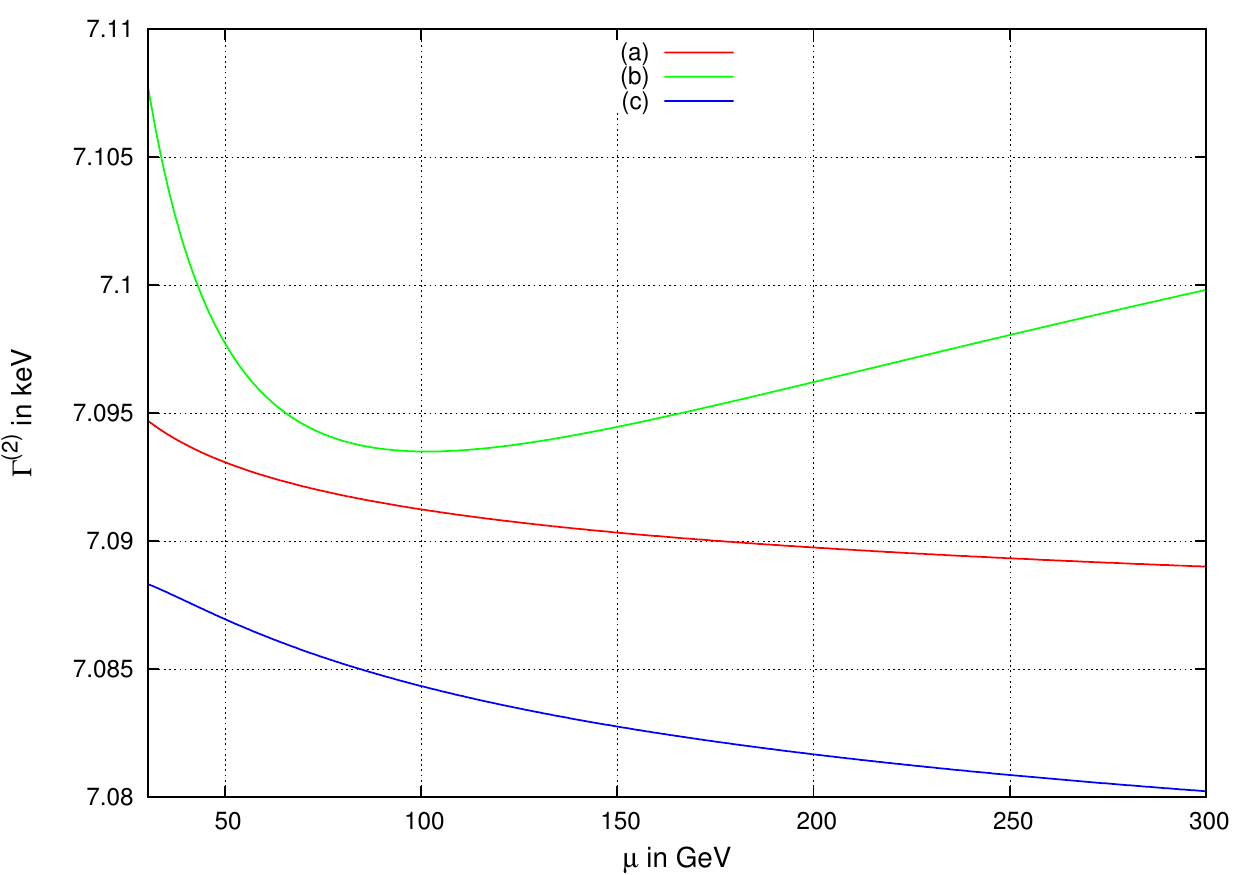}
\caption[Scale variation of the NLO decay width $\Gamma^{(2)}$]{\textbf{Scale variation of the NLO decay width $\boldsymbol{\Gamma^{(2)}}$}\\
in the renormalization schemes (a), (b) and (c) for $30\,\mathrm{GeV}<\mu<300\,\mathrm{GeV}$.}
\label{fig:varscale}
\end{center}
\end{figure}\noindent
The LO decay width $\Gamma^{(1)}$ and the NLO decay width $\Gamma^{(2)}$ can be separated into contributions from the $W$~boson, top quark and bottom quark amplitudes as well as their interferences. The corresponding values are shown in Table~\ref{tab:width}, from which it becomes clear that the bottom quark amplitude has to be taken into account, since its interference with the $W$ amplitude is of the same order of magnitude as the self-interference of the top quark amplitude at one loop. Moreover, the combination $\Gamma^{(2)}_{Wb}$ exceeds the top quark self-interference $\Gamma^{(2)}_{tt}$ at two loops.\\
\begin{table}[tb]
\caption[Various contributions to the numerical result of the LO decay width $\Gamma^{(1)}$ and the NLO decay width $\Gamma^{(2)}$]{\textbf{Various contributions to the numerical result of the LO decay width $\boldsymbol{\Gamma^{(1)}}$ and the NLO decay width $\boldsymbol{\Gamma^{(2)}}$} in the renormalization schemes (a), (b) and (c), evaluated for $\mu=m_H$. In case of $\Gamma^{(1)}_{ij}$, the subscripts $i$ and $j$ indicate the interference of the one-loop amplitudes $A^{(1)}_W$, $A^{(1)}_t$ and $A^{(1)}_b$, whereas $\Gamma^{(2)}_{ij}$ describes the interference of the one-loop amplitude $A^{(1)}_i$ with the two-loop amplitude $A^{(2)}_j$. All values are given in keV.}
\setlength{\tabcolsep}{0.8cm}
\begin{tabularx}{\textwidth}{lrrr}
\toprule
Partial width & \multicolumn{1}{c}{(a)} & \multicolumn{1}{c}{(b)} & \multicolumn{1}{c}{(c)} \\
\midrule
$\Gamma^{(1)}_{WW}$ & $7.86845996$ & $7.86845996$ & $7.86845996$ \\
$\Gamma^{(1)}_{Wt}$ & $-0.83636436$ & $-0.80736905$ & $-0.84015333$ \\
$\Gamma^{(1)}_{Wb}$ & $0.02216139$ & $0.01294390$ & $0.00908488$ \\
$\Gamma^{(1)}_{tt}$ & $0.02222498$ & $0.02071068$ & $0.02242680$ \\
$\Gamma^{(1)}_{tb}$ & $-0.001177803$ & $-0.00066408$ & $-0.00048502$ \\
$\Gamma^{(1)}_{bb}$ & $0.00002103$ & $0.00000717$ & $0.00000325$ \\
\midrule
$\Gamma^{(1)}$ & $7.07532519$ & $7.09408860$ & $7.05933655$ \\
\midrule
$\Gamma^{(2)}_{Wt}$ & $0.02213199$ & $-0.00078617$ & $0.02467587$ \\
$\Gamma^{(2)}_{Wb}$ & $-0.00588750$ & $0.00073044$ & $0.00176120$ \\
$\Gamma^{(2)}_{tt}$ & $-0.00117624$ & $0.00004033$ & $-0.00131738$ \\
$\Gamma^{(2)}_{tb}$ & $0.00031290$ & $-0.00003747$ & $-0.00009403$ \\
$\Gamma^{(2)}_{bt}$ & $0.00003117$ & $-0.00000065$ & $0.00001425$ \\
$\Gamma^{(2)}_{bb}$ & $-0.00001592$ & $-0.00000081$ & $0.00000078$ \\
\midrule
$\Gamma^{(2)}$ & $7.09072159$ & $7.09403427$ & $7.08437723$ \\
\bottomrule
\label{tab:width}
\end{tabularx}
\end{table}\noindent
Our on-shell results are in agreement with the numerical findings of Ref.~\cite{Spira:1991}. Furthermore, we performed a detailed numerical comparison with Ref.~\cite{Bonciani:2015} and found agreement to high precision.\\
We observe that the NLO results for the decay width are consistent between the three schemes. The relative size of the NLO correction is 2$\permille$ in scheme (a), below $10^{-5}$ in scheme (b) and 3$\permille$ in scheme (c). The very small corrections in scheme (b) are however in large part due to numerical cancellations between a priori unrelated terms. The spread between the different schemes is 1.3$\permille$ at $\mu=m_H$, and variations of the renormalization scale change the predictions in either given scheme by at most 0.4$\permille$.

\section{Conclusions}
\label{sec:conc2}

In this chapter, we have revisited the QCD corrections to the rare loop-induced Higgs boson decay $H\to Z\gamma$.
The relevant two-loop three-point integrals with two different external masses and one internal mass were derived analytically, using a reduction to MIs, which were then computed using differential equations. These integrals are also an important constituent of the two-loop amplitudes for Higgs-plus-jet production in gluon fusion with 
full dependence on the internal quark masses.\\
By expanding the one-loop and two-loop matrix elements in the OS scheme for $H\to Z\gamma$ in the limit of small quark masses, we noted the absence of double-logarithmic contributions, which is in contrast to $H\to\gamma\gamma$~\cite{Spira:1995,Akhoury:2001}; single-logarithmic terms are resummed by the introduction of a running Yukawa coupling.\\
We investigated the dependence of the corrections on the renormalization scheme used for the quark mass and Yukawa coupling. We observe that the results for the decay rate  in OS and $\MS$ schemes, as well as in a hybrid scheme with on-shell mass and $\MS$
Yukawa coupling, are well consistent with each other, and that corrections are in the sub-per-cent range in all three schemes (being smallest in the hybrid scheme). We confirm the previously available numerical on-shell result~\cite{Spira:1991} and agree with an independent calculation~\cite{Bonciani:2015}. The residual QCD  uncertainty on the $H\to Z\gamma$ decay rate is around 1.7$\permille$ from the combination of scale variation and  spread between the different renormalization schemes.

\chapter{The Workflow of Multi-Loop Calculations, Part III:\\Calculation of Master Integrals through Series Expansions from Differential Equations}
\chaptermark{Multi-Loop Calculations, Part III: Master Integrals through Series Expansions}
\label{chap:workflow3}

In the previous chapters, we hinted at the fact that the class of MPLs described in Section~\ref{sec:mpl} is not sufficient for the description of Higgs-plus-jet production with full quark mass dependence at two loops. Therefore, we begin this chapter with Section~\ref{sec:beyondmpl}, where we describe integration procedures in terms of functions that are close to, but different from MPLs. In Section~\ref{sec:elliptic}, we present elliptic integrals as a special class of functions beyond MPLs, since they are known to occur in the evaluation of certain Feynman integrals at the two-loop level, from the perspective of their differential equations. Currently there is no straightforward way to integrate differential equations for MIs in terms of elliptic integrals, so that we elaborate on an alternative approach relying on series expansions in a single variable in Section~\ref{sec:seriesexp}. We have used this approach extensively in the context of Higgs-plus-jet production in Chapter~\ref{chap:hj}, and it can be understood as a useful tool for future computations of multi-loop amplitudes involving integrals beyond MPLs. Before doing so, we need to introduce the foundations of complex analysis in Section~\ref{sec:complexana}. Finally, we explain how to cast multi-scale integrals in a form suitable for series expansions in a single variable in Section~\ref{sec:onedim}, and how to connect multiple series expansions in order to cover the whole phase space in Section~\ref{sec:matching}.

\section{Non-Elliptic Integrals beyond Multiple Polylogarithms}
\label{sec:beyondmpl}

It is actually well-known that mathematical structures beyond MPLs can appear in two-loop computations of Feynman integrals, especially when carrying out integrals with massive propagators. In this respect, the two-loop massive sunrise graph is by far the most extensively studied example, and its evaluation is known to involve iterated integrals over elliptic kernels, leading to integrals over elliptic integrals.\\
Let us postpone the discussion on elliptic integrals to the next section and start by elaborating on the gap between them and MPLs. This gap exists due to the restriction of the identities~\eqref{trafo1} and \eqref{trafo2} derived from the coproduct-augmented symbol formalism in Section~\ref{sec:symbol} to rational dependence on kinematic invariants in a multi-scale problem. In the context of integrating differential equations, we rephrased this such that the integration procedure described in Section~\ref{sec:int} can only be carried out if there is at most one variable whose occurence in the alphabet is of non-linear kind. We would like to emphasize that these constraints do not emerge from the definition of MPLs in Eq.~\eqref{defmpl}, which permits algebraic dependence on kinematic invariants within the index vector $\vec{w}$. This explains why integrations can still be performed in the single-variable case with root-valued indices. In other words, the concept of generalized weights is at the moment exclusively understood for one-variable problems~\cite{vonManteuffel:2013}.\\
As a first natural step, one would attempt to remove the square roots emerging from partial fractioning the alphabet through a change of variables. However, this is not feasible if the number of independent square roots is greater than the independent number of variables of the process under consideration. This is exactly what we observe in the two-loop corrections to Higgs-plus-jet production with full quark mass dependence, process~$(c)$. Hence, we conclude that it is not possible to remove the square roots therein, although it cannot be proven rigorously.\\
In such a case, the square roots are inherent to the problem and have to be dealt with. One way to do this involves recycling the exceptional properties of the coproduct and symbol formalism in a different sense. Ref.~\cite{Duhr:2011} provides a procedure which, starting from the total differential of Eq.~\eqref{totdiff}, projects onto a real-valued functional basis of classical polylogarithms~$\mathrm{Li}_n$ and their extension~$\mathrm{Li}_{2,2}$ defined in Eqs.~\eqref{defpolylog} and \eqref{defli22}, respectively. The entity of this basis forms an ansatz up to weight four, whose coefficients can be determined by adding information of both the total differential and the symbol entries associated with these functions. From the relations~\eqref{defli22} and \eqref{gtolrules} it is clear that such a basis must exist and that it is equivalent to the one in terms of MPLs. The described method is advantegeous over the conventional integration to MPLs in the sense that the mentioned transformation identities do not have to be derived, since the integration to a minimal functional basis is directly performed. In addition, the classical polylogarithms can be equally defined through well-converging series representations, leading to much faster and reliable numerical evaluations. The feasibility of this method is only constrained by the question of whether the available computational resources are able to handle the potentially huge combinatorics of the problem under consideration. The combinatorics scale both with the number of letters $l_k$ entering the alphabet~\eqref{defdlog} and with the number of square roots occuring therein. The ansatz in terms of classical polylogarithms has to be extended considerably if the alphabet involves square roots, since one has to allow for half-integer powers of the corresponding functional arguments on top of the integer ones.\\
In practice, this approach has proven beneficial in a number of calculations, e.g. in Refs.~\cite{Bonciani:2015,Gehrmann:2014b}, the former of which corresponds to the calculation of the MIs of process~$(b)$ in Chapter~\ref{chap:hza}. A collaborator of ours~\cite{AvM1} has been able to rederive these results starting from the total differential in Eq.~\eqref{canon}, retaining the full dependence on the undetermined constants in the process of integration at every order in $\e$. In all mentioned examples, square roots could be removed and were not part of the alphabet, which however turns out to be impossible in the context of Higgs-plus-jet production, as stated. Combining this with the high number of letters appearing in the total differential~\eqref{hjdeq3} significantly complicates the application of the described method to the Higgs-plus-jet case. This becomes apparent through two observations: First, our collaborator~\cite{AvM2} tackled the integration of the canonical basis of the sector $A_{5,182}$ defined in Appendix~\ref{sec:hjcan}\footnote{The corresponding Laporta integrals $I_{39}$--$I_{42}$ are given in Appendix~\ref{sec:hjlaporta} and depicted in Fig.~\ref{fig:hjmaster}.} and its subtopology tree in terms of classical polylogarithms. The attempt has been successful up to weight three and the weight-four result is within reach, nevertheless the efforts to accomplish this have shown that applying this method to all MIs of process $(c)$, particularly to more complicated sectors, is not realistic. Second, this method has been used for the computation of all planar MIs in the Euclidean region contributing to the two-loop corrections to Higgs-plus-jet production with full quark mass dependence in Ref.~\cite{Bonciani:2016}. Therein, integrating in this manner turned out to be feasible up to weight two, where nothing but logarithms and dilogarithms appear within the functional basis. Moreover, only 20 out of around 50 letters contribute to the alphabet at this point and thus the combinatorics is under control. This is not the case starting from weight three, where Ref.~\cite{Bonciani:2016} uses a parametric integral representation for numerical integration, so that fully analytic results for the MIs of process~$(c)$ are still missing beyond weight two.\\
In conclusion, we have seen that, if we manage to decouple linear first-order differential equations or even to cast them into canonical form, it might be far from trivial to integrate them, particularly if non-removable square roots are involved. Therefore, we turn to a different approach described in Section~\ref{sec:seriesexp}, which is designed to produce fully analytic results. This alternative formalism is further motivated by the appearance of elliptic integrals within the set of MIs required for the two-loop corrections to the amplitude of Higgs-plus-jet production with full quark mass dependence. Hence, we will elaborate on the definition and the properties of elliptic functions in the next section.

\section[Excursus on a Special Class of Functions: Elliptic Integrals]{Excursus on a Special Class of Functions:\\Elliptic Integrals}
\label{sec:elliptic}

If the underlying class of functions is not of interest, numerical evaluations of elliptic Feynman integrals can be a powerful tool. In this context, the application of the sector decomposition approach~\cite{Hepp:1966,Roth:1996,Binoth:2000,
Heinrich:2008} to a basis of finite integrals~\cite{vonManteuffel:2014b} is particularly well-suited, since all poles are explicit and only the finite-order term of the $\e$-expansion has to be determined numerically. This was successfully used for evaluating the differential cross section of the NLO corrections to Higgs-pair-production with full top mass dependence~\cite{Borowka:2016a,Borowka:2016b}. From this computation, it is not even known if elliptic integrals occur, since all integrals were treated the same from the numerical point of view. The same holds for the very recently calculated differential cross section of the NLO corrections to Higgs-plus-jet production with fixed ratio~$m_H^2/m_t^2$~\cite{Jones:2018}. Other numerical approaches point in the direction of solving differential equations through Runge-Kutta-type methods~\cite{Runge:1895,Kutta:1901}, like in the numerical evaluation of the MIs required for the cross section of top-quark pair production at NNLO~\cite{Baernreuther:2013}. Two collaborators of ours~\cite{Hoff1} took a similar path in the context of the planar two-loop MIs necessary for the NLO corrections to Higgs-plus-jet production with full quark mass dependence, process~$(c)$. More precisely, in the language of the variables introduced in Eq.~\eqref{deqmandelstam}, they integrated the differential equations of the Laporta basis defined in Appendix~\ref{sec:hjlaporta} with respect to the internal mass~$m$. This was done by starting from the boundary value in~$m\to\infty$, where all integrals become massive tadpoles and can be evaluated straightforwardly. However, as in many numerical integration routines, issues with the runtime and with numerical stabilities were encountered, so that we decided to turn to an analytic integration procedure instead. Before describing this method in detail, let us elaborate on different attempts of evaluating elliptic Feynman integrals analytically.\\
Being the simplest possible example of such a case, the equal-mass two-loop sunrise depicted in Fig.~\ref{fig:elliptic}$(a)$ was studied in the literature in great detail. This was done especially from the perspective of the differential equations and extended to arbitrary masses of the propagators~\cite{Kreimer:1991,Berends:1993,Berends:1994,Bauberger:1994a,Bauberger:1994b,
Caffo:1998b,Laporta:2004,Tarasov:2006,Adams:2013,Remiddi:2013,MullerStach:2011}. For a long time, it was unclear how to represent the analytic result of the massive two-loop sunrise in terms of a well-defined class of functions similar to MPLs. However, the recent rediscovery of the \textit{Elliptic Dilogarithm} by particle physicists compensated for this~\cite{Bloch:2013,Adams:2014}. The kite integral shown in Fig.~\ref{fig:elliptic}$(b)$ is the simplest parent topology of the sunrise integral and was subject of further studies~\cite{Remiddi:2016,Adams:2016}. Although the homogeneous solution of the kite integral is not of elliptic nature, iterated integrals over elliptic kernels emerge from integrating the inhomogeneous part of the differential equations. This is due to the fact that the sunrise integral enters the inhomogeneous differential equations as subtopology. Note that this computation required generalizing Elliptic Dilogarithms to \textit{Elliptic Polylogarithms}~\cite{Brown:2011,Adams:2015a,
Adams:2015b,Remiddi:2017,Broedel:2017a,Broedel:2017b}. Very recently, they have been used to compute an elliptic phase space integral contributing to triple real corrections to Higgs boson production at $\mathrm{N^3LO}$, thus leading to the first analytic calculation of a hadron collider cross section involving elliptic integrals~\cite{Mistlberger:2018}. Equivalent formulations of this generalization are referred to as \textit{Elliptic Multiple Zeta Values}~\cite{Broedel:2014,Broedel:2015,Broedel:2017c} or \textit{Iterated Integrals of Modular Forms}~\cite{Adams:2017b}. Other tools to determine the solutions of elliptic differential equations are provided by computing the maximal cut of Feynman integrals in specific integral representations~\cite{Primo:2016,Primo:2017,Frellesvig:2017,
Harley:2017,Hidding:2017}.
\begin{figure}[tb]
\begin{center}
\includegraphics[width=\textwidth]{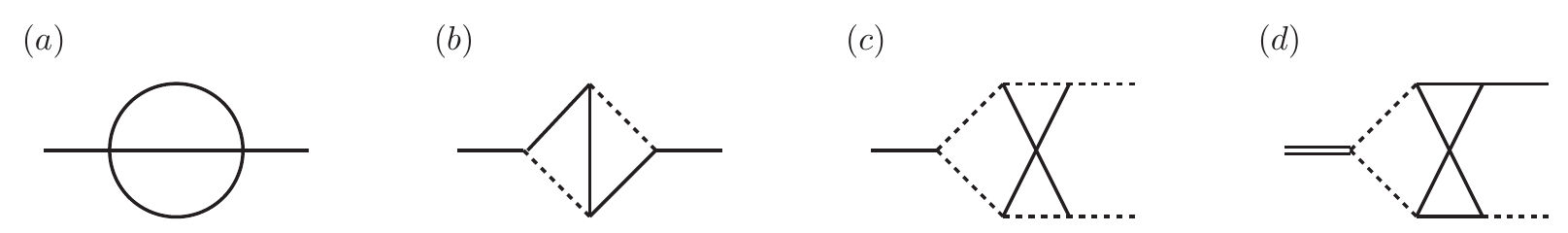}
\caption[Example diagrams known to evaluate to elliptic integrals]{\textbf{Example diagrams known to evaluate to elliptic integrals.}\\
Dashed lines are massless, whereas internal solid lines denote propagators with mass $m$. Double and solid external lines correspond to virtualities $m_H^2=(q_1+q_2+q_3)^2$ and $u=(q_2+q_3)^2$, respectively.\\
$(a)$ Two-loop sunrise diagram in the equal-mass case,\\
$(b)$ two-loop kite diagram in the equal-mass case,\\
$(c)$ two-loop crossed-ladder vertex diagram with two massive exchanges,\\
$(d)$ two-loop crossed-ladder vertex diagram with four massive exchanges.}
\label{fig:elliptic}
\end{center}
\end{figure}\noindent
All these ideas have been developed only recently and are in principle applicable to any Feynman integral that can be expressed through the mentioned class of functions. However, it is at the moment unclear if these classes of functions are sufficient to cover all possible two-loop Feynman integrals. Moreover, the generic combinations of the known elliptic functions relevant to the computation of Feynman diagrams are still widely unknown and subject of current research. Furthermore, almost all results available in the literature refer to the Euclidean region. Although first steps were made in Refs.~\cite{Remiddi:2017,Primo:2017,Bogner:2017b,vonManteuffel:2017b}, the analytic continuation of the newly introduced classes of functions is still not well-understood. This holds as far as two-loop four point graphs evaluating to elliptic functions are concerned, which is in contrast to the fully established concept of analytic continuation of two-loop four point functions that evaluate to MPLs~\cite{Gehrmann:2002}. In Section~\ref{sec:seriesexp}, we will therefore follow a different analytic approach, in which the integration is performed directly in the physical region so that the results do not have to be continued analytically. Before doing so, we introduce the notion of complete elliptic integrals and elaborate on how to identify Feynman graphs including elliptic functions prior to integration.

\subsection{The Massive Two-Loop Sunrise and Complete Elliptic Integrals}
\label{sec:defell}

We point out that the review on the definitions and properties of elliptic functions in this section is far from being complete and comprehensive. In the following, we limit our discussion to functions, which will be of use at a later point of this thesis.\\
Let us illustrate how elliptic integrals arise naturally from computations of Feynman diagrams by inspecting the imaginary part of the two-loop sunrise diagram in Fig.~\ref{fig:elliptic}$(a)$, but for arbitrary propagator masses $m_1$, $m_2$ and $m_3$. Due to the optical theorem, this is equal to the evaluation of the three-particle phase space with masses $m_1$, $m_2$ and $m_3$ in the final state. By denoting the total center-of-mass energy $\sqrt{u}$, the three-particle phase space in $D$ space-time dimensions beyond the threshold $\sqrt{u}>m_1+m_2+m_3$ can be rephrased as the factorization of two two-particle phase space contributions~\cite{Tancredi:2014},
\begin{align}
\Phi_3\left(D;u,m_1^2,m_2^2,m_3^2\right) &\propto \int_{(m_2+m_3)^2}^{(\sqrt{u}-m_1^2)^2} \d b \, \Phi_2\left(D;u,b,m_1^2\right) \, \Phi_2\left(D;b,m_2^2,m_3^2\right) \nonumber \\
&= \int_{(m_2+m_3)^2}^{(\sqrt{u}-m_1^2)^2} \frac{\d b}{\sqrt{R_2\left(u,b,m_1^2\right) \, R_2\left(b,m_2^2,m_3^2\right)}} \nonumber \\
&\qquad \times \left(\frac{R_2\left(u,b,m_1^2\right) \, R_2\left(b,m_2^2,m_3^2\right)}{u\,b}\right)^{\frac{D}{2}-1} \,,
\label{3pps1}
\end{align}
up to a normalization constant depending on $D$, where the K\"allen function
\begin{equation}
R_2\left(a,b,c\right) \equiv a^2 + b^2 + c^2 - 2 \, a \, b - 2 \, a \, c - 2 \, b \,c
\end{equation}
was introduced. Upon integration in $D=4$ dimensions, Eq.~\eqref{3pps1} turns into
\begin{equation}
\Phi_3\left(4;u,m_1^2,m_2^2,m_3^2\right) \propto \int_{b_2}^{b_3} \frac{\d b}{b} \, \sqrt{R_4\left(b;b_1,b_2,b_3,b_4\right)}
\label{3pps2}
\end{equation}
with
\begin{align}
R_4\left(b;b_1,b_2,b_3,b_4\right) &\equiv R_2\left(u,b,m_1^2\right) \, R_2\left(b,m_2^2,m_3^2\right) \nonumber \\
&= \left(b-b_1\right) \, \left(b-b_2\right) \, \left(b-b_3\right) \, \left(b-b_4\right) \,.
\end{align}
Therein, the definitions and inequalities
\begin{equation}
b_1 = \left(m_2-m_3\right)^2 \leq b_2 = \left(m_2+m_3\right)^2 \leq b_3 = \left(\sqrt{u}-m_1\right)^2 \leq b_4 = \left(\sqrt{u}+m_1\right)^2
\end{equation}
have been applied. By explicitly performing the differentiation within the IBP-like identity
\begin{equation}
\int_{b_2}^{b_3} \d b \, \frac{\d}{\d b} \, \left(b^n \, \sqrt{R_4\left(b;b_1,b_2,b_3,b_4\right)}\right) = 0 \,,
\end{equation}
one obtains a relation of up to five integrals of the integral family
\begin{equation}
I(n;b_1,b_2,b_3,b_4) \equiv \int_{b_2}^{b_3} \d b \, \frac{b^n}{\sqrt{R_4\left(b;b_1,b_2,b_3,b_4\right)}} \,,
\end{equation}
which means that the three-particle phase space in Eq.~\eqref{3pps2} can be expressed in terms of four MIs belonging to this family. From that set, we choose the following integrals as MIs:
\begin{align}
I(-1;b_1,b_2,b_3,b_4) &= \int_{b_2}^{b_3} \frac{\d b}{b \, \sqrt{R_4\left(b;b_1,b_2,b_3,b_4\right)}} \,, \nonumber \\
I(0;b_1,b_2,b_3,b_4) &= \int_{b_2}^{b_3} \frac{\d b}{\sqrt{R_4\left(b;b_1,b_2,b_3,b_4\right)}} \,, \nonumber \\
I(1;b_1,b_2,b_3,b_4) &= \int_{b_2}^{b_3} \frac{b \, \d b}{\sqrt{R_4\left(b;b_1,b_2,b_3,b_4\right)}} \,, \nonumber \\
I(2;b_1,b_2,b_3,b_4) &= \int_{b_2}^{b_3} \frac{b^2 \, \d b}{\sqrt{R_4\left(b;b_1,b_2,b_3,b_4\right)}} \,.
\end{align}
Through the change of variables from $\left(b,b_1,b_2,b_3,b_4\right)$ to $\left(x,b_1,b_2,b_3,b_4\right)$ with
\begin{equation}
x^2 = \frac{\left(b_3-b_1\right) \, \left(b-b_2\right)}{\left(b_3-b_2\right) \, \left(b-b_1\right)} \,,
\end{equation}
these MIs can be expressed in terms of a different, but equivalent integral representation:
\begin{align}
I(-1;b_1,b_2,b_3,b_4) &= \frac{2}{\sqrt{\left(b_3-b_1\right) \, \left(b_4-b_2\right)} \, b_1 \, b_2} \left[ b_2 \, K(w) - \left(b_2-b_1\right) \, \Pi\left(a_2;w\right) \right] \,, \nonumber \\
I(0;b_1,b_2,b_3,b_4) &= \frac{2}{\sqrt{\left(b_3-b_1\right) \, \left(b_4-b_2\right)}} \, K(w) \,, \nonumber \\
I(1;b_1,b_2,b_3,b_4) &= \frac{2}{\sqrt{\left(b_3-b_1\right) \, \left(b_4-b_2\right)}} \left[b_1 \, K(w) - \left(b_2-b_1\right) \, \Pi\left(a_1;w\right) \right] \,, \nonumber \\
I(2;b_1,b_2,b_3,b_4) &= \frac{2}{\sqrt{\left(b_3-b_1\right) \, \left(b_4-b_2\right)}} \left[ \left( b_1^2 + b_1 \, \left(b_2+b_3\right) - b_2 \, b_3 \right) \, K(w) \right. \\
&\qquad\qquad\qquad\qquad\qquad\quad \left. - \left(b_3-b_1\right) \, \left(b_4-b_2\right) \, E(w) \right. \nonumber \\
&\qquad\qquad\qquad\qquad\qquad\quad \left. + \left(b_2-b_1\right) \, \left(b_1+b_2+b_3+b_4\right) \, \Pi\left(a_1;w\right) \right] \,. \nonumber
\end{align}
We have done nothing other than changing to a new basis, which is advantageous in the sense that their integral representations defined by
\begin{align}
K(w) &= \int_0^1 \frac{\d x}{\sqrt{\left(1-x^2\right) \, \left(1-w^2\,x^2\right)}} \,, \quad 0 < w < 1 \,, \nonumber \\
E(w) &= \int_0^1 \d x \, \sqrt{\frac{1-w^2\,x^2}{1-x^2}} \,, \quad 0 < w < 1 \,, \nonumber \\
\Pi(a;w) &= \int_0^1 \frac{\d x}{\sqrt{\left(1-x^2\right) \, \left(1-w^2\,x^2\right)}\,\left(1-a\,x^2\right)} \,, \quad 0 < w < 1 \,, \quad 0 < a < 1
\label{defell}
\end{align}
depend on at most two out of the following three functional arguments:
\begin{align}
w^2 &= \frac{\left(b_4-b_1\right) \, \left(b_3-b_2\right)}{\left(b_4-b_2\right) \, \left(b_3-b_1\right)} \,, \nonumber \\
a_1 &= \frac{b_3-b_2}{b_3-b_1} \,, \nonumber \\
a_2 &= \frac{b_1 \, \left(b_3-b_2\right)}{b_2 \, \left(b_3-b_1\right)} \,.
\label{argell}
\end{align}
The functions $K(w)$, $E(w)$ and $\Pi(a;w)$ are referred to as the \textit{Complete Elliptic Integrals of first, second and third kind}, respectively. They emerge from the \textit{Incomplete Elliptic Integrals} with arbitrary upper integration limit in the special case, where that limit is equal to~$1$. The linear, second-order differential equations of the complete elliptic integrals of first and second kind read
\begin{align}
\left( w\,\left(1-w^2\right) \, \frac{\d^2}{\d w^2} + \left(1-3 \, w^2\right) \frac{\d}{\d w} - w \right) \, K(w) &= 0 \,, \nonumber \\
\left( w\,\left(1-w^2\right) \, \frac{\d^2}{\d w^2} + \left(1-w^2\right) \frac{\d}{\d w} + w \right) \, E(w) &= 0 \,.
\label{deqell}
\end{align}
Let us perform another change of variable according to
\begin{equation}
w \to z=w^2 \,.
\label{varchangeell}
\end{equation}
With the help of the chain rule, the differential equations~\eqref{deqell} turn into
\begin{align}
\left( z \, \left(1-z\right) \, \frac{\d^2}{\d z^2} + \left(1-2 \, z\right) \frac{\d}{\d z} - \frac{1}{4} \right) \, K\left(\sqrt{z}\right) &= 0 \,, \nonumber \\
\left( z \, \left(1-z\right) \, \frac{\d^2}{\d z^2} + \left(1-z\right) \frac{\d}{\d z} + \frac{1}{4} \right) \, E\left(\sqrt{z}\right) &= 0 \,.
\end{align}
The equations in the new variable $z$ can be identified with the differential equation of the hypergeometric function $_2F_1(a,b;c;z)$ in the special cases
\begin{equation}
K\left(\sqrt{z}\right) = \frac{\pi}{2} \, _2F_1\left(\frac{1}{2},\frac{1}{2};1;z\right) \,, \qquad E\left(\sqrt{z}\right) = \frac{\pi}{2} \, _2F_1\left(-\frac{1}{2},\frac{1}{2};1;z\right) \,,
\end{equation}
which will be useful in the next subsection. This differential equation reads
\begin{equation}
\left( z \, \left(1-z\right) \, \frac{\d^2}{\d z^2} + \left( c - \left( a+b+1 \right) \, z \right) \frac{\d}{\d z} - a \, b \right) \, _2F_1\left(a,b;c;z\right) = 0 \,.
\label{deqhyp}
\end{equation}
In a final remark, we would like to comment on two interesting limits of these results:
\begin{itemize}
\item[\textbf{i)}] \textbf{Equal-Mass Limit}\\
In the limit in which all internal masses coincide, $m_1=m_2=m_3\equiv m$, the arguments of the complete elliptic integrals become
\begin{equation}
w^2 \to \tilde{w}^2 \equiv \frac{\left(\sqrt{u}-3\,m\right) \, \left(\sqrt{u}+m\right)^3}{\left(\sqrt{u}+3\,m\right) \, \left(\sqrt{u}-m\right)^3} \,, \quad a_1 \to \tilde{a}_1 \equiv 1-\frac{4\,m^2}{\left(\sqrt{u}-m\right)^2} \,, \quad a_2 \to 0 \,,
\end{equation}
telling us that the complete elliptic integral of third kind,
\begin{equation}
\Pi\left(0;\tilde{w}\right) = K\left(\tilde{w}\right) \,,
\end{equation}
reduces to the complete elliptic integral of first kind. Further investigation reveals that $\Pi\left(\tilde{a}_1;\tilde{w}\right)$ can be written as a linear combination of the elliptic integrals $K\left(\tilde{w}\right)$ and $E\left(\tilde{w}\right)$ of first and second kind in that limit, thus leaving us with the well-known number of two MIs in the equal-mass case.
\item[\textbf{ii)}] \textbf{Limit of One Vanishing Internal Mass}\\
Let us assume that one of the masses vanishes, e.g. $m_3\to 0$. In this case, the arguments in Eq.~\eqref{argell} turn into
\begin{equation}
w^2 \to 1 \,, \quad a_1 \to 1 \,, \quad a_2 \to 1 \,,
\end{equation}
so that the integral representations of the complete elliptic integrals in Eq.~\eqref{defell} trivially evaluate to rational and logarithmic functions, i.e. they can be represented through MPLs in this limit. This is in agreement with our result for sector $A\mathrm{x}13_{3,38}$, which is defined in Appendix~\ref{sec:hjlaporta} and coincides with this limit in the case $m_1=m_2\equiv m$.
\end{itemize}

\subsection{Identification of Elliptic Integrals}
\label{sec:identell}

Identifying elliptic integrals is a non-trivial task and there is no straightforward way to prove elliptic behavior prior to integration. It is clear that integrals with three-particle massive cuts involve the sunrise integral as subtopology and therefore have to be integrated over elliptic functions at some point. For a long time, it was common sense to turn the argument around: As long as no three-particle massive cut is present in the process under consideration, elliptic integrals are unlikely to occur. Nevertheless, recent results have shown that this rule of thumb does not apply to a number of cases. Two examples of non-planar three-point functions, which are known to evaluate to elliptic integrals despite involving only two-particle massive cuts, are depicted in the last two diagrams of Fig.~\ref{fig:elliptic}: Figure~$(c)$ is required for the computation of the two-loop electroweak form factor~\cite{Aglietti:2007}, whereas Fig.~$(d)$ is relevant to the two-loop amplitude for Higgs-plus-jet production~\cite{Primo:2016,Harley:2017}, process~$(c)$, and can be embedded into integral family~$C$ of Table~\ref{tab:hjtopo} through the sector~$C_{6,246}$ with two MIs. In fact, by setting the virtualities of the external legs of Fig.~\ref{fig:elliptic}$(d)$ to the same value $m_H^2=u$, it can be shown that the elliptic nature of the integral is preserved~\cite{AvM3}. In order to describe the current lack of understanding within the particle physics community with respect to the question of how to distinguish elliptic integrals from non-elliptic ones based on a fundamental mechanism, two more special cases of Fig.~\ref{fig:elliptic}$(d)$ are exceptionally suited:
\begin{itemize}
\item We can simplify the equal-mass case and set the mass of the external leg adjacent to the internal massive loop to zero, i.e. $u=0$. The corresponding integral contributes to the two-loop amplitudes for $t\bar{t}$ and $\gamma\gamma$ production and was analyzed in Ref.~\cite{vonManteuffel:2017b}, according to which it remains elliptic.
\item Let us go in the opposite direction by starting from Fig.~\ref{fig:elliptic}$(d)$ and setting the virtuality of the non-adjacent external leg to zero, i.e. $m_H^2=0$, with the resulting integral being relevant to the two-loop amplitude for Higgs production. This integral was computed in Ref.~\cite{Anastasiou:2006} in terms of HPLs only, which means that the elliptic structure disappears in that case.
\end{itemize}
Consequently, the question of \textit{why} a certain integral turns out to be elliptic and other ones of similar mass configuration do not remains an open issue. At the moment, the only question that can be answered is \textit{whether} or not a given integral is composed of elliptic functions. Let us dicuss this in the context of the first planar four-point function, which has been shown to be elliptic in spite of lacking a three-particle massive cut. In fact, this integral is required for the two-loop amplitude of Higgs-plus-jet production and its evaluation will be discussed in Chapter~\ref{chap:hj}. Our choice of the Laporta basis integrals $I_{59}$--$I_{62}$ of the corresponding sector~$A_{6,215}$ is defined in Appendix~\ref{sec:hjlaporta} and depicted in Fig.~\ref{fig:hjmaster}. With this example in mind, we illustrate how elliptic integrals can be identified prior to integration in what follows:
\begin{itemize}
\item[\textbf{a)}] \textbf{Coupled Differential Equations}\\
A $n$-th order differential equation of a single integral can be cast into $n$ coupled first-order differential equations and according to Eq.~\eqref{deqell}, a single elliptic integral fulfills at least a second-order differential equation. Hence, the elliptic nature of the integral becomes manifest in the coupling at the level of first-order differential equations, i.e. through the fact that other MIs of the same sector appear on the right-hand side of the equations in $D=4$. With the guidelines presented in Section~\ref{sec:basischoice}, we were able to decouple all linear first-order differential equations of the planar MIs relevant to the two-loop amplitude for Higgs-plus-jet production in a reasonable amount of time. The only exception was the mentioned sector~$A_{6,215}$, where we spent a disproportionate amount of time scanning integrals over a large range of the values $r$ and $s$ defined in Eq.~\eqref{rs} with respect to their decoupling properties. Although this was only a first indication that a decoupling might not be possible, a more concrete sign was provided in conjunction with guideline~g) in Section~\ref{sec:basischoice}, where we suggested decoupling differential equations by applying the procedure described in Ref.~\cite{Tancredi:2015}. We recall that we found relations in $D=2$ designed to decouple the differential equations in $D=4$, which failed solely in the case of sector~$A_{6,215}$.
\item[\textbf{b)}] \textbf{Factorization of Picard-Fuchs operators}\\
It is instructive to study the factorization properties of the Picard-Fuchs operator~\cite{Adams:2013,Adams:2017a,MullerStach:2012}, which corresponds to the entity of differential operators in the higher-order differential equation of a single integral belonging to a given sector of $n$~MIs. In this method, the Picard-Fuchs operator of order~$n$ factorizes into blocks of operators of order~$n_i$. The blocks of differential operators of order~$n_i$ fulfill the restriction $n=\sum_i n_i$ and exactly reflect the maximal decoupling behavior of the sector under consideration. We verified this approach by analyzing the differential operators of all planar two-loop MIs relevant to Higgs-plus-jet production and found that all sectors factorize into blocks with $n_i=1$. As expected, sector~$A_{6,215}$ with four MIs is the only exception and factorizes into blocks of $2\times 1\times 1$~\cite{Hoff2}.
\item[\textbf{c)}] \textbf{Mapping onto Known Elliptic Differential Equations}\\
In some cases, a change of variables might be found such that the differential equation of an integral under consideration can be mapped onto a differential equation of an integral known to evaluate to elliptic functions. For example, this has proven beneficial in Ref.~\cite{Aglietti:2007}, where the second-oder differential equation of Fig.~\ref{fig:elliptic}$(c)$ was mapped onto the differential equation of the equal-mass sunrise diagram. We follow a similar procedure for the corner integral~$I_{59}$ of sector~$A_{6,215}$, for which we consider the set of equations~\eqref{deqmandelstam}. For special kinematic values of the variables $s$, $t$ and $u$, the differential equation in $w\equiv m^2$ can be cast into a form similar to Eq.~\eqref{deqell}:
\begin{equation}
\left[ w \, \left(1-w^2\right) \, \frac{\d^2}{\d w^2} + \left(1 - 3 \, w^2 \right) \, \frac{\d}{\d w} - \frac{3}{4} \, w \right] \, I_{59}(w) = 0 \,.
\end{equation}
By changing the variable according to Eq.~\eqref{varchangeell}, we can perform the same transition as from Eq.~\eqref{deqell} to Eq.~\eqref{deqhyp}. The equivalent of Eq.~\eqref{deqhyp} is solved by the hypergeometric functions
\begin{equation}
_2F_1\left(\frac{3}{4},\frac{1}{4};1;z\right) \,, \qquad _2F_1\left(\frac{1}{4},\frac{3}{4};1;z\right)
\label{solhyp0}
\end{equation}
with argument $z\equiv m^4$. Reference~\cite{NIST} provides the relation
\begin{equation}
_2F_1\left(a,b;2\,b;z\right) = \left(1-\frac{z}{2}\right)^{-a} \, _2F_1\left(\frac{a}{2},\frac{a}{2}+1;b+\frac{1}{2};\left(\frac{z}{2-z}\right)^2\right) \,,
\end{equation}
which in our case turns into
\begin{equation}
_2F_1\left(\frac{1}{4},\frac{3}{4};1;z'\right) = \left(1\pm\sqrt{z'}\right)^{-\frac{1}{2}} \, _2F_1\left(\frac{1}{2},\frac{1}{2};1;2\,\left(1-\frac{1}{1\pm\sqrt{z'}}\right)\right)
\label{trafohyp}
\end{equation}
with $a=b=\nicefrac{1}{2}$ and $z'\equiv \left(\frac{z}{2-z}\right)^2$. Equation~\eqref{trafohyp} establishes a connection between the first solution in Eq.~\eqref{solhyp0} and the hypergeometric differential equation~\eqref{deqhyp} known to be fulfilled by the complete elliptic integral of first kind. Since the elliptic behavior is present in at least one point of the phase space, the integral must be considered elliptic on the whole phase space.
\item[\textbf{d)}] \textbf{Maximal Cut}\\
Another increasingly used tool consists in computing the maximal cut of a given integral. Although this approach is completely independent of the method of differential equations, it provides the solution of the homogeneous differential equation given that the integral under consideration is finite in a fixed, even number of dimensions. The corner integral~$I_{59}$ of sector~$A_{6,215}$ is finite in $D=4$ dimensions, so that this method could be applied successfully in Refs.~\cite{Bonciani:2016,Primo:2016,Frellesvig:2017,Harley:2017}. However, the integration does not commute with the $\e$-expansion if the integral diverges in a given number of fixed, even space-time dimensions. In such a case, the integral has to be evaluated by retaining the full dependence on~$D$, which turns out to be infeasible for many cases in practice.
\end{itemize}

\section{The Foundations of Complex Analysis}
\label{sec:complexana}

\begin{center}
\textit{``The imaginary numbers are a wonderful flight of God's spirit;\\
they are almost an amphibian between being and not being."} \cite{Arfken:2012}
\end{center}
\hfill \textsc{Gottfried Wilhelm von Leibniz}, 1702\\\\
In this section, we introduce some of the most useful tools of analysis and combine it with the machinery of complex numbers in order to form the powerful branch of complex analysis. This serves as mathematical basis for the next section, in which we explain how to obtain series expansions of Feynman integrals from their differential equations.\\
In the previous chapters of this thesis, the importance of complex numbers within particle physics applications has appeared in several places:
\begin{itemize}
\item In Section~\ref{sec:rules}, the Feynman rules for the propagators contain infinitesimal imaginary parts, which shift the poles off the real axis and enable the computation of Feynman integrals through deformation of the integration contour.
\item In Chapters~\ref{chap:hbb} and \ref{chap:hza}, the form factors and amplitudes are initially computed in the non-physical Euclidean region, where they are real. However, physical results in the Minkowskian region require continuing the non-physical expressions analytically, which we explained in detail for process~$(b)$ in Section~\ref{sec:results}. This implies that the classes of functions representing the results might in general develop imaginary parts, as described in Section~\ref{sec:mplproperties} in case of the MPLs defined in Eq.~\eqref{defmpl}. Ultimately, these imaginary parts vanish in physical observables like cross sections or decay rates since they arise from the absolute squares of the complex-valued amplitudes, as stated in~Eq.~\eqref{cs2}.
\item In Section~\ref{sec:defell}, we applied the Cutkosky rules to relate the imaginary part of a two-loop two-point function to the evaluation of the three-particle phase-space at tree level. The generalization thereof, known as the optical theorem, can be expressed as a link between the imaginary part of a forward-scattering amplitude and the total cross section that it corresponds to.
\end{itemize}
To sum up, this list, though being far from complete, makes clear that today's particle physics phenomenology would not exist without the theory of functions of complex variables, whose key ideas will be specified in the following subsections.

\subsection{Complex Variables and Functions}

Let us recall that complex numbers
\begin{equation}
z=x+i\,y
\label{cartesian}
\end{equation}
are defined as ordered pairs of real numbers~$x$ and $y$, and that the \textit{complex conjugate} of~$z$ is given by $z^*=x-i\,y$. The \textit{polar representation}, as opposed to the \textit{Cartesian representation} in Eq.~\eqref{cartesian}, makes use of the the magnitude~$r$ of complex numbers, referred to as \textit{modulus}, and of the \textit{argument}~$\theta$:
\begin{equation}
z = r \, e^{i\,\theta} = r \, \left( \cos\theta + i \, \sin\theta \right) \,.
\end{equation}
For real~$\theta$, $e^{i\,\theta}$ is obviously situated on the unit circle at an angle~$\theta$ from the real axis.\\
The main focus of this section are \textit{complex functions},
\begin{equation}
f(z) = u + i\,v \,, \qquad u(x,y)\in\mathbb{R} \,, \quad v(x,y)\in\mathbb{R} \,,
\label{defcomplex}
\end{equation}
which coincide with real functions~$f(x)$ for real argument~$z=x$. One distinctive property of complex functions becomes clear when roots like $z^{1/m}$ are computed, leading to $m$~different complex values, so that the function is called \textit{multi-valued} in the complex domain. This is due to the fact that
\begin{equation}
e^{2\,i\,\pi\,n} = 1 \,, \qquad \forall \, n\in\mathbb{Z} \,,
\end{equation}
which turns the logarithm into a multi-valued function upon extension to complex values:
\begin{equation}
\log z = \log\left(r\,e^{i\,\theta}\right) = \log r + i\,\left(\theta + 2\,\pi\,n\right) \,.
\label{complexlog}
\end{equation}

\subsection{Cauchy-Riemann Conditions}

As a next step, let us elaborate on the differentiation of complex functions. The derivative of~$f(z)$ is defined in the same way as for real functions,
\begin{equation}
\lim_{\delta z\to 0} \frac{f(z+\delta z) - f(z)}{z+\delta z-z} = \lim_{\delta z\to 0} \frac{\delta f(z)}{\delta z} = \frac{\d f}{\d z} = f'(z) \,.
\end{equation}
For real functions, this definition is valid provided that the left- and right-hand limits $x\to x_0^-$ and $x\to x_0^+$, respectively, are equal for the derivative $\d f/\d x$. Inevitably, this requirement has to be extended for complex functions, where the limit must be independent of the direction of approach in the complex plane. Starting from the definition~\eqref{defcomplex}, it can be shown that this constraint can be rephrased through the \textit{Cauchy-Riemann conditions},
\begin{equation}
\frac{\p u}{\p x} = \frac{\p v}{\p y} \,, \qquad \frac{\p u}{\p y} = -\frac{\p v}{\p x} \,,
\label{cauchyriemann}
\end{equation}
stating that they must hold in order for $\d f/\d z$ to exist. Conversely, the derivative $\d f/\d z$ exists if Eqs.~\eqref{cauchyriemann} are satisfied and if the partial derivatives of $u(x,y)$ and $v(x,y)$ are continuous. If $f'(z)$ does not exist in a point $z=z_0$, then $z_0$ is labeled a \textit{singular point}, which will be discussed in Section~\ref{sec:singularities}.

\subsection{Analytic Functions}

A real function~$f(x)$ is said to be \textit{real analytic} in a region, if it is differentiable and single-valued in that region. The complex equivalent of an analytic function~$f(z)$ in the complex plane is called \textit{complex analytic}\footnote{In the literature, analyticity in the complex plane is often distinguished by using the terms \textit{holomorphic} or \textit{regular}. In the following, we will use the term `analytic' by referring to complex functions in general.}. Note that a multi-valued function, which is single-valued in a specific region, can also be analytic in that region. A complex function that is analytic everywhere in the finite complex plane is known as an \textit{entire} function. In fact, extending these definitions to the complex plane entails much more far-reaching implications compared to the purely real case. For instance, an analytic function has the global characteristic of possessing all higher-order derivatives, as opposed to its real counterpart, which will be proven through \textit{Cauchy's integral formula} in Section~\ref{sec:cauchyformula}.\\
Let us give a few examples for the definitions from above: $z^k$ ($k\in\mathbb{N})$ and $e^z$ are entire functions, which can be shown through the application of the Cauchy-Riemann conditions. Similarly, $z^*$, though being continuous in the whole complex plane, is disproven to be analytic in the exact same domain. Finally, the function $\log z$ illustrates a differentiable function in every point apart from $z=0$, however it is multi-valued and thus not analytic in that region. In Section~\ref{sec:singularities}, we will see how to constrain $\log z$ to a single-valued function in a specific domain, resulting in an analytic function in that domain. Alternatively, analytic functions can be defined through the existence of their Taylor series, which we elaborate on in Section~\ref{sec:laurent}.

\subsection{Cauchy's Integral Theorem}

After establishing the differentiation of complex functions, let us turn to their integration. \textit{Cauchy's integral theorem} states that
\begin{equation}
\oint_C f(z) \, \d z = 0
\end{equation}
provided that the \textit{integration contour}~$C$ is closed within a so-called \textit{simply connected} region in the complex plane, wherein $f(z)$ is an analytic function at all points. In this respect, an integration contour is simply a path describing the integral of a variable in the complex plane, and a closed contour is indicated by the symbol~$\oint$. In addition, a region is simply connected if every closed curve within that region can be shrunk continuously to a point. By extending the theorem from simply to multiply connected regions, in which an imaginary barrier is created such that simply connected subregions are constructed, one arrives at a more general formulation of Cauchy's integral theorem:\\\\
\textit{The integral of an analytic function over a closed path has a value that remains unchanged over all possible continuous deformations of the contour within the region of analyticity.}\\\\
An example for such a construction is depicted in Fig.~\ref{fig:laurent}$(b)$ of Section~\ref{sec:laurent}. Based on this observation, one can show that
\begin{equation}
\oint_C \left(z'-z\right)^n \, \d z' =
\begin{cases}
0 \,, \quad &n\neq -1 \,, \\
2\,\pi\,i \quad &n=-1 \,,
\end{cases}
\label{cauchytheorem}
\end{equation}
given that the path~$C$ is counterclockwised closed. A rigorous proof of Cauchy's integral theorem is beyond the scope of this thesis and can be found in Refs.~\cite{Arfken:2012,Fischer:1994}, for example.

\subsection{Cauchy's Integral Formula}
\label{sec:cauchyformula}

In a more practical sense, the statement of Eq.~\eqref{cauchytheorem} can be rephrased as
\begin{equation}
\frac{1}{2\,\pi\,i} \oint_C \frac{f(z')}{z'-z} \, \d z' = f(z) \,,
\label{cauchyformula}
\end{equation}
known as \textit{Cauchy's integral formula}. Therein, $f(z')$ is an analytic function on the closed contour~$C$ and within the region bounded by~$C$, where $C$ is meant to be traversed in the counterclockwise direction. The integral is well-defined since $z$ is any point in the interior region bounded by~$C$, whereas~$z'\neq z$ lies on the contour~$C$. Equation~\eqref{cauchyformula} can be proven by introducing polar coordinates~$z'=z+r\,e^{i\,\theta}$ and analyzing the limit~$r\to 0$.\\
Cauchy's integral formula allows expressing the first derivative of~$f(z)$ at the point $z=z_0$ as
\begin{equation}
f'(z_0) = \frac{1}{2\,\pi\,i} \, \oint \frac{f(z')}{(z'-z_0)^2} \, \d z' \,.
\end{equation}
Iterating the operation of differentiation $i$~times leads to
\begin{equation}
f^{(i)}(z_0) = \frac{i!}{2\,\pi\,i} \oint \frac{f(z')}{(z'-z_0)^{i+1}} \, \d z' \,,
\label{cauchyderivative}
\end{equation}
thereby confirming the existence of derivatives of a complex analytic function to all orders.

\subsection{Laurent Series}
\label{sec:laurent}

Equipped with the theoretical foundations of the previous sections, we begin with the description of more practical tools, which will be used in the remainder of this thesis.\\
The Cauchy integral formula derived in the last section opens up the way for the derivation of \textit{Taylor's series} for functions of a complex variable: Let us suppose that we attempt an expansion of~$f(z)$ around $z=z_0$ under the condition that~$z=z_1$ is the nearest point in the complex plane, for which $f(z)$ is not analytic. We then construct a circle~$C$ centered at $z=z_0$ with radius smaller than $|z_1-z_0|$, as depicted in Fig.~\ref{fig:laurent}$(a)$. This implies that $f(z)$ is analytic on and within~$C$, so that its value is given by Eq.~\eqref{cauchyformula}, which we rewrite in the following way:
\begin{equation}
f(z) = \frac{1}{2\,\pi\,i} \oint_C \frac{f(z')}{z'-z} \, \d z' = \frac{1}{2\,\pi\,i} \oint_C \frac{f(z)}{\left(z'-z_0\right) \, \left(1-\frac{z-z_0}{z'-z_0}\right)} \, \d z' \,.
\label{taylorderive}
\end{equation}
\begin{figure}[tb]
\begin{center}
\hspace{0.3cm}
\begin{tikzpicture}
\def\gap{0.2}
\def\bigradius{3}
\def\littleradius{0.5}

\draw [help lines,->] (-0.5*\bigradius, 0) -- (1.25*\bigradius,0);
\draw [help lines,->] (0, -0.5*\bigradius) -- (0, 1.25*\bigradius);
\draw (1,1) circle (1.2);
\draw (1,1) circle (2);
\foreach \Point in {(1,1), (2,0.35), (2.4,2.4)}{
    \node at \Point {\textbullet};
    }
\draw[thick,->] (1,1) -- (2.15,0.7);
\draw[thick,->] (1,1) -- (2.8,1.9);

\node[font=\footnotesize] at (3.7,-0.3){$\mathrm{Re}\,z$};
\node[font=\footnotesize] at (0,4) {$\mathrm{Im}\,z$};
\node[font=\scriptsize] at (1,2.4) {$C$};
\node[font=\scriptsize] at (0.7,1) {$z_0$};
\node[font=\scriptsize] at (2.2,0.2) {$z'$};
\node[font=\scriptsize] at (2.6,2.6) {$z_1$};
\node[font=\scriptsize] at (1.4,1.6) {$|z_1\text{--}z_0|$};
\node[font=\scriptsize] at (1.4,0.6) {$|z'\text{--}z_0|$};
\node[font=\footnotesize] at (-2,4) {$(a)$};
\end{tikzpicture}
\hspace{1cm}
\begin{tikzpicture}
[decoration={markings,
mark=at position 7.5cm with {\arrow{>}},
mark=at position 13cm with {\arrow{<}}
}
]

\def\gap{0.2}
\def\bigradius{3}
\def\littleradius{0.5}

\draw [help lines,->] (-0.5*\bigradius, 0) -- (1.25*\bigradius,0);
\draw [help lines,->] (0, -0.5*\bigradius) -- (0, 1.25*\bigradius);
\draw (1,1) circle (1);
\draw (1,1) circle (2);
\foreach \Point in {(1,1), (-0.1,0.5), (-0.37,2), (1,2.5)}{
    \node at \Point {\textbullet};
    }
\draw[thick,dashed,-] (1,1) -- (3.75,1);
\draw[thick,->] (1,1) -- (1.8,0.4);
\draw[thick,->] (1,1) -- (2.8,1.9);

\path[draw,postaction=decorate,thick] (2.25,1.25) -- +(0.5,0) arc (10:355:1.75) -- +(-0.5,0) arc (-10:-350:1.25);

\node[font=\footnotesize] at (3.7,-0.3){$\mathrm{Re}\,z$};
\node[font=\footnotesize] at (0,4) {$\mathrm{Im}\,z$};
\node[font=\scriptsize] at (0.8,2.4) {$z$};
\node[font=\scriptsize] at (0.7,1) {$z_0$};
\node[font=\scriptsize] at (-0.35,0.5) {$z'_2$};
\node[font=\scriptsize] at (-0.1,2) {$z'_1$};
\node[font=\scriptsize] at (3,2) {$R$};
\node[font=\scriptsize] at (1.5,0.4) {$r$};
\node[font=\scriptsize] at (0.4,-0.4) {$C_1$};
\node[font=\scriptsize] at (1.3,-0.5) {$C_2$};
\node[font=\footnotesize] at (-2,4) {$(b)$};
\end{tikzpicture}
\caption[Domains for the derivation of the Taylor and Laurent series]{\textbf{Domains for the derivation of the Taylor and Laurent series.}\\
$(a)$ Circular domains for the derivation of the Taylor expansion with the expansion point~$z_0$, the point~$z'$ located on the integration contour~$C$ and the non-analytic point~$z_1$ closest to~$z_0$.\\
$(b)$ Annular region between two circles of radii~$r$ and $R$, in which the considered function~$f(z)$ is analytic, for the derivation of the Laurent expansion with the expansion point~$z_0$. $z'_1\equiv z'(C_1)$ and $z'_2\equiv z'(C_2)$ are understood to be on the integration contours~$C_1$ and $C_2$, respectively, and the point~$z$ lies within. The dashed line denotes a barrier to produce a simply connected region.}
\label{fig:laurent}
\end{center}
\end{figure}\noindent
Since $z$ is interior to the contour~$C$ and $z'$ lies on the contour, the ratio $t\equiv |z-z_0|/|z'-z_0|$ of the radii of the circles centered around~$z_0$ is smaller than one and allows applying the summation formula for geometric series:
\begin{equation}
\frac{1}{1-t} = \sum_{i=0}^{\infty} t^i \,.
\label{geometric}
\end{equation}
With this relation, Eq.~\eqref{taylorderive} becomes
\begin{align}
f(z) &= \frac{1}{2\,\pi\,i} \sum_{i=0}^\infty (z-z_0)^i \oint_C \frac{f(z')}{(z'-z_0)^{i+1}} \d z' \,, \nonumber \\
&= \sum_{i=0}^\infty \frac{f^{(i)}(z_0)}{i!} \, \left(z-z_0\right)^i \,.
\label{taylor}
\end{align}
Therein, we interchanged the order of integration and summation in the first step due to the uniform convergence of Eq.~\eqref{geometric} and made use of Eq.~\eqref{cauchyderivative} in the second step.\\
Equation~\eqref{taylor} defines the Taylor expansion associated with the analytic function~$f(z)$. It allows rephrasing the definition of an analytic function as requiring a Taylor series of above type to exist and to converge pointwise to~$f(z)$ for $z$ in a neighborhood of~$z_0$. In this context, we would like to emphasize that the derivation of the Taylor series up to Eq.~\eqref{taylor} implies that such an expansion converges for $|z-z_0|<|z_1-z_0|\equiv r$, i.e. $r$~is given by the distance from~$z_0$ to the one non-analyticity~$z_1$ of~$f(z)$ closest to~$z_0$. Since $r$~forms a circular disk of convergence in the complex plane, it is referred to as the \textit{circle of convergence} or \textit{radius of convergence} of the Taylor series.\\
At several places in this thesis, we have seen that Feynman integrals are in general ill-defined, divergent quantities. Their representation in terms of a series expansion in the dimensional regulator~$\e$ like in Eq.~\eqref{laurent} requires extending the Taylor series to negative powers of $z-z_0$. As a starting point for the derivation of such \textit{Laurent series}, we assume a function~$f(z)$ to be analytic in the annular region between an inner circle of radius~$r$ and an outer circle of Radius~$R$, as shown in Fig.~\ref{fig:laurent}$(b)$. Therein, $z$ is a point in the annular region, which is converted into a simply connected region through the introduction of a barrier intercepting the contours~$C_1$ and $C_2$ with radii~$r_1$ and $r_2$, respectively. With these assumptions, we can express $f(z)$ with the help of Cauchy's integration formula,
\begin{equation}
f(z) = \frac{1}{2\,\pi\,i} \oint_{C_1} \frac{f(z')}{z'-z} \, \d z' - \frac{1}{2\,\pi\,i} \oint_{C_2} \frac{f(z')}{z'-z} \, \d z'
\end{equation}
where the explicit minus sign in front of the second integral has been introduced in order to traverse both $C_1$ and $C_2$ in the counterclockwise sense. The exact same line of argument as for the derivation of Taylor's series up to Eq.~\eqref{taylor} results in a Taylor-like series plus an additional series, which can be converted into an equivalent series running over negative indices. The combination of both leads to
\begin{align}
f(z) &= \sum_{i=-\infty}^\infty a_i \, \left(z-z_0\right)^i \,, \nonumber \\
a_i &= \frac{1}{2\,\pi\,i} \, \oint_C \frac{f(z')}{(z'-z_0)^{i+1}} \, \d z' \,,
\label{laurentcauchy}
\end{align}
where $C$ is understood as any contour within the annular region $r<|z-z_0|<R$ encircling~$z_0$ in counterclockwise direction. If $f(z)$ is analytic within that annular region, then Eq.~\eqref{laurentcauchy} defines the Laurent series associated with~$f(z)$. Comparing this definition with Eq.~\eqref{taylor} reveals that the coefficients coincide with the ones of the Taylor expansion for $i\geq 0$, which however are rarely determined from contour integrals in practice. In Section~\ref{sec:seriesexp}, for example, we will construct an ansatz of Laurent series type in order to solve linear higher-order differential equations at fixed order in the dimensional regulator~$\e$.

\subsection{Singularities}
\label{sec:singularities}

The remarkable property of Laurent series is that they allow the representation of functions in so-called isolated singular points. A function~$f(z)$ is said to have an \textit{isolated singularity} in $z_0$ if it is not analytic in that point, but in its vicinity, and the function is called \textit{meromorphic} if it is analytic throughout the finite complex plane except for isolated singularities. The Laurent expansion around an isolated singular point will exist and will be of one of the following types:
\begin{itemize}
\item The most negative power of $(z-z_0)^i$ in the Laurent series~\eqref{laurentcauchy} is a finite negative integer~$n$. In this case, the singularity is referred to as \textit{pole of order}~$n$ and corresponds to what we have seen for the Laurent expansion of the Feynman integrals in~$\e$ in Eq.~\eqref{laurent}. Poles of order one are known as \textit{simple poles}. Without having the Laurent expansion to hand, the order of the pole can be inferred from the relation
\begin{equation}
\lim_{z\to z_0} (z-z_0)^i \, f(z)
\label{pole}
\end{equation}
as the smallest integer $i=n$, for which this limit exists.
\item The Laurent series of $f(z)$ around $z-z_0$ continues to infinite negative powers of $z-z_0$. The associated `pole of infinite order' is then denoted by the term \textit{essential singularity}. This kind of singularities are often identified directly from their Laurent expansion.
\end{itemize}
It remains to comment on the peculiarities of the point at infinity: In complex analysis, infinity is treated as a single point, and the behavior in its vicinity is studied by performing a change of variable from~$z$ to $w=1/z$ and analyzing the limit~$w\to 0$. Consequently, entire functions like~$e^z$ are said to have a singular point at $z=\infty$, and through the above distinction one can determine whether it is a pole or an essential singularity. In case of $e^z$, we have
\begin{equation}
e^z = e^{1/w} = \sum_{i=0}^\infty \frac{1}{i!} \, \left(\frac{1}{w}\right)^i \,,
\end{equation}
revealing that we are dealing with an essential singularity in the limit $w\to 0$ or equivalently in $z\to \infty$. This is in agreement with Picard's theorem, which states that any entire function that is not a polynomial has an essential singularity at infinity. The generalization of Cauchy's integral theorem to functions with isolated singularities is given by the \textit{residue theorem}, whose description is however not relevant for this thesis. \\
Apart from isolated singularities, there is a different kind of singularity uniquely associated with multi-valued functions. Ambiguities can arise from integration paths forming a closed loop around singularities of multi-valued functions. The determination of these singularities, known as \textit{branch points}, is useful in order to reduce the ambiguity with respect to the function values to the maximum possible extent. The \textit{order} of a branch point is obtained by the number of rotations required to return to the original function value. As an example, let us consider $f(z)=\sqrt{z}$, which is non-analytic at the point $z=0$. Let us draw a unit circle in the complex plane and encircle this point by starting from $z=1$, where the function value is given by~$f(z)=1$. After one counterclockwise rotation, we end up at the same point $z=1$, however with the function value $f(z)=-1$, which means that there is a branch point at $z=0$. A second rotation brings us back to the original function value, so that the branch point at $z=0$ is of order two.\\
The ambiguity of multi-valued functions can be removed by converting them into single-valued ones. For this purpose, we connect two branch points, thereby creating a line in the complex plane referred to as \textit{branch cut}\footnote{For an odd number of branch points, one of them is connected with the branch point at~$\infty$.}. The precise path between the two endpoints can be chosen freely in principle, but is usually taken to be a straight line. Any evaluation path is now prohibited to cross branch cuts, so that the original multi-valued function is restricted to be single-valued in the region bounded by them, and the corresponding single-valued functions are called \textit{branches} of the original function. In the example $f(z)=\sqrt{z}$ from above, a branch cut is created by connecting the points $z=0$ and $z=\infty$, which results in two branches. As indicated in Fig.~\ref{fig:branchcut}, any possible integration contour~$C$ is required to encircle the branch point twice, so that the ambiguity with respect to the sign of the function value at $z=1$ is removed.\\
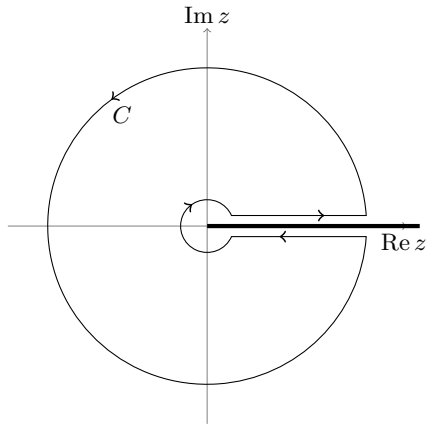
\begin{figure}[tb]
\begin{center}
\begin{tikzpicture}[scale=0.7]
\def\gap{0.4}
\def\bigradius{3}
\def\littleradius{0.5}

\draw [help lines,->] (-1.25*\bigradius, 0) -- (1.25*\bigradius,0);
\draw [help lines,->] (0, -1.25*\bigradius) -- (0, 1.25*\bigradius);
\draw[decoration={ markings,
  mark=at position 0.2455 with {\arrow[line width=0.7pt]{>}},
  mark=at position 0.765 with {\arrow[line width=0.7pt]{>}},
  mark=at position 0.87 with {\arrow[line width=0.7pt]{>}},
  mark=at position 0.97 with {\arrow[line width=0.7pt]{>}}},
  postaction={decorate}]
  let
     \n1 = {asin(\gap/2/\bigradius)},
     \n2 = {asin(\gap/2/\littleradius)}
  in (\n1:\bigradius) arc (\n1:360-\n1:\bigradius)
  -- (-\n2:\littleradius) arc (-\n2:-360+\n2:\littleradius)
  -- cycle;
\draw[ultra thick,-] (0,0) -- (4,0);

\node[font=\footnotesize] at (3.7,-0.3){$\mathrm{Re}\,z$};
\node[font=\footnotesize] at (0,4) {$\mathrm{Im}\,z$};
\node[font=\footnotesize] at (-1.6,2.1) {$C$};
\end{tikzpicture}
\caption[Integration contour~$C$ for the function $f(z)=\sqrt{z}$]{\textbf{Integration contour~$\boldsymbol{C}$ for the function $\boldsymbol{f(z)=\sqrt{z}}$} with branch points at $z=0$ and $z=\infty$. The thick line denotes the branch cut connecting the branch points.}
\label{fig:branchcut}
\end{center}
\end{figure}\noindent
In order to pass from one branch to another, one would have to cross the cut line, along which the branches are `glued together'. Across this line, the function is evidently not continuous and by construction, all the branches are equally legitimate. From a practical point of view, however, it might be convenient to agree on a branch to be used. Such a branch is often known as \textit{principal branch}, and the value~$f(z)$ on that branch is called \textit{principal value}. Commonly, the principal branch for our example $f(z)=\sqrt{z}$ is chosen to be the positive one for real, positive~$z$. Another well-known example is given by the complex logarithm, which plays an important role in the remainder of this thesis and has a branch point in $z=0$. This branch point is of infinite order, corresponding to the infinite number of its multiple values within Eq.~\eqref{complexlog}. Typically, the branch with $n=0$ is selected as the principal branch of the logarithm.\\
Let us close this section by stating that branch points are often called \textit{thresholds} in physical applications. They appear for specific limits of kinematic invariants within multi-scale processes and separate multiple kinematic regions, thereby manifesting themselves both at the level of the Feynman integrals and of the amplitude. Provided that we remain within the range of allowed values in the physical region, processes may involve thresholds of massless type, which correspond to the limits of the external invariants either tending to zero or infinity. Massive final-state particles open up further thresholds, at which the external invariants are equal to the invariant mass of the final state. Finally, processes involving massive propagators imply additional massive thresholds due to the creation of virtual particles running in the loops. \textit{Pseudo-thresholds} are distinguished from actual thresholds by the property that the integral under consideration is analytic in the corresponding limit, although the differential equations of the integrals have poles in that limit. This is precisely the reason why we have been able to derive the boundary conditions~\eqref{hzabc} for the two-loop corrections to $H\to Z\gamma$ by imposing regularity in the pseudo-thresholds, as described in Section~\ref{sec:boundary}. In contrast, this would have not worked in the actual threshold $s=4\,m_q^2$, where the integrals diverge and the procedure is not feasible. For Higgs-plus-jet production with full quark mass dependence, we will substantiate similar statements in Chapter~\ref{chap:hj} in the context of the computation of the MIs.

\section{Series Expansions from Differential Equations}
\label{sec:seriesexp}

After motivating the necessity of deriving series expansions from first- and second-order differential equations in Sections~\ref{sec:beyondmpl} and \ref{sec:elliptic}, respectively, we have introduced the underlying mathematical concepts in Section~\ref{sec:complexana}. With these preliminary insights, we are capable of describing the computation of Feynman integrals through series expansions obtained by differential equations, which has rarely been used in physical applications so far. Such power series approach was applied in case of non-elliptic one-variable integrals~\cite{Mueller:2015,Kniehl:2017} and for the calculation of specific limits of multi-scale problems, in which the elliptic nature of the integrals disappears~\cite{Melnikov:2016,Kudashkin:2017,Davies:2018}. All mentioned applications have in common that they were used to compute expansions in a single small parameter around a singular limit. This approach has been systematized for arbitrary two-scale problems with rational dependence on the kinematic invariants in Ref.~\cite{Lee:2017b}, in which multiple expansions around singular points are suggested. Two of these expansions are then connected by a numerical matching procedure in the only appearing variable of the problem in order to obtain expressions over the whole phase space. As an example, Ref.~\cite{Lee:2017b} applies this method to a four-loop sunset graph with two massless and three equally massive propagators. In order to expand elliptic integrals in a single variable, a similar approach served to compute two-loop sunrise of arbitrary masses and the two-loop crossed-ladder vertex diagram with two massive exchanges of Fig.~\ref{fig:elliptic}$(c)$\cite{Caffo:1998b,Aglietti:2007}.\\
Our goal is to generalize the approach such that it can be applied to elliptic integrals of multiple scales with algebraic dependence on the kinematic invariants. Therefore, we aim at solving up to linear homogeneous second-order differential equations of a single integral with the so-called~\textit{Frobenius method} in Section~\ref{sec:frobenius}. Subsequently, we apply it to the hypergeometric differential equation in Section~\ref{sec:frobeniusexp} and complete the picture by supplementing the inhomogeneous solution in Section~\ref{sec:frobenius2}. Finally, we explain how the power series solution associated with the inhomogeneous second-order differential equations can be simplified and used for solving canonical first-order differential equations in Section~\ref{sec:frobenius2simp}.

\subsection{The Homogeneous Second-Order Differential Equation}
\label{sec:frobenius}

In this section, we develop the mechanics of the Frobenius method for the solution of a linear, second-order differential equation of a single integral, that is coupled to other integrals of the same sector at the level of the first-order differential equations~\cite{Arfken:2012,Smirnow:1976,Dallmann:1992}. The approach can be generalized to arbitrary order of the differential equation~\cite{Frobenius:1873,Coddington:1955}, which is however not required for the computation of the planar two-loop MIs for Higgs-plus-jet production. Conversely, the corresponding solution of linear first-order differential equations of a single integral can be obtained through a trivial simplification of the procedure in the second-order case.\\
Let us start by considering a generic linear, homogeneous, second-order differential equation of the function~$M^{(h)}$ depending on the variable~$\lambda$ in the complex plane:
\begin{equation}
\frac{\d^2 M^{(h)}(\lambda)}{\d \lambda^2} + p(\lambda) \, \frac{\d M^{(h)}(\lambda)}{\d \lambda} + q(\lambda) \, M^{(h)}(\lambda) = 0 \,.
\label{deqfrobeniushom}
\end{equation}
It can be shown that a differential equation of second order possesses two linear-independent solutions $y_1(\lambda)$ and $y_2(\lambda)$\footnote{With the help of the Wronskian in Eq.~\eqref{wronskian}, this can by done by proving that any third solution must be linear dependent, i.e. that a second-order differential equation has at most two independent solutions.}, so that the most general structure of the homogeneous solution can be expressed as
\begin{equation}
M^{(h)}(\lambda) = c_1 \, y_1(\lambda) + c_2 \, y_2(\lambda) \,,
\label{solhomfrobenius}
\end{equation}
where the constants~$c_1$ and $c_2$ have to be determined by appropriate boundary conditions.\\

\textbf{Regular Singularities}

We attempt to identify the two solutions through a Laurent series ansatz of the kind~\eqref{laurentcauchy} by expanding around the point~$\lambda_0$,
\begin{equation}
M^{(h)}(\lambda) = \sum_{i=0}^\infty a_i \, \left(\lambda-\lambda_0\right)^{s+i} \,, \quad a_0 \neq 0 \,,
\label{ansatzfrobenius}
\end{equation}
where~$s$ takes a value that remains to be determined. There are restrictions on the analytic behavior of the function~$M^{(h)}(\lambda)$ in the vicinity of this point in order for the attempt to be successful, namely the point~$\lambda_0$ is forbidden to coincide with an essential singularity. Moreover, it cannot be a pole of any order, but has to be a \textit{regular singular} point of the differential equation. In order to understand what this means, let us have a look at a generic linear, homogeneous differential equation of order~$n$,
\begin{equation}
\sum_{i=0}^n p_i(\lambda) \, \frac{\d^i M^{(h)}(\lambda)}{\d \lambda^i} = 0 \,,
\end{equation}
with meromorphic functions~$p_i(\lambda)$, which can always be cast in the form with $p_n(\lambda)=1$. This equation has only regular singularities in the expansion point~$\lambda_0$ provided that $p_{n-i}(\lambda)$ has at most a pole of order~$i$ in $\lambda_0$. In the second-order equation of Eq.~\eqref{deqfrobeniushom}, this translates into the constraints that $p(\lambda)$ and $q(\lambda)$ have at most poles of order one and two in $\lambda_0$, respectively. Such a differential equation, which has at most regular singularities in the whole finite complex plane, is referred to as~\textit{Fuchsian}. An example for a Fuchsian equation is given by the hypergeometric differential equation in Eq.~\eqref{deqhyp} with regular singularities in $z=0,1,\infty$. Beyond that, we will shortly see that the requirement of Fuchsian differential equations is fulfilled by all applications presented in Chapter~\ref{chap:hj}.\\

\textbf{Finding Two Solutions}

Let us return to the ansatz in Eq.~\eqref{ansatzfrobenius} and substitute it into the homogeneous differential equation~\eqref{deqfrobeniushom}. Consequently, we obtain an algebraic equation, in which the coefficients of different powers of~$\lambda$ must vanish independently due to the uniqueness of power series. In the resulting equations, the coefficient of the lowest power in $\lambda$ leads to the so-called \textit{indicial equation}, which allows us to determine the two solutions $s_1$ and $s_2$ of the exponent~$s$ under the condition that $a_0\neq 0$ is the lowest non-vanishing term of the series. In theory, the equations $\lambda^{s+i}$ ($i\geq 1$) of higher order then produce a recurrence relation so that the coefficients~$a_i$ ($i\geq 1$) can be fixed in terms of $a_0$, which might lead to a closed analytical form in simple cases. In practice, we find it more convenient to identify the coefficients $a_i$ by inserting the power series ansatz with fixed $s=s_i$ ($i=1,2$) into Eq.~\eqref{deqfrobeniushom}. It remains to comment on two issues:
\begin{itemize}
\item In this procedure, $a_0$ remains undetermined and can be safely set to one, since it is understood to mimic the boundary constants $c_1$ and $c_2$ already introduced in Eq.~\eqref{solhomfrobenius}.
\item If the considered Feynman integral is regular in the expansion point~$\lambda_0$, then $\lambda_0$ is an \textit{ordinary point} of the differential equation. As a consequence, one of the roots of the indicial equation must evaluate to zero, so that the Laurent series simplifies to a Taylor series of the form~\eqref{taylor}.
\item In case of a homogeneous, first-order differential equation, this procedure trivially yields an indicial equation with only one possible value for~$s$.\\
\end{itemize}

\textbf{Linear Independence of the Two Solutions}

As a next step, we have to verify whether the two homogeneous solutions~$y_1$ and $y_2$ are linearly independent. A criterion of linear independence of a set of $n$ solutions can be inferred from the non-vanishing value of a determinant known as \textit{Wronskian}, which is defined by
\begin{align}
W\left[y_1(\lambda),\dots,y_n(\lambda)\right]&=
\begin{vmatrix}
y_{1}(\lambda)&y_{2}(\lambda)&\cdots &y_{n}(\lambda)\\y_{1}^{(1)}(\lambda)&y_{2}^{(1)}(\lambda)&\cdots &y_{n}^{(1)}(\lambda)\\\vdots &\vdots &\ddots &\vdots \\y_{1}^{(n-1)}(\lambda)&y_{2}^{(n-1)}(\lambda)&\cdots &y_{n}^{(n-1)}(\lambda)
\end{vmatrix}
\stackrel{!}{\neq} 0 \,, \nonumber \\
y^{(i)}(\lambda) &\equiv \frac{\d^i y(\lambda)}{\d \lambda^i} \,.
\end{align}
In our case, $n=2$ and this condition turns into
\begin{equation}
W\left[y_1(\lambda),y_2(\lambda)\right]=
\begin{vmatrix}
y_{1}(\lambda)&y_{2}(\lambda)\\ \d y_1(\lambda)/\d \lambda&\d y_2(\lambda)/\d \lambda
\end{vmatrix}
\stackrel{!}{\neq} 0 \,.
\label{wronskian}
\end{equation}
In general, the approach described so far guarantees two linear independent power series if the roots $s_1$ and $s_2=s_1+n\geq s_1$ of the indicial equation are not separated by an integer $n\in\mathbb{N}_0$. If only one linear independent solution
\begin{equation}
y_1(\lambda) = \lambda^{s_1} \sum_{i=0}^\infty a_i \, \lambda^i
\label{solfrobenius}
\end{equation}
is found with the help of the Frobenius method, the second solution~$y_2$ can be determined by requiring a non-vanishing Wronskian, which leads to the formula
\begin{equation}
y_2(\lambda) = y_1(\lambda) \, \int^\lambda \frac{\exp\left[-\int^{\lambda_2} p(\lambda_1) \, \d\lambda_1\right]}{\left[y_1\left(\lambda_2\right)\right]^2} \, \d\lambda_2 \,.
\label{liformula1}
\end{equation}
By taking the specific structure of the power series~\eqref{solfrobenius} into account, Eq.~\eqref{liformula1} can be rephrased as
\begin{equation}
y_2(\lambda) = y_1(\lambda) \, \log\left(\lambda-\lambda_0\right) + \sum_{i=-n}^\infty b_i \, (\lambda-\lambda_0)^{s_1+i} \,.
\label{liformula2}
\end{equation}
In fact, we have observed an exception to this rule: By computing the Wronskian, we have verified that our solutions are linearly independent, although the roots $s_{1,2}$ of the indicial equation are separated by an integer~$n$. These cases have in common that the first coefficient $a_1$ of either $y_1$ or $y_2$ vanishes; we therefore conjecture that this particular behavior restores the linear independence of the two solutions. In the context of Higgs-plus-jet production presented in Chapter~\ref{chap:hj}, this observation applies to all homogeneous solutions of second-order differential equations, in which the roots $s_{1,2}$ of the indicial equation are separated by a non-vanishing integer $n\neq 0$. In the case $n=0$, the indicial polynomial is solved by a double-root $s_1=s_2$, thus requiring the application of Eq.~\eqref{liformula2}.\\

\textbf{The Prescription in a Nutshell}

To sum up, we have described a method to find two solutions of the linear, homogeneous, second-order differential equation~\eqref{deqfrobeniushom} under the condition that the point of expansion~$\lambda_0$ is at worst a regular singularity of the differential equation.  Substituting the power series ansatz~\eqref{ansatzfrobenius} results in the indicial equation, which allows us to determine at least one solution. A second, linearly independent solution is provided by Eq.~\eqref{liformula2}. By requiring a non-vanishing Wronskian as in Eq.~\eqref{wronskian}, the linear independence of the two solutions can be verified. The full homogeneous solution is obtained by the sum of the two linearly independent solutions as indicated in Eq.~\eqref{solhomfrobenius} and its radius of convergence is given by the distance to the singularity closest to~$\lambda_0$.

\subsection{Example: The Hypergeometric Differential Equation}
\label{sec:frobeniusexp}

In order to clarify the steps we described so far, let us consider the hypergeometric equation~\eqref{deqhyp} with $z=\lambda$,
\begin{equation}
\left( \lambda \, \left(1-\lambda\right) \, \frac{\d^2}{\d \lambda^2} + \left( c - \left( a+b+1 \right) \, \lambda \right) \frac{\d}{\d \lambda} - a \, b \right) \, M^{(h)}(\lambda) = 0 \,,
\label{deqhyp2}
\end{equation}
and try to derive a series expansion around the point $\lambda_0=0$. In order to verify that this is a regular singular point, we cast the differential equation into the generic form~\eqref{deqfrobeniushom} with
\begin{equation}
p(\lambda) = \frac{c-\left(a+b+1\right)\,\lambda}{\lambda\,\left(1-\lambda\right)} \,, \quad q(\lambda) = -\frac{a\,b}{\lambda\,\left(1-\lambda\right)} \,.
\end{equation}
By means of Eq.~\eqref{pole}, we can show that the limits
\begin{align}
\lim_{\lambda\to \lambda_0} (\lambda-\lambda_0) \, p(\lambda)  &= \lim_{\lambda\to 0} \frac{\lambda\,\left(c-\left(a+b+1\right)\,\lambda\right)}{\lambda\,\left(1-\lambda\right)} = c \,, \nonumber \\
\lim_{\lambda\to \lambda_0} (\lambda-\lambda_0)^2 \, q(\lambda)  &= \lim_{\lambda\to 0} \frac{-\lambda^2\,a\,b}{\lambda\,\left(1-\lambda\right)} = 0
\end{align}
exist, so that $\lambda=0$ is a regular singular point. As a next step, we make an ansatz of the kind~\eqref{ansatzfrobenius} with $a_0 \neq 0$,
\begin{align}
M^{(h)}(\lambda) &= \sum_{i=0}^\infty a_i \, \lambda^{s+i} \,, \nonumber \\
\frac{\d M^{(h)}(\lambda)}{\d\lambda} &= \sum_{i=0}^\infty a_i \, (s+i) \, \lambda^{s+i-1} \,, \nonumber \\
\frac{\d^2 M^{(h)}(\lambda)}{\d\lambda^2} &= \sum_{i=0}^\infty a_i \, (s+i) \, (s+i-1) \, \lambda^{s+i-2} \,,
\end{align}
and substitute it into the hypergeometric equation~\eqref{deqhyp2}. After a few simplifications, we obtain
\begin{align}
0 = \, &a_0 \, \left(s\,(s-1)+c\,s\right) \, \lambda^{s-1} + \sum_{i=1}^\infty a_i \, (s+i) \, (s+i-1) \, \lambda^{s+i-1} \nonumber \\
&- \sum_{i=1}^\infty a_{i-1} \, (s+i-1) \, (s+i-2) \, \lambda^{s+i-1} + c \, \sum_{i=1}^\infty a_i \, (s+i) \lambda^{s+i-1} \nonumber \\
&- (a+b+1) \, \sum_{i=1}^\infty a_{i-1} \, (s+i-1) \lambda^{s+i-1} - a \, b \, \sum_{i=1}^\infty a_{i-1} \, \lambda^{s+i-1} \,.
\label{indicialhyp}
\end{align}
Since all powers of~$\lambda$ are linearly independent, their coefficients must vanish separately. The first-term on the right-hand side then yields the indicial equation,
\begin{equation}
a_0 \, \left(s\,(s-1)+c\,s\right) = 0 \,,
\end{equation}
with the two possible solutions
\begin{equation}
s_1 = 0 \,, \quad s_2 = 1-c
\end{equation}
due to $a_0 \neq 0$. The remaining part of Eq.~\eqref{indicialhyp} results in a recurrence relation, which enables us to determine the higher-order coefficients of the series expansion:
\begin{equation}
a_i = \frac{\left(s+i+a-1\right)\,\left(s+i+b-1\right)}{\left(s+i\right)\,\left(s+i+c-1\right)} \, a_{i-1} \,, \quad i\in\mathbb{N} \,.
\label{recurrencehyp}
\end{equation}
Starting from this recurrence relation, one may compute the first few coefficients~$a_i$ in terms of~$a_0$ and guess the corresponding relation for generic~$i\geq 1$, which can ultimately be proven through mathematical induction. Note that this step is a question of convenience, but not required for the success of the method.\\
At this point, we have to distinguish two cases: If $c$ is not an integer, then $s_1$ and $s_2$ do not differ by an integer and the full solution is constructed by the sum of two linearly independent power series according to
\begin{equation}
M^{(h)}(\lambda) = c_1 \sum_{i=0}^\infty a_i \, \lambda^i + c_2 \sum_{i=0}^\infty a_i \, \lambda^{1-c+i} \,,
\label{solhyp}
\end{equation}
where the coefficients $a_i$ are given by Eq.~\eqref{recurrencehyp} together with $a_0=1$ and the boundary constants $c_1$ and $c_2$ remain to be determined. If~$c$ is an integer, then $s_1=0$ and $s_2=1-c$ differ by an integer and the one power series within Eq.~\eqref{solhyp}, whose value of $s$ within the set~$(s_1,s_2)$ is smaller, has to be replaced by Eq.~\eqref{liformula2}. Note that the sign of the expression $s_1-s_2=c-1$ depends on the value of the parameter~$c$.

\subsection{The Inhomogeneous Second-Order Differential Equation}
\label{sec:frobenius2}

In view of the fact that we would like to apply the method to compute Feynman integrals, we have to consider an inhomogeneous differential equation of the form
\begin{equation}
\frac{\d^2 M(\lambda)}{\d \lambda^2} + p(\lambda) \, \frac{\d M(\lambda)}{\d \lambda} + q(\lambda) \, M(\lambda) = F(\lambda) \,,
\label{deqfrobenius}
\end{equation}
which turns into the homogeneous equation~\eqref{deqfrobeniushom} in the special case $F(\lambda)=0$. Therein, the inhomogeneity~$F(\lambda)$ represents a linear combination of all subsector integrals plus integrals of the sector under consideration that are different from~$M(\lambda)$.\\
If we manage to find one particular solution~$M^{(p)}(\lambda)$, which solves the inhomogeneous equation~\eqref{deqfrobenius}, then it contributes together with the homogeneous solution~$M^{(h)}(\lambda)$ to the full solution as follows:
\begin{equation}
M(\lambda) = M^{(h)}(\lambda) + M^{(p)}(\lambda) \,.
\label{solfrobenius2}
\end{equation}
Given that the homogeneous solution is known, it is straightforward to obtain a particular solution through the so-called \textit{variation of constants}. The procedure starts by assuming that the particular solution can be expressed in terms of the two homogeneous solutions according to
\begin{equation}
M^{(p)}(\lambda) = u_1(\lambda) \, y_1(\lambda) + u_2(\lambda) \, y_2(\lambda) \,,
\label{partsol1}
\end{equation}
where the coefficients are functions of the independent variable~$\lambda$. By imposing that this ansatz fulfills the inhomogeneous differential equation~\eqref{deqfrobenius} and that the homogeneous equation~\eqref{deqfrobeniushom} is satisfied by both $y_1$ and $y_2$, one arrives at a set of two simultaneous algebraic equations in the variables~$u'_1$ and $u'_2$:
\begin{align}
y_1(\lambda) \, u'_1(\lambda) + y_2(\lambda) \, u'_2(\lambda) &= 0 \,, \nonumber \\
y'_1(\lambda) \, u'_1(\lambda) + y'_2(\lambda) \, u'_2(\lambda) &= F(\lambda) \,.
\end{align}
Evidently, the determinant of the coefficients of these equations is given by the Wronskian in Eq.~\eqref{wronskian}, which can be used to rephrase Eq.~\eqref{partsol1} as
\begin{equation}
M^{(p)}(\lambda) = y_2(\lambda) \int^\lambda \frac{y_1(\lambda') \, F(\lambda')}{W\left[y_1(\lambda'),y_2(\lambda')\right]} \, \d\lambda' - y_1(\lambda) \int^\lambda \frac{y_2(\lambda') \, F(\lambda')}{W\left[y_1(\lambda'),y_2(\lambda')\right]} \, \d\lambda' \,,
\label{partsol2}
\end{equation}
providing a systematic way to compute an inhomogeneous solution of a second-order differential equation through the previously determined homogeneous ones. Although we have used the variation of constants in the context of Higgs-plus-jet production, we have found it more convenient to explore the particular structure inspired by the power series representation of all quantities within Eq.~\eqref{partsol2}. By recalling the universal leading-power behavior $\lambda^{s_1}$ and $\lambda^{s_2}$ of the homogeneous solutions $y_1$ and $y_2$, respectively, we can show that their leading-order contributions within Eq.~\eqref{partsol2} cancel. As a consequence, the power series structure of the particular solution~\eqref{partsol2} is fully determined by the one of the inhomogeneity~$F(\lambda)$, which we assume to be
\begin{equation}
F(\lambda) = \sum_{i\in\mathbb{N}} \sum_{j\in\mathbb{Z}} \sum_{f\in \{0\bigcup\mathcal{F}\}} b_{i,j+f} \, (\lambda-\lambda_0)^{j+f} \, \log^i(\lambda-\lambda_0) \,.
\end{equation}
Therein, the set~$\mathcal{F}$ is composed of the lowest non-integer exponents, which appear within $F(\lambda)$ and are not related by an integer. It reflects that the non-homogeneous term might involve algebraic expressions, which manifest themselves as branch points in~$\lambda=\lambda_0$. Hence, the most general structure of the particular solution can be predicted by
\begin{equation}
M^{(p)}(\lambda) = \sum_{i\in\mathbb{N}} \sum_{j\in\mathbb{Z}} \sum_{f\in \{0\bigcup\mathcal{F}\}} a_{i,j+f} \, (\lambda-\lambda_0)^{j+f} \, \log^i(\lambda-\lambda_0) \,.
\label{partsol3}
\end{equation}
We emphasize that the logarithmic contributions therein also cover the case in which a second homogeneous solution $y_2$ must be computed by means of Eq.~\eqref{liformula2}. By exploiting the representation in terms of series expansions a priori, the series form~\eqref{partsol3} of the particular solution can be used instead of the integral form~\eqref{partsol2} inferred by the variation of constants. According to our observations, this results in a substantial speed-up and is particularly useful in cases, in which the integration step within the variation of constants cannot be carried out. Note that the series form allows deriving a particular solution without being aware of the homogeneous solution. Moreover, knowing the precise structure of the inhomogeneous term might reduce the complexity of Eq.~\eqref{partsol3} considerably, as discussed in the following.

\subsection{Series Form of Inhomogeneous First- and Second-Order Solutions}
\label{sec:frobenius2simp}

All considerations made so far are valid for a certain order in the dimensional parameter~$\e$. The most striking property of the first-order differential equations~\eqref{canonicaldeq} in canonical form is that this parameter is factorized, so that the integration can be carried out order by order in~$\e$. The result for a Feynman integral~$M^{(k)}$ at a given order of the Laurent expansion
\begin{equation}
M^{(p)}(D;\lambda) = \sum_{k=0}^\infty \e^k \, M^{(p,k)}(\lambda)
\label{laurent3}
\end{equation}
is then obtained by exclusively using expressions of the previous order. With a canonical form at hand, the homogeneous equation is therefore considered solved and the full solution is uniquely determined by the inhomogeneous solution:
\begin{equation}
M(D;\lambda) = M^{(p)}(D;\lambda) \,.
\end{equation}
Therefore, the inhomogeneous solution remains to be computed, whose most general structure can be described by Eq.~\eqref{partsol3} as well. Consequently, we are able to make an ansatz of type~\eqref{partsol3} both for the inhomogeneous solution of the second-order differential equation~\eqref{deqfrobenius} in the elliptic case and for the full solution of canonical differential equations in the non-elliptic case. In the following, we will reveal possible simplifications of this ansatz under specific circumstances, which are useful in the context of Higgs-plus-jet production with full quark mass dependence.\\
The first and most obvious simplification consists in reducing the non-integer set~$\mathcal{F}$ within Eq.~\eqref{partsol3}. In order to calculate the two-loop MIs required for Higgs-plus-jet production, we deal with expressions beyond rational dependence on the kinematic invariants. Like for most known applications in particle physics phenomenology, these non-rational terms appear at most in the shape of square roots. Hence, we fix $\mathcal{F}=\{\nicefrac{1}{2}\}$ in the following, leading to
\begin{align}
M^{(p,k)}(\lambda) &= \sum_{i\in\mathbb{N}} \sum_{j\in\mathbb{Z}} \sum_{f\in \{0,\frac{1}{2}\}} a_{i,j+f} \, (\lambda-\lambda_0)^{j+f} \, \log^i(\lambda-\lambda_0) \nonumber \\
&= \sum_{i\in\mathbb{N}} \sum_{j\in\mathbb{Z}} \left( a_{i,j} + a_{i,j+\frac{1}{2}} \, \sqrt{\lambda-\lambda_0} \right) \, (\lambda-\lambda_0)^j \, \log^i(\lambda-\lambda_0) \,.
\label{partsol4}
\end{align}
As a next step, let us recall the Fuchsian nature of the homogeneous differential equation~\eqref{deqfrobeniushom} in the framework of the Frobenius method. In fact, this characteristic also applies to all inhomogeneous terms of the form~\eqref{partsol2} that we encountered in connection with the second-order differential equations~\eqref{partsol3} of the elliptic sectors $A_{6,215}$ and $A_{7,247}$ defined in Appendix~\ref{sec:hjlaporta}. Combining this with the fact that canonical differential equations are Fuchsian by definition entails that all differential equations of the planar two-loop MIs required for Higgs-plus-jet production have this property. This can be easily verified at the level of the full system of first-order differential equations~\eqref{hjdeq3} in matrix form, which means that their inhomogeneities of the kind~\eqref{partsol2} have at most simple poles. The ansatz~\eqref{partsol3} as the corresponding integrated quantity can therefore not involve negative indices~$j$, so that Eq.~\eqref{partsol4} turns into
\begin{equation}
M^{(p,k)}(\lambda) = \sum_{i=0}^\infty \sum_{j=0}^\infty \left( a_{i,j} + a_{i,j+\frac{1}{2}} \, \sqrt{\lambda-\lambda_0} \right) \, (\lambda-\lambda_0)^j \, \log^i(\lambda-\lambda_0) \,.
\label{partsol5}
\end{equation}
We can take this even further by exploring another powerful property of canonical differential equations: The coefficient at order~$k$ of the Laurent expansion in~$\e$ is given in terms of pure functions of uniform transcendental weight, allowing us to predict the structure of the result. At given weight~$k$, the power~$i$ of the logarithm can therefore be at most~$i\leq k$. This statement is precisely correct for the divergent logarithmic terms within Eq.~\eqref{partsol5}, i.e. for the ones with $j=0$, and can be extended if the logarithms are accompanied by a power series in $\lambda-\lambda_0$: Since the logarithm to the power of one could equivalently be represented through a power series, we assign a weight of one to the non-logarithmic power series within Eq.~\eqref{partsol5}, i.e. to the one with $i=0$, resulting in the requirement that $i\leq k-1$ if $j\geq 1$. In conclusion, we have
\begin{equation*}
i \leq
\begin{cases}
k \,, \quad &j=0 \,,\\
k-1 \,, \quad &j\geq 1 \,,
\end{cases}
\end{equation*}
leading to the most general ansatz for the series expansions used to solve the first-order and inhomogeneous second-order differential equations for process~$(c)$ presented in Chapter~\ref{chap:hj}:
\begin{align}
M^{(p,k)}(\lambda) &= \sum_{i=0}^{k-1} \sum_{j=1}^p \left( a_{i,j} + \frac{a_{i,j-\frac{1}{2}}}{\sqrt{\lambda-\lambda_0}} \right) \, (\lambda-\lambda_0)^j \, \log^i(\lambda-\lambda_0) \nonumber \\
&\quad\,+ \sum_{i=0}^k a_{i,0} \, \log^i(\lambda-\lambda_0) \,.
\label{partsol6}
\end{align}
Note that the upper bound of the summation in~$j$ is set to a finite integer~$p\geq 1$ in practice. On the one hand, this value has to be sufficiently small so that the determination of the coefficients $a_{i,j}$ is feasible with available computational resources. On the other hand, it must be chosen large enough in view of a desired target precision.\\
There exist two more possibilities to simplify Eq.~\eqref{partsol6} with respect to regularity conditions\footnote{The term \textit{regular} requires the property of single-valuedness, which is satisfied by converting roots and logarithms into single-valued functions as described in Section~\ref{sec:singularities}.}:
\begin{itemize}
\item Let us assume that the considered Feynman integral is known to be regular in the expansion point~$\lambda_0$, which can be deduced from physical arguments as explained in Section~\ref{sec:boundary}. In this case, logarithmically divergent terms cannot occur in its power series representation, so that Eq.~\eqref{partsol6} turns into
\begin{equation}
M^{(p,k)}(\lambda) = a_{0,0} + \sum_{i=0}^{k-1} \sum_{j=1}^p \left( a_{i,j} + \frac{a_{i,j-\frac{1}{2}}}{\sqrt{\lambda-\lambda_0}} \right) \, (\lambda-\lambda_0)^j \, \log^i(\lambda-\lambda_0) \,.
\label{partsol7}
\end{equation}
\item Moreover, let us assume that the considered Feynman integral is regular in the expansion point~$\lambda_0$ and that its differential equation contains only regular subsector integrals. In this case, logarithms cannot propagate into the solution, which translates into the requirement $i=0$ within Eq.~\eqref{partsol7}:
\begin{equation}
M^{(p,k)}(\lambda) = a_{0,0} + \sum_{j=1}^p \left( a_{0,j} + \frac{a_{0,j-\frac{1}{2}}}{\sqrt{\lambda-\lambda_0}} \right) \, (\lambda-\lambda_0)^j \,.
\label{partsol8}
\end{equation}
\end{itemize}
We have finally arrived at Eqs.~\eqref{partsol6}, \eqref{partsol7} and \eqref{partsol8}, which are sufficient to cover all solutions of the first-order and inhomogeneous second-order differential equations of the planar MIs required for Higgs-plus-jet production with full quark mass dependence. Note that the integration constant~$a_{0,0}$ has to be determined through appropriate boundary conditions in case of canonical first-order differential equations, as described in Section~\ref{sec:boundary}. In case of second-order differential equations, the boundary values have already been taken care of by the homogeneous solution, so that $a_{0,0}$ can safely be set to zero. The full solution for any Feynman integral $M(\lambda)$ is then obtained by supplementing the sum of the homogeneous and particular solution~\eqref{solfrobenius2} at a given weight~$k$ with the Laurent series introduced in Eq~\eqref{laurent3}:
\begin{align}
M(D,\lambda) &= \sum_{k=0}^4 \e^k \, M^{(k)}(\lambda) \nonumber \\
&= \sum_{k=0}^4 \e^k \, \left( M^{(h,k)}(\lambda) + M^{(p,k)}(\lambda) \right) \,.
\label{laurent4}
\end{align}
Therein, $M^{(h,k)}$ vanishes or is given by Eq.~\eqref{solhomfrobenius} in case of first-order canonical or second-order equations, respectively, whereas $M^{(p,k)}$ is given by Eqs.~\eqref{partsol6}, \eqref{partsol7} or \eqref{partsol8}, depending on the circumstances\footnote{In the following, we omit the superscript $(p)$ whenever we deal with canonical integrals.}. Moreover, the upper bound of~$k$ is determined by the fact that computing the finite part of two-loop observables requires the evaluation of the Laurent series in~$\e$ up to weight four, as explained in Section~\ref{sec:int}. Vice versa, the lower bound is given by exploring the definitions of the MIs in Appendix~\ref{sec:hjcan}, which are normalized such that their Laurent expansion starts at order~$\e^0$.\\
Before providing an example for the application of the derived ansatz, we would like to comment on a possible variation: All Eqs.~$\eqref{partsol3}$--$\eqref{partsol8}$ could have been multiplied by the factor
\begin{equation}
\sum_{m=0}^l \left(\lambda-\lambda_0\right)^{-m\,\e} \,,
\end{equation}
where~$l$ denotes the loop number of the considered integral. This would give access to the different $\lambda$-branches of order $\lambda^{-2\e},\lambda^{-\e},\lambda^0$ in the vicinity of the expansion point~$\e$, corresponding to contributions of distinct regions in the language of the \textit{strategy-of-regions}~\cite{Beneke:1997,Smirnov:2002}, which will be applied to a simple example in Section~\ref{sec:thresholdexp}. Beyond that, this approach would resum some of the divergent logarithmic expressions appearing in Eqs.~\eqref{partsol3}--\eqref{partsol8}, that emerge from integrating terms of the form~$1/\lambda$ (as opposed to explicit logarithms introduced by subsector results). However, this resummation absorbs the logarithms only for a limited number of two-point functions with one variable. As soon as it has to be carried out systematically for three- and four-loop functions with multiple scales, the pattern is unclear and a naive factorization as in the one-variable case results in expression swell. We prefer to avoid this effect and retain the explicit dependence on the logarithmic functions as shown in Eqs.~$\eqref{partsol3}$--$\eqref{partsol8}$.

\subsection{Example: The One-Loop Massive Bubble near Origin}
\label{sec:frobenius2exp}

Let us reconsider the one-loop massive bubble from Section~\ref{sec:oneloopsunrise}, but this time from the viewpoint of the power series approach. The corresponding Laporta integral~$I_{2}$ is shown in Fig.~\ref{fig:bubble} and defined in Appendix~\ref{sec:hjlaporta}, according to which its integral representation in the Euclidean region is given by:
\begin{equation}
I_2 = \int \frac{\d^D k}{(2\pi)^D} \, \frac{1}{(k^2+m_q^2)^2 \, ((k-q_1-q_2)^2+m_q^2)} \, \int \frac{\d^D l}{(2\pi)^D} \frac{1}{(l^2+m_q^2)^2} \,.
\end{equation}
As in Section~\ref{sec:oneloopsunrise}, the second integral corresponds to a factorized tapole in order to remain in the two-loop notation. Through the procedure described in Section~\ref{sec:deq1}, we switch to the canonical equivalent $M_2$ defined in Appendix~\ref{sec:hjcan}, which is given in terms of the variable $x=s/m_q^2$:
\begin{equation}
M_2 = m_q^2 \, \e^2 \, \sqrt{x\,(x-4)} \, I_2 \,.
\end{equation}
This yields the canonical differential equation
\begin{align}
\frac{\p M_2}{\p x} &= \e \left( \frac{M_1}{\sqrt{x\,(x-4)}}  - \frac{M_2}{x-4} \right) \,, \nonumber \\
\leftrightarrow \frac{\p M_2^{(k+1)}}{\p x} &= \frac{M_1^{(k)}}{\sqrt{x\,(x-4)}}  - \frac{M_2^{(k)}}{x-4} \,,
\label{sunrisedeq}
\end{align}
where we substituted the Laurent series~\eqref{laurent4} in the second step. By neglecting the normalization factor $S_\e$, we can identify the only occuring subsector with the double tadpole
\begin{equation}
M_1 = \e^2 \, I_1 = 1
\end{equation}
as indicated in Eq.~\eqref{tadpole}. The last missing ingredient required for solving the differential equation~\eqref{sunrisedeq} is the boundary condition. The boundary value of $M_2$ at the point $x_0=0$ has already been determined in Eq.~\eqref{exampleboundary},
\begin{equation}
\lim_{x\to 0} M_2^{(k)} = 0 \qquad \forall \, k \,,
\end{equation}
which reveals that $M_2$ is not only regular in the expansion point~$x_0=0$, but even vanishes at all $\e$-orders. On top of that, the differential equation~\eqref{sunrisedeq} involves only subsectors that are regular in this point, so that we can use the most simplified version~\eqref{partsol8} of the power series ansatz at given weight~$k$:
\begin{equation}
M_2^{(k)}(x) = a_{0,0} + \sum_{j=1}^p \left( a_{0,j} + \frac{a_{0,j-\frac{1}{2}}}{\sqrt{x}} \right) \, x^j \,.
\end{equation}
A powerful characteristic of the series approach is that the structure of the series expansion allows fixing the boundary constant prior to integration due to
\begin{equation}
\lim_{x\to 0} M_2^{(k)} = a_{0,0} = 0 \qquad \forall \, k \,,
\end{equation}
thereby reducing the complexity of the ansatz and the size of intermediate expressions in the process of determining the coefficients $a_{0,j}$ ($j\geq 1$). In case of canonical differential equations, however, $a_{0,0}$ does not enter the differential equation since the only unknown quantity is the derivative
\begin{equation}
\frac{\p M_2^{(k)}(x)}{\p x} = \sum_{j=1}^p \left( j \, a_{0,j} + \left(j-\frac{1}{2}\right) \, \frac{a_{0,j-\frac{1}{2}}}{\sqrt{x}} \right) \, x^{j-1} \,.
\end{equation}
Substituting this identity into the differential equation~\eqref{sunrisedeq} at weight $k=0$ and using $M_1^{(0)}=1$ as well as $M_2^{(0)}=0$ yields the purely algebraic relation
\begin{align}
\sum_{j=1}^p \left( j \, a_{0,j} + \left(j-\frac{1}{2}\right) \, \frac{a_{0,j-\frac{1}{2}}}{\sqrt{x}} \right) &= \frac{1}{\sqrt{x\,(x-4)}} \nonumber \\
&= \frac{i}{2\,\sqrt{x}} \left(1+\frac{x}{8}+\frac{3\,x^2}{128}+\frac{5\,x^3}{1024}\right) + \mathcal{O}\left(x^4\right)
\end{align}
for the weight-one coefficients $a_{0,j}$ ($j\geq 1$). Therein, we have expanded the square root expression on the right-hand side around $x=0$ and simplified it for values~$x<0$ in the Euclidean region. As a next step, we equate coefficients of the powers in~$x$ up to~$p=4$, so that
\begin{align}
a_{0,1} &= a_{0,2} = a_{0,3} = a_{0,4} = 0 \,, \nonumber \\
a_{0,\frac{1}{2}} &= i \,, \quad a_{0,\frac{3}{2}} = \frac{i}{24} \,, \quad  a_{0,\frac{5}{2}} = \frac{3\,i}{640} \,, \quad  a_{0,\frac{7}{2}} = \frac{5\,i}{7168} \,,
\end{align}
which results in the following weight-one coefficient $M_2^{(1)}(x)$ of the Laurent expansion in~$\e$:
\begin{equation}
M_2^{(1)}(x) = i\,\sqrt{x} \, \left( 1 + \frac{x}{24} + \frac{3\,x^2}{640} + \frac{5\,x^3}{7168} \right) + \mathcal{O}\left(x^4\right) \,.
\end{equation}
By using this expression and $M_2^{(0)}=0$ as input, the procedure can be iterated for $k=1$ within the differential equation~\eqref{sunrisedeq}, leading to the weight-two result~$M_2^{(2)}(x)$, and so on. The result for the expansion of the massive one-loop bubble around the origin up to weight three can then be written as
\begin{align}
M_2(x) &= i\,\sqrt{x} \, \left[ \,\e \left(1 + \frac{x}{24} + \frac{3\,x^2}{640} + \frac{5\,x^3}{7168} + \mathcal{O}\left(x^4\right) \right) \right. \nonumber \\
&\qquad\qquad\quad +\e^2 \left( \frac{x}{6} + \frac{7\,x^2}{240} + \frac{149\,x^3}{26880} + \mathcal{O}\left(x^4\right) \right) \nonumber \\
&\qquad\qquad\quad \left. +\e^3 \left( \frac{x^2}{60} + \frac{17\,x^3}{3360} + \mathcal{O}\left(x^4\right) \right) \right] + \mathcal{O}\left(\e^4\right)
\end{align}
and is in agreement with the exact expression quoted in Eq.~\eqref{sunriseexact}. This can be verified by converting the exact result to classical polylogarithms as shown in Eq.~\eqref{gtolrules}, e.g. by using the package \textsc{Gtolrules}. Subsequently, one has to perform the change of variables $\tilde{x}\to x$ with the help of relation~\eqref{landau1} and expand the resulting expression around~$x=0$.

\section{One-Dimensional Parametrization}
\label{sec:onedim}

All statements made in this chapter so far have been formulated in the single-variable case, i.e. for integrals depending on at most two scales, however we will obviously deal with a multivariate problem in Chapter~\ref{chap:hj}. Therefore, we introduce a method designed to parametrize the kinematic invariants of the process such that we are left with a one-variable problem, which enables us to apply the method of series expansions outlined in the previous section for the computation of multi-scale integrals.\\
Let us start by recalling that a process with $N+1$ scales can be described using $N$~independent variables $(x_1,\dots,x_N)\equiv \vec{x}$. We choose to parametrize the kinematic invariants with a scalar parameter~$\lambda$ as follows:
\begin{equation}
\vec{x} \to \vec{\gamma}\left(\vec{x},\lambda\right) \,, \quad \vec{\gamma}\left(\vec{x},0\right) = \vec{x}_0 \,, \quad \vec{\gamma}\left(\vec{x},1\right) = \vec{x} \,.
\label{onedim}
\end{equation}
In other words, the parametrization $\vec{\gamma}\left(\vec{x},\lambda\right)\equiv(\gamma_1\left(\vec{x},\lambda\right),\dots,\gamma_N\left(\vec{x},\lambda\right))$ has to be such that it restores the full dependence on $\vec{x}$ in the limit $\lambda\to 1$. The power series approach can then be carried out around the expansion point~$\vec{x}_0$, which translates into an expansion around the scalar parameter $\lambda=0$. Subsequently, the full dependence on the kinematic invariants is recovered in the limit $\lambda\to 1$, provided that the radius of convergence is sufficiently large. This can be understood as a way to parametrize the integration path~$\vec{\gamma}$ through $\lambda=[0,1]$. In this process, the integrals become functions of~$\lambda$ and their variation is analyzed in the direction specified by $\vec{\gamma}$. This can be done at the level of the differential equation in~$\lambda$, which is calculated from the ones in the kinematic invariants $(x_1,\dots,x_N)$ according to
\begin{equation}
\frac{\p M(D,\lambda)}{\p \lambda} = \sum_{i=1}^N \frac{\p \gamma_i(\vec{x},\lambda)}{\p\lambda} \, \left(\left.\frac{\p M(D,\vec{x})}{\p x_i} \right|_{\vec{x}\to \vec{\gamma}(\vec{x},\lambda)} \right) \,.
\label{deqlambda}
\end{equation}
Alternatively, this equation can be obtained from the total differential~\eqref{totdiff} by introducing the parametrization $\vec{x}\to \vec{\gamma}(\vec{x},\lambda)$ and subsequently differentiating with respect to~$\lambda$.\\
The simplest possibility is to construct a linear parametrization:
\begin{equation}
\vec{x} \to \vec{\gamma}\left(\vec{x},\lambda\right) = \lambda\,\left(\vec{x}-\vec{x_0}\right) + \vec{x_0} \,, \quad \vec{\gamma}\left(\vec{x},0\right) = \vec{x}_0 \,, \quad \vec{\gamma}\left(\vec{x},1\right) = \vec{x} \,.
\label{onedimlinear}
\end{equation}
The special case $N=3$, $\vec{x}_0=0$ has been applied in Ref.~\cite{Bonciani:2016} to derive one-dimensional integral representations for the planar two-loop MIs of Higgs-plus-jet production in the Euclidean region, which are integrated numerically. Moreover, the case $\vec{x}_0=0$ for generic~$N$ has proven benefical in analyzing the properties of the Picard-Fuchs operator in Ref.~\cite{Adams:2017a}. Beyond that, an approach similar in spirit has been suggested by Ref.~\cite{Papadopoulos:2014a}, where a parameter~$\tilde{\lambda}$ is introduced to parametrize the `off-shellness' of a massive external leg, corresponding to the linear parametrization of just one of the kinematic invariants~$x_m$ in our case. The differential equations in this parameter are then solved in terms of MPLs, whose argument
eventually tends to one by taking the limit $\tilde{\lambda}\to 1$. The method has been applied to compute four- and five-point functions with massless propagators~\cite{Papadopoulos:2014b,Papadopoulos:2015}, however it is not clear at this point how to extend the method for integrals with massive propagators and thus to elliptic integrals.\\
As a next step, let us analyze the impact of such a parametrization on the power series approach described in the previous section: The coefficients~$a_i$ and $a_{i,j}$ ($j\geq 1$) in the ansatz~\eqref{solhomfrobenius} and \eqref{partsol6} for solving a homogeneous second-order and an inhomogeneous first- or second-order differential equation, respectively, become functions of the kinematic invariants~$\vec{x}$. On top of that, we can fix $\lambda_0=0$, leading to
\begin{align}
M^{(h,k)}(\lambda) &= \sum_{i=0}^p a_i(\vec{x}) \, \lambda^{s+i} \,, \quad a_0 \neq 0 \,, \\
M^{(p,k)}(\lambda) &= \sum_{i=0}^{k-1} \sum_{j=1}^p \left( a_{i,j}(\vec{x}) + \frac{a_{i,j-\frac{1}{2}}(\vec{x})}{\sqrt{\lambda}} \right) \, \lambda^j \, \log^i(\lambda) \nonumber \\
&\quad\,+ \sum_{i=0}^k a_{i,0} \, \log^i(\lambda) \,.
\label{partsol9}
\end{align}
More precisely, the coefficients $a_{i,j}$ ($i,j\geq 1$) depend not only algebraically on~$\vec{x}$, but may contain logarithms of $\vec{x}$ introduced by subsector results at the level of the differential equations. In fact, this is not the case for the coefficients~$a_{i,0}$ ($i\geq 1$) of the divergent logarithmic terms within Eq.~\eqref{partsol9}, which remain purely algebraic or transcendental numbers. As explained in Section~\ref{sec:frobenius2simp}, these divergent logarithms emerge from integrating terms of the form~$1/\lambda$, i.e. they do not contain information about the behavior in the vicinity of a singular point. Therefore, this information has to be supplemented by adjusting the argument of the logarithm if the expansion is performed around a regular singular point~$x_{m,0}$ in the variable~$x_m$. In practice, this means that Eq.~\eqref{partsol9} turns into
\begin{align}
M^{(p,k)}(\lambda) &= \sum_{i=0}^{k-1} \sum_{j=1}^p \left( a_{i,j}(\vec{x}) + \frac{a_{i,j-\frac{1}{2}}(\vec{x})}{\sqrt{\lambda}} \right) \, \lambda^j \, \log^i(\lambda) \nonumber \\
&\quad\,+ \sum_{i=0}^k a_{i,0} \, \log^i\left[\gamma(x_m,\lambda)-x_{m,0}\right] \,.
\label{partsol10}
\end{align}
All statements made so far for the most general ansatz~\eqref{partsol9} equally hold for its simplified versions~\eqref{partsol7} and \eqref{partsol8}, corresponding to the case where the integrals of the sector under consideration are regular in the expansion point and to the case where also their subsectors are regular in that point, respectively. For the sake of completeness, let us specify these two simplifications for a parametrization of the kind~\eqref{onedim}:
\begin{align}
M^{(p,k)}(\lambda) &= a_{0,0} + \sum_{i=0}^{k-1} \sum_{j=1}^p \left( a_{i,j}(\vec{x}) + \frac{a_{i,j-\frac{1}{2}}(\vec{x})}{\sqrt{\lambda}} \right) \, \lambda^j \, \log^i(\lambda) \,, \label{partsol11} \\
M^{(p,k)}(\lambda) &= a_{0,0} + \sum_{j=1}^p \left( a_{0,j}(\vec{x}) + \frac{a_{0,j-\frac{1}{2}}(\vec{x})}{\sqrt{\lambda}} \right) \, \lambda^j \,. \label{partsol12}
\end{align}
Let us emphasize that we checked the validity of Eqs.~\eqref{partsol10}, \eqref{partsol11} and \eqref{partsol12} in depth: First, we derived series expansions from differential equations for the two-loop three-scale integrals required for the $H\to Z\,\gamma$ decay rate. Subsequently, we compared these power series to the ones obtained by expanding the exact results presented in Chapter~\ref{chap:hza}. As in the example of the previous section, this can be done by converting the MPLs to classical polylogarithms by means of Eq.~\eqref{gtolrules}, e.g. with the help of the package \textsc{Gtolrules}. Subsequently, the change of variables $\tilde{x}\to x$ and $\tilde{h}\to h$ has to be carried out using Eq.~\eqref{landau1}, which allows us to parametrize and expand the resulting expression in $\lambda$.\\
In a last comment, we point out that a deliberate choice of the parametrization can enlarge the radius of convergence in the multi-dimensional phase space spanned by the kinematic invariants considerably, so that the series expansion in $\lambda$ can be used over a much a wider range of the phase space. More precisely, the symmetry of the definition of the radius of convergence in $\lambda$-space as the distance to the closest singular point may be broken in $x_m$-space. Pictorially, this means that the circle of convergence associated with the parameter~$\lambda$ in the complex plane can be distorted, e.g. to an egg shaped curve in $x_m$-space as shown in Fig.~\ref{fig:egg}, by choosing an asymmetric parametrization in $\lambda$. We will explore this possibility by performing an exponential-type parametrization in order to derive series expansions for the planar two-loop MIs of Higgs-plus-jet production beyond the threshold induced by the quark mass in Chapter~\ref{chap:hj}.\\
To sum up, combining the method of series expansions in a single variable with this simple yet elegant method leads to a powerful tool, which will be used throughout the remainder of this thesis.
\begin{figure}[tb]
\begin{center}
\begin{tikzpicture}
\def\eggheight{0.3cm}
\draw [->,thick] (-3, 0) -- (5,0);
\draw [->,thick] (0, -2.5) -- (0, 2.5);
\draw[looseness=0.75,rotate=-90,color=blue,thick] (-2,0) arc (180:360:2) to[out=90,in=0] (0,4) to[out=180,in=90] (-2,0) -- cycle;

\node[circle, fill=black,inner sep=0pt, minimum size=7pt] at (0,0) {};

\node at (5.7,0){$\mathrm{Re}\,x_m$};
\node at (0,2.7) {$\mathrm{Im}\,x_m$};
\node[font=\small] at (-0.4,-0.3) {$x_{m,0}$};
\end{tikzpicture}
\caption[Distorted shape of `circle of convergence' in complex $x_m$-plane]{\textbf{Distorted shape of `circle of convergence' in complex $\boldsymbol{x_m}$-plane}\\
in case of an asymmetric parametrization $\gamma(x_m,\lambda)$ with respect to~$\lambda$.\\
The shape is centered around $\gamma(x_m,0)=x_{m,0}$.}
\label{fig:egg}
\end{center}
\end{figure}
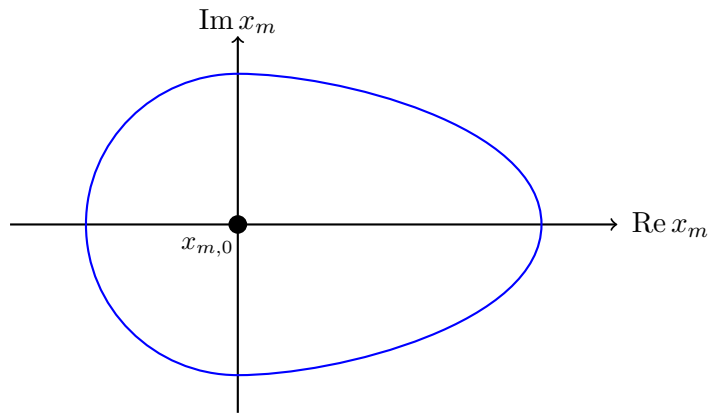

\section{The Matching Procedure}
\label{sec:matching}

In order to cover the whole phase space, multiple series expansions at different phase space points have to be computed. In the following, we elaborate on a deliberate choice of expansion points and on possibilities to connect the expansion around these points by means of a single-variable integral depending on the ratio~$y$. The multi-variate case can be obtained from a generalization of these statements.

\subsection{Partitioning of the Phase Space through Singular Points}

In general, it makes sense to derive one power series around every non-analyticity of the phase space, since their radii of convergence are such that every expansion covers all phase space points within the distance to the closest non-analyticity. In case of two singular points $y_2$ and $y_3$ over the physical range of $y$-values, this is depicted in Fig.~\ref{fig:matching}, telling us that the combination of both expansions would cover the range
\begin{equation}
y_2-r_2\leq y \leq y_+=y_3+r_3 \,,
\label{convergence}
\end{equation}
where $r_2$ and $r_3$ are the radii of convergence associated with the series expansions around $y_2$ and $y_3$, respectively. If the region quoted in Eq.~\eqref{convergence} covers all $y$-values of phenomenological interest, the only missing piece consists in determining the boundary conditions at every singular point. Should the mentioned region be insufficient, it can be augmented by an arbitrary number of series expansions around regular points to the left- or rightmost end of the $y$~axis. In Fig.~\ref{fig:matching}, this is indicated by the additional expansion around the point $y_1$, which extends the region in Eq.~\eqref{convergence} to the left:
\begin{equation}
y_-=y_1-r_1\leq y \leq y_+=y_3+r_3 \,.
\end{equation}
Note that one expansion point, e.g. $y_1$, can be set to zero without loss of generality due to the possibility of redefining the $y$~axis. Let us stress that an additional expansion around a regular point can also be carried out in between two singular points in order to increase the precision of the result in that region.\\
In general, the boundary conditions in all regular and singular points emerging from this selection procedure have to be determined separately, which requires large effort and manual interaction. This statement holds in particular when it comes to singular points, in which the integral under consideration diverges, so that the boundary conditions cannot be derived from the differential equations. Instead, they have to be supplied by explicit calculations corresponding to the highly non-trivial derivation of threshold expansions~\cite{Beneke:1997,Smirnov:2002} in the particle physics language.
\begin{figure}[tb]
\begin{center}
\begin{tikzpicture}[
    thick      ]
    
\pgfdeclarelayer{bg}    
\pgfsetlayers{bg,main}

\node[circle, draw=black, fill=red,inner sep=0pt, minimum size=10pt,label=below:{\small{$y_1=0$}}] (a) at (0,0) {};
\node[circle, draw=black, fill=blue,inner sep=0pt, minimum size=10pt,label=below:{\small{$y_2$}}] at (2,0) {};
\node[circle, draw=black, fill=green,inner sep=0pt, minimum size=10pt,label=below:{\small{$y_3$}}] at (6,0) {};
\draw[red, very thick, <->] (-2,0.4) -- (2,0.4);
\draw[blue, very thick, <->] (0,0.6) -- (4,0.6);
\draw[green, very thick, <->] (2,0.4) -- (10,0.4);
\draw (1,0.2) -- + (0,-0.4) node[below] {\small{$m_1$}};
\draw (3,0.2) -- + (0,-0.4) node[below] {\small{$m_2$}};
\draw (-2,0.2) -- + (0,-0.4) node[below] {\small{$y_-$}};
\draw (10,0.2) -- + (0,-0.4) node[below] {\small{$y_+$}};

\begin{pgfonlayer}{bg}
\draw[thick,->] (-3,0) -- (11,0) node[anchor=west] {\large{$y$}};
\end{pgfonlayer}
\end{tikzpicture}
\caption[Partitioning of the phase space for a one-variable integral]{\textbf{Partitioning of the phase space for a one-variable integral}\\
depending on real values of~$y$ in the physical region. The colored dots denote the expansion points $y_1$, $y_2$ and $y_3$, one of which can be set to zero without loss of generality. The arrows indicate the radii of convergence associated with the series expansions around points of same color and $m_1$~and $m_2$ suggest matching points in between the series expansions.}
\label{fig:matching}
\end{center}
\end{figure}

\subsection{Example: The One-Loop Massive Bubble near Threshold}
\label{sec:thresholdexp}

Let us illustrate this with the help of the simplest possible Laporta integral~$I_2$ from within sector $A_{3,35}$ in the context of Higgs-plus-jet production, which is shown in Fig.~\ref{fig:bubble} and defined in Appendix~\ref{sec:hjlaporta}. We would like to expand its integral representation, given in the Minkowski region by
\begin{equation}
I_2 = \int \frac{\d^D k}{(2\pi)^D} \, \frac{1}{(k^2-m_q^2)^2 \, ((k-q_{12})^2-m_q^2)} \, \int \frac{\d^D l}{(2\pi)^D} \frac{1}{(l^2-m_q^2)^2} \,,
\end{equation}
around the threshold $q_{12}^2\equiv (q_1+q_2)^2=4\,m_q^2$, whose existence follows from the two-particle cut of the internal quark mass lines. By applying IBP reduction techniques, we can rephrase $I_2$ in terms of the double-tadpole~$I_1$ defined in Eq.~\eqref{tadpole} and of the corner integral $K_2$ of the sector $A_{3,35}$:
\begin{align}
I_2 &= \frac{I_1-\left(D-3\right)\,K_2}{q_{12}^2-4\,m_q^2} \,, \\
K_2 &= \frac{4-D}{2} \, \int \frac{\d^D k}{(2\pi)^D} \, \frac{1}{(k^2-m_q^2) \, ((k-q_{12})^2-m_q^2)} \nonumber \\
&= \frac{4-D}{2} \, \int \frac{\d^D k}{(2\pi)^D} \, \frac{1}{(k^2+k\cdot q_{12}-y) \, (k-k\cdot q_{12}-y)} \,.
\end{align}
In the last step, we redefined the loop momentum flow in order to make explicit the dependence of the propagators on the expansion parameter $y=m_q^2-q_{12}^2/4$, corresponding to the variable transformation $(q_{12}^2,m_q^2)\to (q_{12}^2,y)$. The integral~$K_2$ can be evaluated by means of the strategy-of-regions~\cite{Beneke:1997,Smirnov:2002}, in which asymptotic expansions are derived systematically based on the hierarchy of certain momentum components. In that approach, the integral~$K_2$ receives contributions from the so-called \textit{hard region} arising from large loop momenta~$k$, which are obtained by a naive Taylor expansion:
\begin{align}
K_2^H &= \frac{4-D}{2} \, \int \frac{\d^D k}{(2\pi)^D} \, \frac{1}{(k^2+k\cdot q_{12}) \, (k-k\cdot q_{12})} + \mathcal{O}\left(y\right) \nonumber \\
&= \frac{1}{\e} \, \frac{i}{16\,\pi^2\e} \, \left(\frac{16\,\pi}{q_{12}^2}\right)^\e \, \sum_{n=0}^\infty \frac{\Gamma(n+\e)}{n!\,(1-2\,\e-2\,n)} \, \left(\frac{-4\,y}{q_{12}^2}\right)^n \,.
\label{oneloopthreshold1}
\end{align}
Beyond that, the contributions from the \textit{soft} and \textit{ultra-soft regions} corresponding to $k\sim \sqrt{y}$ and $k\sim y/\sqrt{q_{12}^2}$, respectively, vanish due to the appearance of scaleless integrals. The pattern of these contributions suggests that the missing non-zero contribution is obtained from the small-$k_0$ limit within the loop momentum $k=(k_0,\vec{k})$, so that switching to the frame $q_{12}=\{q_0,\vec{0}\}$ might pave the way. In this frame, the Lorentz invariance of the problem breaks down and the integral representation turns into
\begin{align}
K_2^P &= \int \frac{\d^{D-1} \vec{k} \, \d k_0}{(2\pi)^D} \, \frac{1}{(\vec{k}^2-k_0^2+k_0\cdot q_0+y) \, (\vec{k}^2-k_0^2-k_0\cdot q_0+y)} \nonumber \\
&= \int \frac{\d^{D-1} \vec{k} \, \d k_0}{(2\pi)^D} \, \frac{1}{(\vec{k}^2+k_0\cdot q_0+y) \, (\vec{k}^2-k_0\cdot q_0+y)} + \mathcal{O}\left(k_0^2\right) \,.
\end{align}
We assumed $k_0$ to be small in the last step, i.e. at least $|k_0|\leq \sqrt{y}$, which means that $k_0^2 \ll k_0\,q_0$ can be neglected in the denominator. Within this power series, only the leading-order term in $k_0$ survives and can be evaluated with the help of Cauchy's integral formula of Eq.~\eqref{cauchyformula} by closing the integration contour in the upper half-plane for the purpose of definiteness. The remaining integral can be written as
\begin{align}
K_2^P &= \frac{1}{(2\pi)^D} \, \frac{i\,\pi}{\sqrt{q_{12}^2}} \int \frac{\d^{D-1}\vec{k}}{\vec{k}^2+y} \nonumber \\
&= \frac{i\,(4\pi)^\e}{16\,\pi^2} \, \Gamma\left(\e-\frac{1}{2}\right) \, \sqrt{\frac{\pi\,y}{q_{12}^2}} \, y^{-\e}
\label{oneloopthreshold2}
\end{align}
and is derived by imposing $\vec{k}\sim \sqrt{y}$ in order to prevent another vanishing, scaleless integration. Being related to the Coulomb potential, the two requirements $k_0\sim y/\sqrt{q_{12}^2}$ and $\vec{k}\sim \sqrt{y}$ characterize the so-called \textit{potential region}.\\
The sum of Eqs.~\eqref{oneloopthreshold1} and \eqref{oneloopthreshold2} of the contributions from the hard and potential region, respectively, then yields the full result for the one-loop massive bubble near threshold,
\begin{align}
K_2 &= K_2^H + K_2^P \nonumber \\
&= \frac{i\,(4\pi)^\e}{16\,\pi^2} \, \Gamma(\e) \, y^{-\e} \, _2F_1\left(\frac{1}{2},\e,\frac{3}{2};-\frac{q_{12}^2}{4\,y}\right) \,,
\end{align}
successfully reproducing the known analytic expression in terms of the hypergeometric function defined through the differential equation~\eqref{deqhyp}.\\
We have seen that the simplest example appearing in our calculations requires great effort and a case-by-case treatment, which becomes even more obvious when recalling the simplicity of the computation of the same integral in an exact manner and near origin in Sections~\ref{sec:oneloopsunrise} and \ref{sec:frobenius2exp}, respectively. For more complicated integrals, multiple scales might be involved and several cases have to be distinguished. This renders the derivation of a fully automatic setup difficult, so that a more systematic approach to determine the boundary conditions at singular points is desirable.

\subsection{Connecting Multiple Series Expansions}

A remedy is found by realizing that the results which arise from individual series expansions are not independent, but have to coincide in regions where their radii of convergence overlap. In the language of Fig.~\ref{fig:matching}, these regions can be identified with
\begin{align}
\text{Region I}: \quad &y_1 \leq y \leq y_2 \,, \nonumber \\
\text{Region II}: \quad &y_2 \leq y \leq y_3+r_2 \,,
\label{overlap}
\end{align}
where $r_2$ is the radius of convergence associated with the series expansion around the singular point~$y_2$. We then proceed as follows:
\begin{enumerate}
\item First, we choose one of the available power series ought to serve for the explicit determination of the boundary conditions. This can be the series expansion around any regular or singular point, however it should be selected on the condition that its boundary values are derived in the simplest way. Since this statement most likely holds for regular points, whose limits can be analyzed by imposing regularity conditions at the level of the differential equation, regular points should be preferred over singular ones. If no reasonable series expansion exists suitable for the explicit determination of the boundary conditions, it might be more convenient to derive an additional power series around an arbitrary point on the $y$~axis eligible for that task.
\item Next, the boundary conditions are determined in the selected expansion point, e.g. $y_1=0$ in Fig.~\ref{fig:matching}.
\item Retaining the framework of Fig.~\ref{fig:matching}, we demand that the series expansions around the points $y_1$ and $y_2$ coincide in Region~I defined in Eq.~\eqref{overlap}. To warrant sufficient precision, the actual matching point should be chosen such that it is not to close to the borders of the radii of convergence. In Fig.~\ref{fig:matching}, we picked the point~$m_1$ centered in Region~I.
\item Finally, we iterate the procedure of the previous step for all regions of overlapping radii of convergence. In case of three expansion points in Fig.~\ref{fig:matching}, only one further matching at the point~$m_2$ would have to be performed. Note that, unlike the previous matching point~$m_1$, $m_2$ was not chosen in the middle of Region~II indicated in Eq.~\eqref{overlap}, since the convergence of the power series around $y_2$ is unsatisfactory at this point. Hence, a deliberate choice of the matching point might increase the precision of the results considerably.
\end{enumerate}
Let us emphasize that the step of equating two power series, one of which is accompanied with undetermined boundary conditions, translates into the solution of a system of equations with algebraic and transcendental coefficients. Since there are as many boundary constants as linearly independent equations, the system can be inverted. This matching is referred to as \textit{implicit determination of boundary constants} and can be carried out numerically or analytically. Together with the derivation of the individual power series, this opens up three possibilies:
\begin{itemize}
\item[a)] Analytical computation of the series expansions and analytical matching.
\item[b)] Analytical computation of the series expansions and numerical matching.
\item[c)] Numerical computation of the series expansions and numerical matching.
\end{itemize}
Given the complexity of the matching equations and the transcendental functions appearing therein, the analytical invertion of the system in option~a) requires enormous computational resources or may fail at all. In contrast, the analytical calculation of the series expansions is feasible, so that we choose option~b) in the remainder of this thesis. Moreover, option~b) prevents rederiving series expansions in a vast number of phase space points in case of multiple scales, as opposed to option~c).\\
Let us close this section by trivially generalizing the outlined procedure to multi-variate problems. Although we apply the method described in Section~\ref{sec:onedim} to reduce integrals depending on multiple variables to single-variable ones, the matching procedure has to be performed in the multi-dimensional space spanned by the kinematic invariants $\vec{x}=(x_1,\dots,x_N)$. This is due to the fact that $\lambda$ is restricted to the range $[0,1]$ and the one-dimensional parametrization in~$\lambda$ is eventually undone by taking the limit $\lambda\to 1$, in which the full kinematic dependence on $\vec{x}$ is recovered. As a consequence, a process with $N$ independent variables, which are real in the physical region, requires a $N$-dimensional matching. In the geometric language of Fig.~\ref{fig:matching}, this means that we have to consider $N$-dimensional diagrams with $N$-dimensional `figures of convergence'. In Fig.~\ref{fig:matching}, we have $N=1$ and \textit{lines of convergence}, whereas $N=2$ would lead to the conventional term \textit{circle of convergence} originating from the two-dimensional complex plane. Finally, $N=3$ would result in \textit{spheres of convergence}, and so on.\\
If $k$ out of the $N$ kinematic invariants are given by physical values of masses lying within the radius of convergence of the series expansion around~$y_1$, which serves to determine the boundary conditions explicitly, then the dimension of the matching reduces to $N-k$. Note that, on top of regularity in the expansion point, this establishes another criterion for selecting the power series suitable for explicit determination of the boundary values. In case of Higgs-plus-jet production, for example, we are dealing with the three independent ratios defined in Eq.~\eqref{ratios}, i.e. $N=3$. However, by choosing the series expansion which is used to derive the boundary constants explicitly as the one closest to the physical value of the Higgs boson mass, we manage to obtain the reduced dimensionality $N-k=2$ of the matching procedure. In Chapter~\ref{chap:hj}, the details of the individual series expansions and their matching in the context of Higgs-plus-jet production with full quark mass dependence in the multi-variate case will be discussed in detail.

\addtocontents{toc}{\protect\newpage}

\chapter{Two-Loop Corrections to Higgs-plus-Jet Production with Full Quark Mass Dependence}
\chaptermark{Two-Loop Corrections to Higgs-plus-Jet Production}
\label{chap:hj}

In the Standard Model, the amplitude of Higgs-plus-jet production does not exist at tree level and is loop-mediated through a heavy quark. In this chapter, we calculate the two-loop amplitude, which enters the QCD corrections of order~$\alpha_s^2$ to the cross section, in terms of Master Integrals. Furthermore, we describe the computation of the planar Master Integrals as a series expansion in the physical region, the coefficients of which depend on the quark mass.\\
After motivating the importance of quark mass corrections to Higgs-plus-jet production in Section~\ref{sec:hjintro}, we describe the kinematics and establish the integral families in Section~\ref{sec:hjnotation}. These integral families are a substantial ingredient of the amplitude computation and the IBP reduction, which are discussed in Sections~\ref{sec:hjamp} and \ref{sec:hjibp}, respectively. In Section~\ref{sec:hjdeq}, we present the calculation of the planar non-elliptic MIs, before we turn to the two planar elliptic sectors in Sections~\ref{sec:hjdeqell215} and \ref{sec:hjdeqell247} and give an outlook on the evaluation of the non-planar integrals in Section~\ref{sec:hjdeqellC}. Finally, we elaborate on the numerical evaluation of the MIs in Section~\ref{sec:hjnum} and conclude in Section~\ref{sec:hjconclusions}.

\section{Introduction}
\label{sec:hjintro}

As mentioned in Eq.~\eqref{hjchannels}, the three channels $g\,g\to H\,g$, $q\,\bar{q}\to H\,g$ and $q\,g\to H\,q$ contribute to the cross section of Higgs-plus-jet production in the Standard Model at order~$\alpha_s^2$. At the LHC, Higgs-plus-jet production in gluon fusion is the dominant production mode and yields by far the largest contribution. In this context, the CMS collaboration has recently published results for the transverse momentum ($p_T$) distribution of the Higgs boson for $p_T>450\,\mathrm{GeV}$ through the application of boosted techniques~\cite{Butterworth:2008}, where the Higgs boson was identified through its decay to bottom quarks.\\
On the theoretical side, the one-loop amplitudes of the mentioned channels, corresponding to the LO in perturbation theory, were derived long ago including finite quark mass effects~\cite{Ellis:1987,Baur:1989}. The LO contributions to higher-multiplicity processes of the Higgs boson in association with two and three jets were also calculated retaining the full dependence on the quark mass, see Refs.~\cite{DelDuca:2001a,DelDuca:2001b} and~\cite{Campanario:2013b,Greiner:2016}, respectively.\\
For precision studies of the transverse momentum distribution of the Higgs boson and of Higgs-boson-plus-jet production, an effective field theory in the limit of infinite top quark mass is commonly used. In this approach, the heavy quarks are integrated out and the Higgs boson couples directly to the gluons, thereby reducing the number of loops by one and simplifying the calculations considerably. In this way, NLO QCD corrections were obtained in Refs.~\cite{Schmidt:1997,Glosser:2002,deFlorian:1999,Ravindran:2002}, supplemented by an expansion in the inverse top quark mass~\cite{Harlander:2012,Neumann:2014} and combined with the exact tree level and real corrections~\cite{Neumann:2016}. Beyond that, the NLO QCD corrections with higher-order quark mass effects were included in multi-purpose Monte Carlo generators, leading to merged samples matched to parton showers~\cite{Buschmann:2014,Hamilton:2015,Frederix:2016}. Not long ago, these quark mass effects have been used to improve the transverse momentum distribution of the Higgs boson at NNLO in the effective field theory, which was previously derived in Refs.~\cite{Gehrmann:2011,Boughezal:2013,Chen:2014,Boughezal:2015c,Boughezal:2015d}, above the top quark mass threshold~\cite{Chen:2016}.\\
\begin{figure}[tb]
	\begin{center}
	\includegraphics[width=0.9\textwidth]{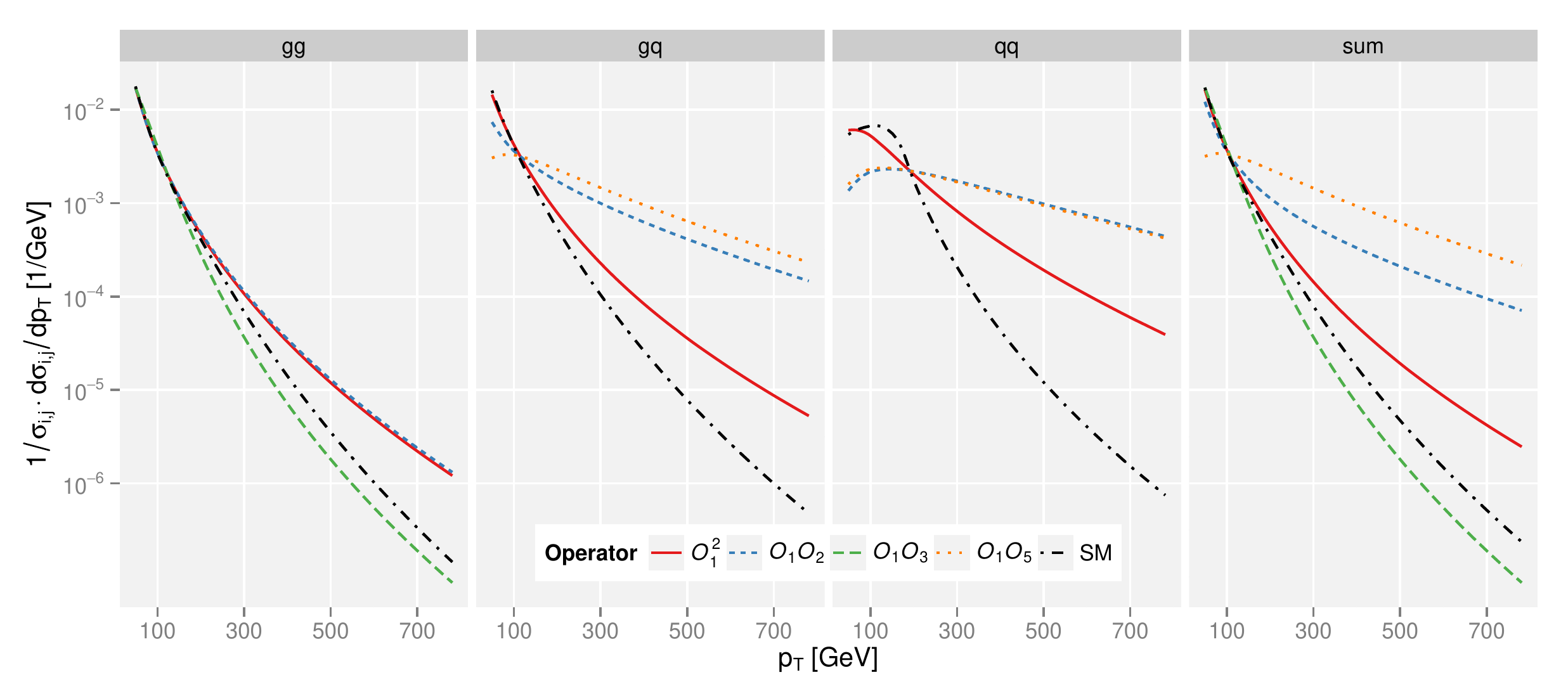}
	\caption[Normalized Higgs transverse momentum distributions for scalar coupling operators]{\textbf{Normalized Higgs transverse momentum distributions for scalar coupling operators}~\cite{Harlander:2013}. The orange, blue and green lines stand for BSM operators, whereas the SM predictions, denoted by black and red lines, differ in the high-$p_T$~regime, corresponding to the estimated deviation of the full theory from the effective field theory description.}
	\label{fig:highpt}
	\end{center}
\end{figure}\noindent
The effective field theory description is expected to be inappropriate at large transverse momenta, where the top quark loop is resolved by the recoiling jet. This can be seen from Fig.~\ref{fig:highpt}, in which the result with full top quark mass dependence deviates from the one in the effective field theory in the high-$p_T$ regime, and it is precisely in this region that deviations from the Standard Model due to new heavy states could become visible~\cite{Grojean:2013,Azatov:2013,Azatov:2016,Grazzini:2016c}. The calculation of NLO QCD corrections with full top quark mass dependence was therefore recognized as high-priority aim in order to produce reliable predictions in the regime of large transverse momenta of the Higgs boson~\cite{Dittmaier:2012,Heinemeyer:2013}.\\
In the meantime, substantial progress was made on the side of the NLO QCD corrections with full top quark mass dependence: Initially, the planar MIs were computed in the Euclidean region in Ref.~\cite{Bonciani:2016} in terms of a parametric integral representation, which is integrated numerically. Therein, it was also shown that the set of two-loop planar MIs contains elliptic integrals. Not long ago, the two-loop amplitudes for the three mentioned channels were approximated by the limit, in which the Higgs transverse momentum is much larger than the top quark mass~\cite{Kudashkin:2017}. Subsequently, these amplitudes were combined with the squared one-loop amplitudes for Higgs production in association with two jets in order to calculate the differential cross sections to $g\,g\to H\,g$, $q\,\bar{q}\to H\,g$ and $q\,g\to H\,q$ at NLO in the limit $p_T\gg m_t$~\cite{Lindert:2018}\footnote{Very recently, a similar calculation has been carried out by combining asymptotic expansions in the low- and high-energy limit, leading to predictions for $p_T<225\,\mathrm{GeV}$ and $p_T>500\,\mathrm{GeV}$ at NLO~\cite{Neumann:2018}.}. The NLO corrections in the regime relevant for the description of the Higgs boson transverse momentum distribution at $p_T\gtrsim 400\,\,\mathrm{GeV}$ turn out to be large and very different from the results obtained for a point-like $Hgg$ coupling. This is due to the fact that the effective field theory description is violated by processes, in which a Higgs boson recoils against one or more jets and acquires a large transverse momentum. However, the $K$-factor\footnote{The $K$-factor is simply the ratio of the NLO cross section and the LO cross section in this case.} is similar in both cases, since the large difference between the LO calculations with full quark mass dependence and in the effective field theory cancels the difference of the NLO corrections to a large extent. The observation of Ref.~\cite{Lindert:2018} has been confirmed by the very recent numerical computation of the NLO corrections to Higgs-plus-jet production with full top quark mass dependence~\cite{Jones:2018}, which proceeds along the lines of the NLO corrections to double-Higgs production including finite top mass effects~\cite{Borowka:2016a, Borowka:2016b}. In Ref.~\cite{Jones:2018}, the squared ratio of the Higgs boson and the top quark mass is set to $m_H^2/m_t^2=12/23$, thereby reducing the number of scales by one, so that the IBP reduction and the computation of the MIs are simplified.\\
At the other end of the spectrum of the internal quark mass, the two-loop amplitudes in the limit of a nearly massless quark were obtained in Refs.~\cite{Melnikov:2016,Melnikov:2017}. As in the high-$p_T$ limit, this was achieved by performing series expansions around one singular point at the level of first-order differential equations, similarly to the procedure described in Section~\ref{sec:seriesexp}. This enabled the computation of the top-bottom interference contribution to the Higgs-plus-jet production cross section at NLO~\cite{Lindert:2017}, which turns out to amount to around ten~percent of the top-mediated Higgs production cross section. Although the interference is mass-suppressed by $m_b^2/m_t^2$ with respect to the pure top quark contribution, it is enhanced by large logarithms of the form $\log^2(m_H^2/m_b^2)$ and thus has to be taken into account. If less inclusive quantities like the transverse momentum distribution of the Higgs boson are considered, the double-logarithmic enhancement depends on~$p_T$. In this context, the origin and structure of terms of the form~$\log^2(p_T^2/m_b^2)$ is unclear, so that an understanding of how to resum them is missing. Empirical studies suggest that bottom quark mass effects lead to differences in the predicted transverse momentum distributions of the Higgs boson, depending on how the $p_T$-dependent logarithms are treated, and cannot be neglected~\cite{Mantler:2012,Grazzini:2013,Banfi:2013,Bagnaschi:2015}.\\
In fact, the results for the NLO corrections to the decay rate of $H\to Z\,\gamma$ presented in Chapter~\ref{chap:hza} exhibit the same behavior~\cite{Gehrmann:2015a,Bonciani:2015}. The MIs evaluated therein are given by up to three-point functions with two massive legs mediated by a quark loop. As explained in Section~\ref{sec:calc}, they form a subset of the two-loop planar MIs of Higgs-plus-jet production belonging to integral family~$A$, as defined in Table~\ref{tab:hjtopo}, and thus are a crucial constituent of the amplitudes derived in the following.\\
Our goal is to provide a method to compute the planar MIs in the physical region and to report on the status of the non-planar MIs relevant to the two-loop corrections to Higgs-plus-jet production. The analytic evaluation of these MIs in the physical region is still not available and the corresponding two-loop amplitudes might be helpful to understand the origin of the behavior of the differential cross section in the high-$p_T$ regime. We emphasize that our results are valid for arbitrary quark masses, so that pure top quark and top-bottom interference contributions can be treated simultaneously. Beyond that, retaining the full dependence on both the Higgs and the quark masses allows us to analyze the variation of the cross section with their ratio and to gain insight into top quark threshold effects. Finally, these MIs serve to demonstrate the success of the power series method described in Chapter~\ref{chap:workflow3}, which is used to compute multi-scale elliptic integrals for the first time analytically in the physical region. This procedure can in principle be applied to any other elliptic integrals with an arbitrary number of scales and thus may be of use in future calculations.\\
Let us point out that the MIs derived in the following are also a crucial ingredient of inclusive Higgs production with full quark mass dependence at NNLO. Inclusive Higgs production is currently known to $\mathrm{N^3LO}$ in the effective field theory~\cite{Anastasiou:2015,Anastasiou:2016} and up to NLO including finite quark mass effects~\cite{Graudenz:1992,Spira:1995}. In Ref.~\cite{Mueller:2015}, the NLO amplitudes with full quark mass dependence were revisited in the limit of a nearly massless quark, thereby suggesting a power series approach for the same calculation at NNLO.

\section{Notation and Conventions}
\label{sec:hjnotation}

\subsection{Kinematics}
\label{sec:hjkinematics}

The Standard Model does not allow a tree-level coupling of the processes
\begin{align}
g(q_1)\,g(q_2)&\to g(q_3)\,H(q_4)\,, \nonumber \\
q(q_1)\,\bar{q}(q_2)&\to g(q_3)\,H(q_4)\,, \nonumber \\
q(q_1)\,g(q_2)&\to q(q_3)\,H(q_4) \,,
\label{hjchannels2}
\end{align}
which are instead mediated through a virtual quark loop. As pointed out in Eq.~\eqref{feynmanampexp}, the corresponding Feynman amplitude~$\mathcal{M}$ obeys a perturbative expansion in the strong coupling constant~$\alpha_s$:
\begin{equation}
\mathcal{M} = \sqrt{\alpha_s} \, \left[\mathcal{M}^{(1)} \, \alpha_s + \mathcal{M}^{(2)} \, \alpha_s^2 + \mathcal{O}(\alpha_s^2) \right] \,.
\label{hjfeynmanamp}
\end{equation}
Similarly to the computation of the $H\to Z\,\gamma$ decay rate presented in Chapter~\ref{chap:hza}, we split the theory of $N_F=6$ active quark flavors into $N_F=N_h+N_l$ with $N_h=2$ heavy and $N_l=4$ light quarks due to the mass hierarchy of the quarks. As a consequence, we consider the full dependence on both the top and bottom quark masses $m_t$ and $m_b$,
\begin{equation}
\mathcal{M}^{(i)} = c_t \, \mathcal{M}_t^{(i)} + c_b \, \mathcal{M}_b^{(i)} \,,
\end{equation}
and neglect the masses $m_u=m_d=m_s=m_c=0$ of the remaining lighter quarks. These light quarks solely contribute to the renormalization of the gauge coupling in the form of $N_l$ and thus $N_F$. In contrast to the two-loop amplitudes necessary for the $H\to Z\,\gamma$ decay width, the two-loop amplitudes for Higgs-plus-jet production correspond to the NLO in~$\alpha_s$ and require a renormalization of the strong coupling constant. However, only the calculation of the unrenormalized two-loop amplitude~$\mathcal{M}^{(2)}$ will be described in Section~\ref{sec:hjamp} in order to draw conclusions about the MIs occuring therein. Note that the one-loop amplitude~$\mathcal{M}^{(1)}$ was calculated long ago~\cite{Ellis:1987,Baur:1989}.\\
The amplitude at any loop order depends on the kinematic invariants introduced in Eq.~\eqref{mandelstam} for four-point functions with three massless external legs $q_1^2=q_2^2=q_3^2=0$ and one massive external leg $q_4^2=m_H^2$ corresponding to the Higgs boson:
\begin{equation}
s\equiv (q_1+q_2)^2 \,, \quad t\equiv (q_1-q_3)^2 \,, \quad u\equiv (q_2-q_3)^2 \,, \quad m_H^2 = s+t+u \,.
\label{mandelstam2}
\end{equation}
Therein, we expressed the Higgs boson mass~$m_H$ through the Mandelstam invariants $s$, $t$ and $u$. Due to momentum conservation $q_1+q_2=q_3+q_4$, we can omit one of the kinematic invariants appearing in Eq.~\eqref{mandelstam2} in favor of the others. In the remainder of this thesis, we choose $s$, $u$ and $m_H^2$ as independent scales, which corresponds to the natural set of invariants in the sense that the planar integral families presented in Section~\ref{sec:integralfamilies} contain cuts in $s$ and $u$, but none in~$t$.\\
The production channels specified in Eq.~\eqref{hjchannels2} require constraining the values of the Mandelstam invariants and of the internal quark mass~$m_q$ to the following kinematical range:
\begin{equation}
s>m_H^2>0 \,, \quad m_H^2-s<u<0 \,, \quad m_q^2 > 0 \,.
\label{srange}
\end{equation}
Consequently, we are dealing with a problem depending on four scales, out of which we can construct three independent ratios thanks to the scaling relation~\eqref{scaling}, as explained in Section~\ref{sec:deq1}. We choose to normalize these ratios to the quark mass:
\begin{equation}
x = \frac{s}{m_q^2} \,, \quad z = \frac{u}{m_q^2} \,, \quad h = \frac{m_H^2}{m_q^2} = \frac{s+t+u}{m_q^2} \,.
\label{ratiodef}
\end{equation}
For this set of independent variables, the restrictions in Eq.~\eqref{srange} translate into the physical region
\begin{equation}
x>h>0 \,, \quad h-x<z<0 \,,
\label{xrange}
\end{equation}
which can be visualized in the $(s,u)$-plane through the so-called \textit{Dalitz plot} depicted in Fig.~\ref{fig:dalitz}.\\
\begin{figure}[tb]
\begin{center}
\scalebox{1.5}{\begin{tikzpicture}
\pgfplotsset{every tick label/.append style={font=\tiny}}
\begin{axis}[
  axis lines=middle,
  xmin=-2,
  xmax=7,
  ymin=-7,
  ymax=2,
  xlabel=$x$,
  ylabel=$z$,
  xtick={-1,0,2,3,5,6},
  ytick={-6,-5,-4,...,1},
  extra x ticks={1,4},
  extra x tick labels={\hspace*{1.2em}$1$,\hspace*{1.2em}$4$},
  every axis x label/.style={
    at={(ticklabel* cs:1.025)},
    anchor=west,
},
every axis y label/.style={
    at={(ticklabel* cs:1.025)},
    anchor=south,
}
]
\addplot[name path=f,blue,very thick,samples=100,domain=0.5:7] {0.5-x} node[left, pos=0.95,font=\scriptsize]{$z=h-x$};
\addplot[name path=g,blue,very thick,samples=100,domain=0.5:7] {0} node[above, pos=0.9,font=\scriptsize]{$z=0$};
\addplot[dashed,blue,very thick,samples=100,domain=-2:0.5] {0.5-x};
\addplot[dashed,blue,very thick,samples=100,domain=-2:0.5] {0};
\addplot +[mark=none,very thick,blue,dashed] coordinates {(0.5,-7) (0.5,2)} node[right, pos=0.95,font=\scriptsize]{$x=h$};
\addplot +[mark=none,very thick,red,dashed] coordinates {(4,-7) (4,2)} node[right, pos=0.95,font=\scriptsize]{$x=4$};

\node[blue,anchor=west,font=\footnotesize] at (axis cs:4.5,-2) {physical};
\node[blue,anchor=west,font=\footnotesize] at (axis cs:4.5,-2.5) {region};

\addplot [
thick,
color=black,
fill=blue, 
fill opacity=0.1
]
fill between[
of=f and g,
soft clip={domain=0:7},
];
\end{axis}
\end{tikzpicture}}
\caption[Dalitz plot for the physical region relevant to Higgs-plus-jet production]{\textbf{Dalitz plot for the physical region relevant to Higgs-plus-jet production} in the plane of the independent variables $x$ and $z$. Blue lines indicate the borders derived in Eq.~\eqref{xrange}, which encircle the physical region denoted by the light blue area. The red line stands for the threshold induced by the massive two-particle cut of the internal quark line.}
\label{fig:dalitz}
\end{center}
\end{figure}
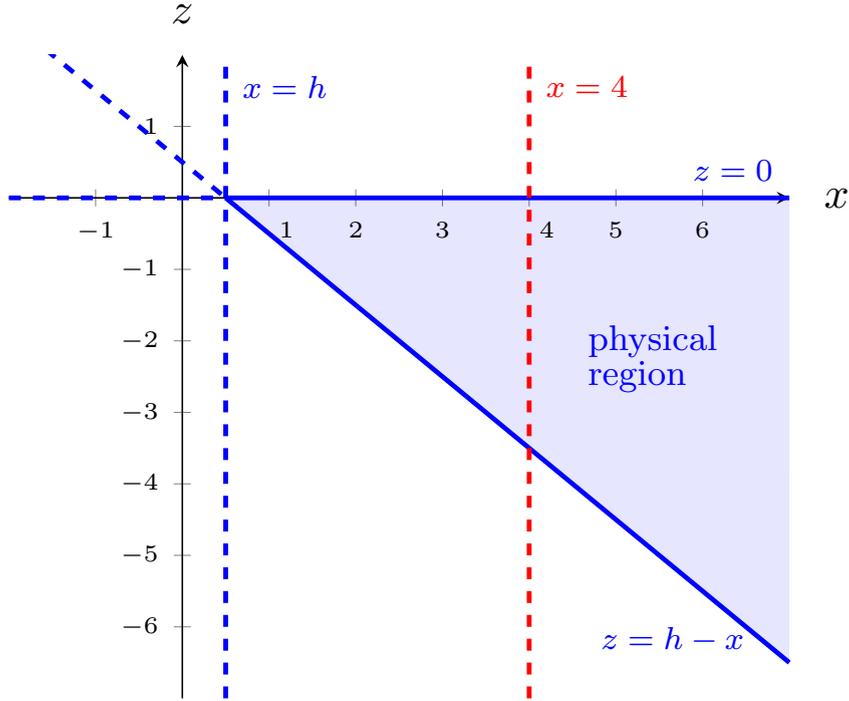\noindent
An estimation of the upper bound of the range in $x$ will be important for evaluating the planar MIs over the whole phase space in Section~\ref{sec:hjdeq}. It can be inferred from the desired upper bound of the transverse momentum~$p_T$ of the Higgs boson by expressing it in terms of the kinematic invariants $s$, $u$ and $m_H^2$:
\begin{equation}
p_T(s,u,m_H^2) = \sqrt{s\,u\,\frac{m_H^2-s-u}{(m_H^2-s)^2}} \,.
\end{equation}
In the high-energy limit, we can impose $s\gg m_H^2$ and $-u\gg m_H^2$, so that
\begin{equation}
p_T(s,u,m_H^2=0) = \sqrt{-\frac{u}{s} (s+u)} \,.
\end{equation}
For a given value of~$s$, the only maximum of this function in the range $-s<u<0$ is at $u=-s/2$, where the function value becomes
\begin{equation}
p_T^\mathrm{max} \equiv p_T(s,u=-\frac{s}{2},m_H^2=0) = \frac{\sqrt{s}}{2} \,.
\end{equation}
If applications of phenomenological interest require predictions for values up to $p_T^\mathrm{max}\approx 800\,\mathrm{GeV}$, then this equation tells us that $x\approx 85$ supposing $m_q=m_t=173\,\mathrm{GeV}$.

\subsection{Integral Families}
\label{sec:integralfamilies}

As explained in Section~\ref{sec:reduction} in great detail, the computation of the scattering amplitudes in terms of MIs and the evaluation of those MIs using differential equations, which is presented in the following sections, requires the definition of integral families. In order to derive a minimal set, we proceed as follows:
\begin{enumerate}
\item First, we determine the number~$\rho$ of distinct propagators of the integral families through Eq.~\eqref{irreducible}. Since we are dealing with two-loop four-point functions, we have $n=4$ and $l=2$ resulting in $\rho=9$. As a consequence, every integral can be written by means of Eq.~\eqref{scalarint3},
\begin{equation}
I(q_1,\dots,q_4) =  \int \frac{\d^Dk\,\d^Dl}{(2\pi)^{2\,D}} \, \frac{1}{D_1^{b_1} \dots D_9^{b_9}} \,, \quad b_j\in\mathbb{Z} \,,
\label{hjscalarint}
\end{equation}
by providing the propagator set $D_1,\dots,D_9$ of the integral families.
\item Second, we generate all two-loop Feynman diagrams contributing to the channels specified in Eq.~\eqref{hjchannels2} with the program~\textsc{Qgraf}. In doing so, we find that they contain at most $\sigma=7$ denominators with positive power. All of these so-called \textit{top-level topologies} are depicted in Fig.~\ref{fig:hjdiag} in the case of gluon fusion, telling us with the help of Eq.~\eqref{irreducible2} that the remaining $\tau=\rho-\sigma=2$ denominators stand for irreducible scalar products. They are represented by admitting negative powers of the corresponding indices $b_j$ within Eq.~\eqref{hjscalarint}, since they cannot be rewritten as a linear combination of the seven denominators with positive exponents.
\begin{figure}[tb]
	\begin{center}
	\includegraphics[width=\textwidth]{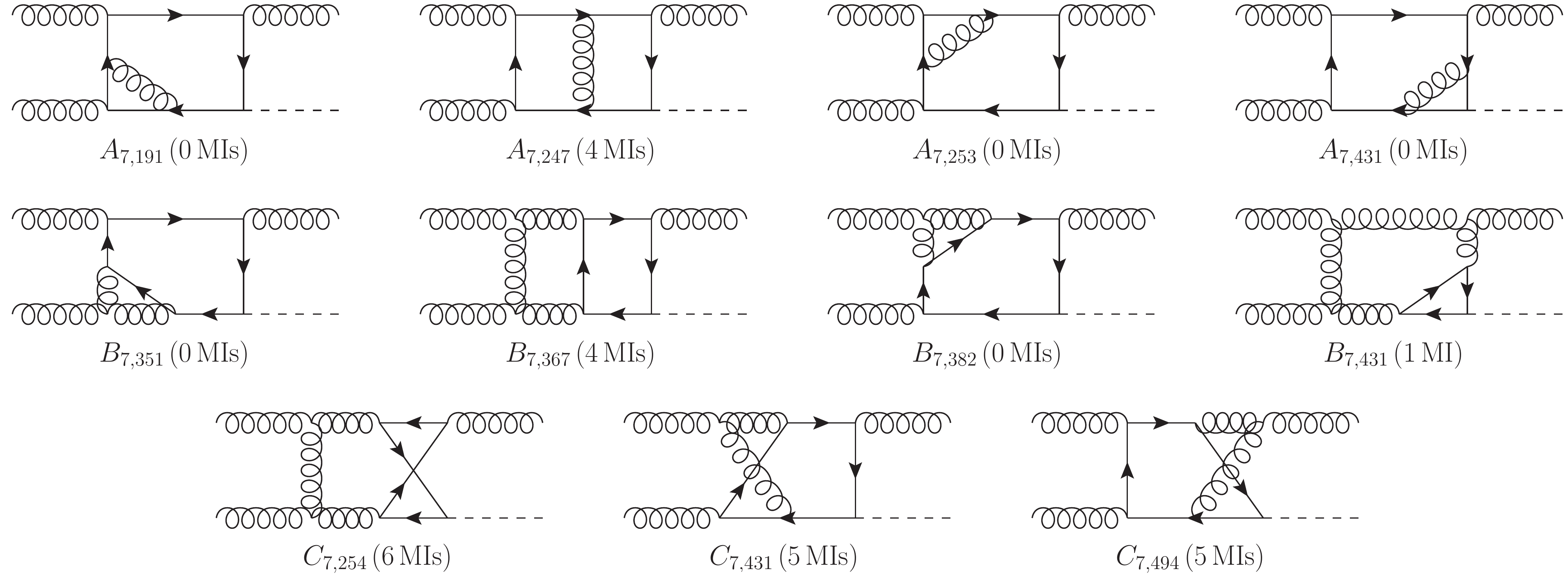}
	\caption[Example diagrams for the two-loop amplitude of Higgs-plus-jet production in gluon fusion]{\textbf{Example diagrams for the two-loop amplitude of Higgs-plus-jet production in gluon fusion.} Dashed, curly and straight lines denote Higgs bosons, gluons and quarks, respectively. Below each diagram we indicate the associated sector embedded in the integral families of Table~\ref{tab:hjtopo} in the notation $\textit{family}_{t,ID}$, which is explained in Appendix~\ref{sec:hjlaporta}. Moreover, we specify the number of MIs of that sector in parentheses.}
	\label{fig:hjdiag}
	\end{center}
\end{figure}\noindent
\item As a next step, the propagators of the integral families are identified by collecting all massive and massless propagators of the top-level topologies and producing as few integral families as possible with nine propagators accommodating them all. In the case of Higgs-plus-jet production, this procedure leads to the integral families defined in Table~\ref{tab:hjtopo}. The integral sectors, onto which the top-level diagrams are mapped, are indicated in Fig.~\ref{fig:hjdiag} below each diagram in the notation $\textit{family}_{t=7,ID}$ introduced in Appendix~\ref{sec:hjlaporta}. Moreover, we include the number of MIs of the corresponding sector in that figure, which arises from solving the IBP identities as described in Section~\ref{sec:hjibp}.
\item Finally, we benefit from a very useful feature of the \textsc{Reduze} code, serving as a check: It allows importing the \textsc{Qgraf} output and applies the crossing relations and sector symmetries specified in Section~\ref{sec:symmetryrel}. Subsequently, the program verifies whether all Feynman diagrams contributing to the amplitude can be expressed in terms of the previously defined integral families.
\end{enumerate}
Note that in general, one has to proceed along the same lines for the Feynman diagrams of the quark channels within Eq.~\eqref{hjchannels2}. However, the required propagators form a subset of the ones collected in the case of gluon fusion, so that the Feynman diagrams involving quarks are also covered by the integral families in Table~\ref{tab:hjtopo}.\\
\begin{table}[tb]
\caption[Integral families for the reduction of the two-loop corrections to Higgs-plus-jet production to Master Integrals]{\textbf{Integral families for the reduction of the two-loop corrections to Higgs-plus-jet production to Master Integrals.} Each of the three families $A$, $B$ and $C$ are given by a set of nine propagators in the Euclidean metric, where the loop momenta are labeled $k$ and $l$, whereas $q_1$, $q_2$ and $q_3$ denote the external momenta of the partons. Note that for the planar families $A$ and $B$, there is only one propagator which contains both loop momenta, whereas the non-planar family $C$ includes three of this kind. In addition, family $A$ is fully symmetric in the exchange~$k\leftrightarrow l$.}
\setlength{\tabcolsep}{0.65cm}
\begin{tabularx}{\textwidth}{lll}
\toprule
\multicolumn{1}{c}{Family $A$} & \multicolumn{1}{c}{Family $B$} & \multicolumn{1}{c}{Family $C$} \\
\midrule
$k^2+m_q^2$ & $k^2$ & $k^2+m_q^2$ \\
$l^2+m_q^2$ & $l^2+m_q^2$ & $l^2+m_q^2$ \\
$(k-l)^2$ & $(k-l)^2+m_q^2$ & $(k-l)^2$ \\
$(k-q_1)^2+m_q^2$ & $(k-q_1)^2$ & $(k-l-q_2)^2$ \\
$(l-q_1)^2+m_q^2$ & $(l-q_1)^2+m_q^2$ & $(k-l-q_2-q_3)^2$ \\
$(k-q_1-q_2)^2+m_q^2$ & $(k-q_1-q_2)^2$ & $(k+q_1)^2+m_q^2$ \\
$(l-q_1-q_2)^2+m_q^2$ & $(l-q_1-q_2)^2+m_q^2$ & $(l+q_1)^2+m_q^2$ \\
$(k-q_1-q_2-q_3)^2+m_q^2$ & $(k-q_1-q_2-q_3)^2$ & $(k-q_2-q_3)^2+m_q^2$ \\
$(l-q_1-q_2-q_3)^2+m_q^2$ & $(l-q_1-q_2-q_3)^2+m_q^2$ & $(l-q_3)^2+m_q^2$ \\
\bottomrule
\label{tab:hjtopo}
\end{tabularx}
\end{table}\noindent
Let us briefly comment on these integral families: Integrals and thus families of integrals can be grouped into \textit{planar} and \textit{non-planar} ones. In our case, families $A$ and $B$ are planar, whereas integral family~$C$ is of non-planar nature. They can be distinguished by the number of propagators involving both loop momenta, where the minimal number is given by one and two in the planar and non-planar case, respectively\footnote{Note that family~$C$ in Table~\ref{tab:hjtopo} contains three of this kind.}. From the geometric perspective, this distinction becomes clear by realizing that the non-planar integrals cannot be drawn in a plane without crossing at least two propagators, as the name suggests. This can easily be verified by Fig.~\ref{fig:hjdiag} and implies that non-planar two-loop four-point integrals with fewer than six propagators do not exist. Consequently, all MIs of family~$C$ are subject to the restriction $t\geq 6$ and their subtopologies are mapped onto either family~$B$ or $C$. Finally, the difference between planar and non-planar integrals becomes manifest at the level of the kinematics: The integral families $A$ and $B$ have a cut in the Mandelstam invariants $s$ and $u$, but not in $t$, leading to the choice of independent scales in the previous section suitable for the computation of the planar MIs in Section~\ref{sec:hjdeq}. In contrast, integral family~$C$ possesses cuts in all three Mandelstam invariants.

\section{Scattering Amplitudes}
\label{sec:hjamp}

This section is designed to compute the unrenormalized two-loop Feynman amplitude for two reasons: First, we would like to determine the range of the values~$r$ and $s$ defined in Eq.~\eqref{rs2} of the integrals appearing therein, which is required to derive the IBP identities in Section~\ref{sec:hjibp}. Second, integrals belonging to specific sectors might not occur in the expression of the amplitude, so that their MIs do not have to be computed.\\
In order to accomplish this, we work in a kinematic configuration different from Eq.~\eqref{srange}, in which all Mandelstam invariants are positive,
\begin{equation}
s>0 \,, \quad u>0 \,, \quad m_H^2>0 \,,
\end{equation}
corresponding to the Higgs decay channels:
\begin{align}
H(q_4) &\to g(q_1)\,g(q_2)\,g(q_3) \,, \nonumber \\
H(q_4) &\to q(q_1)\,\bar{q}(q_2)\,g(q_3) \,.
\label{higgsdecay}
\end{align}
This setup is advantageous in the sense that the computation has to be carried out for two channels only. The kinematic configuration of Eq.~\eqref{xrange} associated with Higgs-plus-jet production can then be obtained by crossing external legs into the appropriate initial or final state and through the proper analytic continuation of the MIs. Note that for the purposes outlined in the beginning of this section, this procedure is not necessary and conclusions at the level of the integrals can be drawn without additional expense.

\subsection{Tensor Decomposition}

Initially, we need to establish the connection between the tensorial quantity $\mathcal{S}$ emerging from the sum of all Feynman diagrams as described in Eq.~\eqref{scatteringamp2} and the Feynman amplitude~$\mathcal{M}$ in Eq.~\eqref{hjfeynmanamp}. A recipe for this is provided by the method of tensor decomposition explained in detail in Section~\ref{sec:tensordecomp}. We start by decomposing the Feynman amplitude of the $ggg$ and $q\bar{q}g$ channels according to Eq.~\eqref{tensor1},
\begin{align}
\mathcal{M}_{ggg} &= f^{abc} \, \mathcal{S}^{ggg}_{\mu\nu\rho} \, \e_1^\mu \, \e_2^\nu \, \e_3^\rho \,, \nonumber \\
\mathcal{M}_{q\bar{q}g} &= T^a_{ij} \, \mathcal{S}^{q\bar{q}g}_\rho \, \e_3^\rho \,,
\end{align}
where we used the shorthand notation $\e_i\equiv \e(q_i)$. Note that the structure constants~$f^{abc}$ and the generators $T^a_{ij}$ of the $SU(3)$ gauge group were defined in Section~\ref{sec:rules}. The tensorial operators $\mathcal{S}$ describe the partonic current and satisfy the same power series as the Feynman amplitude~$\mathcal{M}$, as stated in Eq.~\eqref{scatteringamp1}:
\begin{equation}
\mathcal{S} = \sqrt{\alpha_s} \, \left[ \mathcal{S}^{(1)} \, \alpha_s + \mathcal{S}^{(2)} \, \alpha_s^2 + \mathcal{O}(\alpha_s^2) \right] \,.
\end{equation}
In the following, we discuss the precise structure of the amplitude in each channel along the lines of Ref.~\cite{Gehrmann:2011}.

\newpage

\textbf{The Tensor Structure for the Decay $\boldsymbol{H\to ggg}$}

By using Lorentz and gauge invariance, the most general tensor structure can be predicted to be
\begin{align}
\mathcal{S}^{ggg}_{\mu\nu\rho} \, \e_1^\mu \, \e_2^\nu \, \e_3^\rho &= \sum_{i=1}^3 \sum_{j=1}^3 \sum_{k=1}^3 \mathcal{A}_{ijk} \, q_i\cdot \e_1 \, q_j\cdot \e_2 \, q_k\cdot \e_3 + \sum_{i=1}^3 \mathcal{B}_i \, q_i\cdot \e_1 \, \e_2\cdot\e_3 \nonumber \\
&\quad\, + \sum_{i=1}^3 \mathcal{C}_i \, q_i\cdot \e_2 \, \e_1\cdot\e_3 + \sum_{i=1}^3 \mathcal{D}_i \, q_i\cdot \e_3 \, \e_1\cdot\e_2
\label{Sggg1}
\end{align}
corresponding to the equivalent of Eq.~\eqref{decomposition} for the decay $H\to ggg$. Gluons are massless and have to be described by transverse polarization vectors, so that we can expand the result of Eq.~\eqref{Sggg1} using the conditions $q_i\cdot\e_i=0$ ($i=1,2,3$):
\begin{align}
\mathcal{S}^{ggg}_{\mu\nu\rho} \, \e_1^\mu \, \e_2^\nu \, \e_3^\rho &= \mathcal{A}_{211} \, q_2\cdot \e_1 \, q_1\cdot \e_2 \, q_1\cdot \e_3 + \mathcal{A}_{212} \, q_2\cdot \e_1 \, q_1\cdot \e_2 \, q_2\cdot \e_3 + \mathcal{A}_{231} \, q_2\cdot \e_1 \, q_3\cdot \e_2 \, q_1\cdot \e_3 \nonumber \\
&\quad\, + \mathcal{A}_{232} \, q_2\cdot \e_3 \, q_2\cdot \e_2 \, q_2\cdot \e_3 + \mathcal{A}_{311} \, q_3\cdot \e_1 \, q_1\cdot \e_2 \, q_1\cdot \e_3 + \mathcal{A}_{312} \, q_3\cdot \e_1 \, q_1\cdot \e_2 \, q_2\cdot \e_3 \nonumber \\
&\quad\, + \mathcal{A}_{331} \, q_3\cdot \e_1 \, q_3\cdot \e_2 \, q_1\cdot \e_3 + \mathcal{A}_{332} \, q_3\cdot \e_1 \, q_3\cdot \e_2 \, q_2\cdot \e_3 \nonumber \\
&\quad\, + \mathcal{B}_2 \, \e_2\cdot\e_3 \, q_2\cdot\e_1 + \mathcal{B}_3 \, \e_2\cdot\e_3 \, q_3\cdot\e_1 \nonumber \\
&\quad\, + \mathcal{C}_1 \, \e_1\cdot\e_3 \, q_1\cdot\e_2 + \mathcal{C}_3 \, \e_1\cdot\e_3 \, q_3\cdot\e_2 \nonumber \\
&\quad\, + \mathcal{D}_1 \, \e_1\cdot\e_2 \, q_1\cdot\e_3 + \mathcal{D}_2 \, \e_1\cdot\e_2 \, q_2\cdot\e_3 \,.
\label{Sggg2}
\end{align}
In fact, the remaining set of 14 form factors denoted by $\mathcal{A}_{ijk}$, $\mathcal{B}_i$, $\mathcal{C}_i$ and $\mathcal{D}_i$ is not independent, but can be reduced further by  taking into account that the tensor structure in Eq.~\eqref{Sggg2} fulfills the Slavnov-Taylor identities mentioned in Section~\ref{sec:rules}. In other words, the equation must evaluate to zero if one of the gluon polarization vectors $\e_i$ is replaced with the corresponding momentum $q_i$:
\begin{align}
\mathcal{S}^{ggg}_{\mu\nu\rho} \, q_1^\mu \, \e_2^\nu \e_3^\rho &= 0 \,, \nonumber \\
\mathcal{S}^{ggg}_{\mu\nu\rho} \, \e_1^\mu \, q_2^\nu \e_3^\rho &= 0 \,, \nonumber \\
\mathcal{S}^{ggg}_{\mu\nu\rho} \, \e_1^\mu \, \e_2^\nu q_3^\rho &= 0 \,.
\end{align}
With this knowledge, Eq.~\eqref{Sggg2} simplifies considerably and can be expressed in terms of only four distinct tensor structures $T_{ijk}$ with gauge-independent coefficients $\mathcal{A}_{ijk}$:
\begin{equation}
\mathcal{S}^{ggg}_{\mu\nu\rho} \, \e_1^\mu \, \e_2^\nu \, \e_3^\rho = \mathcal{A}_{211} \, T_{211} + \mathcal{A}_{311} \, T_{311} + \mathcal{A}_{232} \, T_{232} + \mathcal{A}_{312} \, T_{312} \,.
\end{equation}
The tensor structures are given by
\begin{align}
T_{211} &= q_2\cdot\e_1 \, q_1\cdot\e_2 \, q_1\cdot\e_3 - \frac{1}{2} \, s \, \e_1\cdot\e_2 \, q_1\cdot\e_3 - \frac{m_H^2-s-u}{u} \, q_2\cdot\e_1 \, q_1\cdot\e_2 \, q_2\cdot\e_3 \nonumber \\
&\quad\,  + \frac{1}{2} \, \frac{s\,(m_H^2-s-u)}{u} \, \e_1\cdot\e_2 \, q_2\cdot \e_3 \,, \nonumber \\
T_{232} &= q_2\cdot\e_1 \, q_3\cdot\e_2 \, q_2\cdot\e_3 - \frac{1}{2} \, u \, \e_2\cdot\e_3 \, q_2\cdot\e_1 - \frac{s}{m_H^2-s-u} \, q_3\cdot\e_1 \, q_3\cdot\e_2 \, q_2\cdot\e_3 \nonumber \\
&\quad\, + \frac{1}{2} \, \frac{s\,u}{m_H^2-s-u} \, \e_2\cdot\e_3 \, q_3\cdot \e_1 \,, \nonumber \\
T_{311} &= q_3\cdot\e_1 \, q_1\cdot\e_2 \, q_1\cdot\e_3 - \frac{1}{2} \, (m_H^2-s-u) \, \e_1\cdot\e_3 \, q_1\cdot\e_2 - \frac{s}{u} \, q_3\cdot\e_1 \, q_3\cdot\e_2 \, q_1\cdot\e_3 \nonumber \\
&\quad\,  + \frac{1}{2} \, \frac{s\,(m_H^2-s-u)}{u} \, \e_1\cdot\e_3 \, q_3\cdot \e_2 \,, \nonumber \\
T_{312} &= q_3\cdot\e_1 \, q_1\cdot\e_2 \, q_2\cdot\e_3 - q_2\cdot\e_1 \, q_3\cdot\e_2 \, q_1\cdot\e_3 + \frac{1}{2} \, s \, \e_1\cdot\e_3 \, q_3\cdot\e_2 + \frac{1}{2} \, u \, \e_1\cdot\e_2 \, q_1\cdot\e_3  \nonumber \\
&\quad\, - \frac{1}{2} \, u \, \e_1\cdot\e_3 \, q_1\cdot\e_2 + \frac{1}{2} \, (m_H^2-s-u) \, \e_2\cdot\e_3 \, q_2\cdot\e_1 - \frac{1}{2} \, (m_H^2-s-u) \, \e_1\cdot\e_2 \, q_2\cdot\e_3 \nonumber \\
&\quad\, - \frac{1}{2} \, s \, \e_2\cdot\e_3 \, q_3\cdot\e_1 \,.
\end{align}
The scalar form factors depend on the kinematic invariants and fulfill the same perturbative expansion in the strong coupling constant as the Feynman amplitude $\mathcal{M}$ and the partonic current $\mathcal{S}$, as pointed out in Eq.~\eqref{seriesff}:
\begin{equation}
\mathcal{A}_{ijk} = \sqrt{\alpha_s} \, \left[ \mathcal{A}_{ijk}^{(1)} \, \alpha_s + \mathcal{A}_{ijk}^{(2)} \, \alpha_s^2 + \mathcal{O}(\alpha_s^2) \right] \,.
\label{Aggg}
\end{equation}
They are obtained by means of Eq.~\eqref{proj1}, i.e. by applying projectors such that
\begin{equation}
\sum_\mathrm{spins} P\left(\mathcal{A}_{ijk}\right) \, \mathcal{S}^{ggg}_{\mu\nu\rho} \, \e_1^\mu \, \e_2^\nu \e_3^\rho = \mathcal{A}_{ijk} \,,
\label{hjproj1}
\end{equation}
where the sum runs over the polarizations of the external gluons. For the sake of consistency with the cycling gauge fixing condition
\begin{equation}
\e_1\cdot q_2=\e_2\cdot q_3=\e_3\cdot q_1=0 \,,
\label{Hggggauge}
\end{equation}
the polarization sums are given by
\begin{align}
\sum_{spins} \left(\e_1^\mu\right)^* \, \e_1^\nu &= -g^{\mu\nu} + \frac{q_1^\mu\,q_2^\nu + q_1^\nu\,q_2^\mu}{q_1\cdot q_2} \,, \nonumber \\
\sum_{spins} \left(\e_2^\mu\right)^* \, \e_2^\nu &= -g^{\mu\nu} + \frac{q_2^\mu\,q_3^\nu + q_2^\nu\,q_3^\mu}{q_2\cdot q_3} \,, \nonumber \\
\sum_{spins} \left(\e_3^\mu\right)^* \, \e_3^\nu &= -g^{\mu\nu} + \frac{q_3^\mu\,q_1^\nu + q_3^\nu\,q_1^\mu}{q_1\cdot q_3} \,.
\label{Hgggpol}
\end{align}
The explicit form of the projectors is determined by expressing them in the basis of the gauge-invariant tensor structures $T_{ijk}$ and imposing that Eq.~\eqref{hjproj1} is satisfied, leading to
\begin{align}
P\left(\mathcal{A}_{211}\right) &= \frac{1}{D-3} \, \left[ \frac{D\,u}{s^3\,(m_H^2-s-u)} \, T_{211}^\dagger + \frac{(D-4)}{s^2\,u} \, T_{232}^\dagger \right. \nonumber \\
&\qquad\qquad\quad \left. - \frac{(D-4)\,u}{s^2\,(m_H^2-s-u)^2} \, T_{311}^\dagger + \frac{(D-2)}{s^2\,(m_H^2-s-u)} \, T_{312}^\dagger \right] \,, \nonumber \\
P\left(\mathcal{A}_{232}\right) &= \frac{1}{D-3} \, \left[ \frac{(D-4)}{s^2\,u} \, T_{211}^\dagger + \frac{D,(m_H^2-s-u)}{s\,u^3} \, T_{232}^\dagger \right. \nonumber \\
&\qquad\qquad\quad \left. - \frac{(D-4)}{s\,u\,(m_H^2-s-u)} \, T_{311}^\dagger + \frac{(D-2)}{s\,u^2} \, T_{312}^\dagger \right] \,, \nonumber \\
P\left(\mathcal{A}_{311}\right) &= \frac{1}{D-3} \, \left[ -\frac{(D-4)\,u}{s^2\,(m_H^2-s-u)^2} \, T_{211}^\dagger - \frac{(D-4)}{s\,u\,(m_H^2-s-u)} \, T_{232}^\dagger \right. \nonumber \\
&\qquad\qquad\quad \left. + \frac{D\,u}{s\,(m_H^2-s-u)^3} \, T_{311}^\dagger - \frac{(D-2)}{s\,(m_H^2-s-u)^2} \, T_{312}^\dagger \right] \,, \nonumber \\
P\left(\mathcal{A}_{312}\right) &= \frac{1}{D-3} \, \left[ \frac{(D-2)}{s^2\,(m_H^2-s-u)} \, T_{211}^\dagger + \frac{(D-2)}{s\,u^2} \, T_{232}^\dagger \right. \nonumber \\
&\qquad\qquad\quad \left. - \frac{(D-2)}{s\,(m_H^2-s-u)^2} \, T_{311}^\dagger + \frac{D}{s\,u\,(m_H^2-s-u)} \, T_{312}^\dagger \right] \,.
\end{align}

\newpage

\textbf{The Tensor Structure for the Decay $\boldsymbol{H\to q\bar{q}g}$}

Due to the presence of only one external boson, the most general tensor structure in the channels involving quarks is much simpler,
\begin{equation}
\mathcal{S}^{q\bar{q}g}_\rho \, \e_3^\rho = \mathcal{A}_1 \, \bar{u}(q_1) \, \slashed{q}_3 \, v(q_2) \, q_1\cdot\e_3 + \mathcal{A}_2 \, \bar{u}(q_1) \, \slashed{q}_3 \, v(q_2) \, q_2\cdot\e_3 + \mathcal{A}_3 \, \bar{u}(q_1) \, \slashed{q}_3 \, v(q_2) \,,
\label{Sqqg1}
\end{equation}
where we applied $q_3\cdot\e_3$ as before. Moreover, $\bar{u}\equiv u^\dagger\,\gamma_0$ and $v$ denote the four-spinors of the quark and antiquark, respectively. As in the case of $H\to ggg$, the form factors~$\mathcal{A}_i$ depend on the kinematic invariants and can be decomposed as a power series of the form~\eqref{Aggg} in the gauge coupling~$\alpha_s$. They are not independent, but can be related through the Slavnov-Taylor identities:
\begin{equation}
\mathcal{A}_3 = -\left(q_1\cdot q_3 \, \mathcal{A}_1 + q_2\cdot q_3 \, \mathcal{A}_2\right) \,.
\end{equation}
This allows rephrasing Eq.~\eqref{Sqqg1} in terms of only two tensor structures~$T_i$ with scalar coefficients~$\mathcal{A}_i$,
\begin{equation}
\mathcal{S}^{q\bar{q}g}_\rho \, \e_3^\rho = \mathcal{A}_1 \, T_1 + \mathcal{A}_2 \, T_2 \,,
\end{equation}
where the tensor structures are given by
\begin{align}
T_1 &= \bar{u}(q_1) \, \slashed{q}_3 \, v(q_2) \, q_2\cdot\e_3 - \bar{u}(q_1) \, \slashed{\e}_3 \, v(q_2) \, q_2\cdot q_3 \,, \nonumber \\
T_2 &= \bar{u}(q_1) \, \slashed{q}_3 \, v(q_2) \, q_1\cdot\e_3 - \bar{u}(q_1) \, \slashed{\e}_3 \, v(q_2) \, q_1\cdot q_3 \,.
\end{align}
Equivalently to Eq.~\eqref{hjproj1}, the form factors can be extracted by introducing projectors as follows:
\begin{equation}
\sum_{spins} P(\mathcal{A}_i) \, \mathcal{S}^{q\bar{q}g}_\rho \, \e_3^\rho = \mathcal{A}_i \,.
\end{equation}
The polarization sums are carried out by means of the standard relations
\begin{align}
\sum_{spins} u(q_1)\,\bar{u}(q_1) &= \slashed{q}_1 \,, \nonumber \\
\sum_{spins} v(q_2)\,\bar{v}(q_2) &= \slashed{q}_2 \,, \nonumber \\
\sum_{spins} \left(\e_3^\mu\right)^* \, \e_3^\nu &= -g^{\mu\nu} + \frac{q_3^\mu\,q_1^\nu + q_3^\nu\,q_1^\mu}{q_1\cdot q_3} \,,
\label{Hqqgpol}
\end{align}
where the last identity is consistent with the gauge fixing condition
\begin{equation}
\e_3\cdot q_1=0 \,.
\label{Hqqggauge}
\end{equation}
The actual form of the projectors can be determined similarly to the $H\to ggg$ case, yielding
\begin{align}
P(\mathcal{A}_1) &= \frac{1}{D-3} \, \left[ \frac{D-2}{2\,s\,(m_H^2-s-u)^2} \, T_1^\dagger - \frac{D-4}{2\,s\,u\,(m_H^2-s-u)} \, T_2^\dagger \right] \,, \nonumber \\
P(\mathcal{A}_2) &= \frac{1}{D-3} \, \left[ -\frac{D-4}{2\,s\,u\,(m_H^2-s-u)} \, T_1^\dagger + \frac{D-2}{2\,s\,u^2} \, T_2^\dagger \right] \,.
\label{hjproj2}
\end{align}

\subsection{Calculation of the Two-Loop Amplitude}

In order to compute the Feynman amplitude~$\mathcal{M}$ in a given channel, we recycle the Feynman diagrams generated with \textsc{Qgraf} for the derivation of the integral families in Section~\ref{sec:integralfamilies}. At the two-loop level, this results in 49 diagrams for $H\to q\bar{q}g$ and 282 diagrams for $H\to ggg$. Examples for $H\to ggg$ are shown in Fig.~\ref{fig:hjdiag}, where the initial- and final state momenta are crossed into the gluon fusion channel of Higgs-plus-jet production. Let us point out that the number of two-loop diagrams is considerably larger than at one loop, where we find two and twelve diagrams, respectively.\\
Subsequently, we apply the $D$-dimensional projectors indicated in Eqs.~\eqref{hjproj1} and \eqref{hjproj2} and perform the summation over colors and spins with the help of \textsc{Form}. In doing so, we make use of the gauge fixing conditions in Eqs.~\eqref{Hggggauge} and \eqref{Hqqggauge} and the summation formulae for the polarizations of the external particles in Eqs.~\eqref{Hgggpol} and \eqref{Hqqgpol}. Note that we still work in the Feynman-'t Hooft gauge when dealing with internal gluons, so that virtual ghost contributions have to be taken into account, as described in Section~\ref{sec:rules}. After this procedure, we are left with the unrenormalized two-loop form factors for $H\to ggg$ and $H\to q\bar{q}g$, which are given in terms of unreduced scalar integrals in the representation of Eq.~\eqref{hjscalarint}. The complete analytic expression of the two-loop amplitudes exceeds the scope of this thesis, but let us comment on their structure, which is striking in several respects: 
\begin{itemize}
\item The integrals appearing therein cover the range $r\leq 7$, $s\leq 3$, i.e. they have a remarkably simple numerator structure, which simplifies the derivation of the IBP identities for the reduction of the amplitude. The only exception is given by one integral per integral family with $r=6$, $s=4$, corresponding to a factorizing diagram of two one-loop triangles, whose reduction can easily be obtained separately.
\item For the planar integral families $A$ and $B$, all top-level topologies indicated in Fig.~\ref{fig:hjdiag} occur in the final result of the unreduced two-loop amplitude. However, this is not the case for the non-planar sector~$C_{7,431}$ with five MIs, which is depicted in Fig.~\ref{fig:hjdiag}. In fact, the color factor of the corresponding integrals vanishes, which is also true of their subsectors~$C_{6,303}$ and $C_{6,399}$ with two and one MIs, respectively, sparing us the computation of eight non-planar MIs in total.
\item In the context of the color structure, another property of the two-loop amplitudes is worth mentioning: Unlike for many processes with massless propagators, the leading-color part already includes the full dependence on the non-planar integral sectors. In contrast to the effective field theory, the leading-order color coefficient in the full theory receives contributions from both the planar integrals and the non-planar integrals.
\end{itemize}
As a next step, the scalar integrals have to be reduced with the help of IBP relations, which will be outlined in the following section.

\section{Integration-by-Parts Relations}
\label{sec:hjibp}

In this section, we report on technical details and the program packages used to derive the IBP identities in the context of Higgs-plus-jet production with full quark mass dependence. In the following, we will refer both to the reduction of the integrals appearing in the amplitudes and to the IBP relations needed to derive differential equations of the MIs in the next section. As stated previously, these two reductions require different ranges of the values $r$ and $s$. From the last section, we know that the amplitude includes unreduced integrals with $r\leq 7$, $s\leq 3$. For the differential equations, we deduce IBP identities for the range $r\leq 8$, $s\leq 2$, which emerges from the choice of the Laporta MIs presented in Appendix~\ref{sec:hjlaporta} with at most $r=7$, $s=2$. Note that the differentiation operation within the procedure of deriving the differential equations introduces an additional power in the denominator, leading to $r\leq 8$.

\subsection{Program Packages}

We emphasize that the general idea of the method of IBP reduction and its implementations mentioned in the following were described in a more general sense in Sections~\ref{sec:reduction} and \ref{sec:programs}. Therein, we also pointed out that the program~\textsc{Reduze} is our main choice when it comes to IBP reduction, since it includes the highest number of symmetry relations, thus minimizing the number of MIs. Beyond that, it contains multiple useful features, which were mentioned at several points of this thesis.\\
By using \textsc{Reduze}, we managed to derive the full set of IBP relations for all planar sectors $A$ and $B$ as well as for all non-planar sectors $C_{t,ID}$ with $t\leq 6$, except for the sector~$C_{6,238}$ with four MIs. We obtained the same result through the combined use of the programs \textsc{LiteRed} and the \textsc{C++} version of \textsc{Fire}, which is in agreement with the observation made in Ref.~\cite{Melnikov:2016}. Therein, the remaining $t=7$ integrals within integral family~$C$ were reduced with the help of a private \textsc{Form} implementation. We followed a different path by using a private implementation of the recently published reduction code~\textsc{Kira}, which enabled a collaborator~\cite{kira} to derive the full set of IBP identities required for the reduction of all three integral families, both for the integrals appearing in the amplitudes and in the differential equations. Previously, we had obtained the same result by setting the ratio of the Higgs boson and top quark masses to the fixed value $m_H^2/m_t^2=1/2$ for the reduction of the non-planar integrals through the private computer code \textsc{Crusher}~\cite{crusher2}. In this way, the number of scales is reduced by one, similarly to the procedure described in Ref.~\cite{Jones:2018}, which simplifies the derivation a lot.

\subsection{Higher-Sector Relations}

Upon generating IBP identities for a given sector, referred to as \textit{seeds}, and subsequently solving them, it may occur that integrals belonging to that sector drop out of the equation, thus leaving a relation that contains only subsectors. With respect to the subsectors, we refer to these identities as \textit{higher-sector relations}, since they emerge from seeds with a higher number of distinct propagators~$t$. A unique feature of \textsc{Reduze} is that it can be used to derive these extra relations, which allow reducing the number of MIs of a given sector. An example is given by sector~$B_{5,182}$ defined in Appendix~\ref{sec:hjlaporta}, for which the number of MIs could be reduced from three to two by using seeds of the sector $B_{6,190}$.\\
Indeed, distinguishing between 'standard' IBP identities and higher-sector relations seems unnecessary at first, since the latter should be taken into account from the beginning upon solving the system of reduction identities. This would ensure that for every sector, a minimal number of MIs is found, as it is being done in a fully automatic way in the \textsc{Kira} implementation, for example. However, we made a remarkable observation in the context of integral family~$C$, which might prove beneficial in IBP reductions of other processes as well: Upon deriving higher-sector relations from the seed of the unphysical sector $C_{9,511}$, we found an identity relating the MIs of sector $C_{7,254}$. With six MIs, this sector is the one with the highest number of MIs for Higgs-plus-jet production with full quark mass dependence. It is therefore unlikely that all integrals can be decoupled, possibly leading to elliptic integrals or even more complicated structures. Hence, any reduction of the number of MIs is highly appreciated and could simplify the calculation of the MIs considerably. In fact, we have so far only proven the existence of such a relation with the help of \textsc{Reduze} by inserting prime numbers for the kinematic invariants prior to the reduction procedure. Subsequently, the resulting IBP identities have been solved numerically instead of retaining the symbolic dependence on the kinematic invariants. The derivation of the full identity is yet to be accomplished, so that we treat sector $C_{7,254}$ as one with six MIs for the time being.

\subsection{Number of Master Integrals}

As explained in Section~\ref{sec:reduction}, the crucial result of the IBP reduction for the computation of the MIs in the next section is the number of MIs per integral sector. The reduction results of \textsc{Reduze} indicate that there are 70 MIs in family~$A$, 32 in family~$B$ and 26 MIs in family~$C$.\footnote{Note that for now, we do not consider crossings of subsectors, which are necessary for the computation of the planar MIs in Section~\ref{sec:hjdeq} and lead to higher numbers of MIs in total.} Although the full reduction could not be carried out for those sectors of family $C$ with $t=7$, the number of MIs can be obtained by setting the kinematic invariants to prime numbers, thereby limiting the reduction to a specific kinematic point. As mentioned in Section~\ref{sec:hjamp}, eight non-planar MIs do not contribute to the amplitudes, so that the number of MIs within integral family~$C$ decreases to 18. Let us point out that we were able to verify these numbers reliably with the results obtained by \textsc{Crusher} and \textsc{Kira} as soon as we took the symmetry relations into account, that are provided by \textsc{Reduze}. Moreover, the total number of MIs exactly agrees with the one in Ref.~\cite{Jones:2018}. Beyond that, we would like to state that Ref.~\cite{Bonciani:2016} obtains higher numbers of MIs for the planar integral families $A$ and $B$, namely 73 and 50, respectively, which is supposedly due to the usage of \textsc{Fire} instead of \textsc{Reduze}.\\
Eventually, we stress that the code \textsc{Mint}~\cite{Lee:2013b} was used prior to the reduction step. \textsc{Mint} is designed to count the number of MIs per topology by exploring the critical points of the polynomials entering the parametric representation and the Baikov representation, where the relevant topological invariant is the sum of the Milnor numbers of the proper critical points. In the context of massive propagators, however, we encountered difficulties with \textsc{Mint}, leading to a lower number of MIs compared to the results of IBP reduction programs.

\subsection{Technical Details}

We emphasize that the complexity of the considered process is reflected by the mere size of the coefficients appearing in the IBP relations. Due to the presence of a mass parameter in the propagators, they tend to blow up considerably compared to the limit of vanishing internal quark mass. Quantitatively, this becomes clear from two numbers:
\begin{itemize}
\item First, we obtain reduction files of up to a few gigabytes per sector, resulting in reduction relations with a total data size of several hundred gigabytes. These cumbersome expressions prevented us from using the database feature of \textsc{Reduze}, which is designed to store intermediate results at any point of the reduction and continue the calculation at a later, convenient time.
\item Second, we attempted to perform the IBP reduction for the non-planar top-level topologies with \textsc{Reduze} using a machine with four terabytes of memory and failed. We observed that the reduction procedure freezes towards the end of the Laporta algorithm, where only a handful of equations remain to be solved, which however contain the largest possible coefficients. At that point, the available RAM size per core is the crucial quantity, whereas increasing the number of cores is only of use when performing the reduction for many sectors in parallel, e.g. for the ones with lower~$t$.
\end{itemize}
We tried to maximize the probability of obtaining satisfactory results from the reduction procedure by minimizing the coefficients at every intermediate step to the largest possible extent. This includes making use of the recently implemented \textsc{Reduze} feature, which allows us to provide an individual basis selection prior to the IBP reduction, thus minimizing the coefficients from the starting point of the IBP reduction. For example, this enabled the reduction of the top-level topology~$A_{7,247}$ of family~$A$ with four MIs. Beyond that, we ensured constant factorization of the coefficients, thereby performing partial fractioning at the latest possible point and only wherever indispensable, in order to keep the size of the coefficients under control.\\
From these observations, we conclude that this process is at the border of feasibility with conventional IBP reduction techniques. For future computations of similar complexity, completely different methods might be necessary, for example the approach relying on finite fields mentioned in Section~\ref{sec:laporta}.\\
Let us close this section by pointing out that choosing integral families different from the ones in Table~\ref{tab:hjtopo} could have led to a speed-up or to improved reduction results in general. This might apply in particular to the case, in which a higher number of distinct integral families are chosen, so that the topology trees are split and a single family contains a lower number of MIs.

\section[Differential Equations and Master Integrals: The Planar Non-Elliptic Sectors]{Differential Equations and Master Integrals:\\The Planar Non-Elliptic Sectors}
\label{sec:hjdeq}

For reasons explained in great detail in Sections~\ref{sec:beyondmpl} and \ref{sec:elliptic}, we intend to calculate the non-elliptic integrals of families $A$ and $B$ directly in the physical region through series expansions from differential equations. For this purpose, we first derive differential equations in canonical form in Section~\ref{sec:hjdeqcan1}, whenever possible, and shift the discussion on deviations of this form in the shape of elliptic integrals to Sections~\ref{sec:hjdeqell215} and~\ref{sec:hjdeqell247}. In Section~\ref{sec:hjdeqcan2}, we divide the phase space through the selection of suitable expansion points, which are parametrized with the method introduced in Section~\ref{sec:onedim} before actually deriving series expansions around these points in \ref{sec:hjdeqcan3} following the approach explained in Section~\ref{sec:seriesexp}. We finally connect these power series representations in Section~\ref{sec:hjdeqcan4} using the matching procedure described in Section~\ref{sec:matching} after the determination of appropriate boundary conditions. In Section~\ref{sec:hjdeqcan5}, we elaborate on technical details and possible simplifications, and provide alternative ways of dividing the phase space in Section~\ref{sec:hjdeqcan6}.\\
A subset of the integrals presented in the following was computed in the context of the NLO corrections to the $H\to Z\,\gamma$ decay rate in Chapter~\ref{chap:hza} and in Refs.~\cite{Bonciani:2015,Gehrmann:2015a}, namely two-loop three-point functions with two massive external legs within integral family~$A$. The remaining $H\to \gamma\gamma$-like two-loop triangles within family~$A$ with two massless and one massive external leg were calculated in Ref.~\cite{Anastasiou:2006}, some selected three-point functions of which had been known previously~\cite{Bonciani:2003}. We complete these evaluations by providing a basis choice suitable for deriving differential equations in canonical form.

\subsection{Canonical Differential Equations}
\label{sec:hjdeqcan1}

The differential equations for the planar MIs of Higgs-plus-jet production with full quark mass dependence are derived similarly to the ones for the $H\to Z\,\gamma$ decay width, which was described in Section~\ref{sec:hzadeq}. Let us recall in a nutshell the method of differential equations outlined in Chapter~\ref{chap:workflow2}: Initially, we generate differential equations in the internal mass~$m_q$ and the external invariants $s$, $t$ and $u$ for each integral by explicitly carrying out the differentiation at the level of the integrand with the help of Eq.~\eqref{deqmandelstam}. Subsequently, we relate the differential operators in the Mandelstam invariants $s$, $t$ and $u$ to the ones in the three independent ratios $x$, $z$ and $h$ introduced in Eq.~\eqref{ratiodef} by means of Eq.~\eqref{deqratios}:
\begin{align}
\frac{\p}{\p x} &= m_q^2 \, \left(\frac{\p}{\p s} - \frac{\p}{\p t}\right) \,, \nonumber \\
\frac{\p}{\p z} &= m_q^2 \, \left(\frac{\p}{\p u} - \frac{\p}{\p t}\right) \,, \nonumber \\
\frac{\p}{\p h} &= m_q^2 \, \frac{\p}{\p t} \,.
\label{deqratios2}
\end{align}
In doing so, the differential equation in the quark mass turns into the trivial scaling relation~\eqref{scaling}. Eventually, the unreduced integrals arising from the differentiation procedure on the right-hand side are related to the original MIs using the IBP identities derived in Section~\ref{sec:hjibp}. With this, we obtain a system of inhomogeneous differential equations in the three ratios $\vec{x}\equiv (x,z,h)$, which can be summarized as follows:
\begin{equation}
\frac{\p}{\p x_i} \vec{I}(D;\vec{x}) = C^{(i)}(D;\vec{x}) \, \vec{I}(D;\vec{x}) \,.
\label{hjdeq1}
\end{equation}
Therein, $\vec{I}$ is a 151-component vector, whose entries involve all planar two-loop MIs of Higgs-plus-jet production with full quark mass dependence. The number of 151 MIs exceeds the one quoted as a result of the IBP reduction in the previous section, since we treat crossed subsectors required to represent the differential equations of higher sectors as independent from their uncrossed counterparts. In contrast to the exact computation of the integral subset occuring in the $H\to Z\,\gamma$ two-loop amplitude in Chapter~\ref{chap:hza}, we are obliged to proceed in this way. This is due to the fact that the method of solving differential equations through series expansions in the physical region relies on the singularity structure of the crossed integrals, as described in Section~\ref{sec:seriesexp}, which might deviate from the one of their uncrossed equivalent.\\
Our choice of Laporta integrals~$\vec{I}$, which can be understood as a starting point for the derivation of the canonical differential equations, is defined in Appendix~\ref{sec:hjlaporta} and depicted in Fig.~\ref{fig:hjmaster}. It was obtained by applying the guidelines specified in Section~\ref{sec:basischoice}, leading to a triangular form of the matrices $C^{(i)}(4,\vec{x})$ within Eq.~\eqref{hjdeq1}. As a next step, we transform Eq.~\eqref{hjdeq1} into an equation, in which the dependence on the dimensional parameter~$\e$ is fully factorized from the coefficient matrix:
\begin{equation}
\frac{\p}{\p x_i} \vec{M}(D;\vec{x}) = \left( \e \, A^{(i)}(\vec{x}) + A_0^{(i)}(\vec{x}) \right) \, \vec{M}(D;\vec{x}) \,.
\label{hjdeq2}
\end{equation}
This was achieved with the help of the algorithm outlined in Section~\ref{sec:triangulartocanonical}, resulting in the canonical basis~$\vec{M}$ specified in Appendix~\ref{sec:hjcan}. Therein, the homogeneous solutions, also referred to as integrating factors of the differential equations, were chosen such that they are real in the Euclidean region with $x<z<h<0$. Note that the only non-vanishing entries of the matrices~$A_0^{(i)}(\vec{x})$ appear in the differential equations of the elliptic integrals $I_{59}$--$I_{62}$ and $I_{67}$--$I_{70}$, corresponding to the sectors $A_{6,215}$ and $A_{7,247}$. The differential equations of these elliptic integrals cannot be cast into canonical form and will be discussed separately in Sections~\ref{sec:hjdeqell215} and \ref{sec:hjdeqell247}. We refrain from repeating the numerous advantages of differential equations in canonical form here, since this was already done at several points in this thesis, see for example Sections~\ref{sec:canon} and \ref{sec:frobenius2simp}.\\
The representation of Eq.~\eqref{hjdeq2} could be taken one step further for the sub-system of non-elliptic canonical differential equations, in which $A_0^{(i)}(\vec{x})=0$, by making explicit the structure of its total differential:
\begin{equation}
\d \vec{M}(D;\vec{x}) = \e \sum_{k=1}^{N_l} A_k \, \d\log (l_k) \, \vec{M} (D;\vec{x})\,.
\label{hjdeq3}
\end{equation}
Therein, the matrices $A_k$ contain only rational numbers and $N_l$ indicates the number of letters occuring in the alphabet $l_1\,\ldots,l_{N_l}$. The total differential~\eqref{hjdeq3} is obtained by integrating the matrix $A^{(i)}(\vec{x})$ within Eq.~\eqref{hjdeq2} in the respective variables~$x_i$. This results in an arbitrary number of letters $\{\tilde{l}_k\}$, which might however not be independent. In order to achieve a minimal set of letters $\{l_k\}$, one has to find linear dependencies among the letters~$\{\tilde{l}_k\}$ that allow them to be expressed in terms of the alphabet $\{l_k\}$. Although we compute the power series at the level of the differential equations~\eqref{hjdeq2}, the total differential~\eqref{hjdeq3} was derived for the subset of integrals belonging to family~$A$ with $t\leq 5$,
\begin{figure}[H]
\begin{center}
\includegraphics[width=\textwidth]{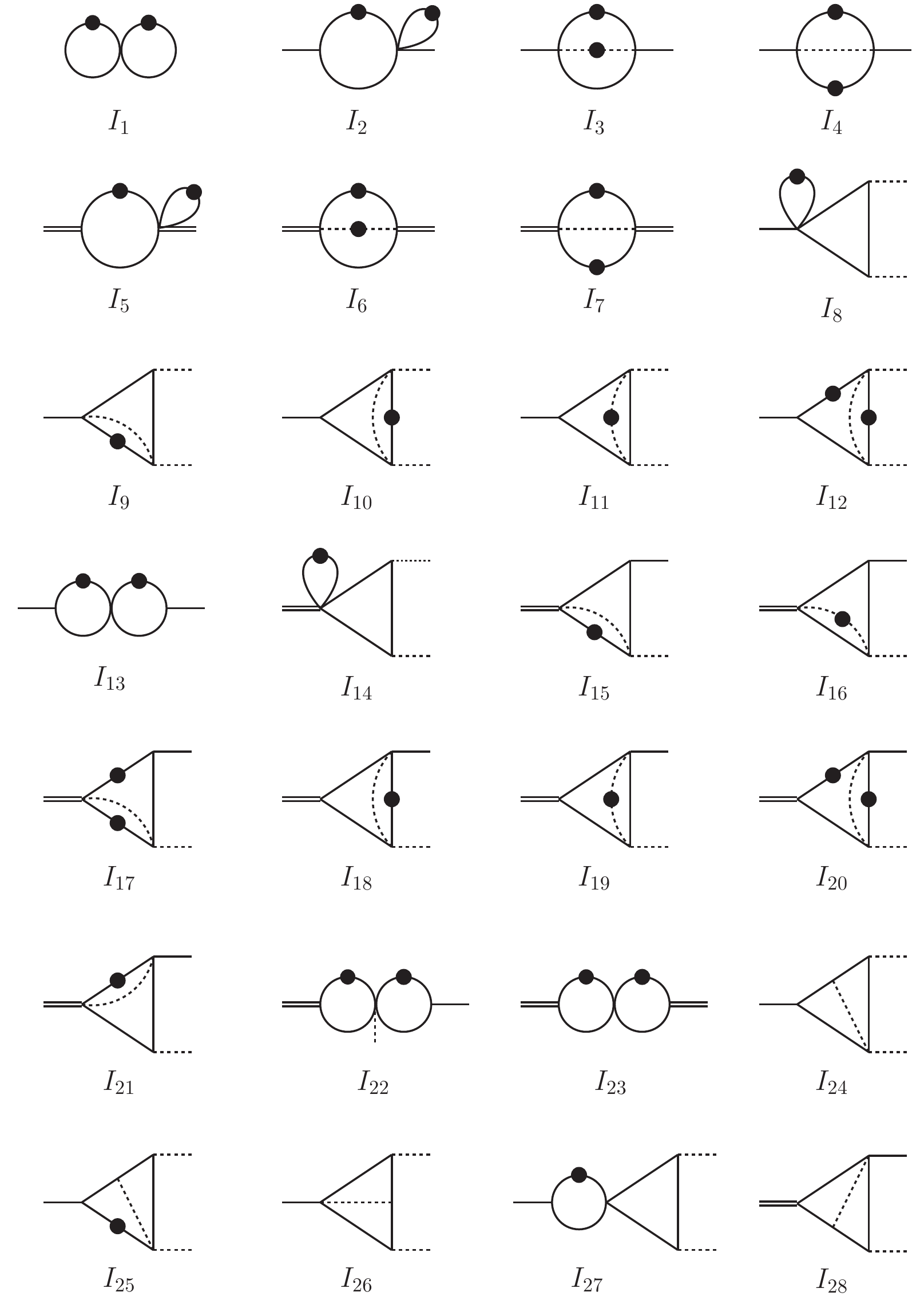}
\caption[Two-loop planar Laporta MIs for the calculation of the amplitude for Higgs-plus-jet production with full quark mass dependence]{\textbf{Two-loop planar Laporta MIs for the calculation of the amplitude for Higgs-plus-jet production with full quark mass dependence.} Dashed lines are massless, whereas internal solid lines denote propagators with mass~$m_q$. Double external lines correspond to $m_H^2$ and solid external lines of up to three-point functions denote virtualities of either~$s$ or $u$, depending on the definition of the Laporta integral in Appendix~\ref{sec:hjlaporta}. Since only uncrossed MIs are depicted, the corresponding crossing is obtained by interchanging $s\leftrightarrow u$.}
\label{fig:hjmaster}
\end{center}
\end{figure}
\begin{figure}[H]
\ContinuedFloat
\captionsetup{list=off,format=cont}
\begin{center}
\vspace{1cm}
\includegraphics[width=\textwidth]{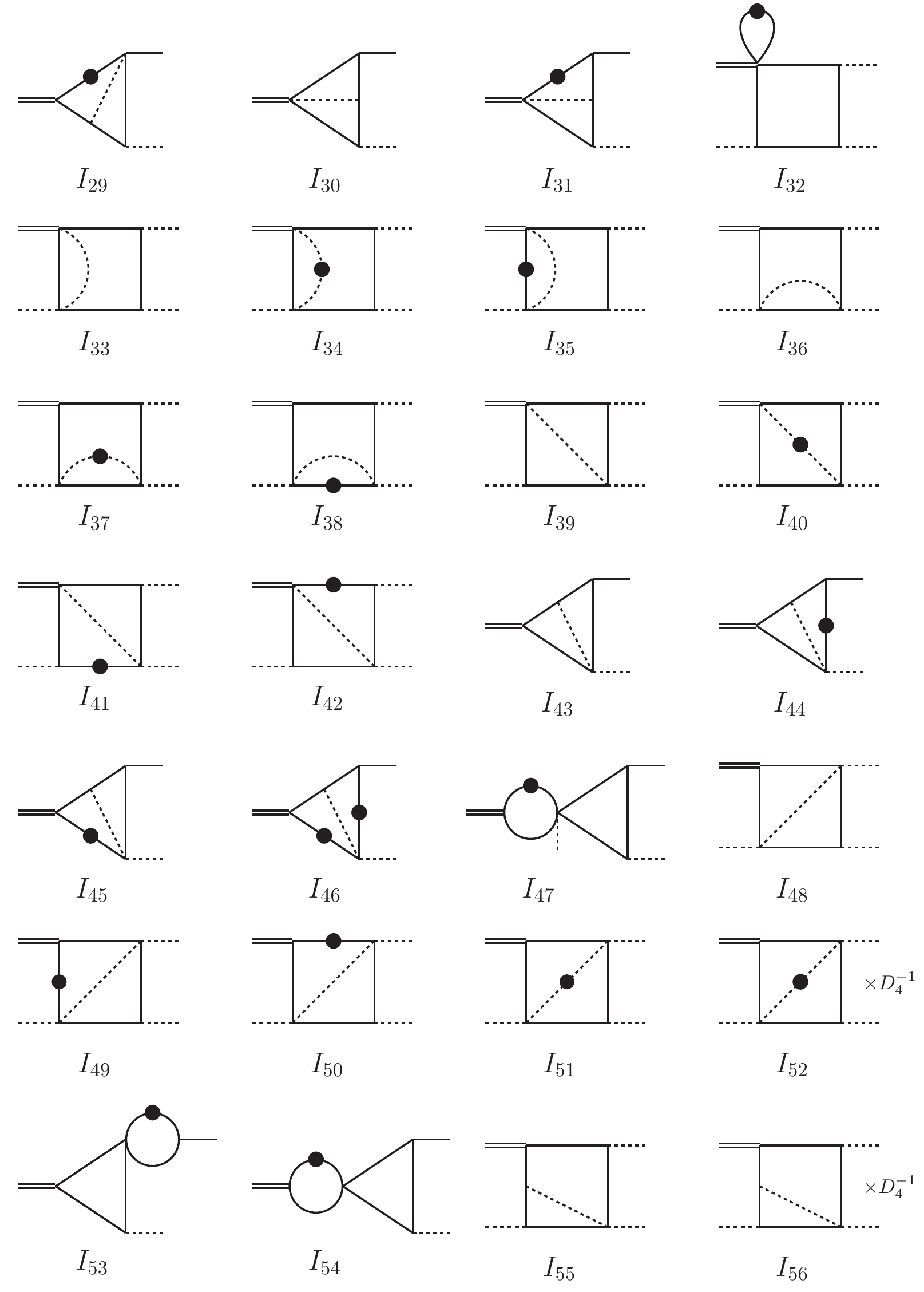}
\caption{Dotted propagators are taken to be squared and scalar products are denoted by~$D_i^{-j}$, where~$i$ refers to the $i$-th propagator of the corresponding integral family provided in Table~\ref{tab:hjtopo} and $j$ indicates the numerator power.}
\end{center}
\end{figure}
\begin{figure}[H]
\ContinuedFloat
\captionsetup{list=off,format=cont}
\begin{center}
\vspace{1.5cm}
\includegraphics[width=\textwidth]{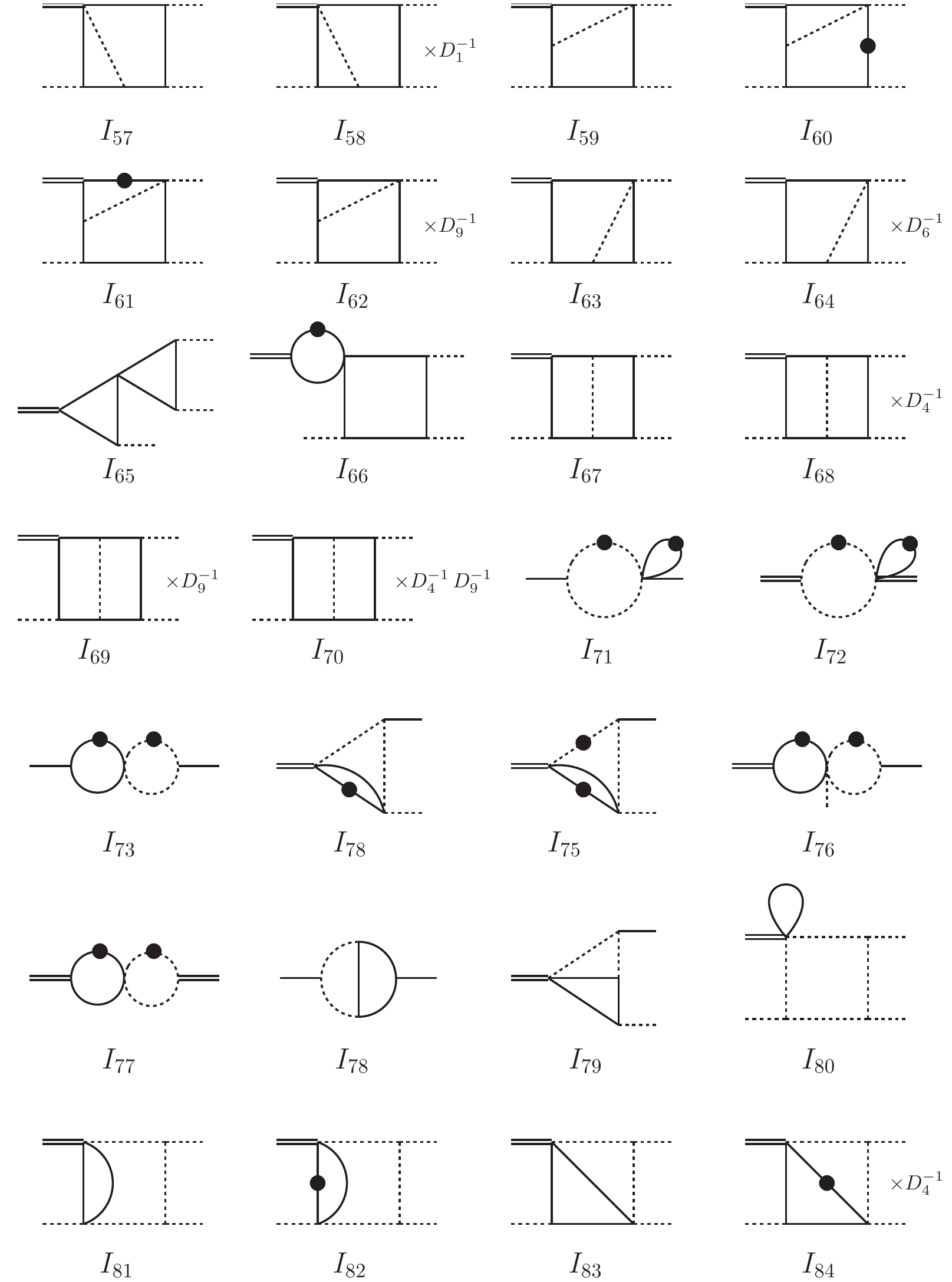}
\caption{}
\end{center}
\end{figure}
\begin{figure}[H]
\ContinuedFloat
\captionsetup{list=off,format=cont}
\begin{center}
\vspace{1.5cm}
\includegraphics[width=\textwidth]{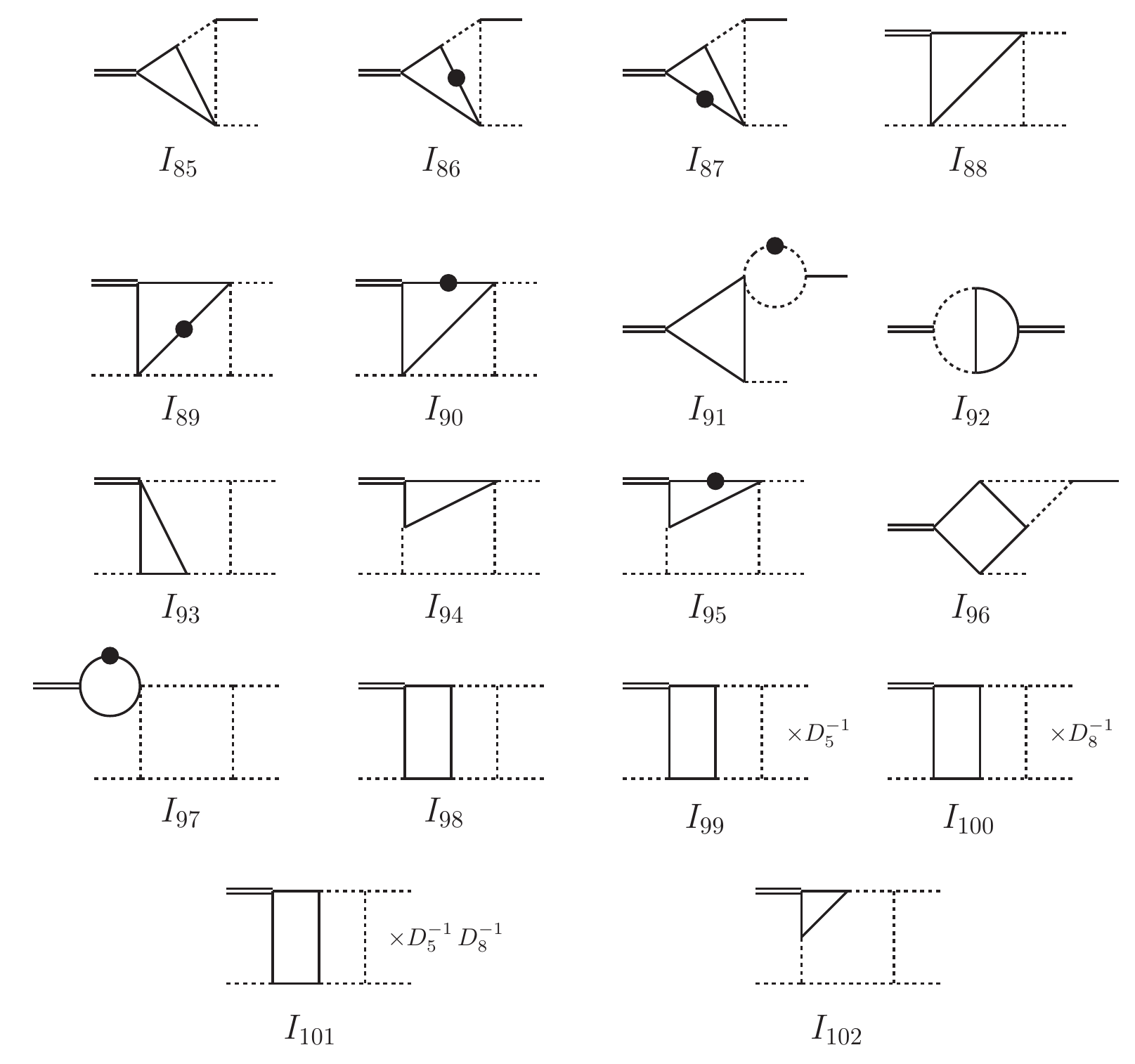}
\caption{}
\end{center}
\end{figure}
leading to an alphabet of around 50 letters. As described in Section~\ref{sec:beyondmpl}, we attempted the direct integration of the differential equations of the sector $A_{5,182}$ to a basis of classical polylogarithms using the symbol and coproduct formalism, thus requiring the total differential of that sector and its subsectors as a crucial ingredient. For this subset, we verified that the differential equations~\eqref{hjdeq2} are reproduced by differentiating Eq.~\eqref{hjdeq3} with respect to the kinematic invariants, implying that the denominators within Eq.~\eqref{hjdeq2} display the structure of the alphabet~$\{l_k\}$, which contains square root expressions. In general, they can be eliminated through a variable transformation, in which the transformed set of variables is equal to these square root expressions. In a problem with three independent variables, this requires that at most three independent square roots occur at the same time in a single differential equation. Since this is not the case here, the square roots have to be dealt with, in contrast to the canonical form~\eqref{canon} of the MIs appearing in the $H\to Z\,\gamma$ decay width. We would like to point out that the full set of letters~$\{l_k\}$ occuring in Eq.~\eqref{hjdeq3} can be found in Appendix~C of Ref.~\cite{Bonciani:2016}.

\subsection{Partitioning of the Phase Space}
\label{sec:hjdeqcan2}

Following the method of deriving series expansions in a multi-variate problem as explained in Sections~\ref{sec:seriesexp}, \ref{sec:onedim} and \ref{sec:matching}, our first task is to find an efficient partitioning of the phase space in the sense that it can be covered by as few series expansions as possible, one of which has to be supplied with appropriate boundary conditions. With a fixed ratio~$h\approx 1/2$ of the squared Higgs and top quark masses, we observe that the physical value of~$h$ is very close to $h=0$ and can therefore be taken care of by expanding around the origin of the $h$~axis, independently of the values of $x$ and $z$. Beyond that, we ought to find power series representations such that the physical region of phenomenological interest is covered, which is given by
\begin{equation}
\frac{1}{2}<x<85 \,, \quad \frac{1}{2}-x<z<0
\label{xrange2}
\end{equation}
according to Fig.~\ref{fig:dalitz} and the considerations made in Section~\ref{sec:hjkinematics}. Let us for the moment assume that we are dealing with small values of $|z|$, so that they lie within the radius of convergence of a series expansion around $z=0$, similarly to $h$. We are then left with the duty of establishing an appropriate partitioning of the domain in $x$, which can be phrased as finding suitable expansion points~$y_i$ in the language of Fig.~\ref{fig:matching}. In order to accomplish this, two important observations have to be taken into account:
\begin{itemize}
\item As indicated in Fig.~\ref{fig:dalitz}, the only threshold of the problem occurs at the point $x=4$ due to the massive two-particle cut of the internal quark lines. The points $x=0$ and $x=h$ correspond to pseudo-thresholds, i.e. the integrals must be regular in spite of singularities of the differential equations at these points.
\item We pointed out in Sections~\ref{sec:boundary} and \ref{sec:singularities} that the pseudo-thresholds are well-suited for the determination of the boundary constants. In the case of Higgs-plus-jet production with full quark mass dependence, this statement is particularly relevant at the origin $x=0$ (and likewise for $z=h=0$), corresponding to the limit of infinite quark mass. At that point, all integrals turn into simple tadpoles and we should try to benefit from this useful observation.
\end{itemize}
As a consequence, the following choice of the first two expansion points becomes obvious, corresponding to $y_1=0$ and $y_2$ in Fig.~\ref{fig:matching}:
\begin{itemize}
\item[\textbf{1)}] \textbf{Linear Series Expansion around the Origin}\\
With the notation introduced in Eq.~\eqref{onedim}, the parametrization reads
\begin{equation}
\vec{x} = \begin{pmatrix}x\\z\\h\end{pmatrix} \to \vec{\gamma}_1(\vec{x},\lambda_1) = \lambda_1 \, \begin{pmatrix}x\\z\\h\end{pmatrix} \,,
\label{param1}
\end{equation}
which is a special case of the linear parametrization in Eq.~\eqref{onedimlinear}. In terms of~$x$, this power series converges in the domain $-4<x<4$ due to the proximity of its expansion point to the singularity $x=4$. Consequently, one and the same series expansion can be evaluated both in the physical region below threshold and in the Euclidean one, where at least one of the kinematic ratios $x$, $z$ and $h$ is negative, provided that the evaluation points lie in the vicinity of the origin $\vec{x}=0$. This is due to the fact that the analytic continuation of the power series in the variables $x$ and $h$ from $0^-$ to $0^+$ is trivial, since it only affects terms of the form $\sqrt{f(\vec{x})}$ and $\log\left(g(\vec{x})\right)$ appearing in the expansion coefficients. These terms have a branch cut for vanishing $f$ and $g$ and develop imaginary parts, depending on the sign of the functions~$f$ and~$g$. Our observations suggest that $f(\vec{x}=0)=g(\vec{x}=0)=0$, implying that there is a branch cut at $\vec{x}=0$, so that the origin is a regular, non-analytic point. Note that the only occuring $\lambda_1$-dependent expressions that involve branch cuts are given by $\sqrt{\lambda_1}$ and $\log\left(\lambda_1\right)$, which however do not entail any ambiguity thanks to the restriction to non-negative values of $\lambda_1$, or more precisely to $\lambda_1\in[0,1]$.
\item[\textbf{2)}] \textbf{Linear Series Expansion around the Threshold}\\
Compared to the power series around the origin, this expansion point is obtained through a simple shift on the $x$~axis from the regular point $x=0$ to the singular point $x=4$:
\begin{equation}
\vec{x} = \begin{pmatrix}x\\z\\h\end{pmatrix} \to \vec{\gamma}_2(\vec{x},\lambda_2) = \lambda_2 \, \begin{pmatrix}x-4\\z\\h\end{pmatrix} + \begin{pmatrix}4\\0\\0\end{pmatrix} \,.
\label{param2}
\end{equation}
The radius of convergence of this series expansion is bounded by the non-analyticity in $\vec{x}=0$, i.e. it is suitable for numerical evaluation in the domain $0<x<8$.
\end{itemize}
In both cases, the series expansions will be carried out around $\lambda_m=0$ and subsequently evaluated at $\lambda_m=1$, thereby reproducing the full dependence on the kinematic variables~$\vec{x}$. This is obviously not sufficient to cover the whole range in~$x$, since the radii of convergence of the two power series do not exhaust the region specified in Eq.~\eqref{xrange2}. Hence, we add a third series expansion around a regular point, corresponding to $y_3$ in Fig.~\ref{fig:matching}:
\begin{itemize}
\item[\textbf{3)}] \textbf{Exponential Series Expansion beyond the Threshold}\\
In this case, we choose a non-linear parametrization in the $xz$-plane according to
\begin{equation}
\vec{x} = \begin{pmatrix}x\\z\\h\end{pmatrix} \to \vec{\gamma}_3(\vec{x},\lambda_3) = \begin{pmatrix}2\,e^{\lambda_3}+4\\z_m\,e^{\alpha\,\lambda_3}\\\lambda_3\,h\end{pmatrix} \,.
\label{param3}
\end{equation}
Although we expand around the point $\lambda_3=0$ as in the previous cases, corresponding to an expansion around $x=6$, this parametrization is different in the sense that the power series is not evaluated at the point~$\lambda_3=1$. Instead, we evaluate it at any point~$\lambda_3$ that carries us to the desired point on the $x$-axis. Remarkably, the non-linear parametrization causes the radius of convergence to be infinite in $\lambda_3$-space, although it is bounded from below by the singularity $x=4$ in $x$-space: The parametrization in Eq.~\eqref{param3} cannot arrive at $x=4$ by setting $\lambda_3$ to very high negative values, but only comes arbitrarily close. For arbitrarily high values of $x$, this means that it is no problem to evaluate the power series, particularly around the upper bound $x=85$ indicated in Eq.~\eqref{xrange2}, which corresponds to $\lambda_3\approx 3.7$. Consequently, we achieved a distortion of the radius of convergence in the transition from $x$- to $\lambda_3$-space, similarly to what was described in Fig.~\ref{fig:egg}.
\end{itemize}
One may wonder why it is necessary to perform a three-dimensional parametrization, corresponding to a series expansion in all three variables $x$, $z$ and $h$, although $h$ will not be varied, but ultimately set to a fixed value. In fact, the exact dependence on $h$ could in principle be retained by limiting oneself to a two-dimensional parametrization $\gamma_m(x,z,\lambda_m)$ that is independent of~$h$. In practice, however, we were not able to derive series expansions associated with such a parametrization, whose coefficients would still depend on~$h$. This is due to the fact that the denominators of the corresponding differential equations in~$\lambda_m$, derived through Eq.~\eqref{deqlambda} with $x_1=x$, $x_2=z$, are given by the letters~$\{l_k\}$ introduced in Eq.~\eqref{hjdeq3}, which have highly non-trivial algebraic dependence on the kinematic invariants. If~$h$ is not accompanied by the parameter~$\lambda_m$ within these denominators, they factorize to a lesser extent, thereby substantially complicating the series expansion of the differential equations. This problem could be resolved by simply setting~$h$ to a rational number describing its physical value prior to the computation of the MIs, e.g. to~$\nicefrac{1}{2}$, thus reducing the number of scales by one. However, this would prevent us from applying boundary conditions at values of~$h$ different from~$\nicefrac{1}{2}$, which will be required in Section~\ref{sec:hjdeqcan4}. Therefore, we include~$h$ in the parametrization procedure and circumvent the mentioned issues without any loss of generality with respect to the numerical evaluation.\\
It remains to comment on the initial assumption of small $z$-values: Thanks to the exponential parametrization of the variable~$z$ in~$\vec{\gamma}_3$, it can be chosen freely and independently of~$x$ using the parameters~$z_m$ and $\alpha$. As a result, the required $z$-domain beyond the threshold in~$x$, given by
\begin{equation}
4<x<85 \,, \quad -\frac{169}{2}<z<0 \,,
\end{equation}
is covered without extra effort and corresponds to the physical region to the right of the red line in Fig.~\ref{fig:dalitz}. Hence, the only remaining part of the physical region is to the left of that red line,
\begin{equation}
0<x<4 \,, \quad -\frac{7}{2}<z<0 \,,
\label{hjregionrem}
\end{equation}
and lies in the gap spanned by the series expansions in $\lambda_1$ and $\lambda_2$. Due to the threshold $z=-x$, their radii of convergence with respect to the variable~$z$ are such that they can be evaluated in the domain $-4<z<0$, thereby exhausting the required region of~$z$ shown in Eq.~\eqref{hjregionrem}. To sum up, the parametrizations $\vec{\gamma}_1$, $\vec{\gamma}_2$ and $\vec{\gamma}_3$ presented in Eqs.~\eqref{param1}, \eqref{param2} and \eqref{param3}, respectively, allow us to cover the complete physical phase space indicated in Eq.~\eqref{xrange2} and depicted in Fig.~\ref{fig:dalitz}.

\subsection{Series Expansions From Canonical Differential Equations}
\label{sec:hjdeqcan3}

Before entering the procedure of deriving series expansions from differential equations, we have to establish those MIs whose differential equations do not contain any subsector contributions. The reason is that the system~\eqref{hjdeq2} of differential equations is solved bottom-up, starting from the topologies with the lowest number $t$ of different propagators and continuously increasing~$t$. However, the differential equations of MIs with the lowest number of distinct propagators do not contain any subsectors, thus cannot be determined through the differential equation approach and have to be computed independently. In the following, we discuss the appearance of such MIs within the system~\eqref{hjdeq2} of differential equations separately for the topology trees of the integral families~$A$ and $B$.\\
In family~$A$, the situation is similar to what was observed for the MIs of the two-loop amplitude for $H\to Z\,\gamma$, where only the double-tadpole result $M_1$ is used as an input of the system of differential equations. The integral representation~\eqref{tadpole} can be evaluated explicitly, so that the result takes the simple form
\begin{equation}
M_1 = 1
\label{tadpoleinput}
\end{equation}
if the normalization factor~$S_\e$ is neglected, i.e. it is independent of the kinematic invariants~$\vec{x}$. An example of a differential equation, in which the double-tadpole appears as subtopology, is given by Eq.~\eqref{sunrisedeq} associated with the canonical integral~$M_2$. The corresponding Laporta integral~$I_2$ is depicted in Fig.~\ref{fig:hjmaster} and emerges from the factorization of the one-loop tadpole and the one-loop massive bubble.\\
The situation is more involved in integral family~$B$, where several MIs without subtopologies occur. First, this concerns the massless bubble integrals $M_{71}$, $M_{72}$ and $M_{235}$, which depend on one of the kinematic invariants in the following way:
\begin{align}
M_{71}(\vec{x}) &= -\frac{\Gamma^2\left(1-\e\right)}{\Gamma\left(1-2\,\e\right)} \, (-x)^{-\e} \,, \nonumber \\
M_{72}(\vec{x}) &= -\frac{\Gamma^2\left(1-\e\right)}{\Gamma\left(1-2\,\e\right)} \, (-h)^{-\e} \,, \nonumber \\
M_{235}(\vec{x}) &= -\frac{\Gamma^2\left(1-\e\right)}{\Gamma\left(1-2\,\e\right)} \, (-z)^{-\e} \,.
\label{masslessbubble}
\end{align}
Beyond that, the differential equation of~$M_{80}$, corresponding to the one-loop massless box with one massive external leg, does not include any integrals different from~$M_{80}$ and is given in terms of hypergeometric functions:
\begin{align}
M_{80}(\vec{x}) &= -2\, \frac{\Gamma^2\left(1-\e\right)}{\Gamma\left(1-2\,\e\right)} \, \left[ \left(\frac{x\,z}{x-h}\right)^{-\e} \, _2F_1\left(-\e,-\e;1-\e;\frac{h-x-z}{h-x}\right) \right. \label{masslessbox} \\
&\qquad\qquad\qquad\qquad + \left(\frac{x\,z}{z-h}\right)^{-\e} \, _2F_1\left(-\e,-\e;1-\e;\frac{h-x-z}{h-z}\right) \nonumber \\
&\qquad\qquad\qquad\qquad - \left. \left(-\frac{x\,z\,h}{(h-x)\,(h-z)}\right)^{-\e} \, _2F_1\left(-\e,-\e;1-\e;\frac{h\,(h-x-z)}{(h-x)\,(h-z)}\right) \right] \nonumber \,.
\end{align}
Both expressions were previously computed in the literature and taken from Refs.~\cite{Baur:1989,Gehrmann:1999,Ellis:2007}. In order to use these results for evaluating differential equations of topologies with higher~$t$, we need to compute their Laurent series in $\e$. This is highly non-trivial for the hypergeometric functions appearing in the expression of $M_{80}$ and requires the use of \textsc{HypExp2}~\cite{Huber:2007}. Since we intend to solve the differential equations with respect to the parameter~$\lambda_m$ introduced in the previous section, we moreover have to introduce the corresponding parametrization~$\vec{\gamma}_m(\vec{x},\lambda_m)$. More precisely, the transition
\begin{equation}
M_{\beta}(\vec{x}) \to M_{\beta,m}(\vec{\gamma}_m(\vec{x},\lambda_m))
\end{equation}
has to be carried out for a given MI~$M_{\beta}$ with~$\beta=\{71,72,80,235\}$. In case of $M_{71}$ and the parametrization~$\vec{\gamma}_1(\vec{x},\lambda_1)$ around the origin~$\vec{x}=0$, the final result suitable for using as an input of the system of differential equations with respect to~$\lambda_1$ reads
\begin{align}
M_{71,1}(\vec{\gamma}_1(\vec{x},\lambda_1)) = &-1 + \log\left(-\lambda_1\,x\right) \, \e + \left[ \frac{\pi^2}{6} - \frac{1}{2} \, \log^2\left(-\lambda_1\,x\right) \right] \, \e^2 \nonumber \\
&+ \left[ 2\,\zeta_3 - \frac{\pi^2}{6} \, \log\left(-\lambda_1\,x\right) + \frac{1}{6} \, \log^3\left(-\lambda_1\,x\right) \right] \, \e^3 \nonumber \\
&+ \left[ \frac{\pi^4}{40} - 2\,\zeta_3 \, \log\left(-\lambda_1\,x\right) + \frac{\pi^2}{12} \, \log^2\left(-\lambda_1\,x\right) - \frac{1}{4} \, \log^4\left(-\lambda_1\,x\right) \right] \, \e^4 \nonumber \\
&+ \mathcal{O}\left(\e^5\right)
\label{masslessbubbleexp}
\end{align}
up to the required order~$\e^4$ of the Laurent series. Interestingly, the series expansion of the massless bubble integrals around the origin introduces logarithms of the kinematic invariants~$\vec{x}$ as opposed to the massive bubbles, which propagate through the full topology tree of integral family~$B$ upon solving the differential equations.\\
With these results in mind, we are in position to derive series expansions for any of the three parametrizations~$\vec{\gamma}_m(\vec{x},\lambda_m)$ specified in the previous section. In order to accomplish this, we proceed as follows:
\begin{enumerate}
\item[\textbf{1.}] \textbf{Compute differential equations in~$\boldsymbol{\lambda_m}$}\\
Initially, we compute the system of differential equations in the parameter $\lambda_m$ from the canonical differential equations~\eqref{hjdeq2} in $x$, $z$ and $h$ by using Eq.~\eqref{deqlambda}. The resulting differential equations are also in canonical form,
\begin{equation}
\frac{\p}{\p \lambda_m} \vec{M}(D;\lambda_m) = \e \, A(\lambda_m) \, \vec{M}(D;\lambda_m) \,,
\label{hjcanlambda}
\end{equation}
so that the power series in $\lambda_m$ can be determined order by order in $\e$ in terms of the previous orders. At every order in $\e$, the only unknown is the derivative on the left-hand side of Eq.~\eqref{hjcanlambda}, whereas all terms on the right-hand side have been identified previously.
\item[\textbf{2.}] \textbf{Substitute known results and solve system of equations}\\
For a given order in~$\e$, we compute the derivative of the ansatz~\eqref{partsol10} up to the desired degree~$p$ and substitute it together with the results of the preceding $\e$-order into the differential equations~\eqref{hjcanlambda}. This applies in particular to the respective $\e$-order of the results in Eqs.~\eqref{tadpoleinput}, \eqref{masslessbubble} and \eqref{masslessbox} after expanding them in~$\e$ and expressing them in terms of~$\lambda_m$, as shown for $M_{71}$ in Eq.~\eqref{masslessbubbleexp}. Next, we calculate the series expansion of the differential equations around $\lambda_m=0$ and collect terms with the same powers of $\lambda_m$ and $\log(\lambda_m)$. The coefficients of these powers must vanish independently, leading to a system of equations, which can be solved algebraically for the unknown coefficients appearing in the ansatz~\eqref{partsol10}.
\item[\textbf{3.}] \textbf{Iterate step 2. at every order in~$\boldsymbol{\e}$}\\
Eventually, we repeat this procedure at every order in the dimensional parameter~$\e$ up to weight $k=4$ and end up with one Laurent series~$M_{n,m}$ in~$\e$ per MI with index~$n$ and expansion point with index~$m$, where $n=\{1,\ldots,102,201,\ldots,249\}$ runs over the indices of the canonical basis integrals specified in Appendix~\ref{sec:hjcan} and $m=\{1,2,3\}$ indicates the use of one of the parametrizations defined in the previous section:
\begin{equation}
M_{n,m} = \sum_{k=0}^4 \e^k \, M_{n,m}^{(k)} \,.
\label{Mnmdef}
\end{equation}
The coefficients $M_{n,m}^{(k)}$ of this Laurent expansion are in turn given by the most general power series representation~\eqref{partsol10} in~$\lambda_m$,
\begin{align}
M_{n,m}^{(k)} &= c_{n,m}^{(k)} + \sum_{i=0}^{k-1} \sum_{j=1}^p \left( a_{i,j}(\vec{x}) + \frac{a_{i,j-\frac{1}{2}}(\vec{x})}{\sqrt{\lambda_m}} \right) \, \lambda_m^j \, \log^i(\lambda_m) \nonumber \\
&\quad\,+ \sum_{i=1}^k a_{i,0} \, \log^i\left[\lambda_2\,(x-4)\right] \,,
\label{Mnmdef2}
\end{align}
where the second line is only present for the expansion in~$\lambda_2$, i.e. $a_{i,0}=0$ if \mbox{$m=\{1,3\}$}. Therein, the coefficients $a_{i,j}$ are fully determined and the only missing ingredient is one boundary constant $c_{n,m}^{(k)}$ per MI with index~$n=\{1,\ldots,102,201,\ldots,249\}$, expansion point with index~$m=\{1,2,3\}$ and weight~$k=\{1,\ldots,4\}$, which have to be fixed through appropriate boundary conditions.
\end{enumerate}
Let us point out that we made an interesting observation by deriving the series expansions around the singular point~$\lambda_2$. Upon solving the system of equations algebraically in step~2, the coefficients $a_{1,0},\ldots,a_{k,0}$ of the divergent logarithmic terms at a given weight~$k$ turn out to be equal to the boundary constants~$c_{n,2}^{(1)},\ldots,c_{n,2}^{(k-1)}$ of lower weights, up to a rational number. More precisely, the system from which these relations emerge can be separated from the remaining equations, i.e. these identities form an independent sub-system of equations. For example, the weight-four result for the one-loop massive bubble integral~$M_2$, which was previously considered in Sections~\ref{sec:oneloopsunrise}, \ref{sec:frobenius2exp} and \ref{sec:thresholdexp}, reads
\begin{align}
M_{2,2}^{(4)} &= \sum_{j=1}^p \left( \frac{a_{i,j-\frac{1}{2}}(x)}{\sqrt{\lambda_2}} \right) \, \lambda_2^j + c_{2,2}^{(4)} - c_{2,2}^{(3)} \, \log\left[\lambda_2\,(x-4)\right] \nonumber \\
&\quad\,+ \frac{1}{2} \, c_{2,2}^{(2)} \, \log^2\left[\lambda_2\,(x-4)\right] - \frac{1}{6} \, c_{2,2}^{(1)} \, \log^3\left[\lambda_2\,(x-4)\right]
\end{align}
upon expansion around the threshold.

\newpage

\subsection{Boundary Conditions and Matching Procedure}
\label{sec:hjdeqcan4}

In the following, we describe how to supply the power series determined in the last section with boundary conditions either through implicit regularity conditions or through the matching procedure described in Section~\ref{sec:matching}.\\
As stated before, the two-loop Laporta integrals~$\vec{I}$ are known to be regular in the origin, i.e. the limit $\lim_{\vec{x}\to\vec{0}} \vec{I}$ is well-defined, which was one reason to perform series expansions around the point~$\lambda_1$. The reason for this is that this limit corresponds to the one of infinite quark mass, where all integrals turn into tadpoles. As a next step, one could enter the differential equations and determine the boundary conditions by imposing regularity at that point, as described in Section~\ref{sec:boundary}. In fact, this procedure is not required in our case, because the canonical basis~$\vec{M}$ defined in Appendix~\ref{sec:hjcan} exhibits a remarkable behavior: Every homogeneous solution or integrating factor~$j_n(\vec{x})$ associated with the integral~$M_n$, which is given by the coefficient of $I_n$ in the definition of~$M_n$, vanishes in the limit~$\vec{x}\to\vec{0}$. Since all~$I_n$ are regular in this point, we conclude that
\begin{equation}
\lim_{\vec{x}\to\vec{0}} \vec{M} = \lim_{\lambda_1\to 0} \vec{M} = 0
\label{hjboundary}
\end{equation}
to all $\e$-orders. Given that the power series in~$\lambda_1$ is of the form
\begin{equation}
M_{n,1}^{(k)} = c_{n,1}^{(k)} + \sum_{i=0}^{k-1} \sum_{j=1}^p \left( a_{i,j}(\vec{x}) + \frac{a_{i,j-\frac{1}{2}}(\vec{x})}{\sqrt{\lambda_1}} \right) \, \lambda_1^j \, \log^i(\lambda_1) \,,
\end{equation}
we obtain with the help of Eq.~\eqref{hjboundary} that the boundary constants of all canonical integrals $M_n$ vanish in the origin for every weight~$k$:
\begin{equation}
c_{n,1}^{(k)} = 0 \qquad \forall \, n,k \,.
\end{equation}
Note that this statement does not apply to the massless bubble integrals $M_{71}$, $M_{72}$ and $M_{235}$ nor to the massless box~$M_{80}$ within family~$B$, which diverge for small values of~$\vec{x}$. However, all of them are fed externally into the system of differential equations, as outlined in the previous section, so that they are excluded from this procedure.\\
Certainly, there is one exception to these findings: The two-loop four-point function~$M_{81}$ belongs to the sector~$B_{5,174}$ with two MIs and diverges in the limit of vanishing~$z$. Therefore, the ansatz for $M_{81}$ takes the more general form
\begin{align}
M_{81,1}^{(k)} &= c_{81,1}^{(k)} + \sum_{i=0}^{k-1} \sum_{j=1}^p \left( a_{i,j}(\vec{x}) + \frac{a_{i,j-\frac{1}{2}}(\vec{x})}{\sqrt{\lambda_1}} \right) \, \lambda_1^j \, \log^i(\lambda_1) \nonumber \\
&\quad\,+ \sum_{i=1}^k a_{i,0} \, \log^i\left[-\lambda_1\,z\right]
\label{hjboundary81}
\end{align}
when expanded around the point~$\lambda_1$. The boundary values~$c_{81,1}^{(k)}$ can be inferred from its differential equations at the point $\lambda_1\to 0$, which is possible despite the singular behavior of $M_{81}$ in $z\to 0$, since it can be traced back to the singular behavior of its subsector~$M_{235}$. In other words, we transform both of them into regular integrals by redefining the basis according to
\begin{equation}
\begin{pmatrix}M_{81}\\M_{235}\end{pmatrix} \to \begin{pmatrix}J_{81}\\J_{235}\end{pmatrix} = (-\lambda_1\,z)^\e \, \begin{pmatrix}M_{81}\\M_{235}\end{pmatrix} \,,
\end{equation}
similarly to the procedure described in the end of Section~\ref{sec:frobenius2simp}. In doing so, the prefactor~$(-\lambda_1\,z)^\e$ absorbs the divergent logarithms~$\log^i\left[-\lambda_1\,z\right]$ within Eq.~\eqref{hjboundary81}, so that we can impose regularity on the differential equations of the new basis, from which we obtain
\begin{equation}
\lim_{\lambda_1\to 0} \lambda_1 \, \frac{\p J_{81}}{\p\,\lambda_1} = J_{81} - 2\,J_{235} \stackrel{!}{=} 0 \,.
\end{equation}
This relation finally enables us to derive the boundary constants of $M_{81,1}$ within Eq.~\eqref{hjboundary81}:
\begin{align}
c_{81,1}^{(1)} &= 0 \,, \nonumber \\
c_{81,1}^{(2)} &= \frac{\pi^2}{6} \,, \nonumber \\
c_{81,1}^{(3)} &= 2\,\zeta_3 \,, \nonumber \\
c_{81,1}^{(4)} &= \frac{\pi^4}{40} \,.
\label{hjboundary81c}
\end{align}
Note that $M_{240}$ corresponds to the $x\leftrightarrow z$-crossing of $M_{81}$ and is treated in the same way, resulting in the boundary constants of Eq.~\eqref{hjboundary81c} as well.\\
As a next step, we aim to determine the remaining boundary constants $c_{n,2}^{(k)}$ and $c_{n,3}^{(k)}$ of the series expansions in $\lambda_2$ and $\lambda_3$ through the matching procedure presented in Section~\ref{sec:matching}. In the language of Fig.~\ref{fig:matching}, this implies that we have to choose two matching points~$m_1$ and $m_2$ lying within the radii of convergence of series expansions with adjacent expansion points~$\lambda_m$ and $\lambda_{m+1}$. This can be achieved by analyzing the impact of the variation of the matching point on the accuracy of the numerical results, which can be verified by comparing with numerical integration routines such as sector decomposition, or in case of the three-point functions appearing in $H\to Z\,\gamma$ with the numerical evaluation of the exact analytical expressions. Such a study reveals that the most accurate numerical predictions are obtained if the series expansion in~$\lambda_2$ is matched to the one in~$\lambda_1$ in the center of the two expansion points with respect to the $x$-axis, i.e. at the point~$x_1=2$. We emphasize that we are not obliged to choose the matching values of~$z$ and $h$ such that the matching is performed at a physical phase space point. Since both expansions are carried out for small values of $z$ and $h$, we should instead keep their absolute values sufficiently small in order to ensure maximally convergent power series throughout the matching procedure. Hence, the first matching point reads
\begin{equation}
\vec{m}_1 = \begin{pmatrix}x_1\\z_1\\h_1\end{pmatrix} = \begin{pmatrix}2\\-10^{-3}\\10^{-2}\end{pmatrix} \,,
\label{hjmatching1}
\end{equation}
where we verified that decreasing the values of $|z_1|$ and $|h_1|$ does not lead to more accurate numerical results. Beyond that, we have chosen a matching point with $z_1\neq -h_1$ in order to circumvent singularities due to denominators of the kind $1/(h+z)$, which might occur in the coefficients of the series expansions. Equation~\eqref{hjmatching1} allows us to determine the boundary constants $c_{n,2}^{(k)}$ by imposing that the two series expansions coincide upon evaluation at that point for all MIs :
\begin{equation}
M_{n,1}^{(k)}\left(\lambda_1=1,\vec{x}=\vec{m}_1\right) = M_{n,2}^{(k)}\left(\lambda_2=1,\vec{x}=\vec{m}_1\right) \,.
\end{equation}
This numerical system of equations is at most triangular with respect to the unknowns~$c_{n,2}^{(k)}$, since boundary constants of low weight~$k$ reappear in the power series result of higher weight, but not vice versa. Let us recall that~$M_{81}$ has a logarithmic divergence at the point~$z\to 0$, thus cannot be matched at the point~$m_1$ specified in Eq.~\eqref{hjmatching1} and has to be treated separately, as above. On the one hand, the integral cannot be matched too close to the origin, on the other we must not choose its matching point too far from the point~$z=0$ due to increasingly unsatisfactory convergence behavior. A compromise is found by matching at the point
\begin{equation}
\vec{m}_{1,81} = \begin{pmatrix}x_1\\z_1\\h_1\end{pmatrix} = \begin{pmatrix}2\\-10^{-1}\\10^{-2}\end{pmatrix} \,.
\end{equation}
Note that this separate treatment is not required for its $x\leftrightarrow z$-crossing $M_{240}$, which has a logarithmic divergence in~$x$ and thus can be matched at the usual point~$\vec{m}_1$ indicated in Eq.~\eqref{hjmatching1}, where~$M_{240}$ converges.\\
With a fully determined power series solution in~$\lambda_2$, we can proceed by matching it to the series expansions in~$\lambda_3$ in a similar way. The choice of the matching point remains to be analyzed in detail, however we expect the point
\begin{equation}
\vec{m}_2 = \begin{pmatrix}x_2\\z_2\\h_2\end{pmatrix} = \begin{pmatrix}6\\-10^{-3}\\10^{-2}\end{pmatrix}
\label{hjmatching2}
\end{equation}
to be appropriate. Due to the particular structure of the parametrization $\vec{\gamma}_3(\vec{x},\lambda_3)$ in Eq.~\eqref{param3}, we have to set $\lambda_3=0$, $z_m=z_2$ and $h=h_2$ within the series expansion in~$\lambda_3$ in order to reach the matching point~$\vec{m}_2$. Note that $\vec{m}_2$ is obtained independently of the parameter~$\alpha$ occuring in Eq.~\eqref{param3}, which can be chosen freely. In this case, the matching point specific to~$M_{81}$ reads
\begin{equation}
\vec{m}_{2,81} = \begin{pmatrix}x_2\\z_2\\h_2\end{pmatrix} = \begin{pmatrix}6\\-10^{-1}\\10^{-2}\end{pmatrix}
\end{equation}
for the same reason as above.\\
To sum up, we have arrived at three series expansions in~$\lambda_1$, $\lambda_2$ and $\lambda_3$ with fully determined coefficients and boundary constants. In order to obtain numerical results to the largest possible precision, we separate the phase space such that the one power series is used for numerical evaluation whose expansion point lies within the region enclosed either by two matching points or by one matching point and the border of the physical region. In our case, this means that physical results are deduced from the series expansion in~$\lambda_1$ in the region $\frac{1}{2}\leq x\leq 2$, the series expansion in~$\lambda_2$ in the region $2\leq x\leq 6$ and the series expansion in~$\lambda_3$ in the region $6\leq x\leq 85$. Note that $h$ is assumed to be in the vicinity of its physical value, i.e.~$h\approx \nicefrac{1}{2}$, and the domain in~$z$ follows from the one in~$x$, yielding the piecewise-defined result for a given MI~$M_n$ with the notation of Eq.~\eqref{Mnmdef}:
\begin{equation}
M_n =
\begin{cases}
\begin{aligned}
&M_{n,1} \,, & \qquad &\frac{1}{2}\leq x\leq 2 \,, & \quad &-\frac{3}{2}\leq z\leq 0\,, & \quad &h\approx \frac{1}{2}\,,\\
&M_{n,2} \,, & \qquad &2\leq x\leq 6 \,, & \quad &-\frac{11}{2}\leq z\leq 0\,, & \quad &h\approx \frac{1}{2} \,,\\
&M_{n,3} \,, & \qquad & 6\leq x\leq 85 \,, & \quad &-\frac{169}{2}\leq z\leq 0 \,, & \quad &h\approx \frac{1}{2}\,.
\end{aligned}
\end{cases}
\end{equation}

\subsection{Technical Details and Speed-Up Possibilities}
\label{sec:hjdeqcan5}

In the following, we will provide the details of the approach presented in the previous sections by commenting on a few useful observations that led to a speed-up of our computations.
\begin{itemize}
\item[\textbf{a)}] \textbf{Simplify ansatz for series expansions around regular expansion points}\\
If the sector under consideration is known to be regular in the expansion point or if furthermore its subsectors are known to be regular in that point, the ansatz~\eqref{partsol10} can be replaced by the simplified version~\eqref{partsol11} or \eqref{partsol12}, respectively. In case of Higgs-plus-jet production with full quark mass dependence, the system of differential equations~\eqref{hjdeq2} is such that the only singularities are introduced through divergent logarithmic terms $\log\left(\lambda_2\,(x-4)\right)$ emerging from the result for massive bubble integrals in~$x$, provided that they are expanded around the threshold~$x=4$.\\
At the level of the Feynman graphs, the initial statement of regularity of a given MI can thus be rephrased as to whether the Feynman graph of that MI can be pinched such that one of the graphs $I_2$, $I_3$ and $I_4$ shown in Fig.~\ref{fig:hjmaster} remains. Similarly, the only convergent logarithmic terms of the form~$\lambda_m\,\log\left(\lambda_m\right)$ are generated through the results for massless bubble integrals in $x$, $z$ or $h$ or massless box integrals, provided that the corresponding component within the parametrization~$\vec{\gamma}_m$ vanishes in the limit $\lambda_m\to 0$. In case of the parametrization $\vec{\gamma}_2$ indicated in Eq.~\eqref{param2}, for example, this implies that we have to verify whether the Feynman graph of a given MI can be pinched such that one of the graphs $I_{72}$, $I_{80}$ or $I_{235}$ depicted in Fig.~\ref{fig:hjmaster} remains, which correspond to the massless bubbles in $z$ and $h$ as well as to the massless box. Note that we explicitly excluded the massless bubble in~$x$ in this case, since the $x$-component of the parametrization~$\vec{\gamma}_2$ does not vanish for $\lambda_2\to 0$. This is different for the parametrization $\vec{\gamma}_1$ given in Eq.~\eqref{param3}, where we would have to check whether massless bubbles in $x$ occur as a subtopology, corresponding to the graph of~$I_{71}$, on top of the ones mentioned for~$\vec{\gamma}_2$.\\
From considerations of this kind, we therefore create a database for every expansion point~$\lambda_m$ in order to simplify the ansatz whenever possible, thereby minimizing the complexity of the algebraic system of equations which has to be solved. This is particularly useful for series expansions around regular points, as the ones in~$\lambda_1$ and $\lambda_3$, where the logarithmic terms within the ansatz can be omitted for many MIs.
\item[\textbf{b)}] \textbf{Set~$\boldsymbol{z_m}$ to numerical value within parametrization $\boldsymbol{\gamma_3(\vec{x},\lambda_3)}$}\\
Let us recall that the matching of the series expansions in $\lambda_2$ and $\lambda_3$ is carried out at the point~$m_2$. As mentioned in the previous section, the $z$-component of $m_2$ simply corresponds to the parameter~$z_m$ and according to Eq.~\eqref{hjmatching2}, it is set to $-10^{-3}$ for the matching. In fact, $z_m$ appears in the parametrization $\gamma_3(\vec{x},\lambda_3)$ and can therefore be set to the numerical matching value before deriving the series expansion in~$\lambda_3$ without any loss of generality, thereby speeding up the calculations considerably. We stress that~$z$ can still be evaluated at the desired evaluation point in the final result of the power series thanks to the parameter~$\alpha$ occuring in Eq.~\eqref{param3}.
\item[\textbf{c)}] \textbf{Use boundary constants to absorb finite numerical terms}\\
When the differential equations of the MIs with respect to~$\lambda_m$ are derived and their series expansion in~$\lambda_m$ is computed, finite logarithms of purely numerical argument may arise. This holds in particular for MIs of integral family~$B$, where the massless one-loop bubble integral in~$x$ appears as subtopology and must be provided to the system of differential equations. This bubble integral is denoted by~$M_{71}$ and introduces terms of the form~$\log\left(\lambda_2(x-4)+4\right)$ due to the structure of the parametrization in Eq.~\eqref{param2}, which turn into
\begin{equation}
\log\left(\lambda_2(x-4)+4\right) = \log(4) + \sum_j^\infty a_j \, \lambda_2^j
\end{equation}
upon expansion in~$\lambda_2$. By comparing to the structure of the result in Eq.~\eqref{Mnmdef2}, it becomes clear that $\log(4)$ is a boundary-like term, which can be absorbed through a redefinition of the boundary constants:
\begin{equation}
c_{n,m}^{(k)} \to c_{n,m}^{(k)} + \log(4) \,.
\end{equation}
In practice, this corresponds to discarding terms of the form~$\log(4)$ within the system of differential equations after inserting the subsector results and expanding in~$\lambda_m$.
\item[\textbf{d)}] \textbf{Determine first non-vanishing $\boldsymbol{\e}$-order in warm-up run with low degree~$\boldsymbol{p}$}\\
In the differential equations approach, the first non-vanishing weight of the integral under consideration is determined by the non-vanishing weights of its subsectors. At a given $\e$-order, it might occur that the subsector results combine such that the result of the considered integral evaluates to zero in a non-trivial way. In order to prevent cancellations between several series expansions of high degree~$p$, we suggest to launch a warm-up run with very low order~$p$ of the power series ansatz~\eqref{partsol10}, which is designed to determine the first non-vanishing weight of a given integral. This can be done around any expansion point from within the parametrizations $\vec{\gamma}_m$, preferably for the one with the smallest coefficients. The result will then be valid for any expansion point and to any degree~$p$ of the expansion. Subsequently, it can be used to set the coefficients of the Laurent expansion in~$\e$ to zero prior to the determination of the series expansion in~$\lambda_m$ with higher degree~$p$. This applies in particular to the boundary constants occuring therein, which would otherwise propagate into the results of higher weights and lead to unnecessary expression swell.
\item[\textbf{e)}] \textbf{Make use of common parametrizations components of previously computed series expansions}\\
Similarly, conclusions about a power series with a given parametrization can be drawn from previous determinations of series expansions around different expansion points. More precisely, if two parametrizations only differ in one variable, as in the case of $\vec{\gamma}_1$ and $\vec{\gamma}_2$, one should not recompute the power series associated with integrals that are not functions of this variable. In such a case, the result of the previous series expansion can be recycled and the boundary constants can be set to zero prior to the matching procedure. In going from the series expansion in~$\lambda_1$ to the one in~$\lambda_2$, for example, this applies to the canonical integrals $M_{1}$, $M_{5}$, $M_{6}$, $M_{7}$, $M_{14}$, $M_{15}$, $M_{16}$, $M_{17}$, $M_{18}$, $M_{19}$, $M_{20}$, $M_{21}$, $M_{23}$, $M_{28}$, $M_{29}$, $M_{30}$, $M_{31}$, $M_{54}$, $M_{72}$, $M_{74}$, $M_{75}$, $M_{77}$, $M_{79}$, $M_{92}$, $M_{201}$, $M_{202}$, $M_{203}$, $M_{204}$, $M_{205}$, $M_{206}$, $M_{207}$, $M_{208}$, $M_{217}$, $M_{218}$.
\item[\textbf{f)}] \textbf{Factorize non-converging homogeneous solutions in the ansatz}\\
Before deriving a power series around a given expansion point, we verify that the homogeneous solution of the differential equations~\eqref{hjdeq1} in the Laporta basis~$\vec{I}$ converges in the vicinity of the expansion point. If this is not the case, then the result for the series expansion will not converge either. This is due to the fact that these integrating factors are part of the definition of the canonical basis~$\vec{M}$ in Appendix~\eqref{sec:hjcan}, so that the canonical differential equations~\eqref{hjdeq2} which we aim to solve also involve these structures.\\
At this point, we have multiple choices: First, we could look for a new canonical basis integral with a different integrating factor. However, we have observed that this is rarely possible, since the homogeneous solutions are inherent to a sector and cannot be removed. Second, as suggested by the convergence behavior of the homogeneous solution in such a case, we could solve the non-canonical differential equations in the Laporta basis~$\vec{I}$ instead. Certainly, this would deprive us of the many praised advantages of the canonical form, see for example Sections~\ref{sec:canon} and \ref{sec:frobenius2simp}. In order to circumvent this, we do not vary the differential equation, but the ansatz~\eqref{partsol10}: We multiply its non-divergent part by the non-converging integrating factor~$j_n(\vec{x})$ associated with the integral~$M_n$, which is given in Appendix~\eqref{sec:hjcan} by the coefficient of $I_n$ in the definition of~$M_n$. Subsequently, we parametrize this prefactor using the parametrization $\vec{\gamma}_m(\vec{x},\lambda_m)$ designed for the given expansion point~$\lambda_m$, so that Eq.~\eqref{partsol10} is replaced by
\begin{align}
M_{n,m}^{(k)}(\lambda_m) &= j_n(\vec{\gamma}_m(\vec{x},\lambda_m)) \, \sum_{i=0}^{k-1} \sum_{j=1}^p \left( a_{i,j}(\vec{x}) + \frac{a_{i,j-\frac{1}{2}}(\vec{x})}{\sqrt{\lambda_m}} \right) \, \lambda_m^j \, \log^i(\lambda_m) \nonumber \\
&\quad\,+ \sum_{i=0}^k a_{i,0} \, \log^i\left[\lambda_2\,\left(x-4\right)\right] \,,
\end{align}
where the second line occurs only for the series expansion in~$\lambda_2$, as before. Note that the same substitution can be made in the simplified Eqs.~\eqref{partsol11} and~\eqref{partsol12}.\\
Next, we proceed in the usual way to determine the coefficients of this new ansatz, leading to results in which the exact dependence on the non-converging integrating factor is retained. That way, we manage to obtain a converging series expansion from canonical differential equations, although the corresponding integrating factor does not converge in the neighborhood of the expansion point. For example, in case of the series expansion in~$\lambda_2$, this procedure was not required at all, whereas in case of the power series in~$\lambda_1$ we had to introduce this approach for the canonical differential equations of $M_{32}$, $M_{34}$, $M_{35}$, $M_{37}$, $M_{38}$, $M_{40}$, $M_{51}$, $M_{55}$, $M_{57}$, $M_{61}$, $M_{63}$, $M_{66}$, $M_{67}$, $M_{84}$, $M_{89}$, $M_{98}$, $M_{224}$, $M_{225}$, $M_{227}$, $M_{228}$, $M_{231}$, $M_{233}$, $M_{243}$, $M_{249}$.
\item[\textbf{g)}] \textbf{Solve algebraic system of equations first for coefficients of divergent logarithmic terms}\\
Let us recall that we eventually take the limit $\lambda_m=1$ for $m=\{1,2\}$, implying that all logarithms of the form $\log(\lambda_m)$ and their coefficients $a_{i,j}$ ($i,j\geq 1$) within the ansatz~\eqref{partsol10} vanish. Upon solving the system of equations for the unknown coefficients of the series expansions, one might therefore assume that it is not required to determine these coefficients $a_{i,j}$ ($i,j\geq 1$), thereby saving a lot of effort. However, the derivative of $\log(\lambda_m)$ leads to the rational expression $1/\lambda_m$, but not the other way around. As a result, the coefficients $a_{i,j}$ ($i,j\geq 1$) of the logarithm propagate upon differentiation of the ansatz into the sub-system of equations which contains the coefficients~$a_{0,j}$ ($j\geq 1$) of the half-integer and integer powers of~$\lambda_m$, and not vice versa. Hence, the determination of the coefficients~$a_{0,j}$ ($j\geq 1$), which will ultimately contribute to the numerical result, requires the coefficients $a_{i,j}$ ($i,j\geq 1$) to be known beforehand, although the latter will not affect the numerical evaluation directly. In turn, this means that the most efficient approach is to solve the sub-system first, in which only the coefficients $a_{i,j}$ ($i,j\geq 1$) of the logarithmic terms appear. With this result, one can enter the system of the remaining equations in order to determine the coefficients $a_{0,j}$ ($j\geq 1$) of the non-logarithmic terms.
\end{itemize}

\subsection{Alternative Partitioning of the Phase Space}
\label{sec:hjdeqcan6}

In the following paragraph, we show that the choice of parametrization in Section~\ref{sec:hjdeqcan2} is not unique and that there are alternative ways to divide the physical phase space depicted in Fig.~\ref{fig:dalitz}, which however turn out to be less efficient.\\
Instead of inferring boundary conditions from small values of~$x$, corresponding to the leftmost end of the physical region in Fig.~\ref{fig:dalitz}, one could do this to the other, rightmost end of the physical region, i.e. for large values of~$x$. Since the physical value of~$h$ is small, we retain the linear parametrization of~$h$ introduced in Section~\ref{sec:hjdeqcan2}, so that
\begin{equation}
\vec{x} = \begin{pmatrix}x\\z\\h\end{pmatrix} \to \vec{\tilde{\gamma}}_{1a}(\vec{x},\tilde{\lambda}_{1a}) = \begin{pmatrix}x/\tilde{\lambda}_{1a}\\z/\tilde{\lambda}_{1a}\\ \tilde{\lambda}_{1a}\,h\end{pmatrix} \,.
\label{paramalt1a}
\end{equation}
In this case, a power series around~$\tilde{\lambda}_{1a}=0$ corresponds to expanding around the point $x=\infty$, $z=-\infty$, $h=0$. Note that we could have equally chosen to derive series expansions around the point $x=\infty$, $z=0$, $h=0$, parametrized by
\begin{equation}
\vec{x} = \begin{pmatrix}x\\z\\h\end{pmatrix} \to \vec{\tilde{\gamma}}_{1b}(\vec{x},\tilde{\lambda}_{1b}) = \begin{pmatrix}x/\tilde{\lambda}_{1b}\\\tilde{\lambda}_{1b}\,z\\ \tilde{\lambda}_{1b}\,h\end{pmatrix} \,,
\label{paramalt1b}
\end{equation}
because we will ultimately have to cover the full range $-\infty<z<0$ to the rightmost end of the physical region in Fig.~\ref{fig:dalitz}, i.e. both for small and large values of~$z$.\\
There is a severe issue with both parametrizations~$\vec{\tilde{\gamma}}_{1a}(\vec{x},\tilde{\lambda}_{1a})$ and $\vec{\tilde{\gamma}}_{1b}(\vec{x},\tilde{\lambda}_{1b})$. In contrast to the linear parametrization $\vec{\gamma}_1(\vec{x},\tilde{\lambda}_1)$ around the origin presented in Eq.~\eqref{param1}, the derivation of the associated boundary conditions is a highly non-trivial task, which has not been taken care of so far\footnote{In the meantime, the boundary conditions corresponding to~$\vec{\tilde{\gamma}}_{1a}(\vec{x},\tilde{\lambda}_{1a})$ in Eq.~\eqref{paramalt1a} became available~\cite{Kudashkin:2017}. However, we are still convinced that the choice of phase space partitioning presented in Section~\ref{sec:hjdeqcan2} is more appropriate for reasons outlined in the following.}. Therefore, we resort to the parametrization
\begin{equation}
\vec{x} = \begin{pmatrix}x\\z\\h\end{pmatrix} \to \vec{\tilde{\gamma}}_1(\vec{x},\tilde{\lambda}_1) = \frac{1}{\tilde{\lambda}_1} \, \begin{pmatrix}x\\z\\h\end{pmatrix} \,,
\label{paramalt1}
\end{equation}
corresponding to a series expansion around $x=\infty$, $z=-\infty$, $h=\infty$, whose boundary conditions are known~\cite{Melnikov:2016}. The results provided therein are given in terms of MPLs with arguments
\begin{equation}
y_h \equiv \frac{t}{m_H^2} = 1-\frac{x+z}{h} \,, \qquad z_h \equiv \frac{u}{m_H^2} = \frac{z}{h}\,,
\label{bcarguments}
\end{equation}
i.e. they are independent of the parameter~$\tilde{\lambda}_1$, since its inverse can be scaled out in all three components of the parametrization~$\vec{\tilde{\gamma}}_1(\vec{x},\tilde{\lambda}_1)$. These arguments do not correspond to our choice of kinematic invariants presented in Eq.~\eqref{ratiodef}, thus requiring transformations of the type~\eqref{trafo2}, which can be derived by applying the symbol and coproduct formalism implemented in the program~\textsc{MPLEval}~\cite{Gehrmann:2013,Weihs:2013}. This procedure changes the arguments of the MPLs given in Eq.~\eqref{bcarguments} to
\begin{equation}
x_h \equiv \frac{s}{m_H^2} = \frac{x}{h} \,, \qquad z_h = \frac{u}{m_H^2} = \frac{z}{h} \,,
\end{equation}
so that they can be used in the framework of our notation introduced in Section~\ref{sec:hjkinematics}. With the help of these boundary conditions, we derived the series expansions in~$\tilde{\lambda}_1$ in the same way as described for the parametrization in~$\lambda_1$ in Sections~\ref{sec:hjdeqcan3} and \ref{sec:hjdeqcan4}. Unfortunately, compared to the phase space partitioning presented in Section~\ref{sec:hjdeqcan2}, covering the full physical phase space by starting from the expansion point~$\tilde{\lambda}_1=0$ is a much more cumbersome task for several reasons:
\begin{itemize}
\item First, we are dealing with considerably larger expressions for the power series associated with the parametrization~$\vec{\tilde{\gamma}}_1(\vec{x},\tilde{\lambda}_1)$ in Eq.~\eqref{paramalt1} than for the one parametrized by $\vec{\gamma}_1(\vec{x},\lambda_1)$ in Eq.~\eqref{param1}. This is due to the fact that the corresponding boundary conditions at $x=\infty$, $z=-\infty$, $h=\infty$ take a much more complicated form than at $\vec{x}=0$.
\item Second, although the radius of convergence of the power series in~$\tilde{\lambda}_1$ is bounded from below only by the threshold at~$x=4$, we observe in practice that the series expansion in~$\tilde{\lambda}_1$ yields reliable numerical predictions only in the region \mbox{$|x_i|>8$}. Presumably, this is due to the convergence slowing down far away from the expansion point, thus requiring the derivation of series expansions to higher degree~$p$, which is simply not feasible\footnote{Let us point out that we will make more precise statements on actual values of the degree~$p$ in Section~\ref{sec:hjnum}.}. Moreover, numerical results that are accurate enough to be used in the matching procedure are only obtained in the domain~$|x_i|>16$, so that the power series in~$\tilde{\lambda}_1$ cannot be connected directly to an expansion around the threshold~$x=4$ with the method described in Section~\ref{sec:matching}. This could be circumvented by providing a supplementary series expansion around a point in between the two expansion points~$\tilde{\lambda}_1$ and $\lambda_2$, whereas the region below the threshold is taken care of in the same way as in Section~\ref{sec:hjdeqcan2}. In fact, since the radius of convergence of the power series around the threshold presented therein is sufficient to cover the complete physical region below the threshold, the series expansion around the origin might not even be necessary in this case.
\item Third, and most importantly, due to the fact that the starting point $h=\infty$ is far from the physical value~$h\approx \nicefrac{1}{2}$, the matching procedure has to be carried out in the three-dimensional space spanned by real values of the kinematic invariants~$x$, $z$ and~$h$. This is in contrast to the approach described in Section~\ref{sec:hjdeqcan2} and \ref{sec:hjdeqcan5}, where we deal with an effective two-dimensional matching in the physical $xz$-plane depicted in Fig.~\ref{fig:dalitz}, since the dependence on $h$ within the parametrizations~$\lambda_m$ ($m=1,2,3$) is identical and close to the physical value of~$h$.\\
One possibility to resolve this issue is to introduce additional linear parametrizations in the three-dimensional space~$\vec{x}$ around a generic, linear expansion point~$\vec{x}_0=(x_0,z_0,h_0)$:
\begin{equation}
\vec{x} = \begin{pmatrix}x\\z\\h\end{pmatrix} \to \vec{\tilde{\gamma}}_2(\vec{x},\tilde{\lambda}_2) = \tilde{\lambda}_2 \, \begin{pmatrix}x-x_0\\z-z_0\\h-h_0\end{pmatrix} + \begin{pmatrix}x_0\\z_0\\h_0\end{pmatrix} \,.
\label{paramalt2}
\end{equation}
This would require the computation of only one extra series expansion, into which arbitrary values of $\vec{x}_0$ could be substituted after its derivation. However in practice, we observe that the computation of this kind of power series is unfeasible, which can be traced back to the large coefficients appearing in the denominators of the differential equations after applying the parametrization~$\vec{\tilde{\gamma}}_2(\vec{x},\tilde{\lambda}_2)$ given in Eq.~\eqref{paramalt2}. One could therefore supply this parametrization with numerical values $x_{i,0}$ from within $x_i=\{x,z,h\}$, thereby multiplying the number of series expansions to be derived by the number of distinct values~$x_{i,0}$ required to cover the whole physical range on the corresponding~$x_i$-axis. This procedure turns out to be particularly inefficient by recalling that it has to be performed in all three directions of~$\vec{x}$: First, in order to cover all physical values of~$x$ between $x=4$ and $x=\infty$, as explained in the previous bullet point; second, to account for small absolute values of~$z$ in addition to the ones close to~$z=-\infty$; and third bearing in mind that the starting point~$h=\infty$ has to be evolved to the physical value~$h\approx \nicefrac{1}{2}$.
\end{itemize}
To sum up, there are multiple reasons why the partitioning of the phase space explained in Section~\ref{sec:hjdeqcan2} is superior to the ones presented here, which require the introduction of additional expansion points in order to cover the complete physical phase space indicated in Eq.~\eqref{xrange2}. Consequently, we use the more efficient procedure described in Section~\ref{sec:hjdeqcan2} in the following.\\
In a final comment, we emphasize that the power series parametrized by~$\vec{\tilde{\gamma}}_2(\vec{x},\tilde{\lambda}_2)$ as described in Eq.~\eqref{paramalt2} cannot only be evaluated for large negative values of~$z$, but also for large positive ones, since we retain the full symbolic dependence on the kinematic invariant~$z$. In this case, one has $x=z=h=\infty$, corresponding to the physical region of the Higgs-decay processes presented in Eq.~\eqref{higgsdecay}. Through comparison with~\textsc{SecDec} results, similarly to what is described in Section~\ref{sec:hjnum} for the parametrizations introduced in Section~\ref{sec:hjdeqcan2}, we verified numerically that the results are correct. As a side-product, we thus calculated the planar MIs of the Higgs boson decaying into three partons in the physical region for high values of the Mandelstam invariants.

\newpage

\section[{Differential Equations and Master Integrals: The Planar Elliptic Sector~$A_{6,215}$}]{Differential Equations and Master Integrals:\\The Planar Elliptic Sector~$\boldsymbol{A_{6,215}}$}
\label{sec:hjdeqell215}

The elliptic sector $A_{6,215}$ has four MIs denoted by $I_{59}$--$I_{62}$, which are defined in terms of integral family~$A$ in Appendix~\ref{sec:hjlaporta} and depicted in Fig.~\ref{fig:hjmaster}. Analytical expressions for these MIs can be found in the literature, which are however limited either to the Euclidean region~\cite{Bonciani:2016} or to the homogeneous solution of the differential equations~\cite{Primo:2016,Frellesvig:2017,Harley:2017}. In the following, we will describe the derivation of their series expansions in the physical region.\\
For this purpose, we start in Section~\ref{sec:hjdeqell1} by reconstructing how to arrive at the basis choice specified in Appendix~\ref{chap:hjlaportacan} before deriving the first- and second-order differential equations in the kinematic invariants~$\vec{x}$. As a next step, we recall the parametrizations defined in Section~\ref{sec:hjdeqcan2}, derive the differential equations with respect to the parameter~$\lambda_m$ and provide a recipe for the determination of the elliptic integrals. In Sections~\ref{sec:hjdeqell3} and \ref{sec:hjdeqell4}, we elaborate on how to calculate the homogeneous and inhomogeneous solutions of the second-order differential equations with respect to~$\lambda_m$ before commenting on the boundary conditions in Section~\ref{sec:hjdeqcan5}.

\subsection{Differential Equations and Basis Choice}
\label{sec:hjdeqell1}

The elliptic sector $A_{6,215}$ has four MIs, which have to be chosen to start with. We accomplish this by following the guidelines outlined in Section~\ref{sec:basischoice} in order to achieve a basis that is as triangular as possible in $D=4$. Beyond that, we establish additional criteria in the elliptic case: First, we require two out of the four basis integrals to decouple at the level of the differential equations, as we will see below. Second, we prefer basis integrals that are finite, i.e. whose Laurent series start at order~$\e^0$. In fact, we observe that the second requirement cannot be met by all four basis integrals at the same time provided that the first condition is satisfied. Therefore, we allow one of the four MIs to start at lower $\e$-order and end up with many different basis combinations, out of which we choose the one whose differential equations have the smallest coefficients:
\begin{equation}
\vec{E}_I \equiv \begin{pmatrix} \e^4\,I_{59}\\ \e^4\,I_{60}\\ \e^3\,I_{61}\\ \e^4\,I_{62}\end{pmatrix} \,.
\label{basisEI}
\end{equation}
The Laporta integrals $I_{59}$--$I_{62}$ occuring therein are defined in Appendix~\ref{sec:hjlaporta} and shown in Fig.~\ref{fig:hjmaster}. If we derive the system of coupled first-order differential equations of this basis with respect to the variables $\vec{x}=(x,z,h)$ in the same way as for the non-elliptic sectors in Section~\ref{sec:hjdeqcan1}, we arrive at
\begin{align}
\frac{\p}{\p x_i} \vec{E}_I(D;\vec{x}) = &\left( C_I^{(i)}(\vec{x}) + \e \, \tilde{C}_I^{(i)}(\vec{x}) \right) \, \vec{E}_I(D;\vec{x}) \nonumber \\
&+ \left( G_I^{(i)}(\vec{x}) + \e \, \tilde{G}_I^{(i)}(\vec{x}) \right) \, \vec{m}_I(D;\vec{x}) + \mathcal{O}\left(\e^2\right) \,.
\end{align}
Therein, the vector~$\vec{m}_I$ contains the Laporta integrals of the subtopologies, whose lowest-order coefficient matrix~$G_I^{(i)}(\vec{x})$ in~$\e$ is proportional to~$\e^0$ thanks to the $\e$-normalization of the basis~$\vec{E}_I$ in Eq.~\eqref{basisEI}. The coupled nature of the equations is reflected by the fact that the matrix~$C_I^{(i)}(\vec{x})$ has non-vanishing entries that cannot all be removed. The structure of this matrix reads
\begin{equation}
C_I^{(i)}(\vec{x}) =
\begin{pmatrix}
a_{11} & a_{12} & 0 & 0 \\
a_{21} & a_{22} & 0 & 0 \\
a_{31} & a_{32} & a_{33} & 0 \\
a_{41} & a_{42} & 0 & a_{44}
\end{pmatrix}
\,,
\end{equation}
revealing that the coupling becomes manifest through the appearance of $I_{59}$ and $I_{60}$ in the differential equations of all components of~$\vec{E}_I$. The diagonal matrix elements~$a_{33}$ and $a_{44}$ can be eliminated by computing the homogeneous solutions of the differential equations of $I_{61}$ and $I_{62}$ as described in Section~\ref{sec:triangulartocanonical} in case of canonical MIs, leading to the basis integrals $M_{61}$ and $M_{62}$ specified in Appendix~\ref{sec:hjcan}. Since the first two columns of $C_I^{(i)}(\vec{x})$ do not vanish, this procedure is not appropriate to determine suitable coefficients $j_{59}$ and $j_{60}$ in the definition of the new basis
\begin{align}
M_{59} &= \e^4 \, j_{59} \, I_{59} \,, \nonumber \\
M_{60} &= \e^4 \, j_{60} \, I_{60} \,,
\end{align}
so that we follow a different approach in this case: As mentioned before, coupled first-order differential equations of two integrals can be cast into one second-order differential equation of one of them. We start by deriving the second-order differential equation
\begin{equation}
\frac{\p^2 I_{59}(\vec{x})}{\p x_i} + p_I(\vec{x}) \, \frac{\p I_{59}(\vec{x})}{\p x_i} + q_I(\vec{x}) \, I_{59}(\vec{x}) = \left( \vec{g}_I(\vec{x})) + \e \, \vec{\tilde{g}}_I(\vec{x})) \right) \, \vec{m}_I(D;\vec{x})) + \mathcal{O}\left(\e^2\right)
\label{deqI59}
\end{equation}
of the corner integral~$I_{59}$ and find that the coefficient functions~$p_I$ and $q_I$ have six singular points including the point at infinity. Note that the hypergeometric differential equation has only three singular points and is known to evaluate to complete elliptic integrals of first and second kind for certain values of the indices~$a$, $b$ and $c$ in Eq.~\eqref{deqhyp}. For the sake of convenience, we would like to reduce the number of singular points within Eq.~\eqref{deqI59} and choose $j_{59}$ such that the second-order differential equation of~$M_{59}$, given by
\begin{equation}
\frac{\p^2 M_{59}(\vec{x})}{\p x_i} + p_M(\vec{x}) \, \frac{\p M_{59}(\vec{x})}{\p x_i} + q_M(\vec{x}) \, M_{59}(\vec{x}) = \e \, \vec{\tilde{g}}_M(\vec{x})) \, \vec{m}_I(D;\vec{x})) + \mathcal{O}\left(\e^2\right) \,,
\label{deqM59}
\end{equation}
involves only five singular points. This procedure can be repeated for $I_{60}$, resulting in $j_{59}=j_{60}=(-x)^{3/2}$ and thus in the new basis
\begin{equation}
\vec{E}_M \equiv \begin{pmatrix} M_{59}\\ M_{60}\\ M_{61}\\ M_{62}\end{pmatrix}
\label{basisEM}
\end{equation}
indicated in Appendix~\ref{sec:hjcan}. The first-order differential equations of the basis~$\vec{E}_M$ can be written as
\begin{align}
\frac{\p}{\p x_i} \vec{E}_M(D;\vec{x}) &= \left( C_M^{(i)}(\vec{x}) + \e \, \tilde{C}_M^{(i)}(\vec{x}) \right) \, \vec{E}_M(D;\vec{x}) + \e \, \tilde{G}_M^{(i)}(\vec{x}) \, \vec{m}_E(D;\vec{x}) + \mathcal{O}\left(\e^2\right) \,, \label{deqEM} \\
C_M^{(i)}(\vec{x}) &=
\begin{pmatrix}
b_{11} & b_{12} & 0 & 0 \\
b_{21} & b_{22} & 0 & 0 \\
b_{31} & b_{32} & 0 & 0 \\
b_{41} & b_{42} & 0 & 0
\end{pmatrix}
\,,
\label{matrixCM}
\end{align}
where we removed the subsector contributions with coefficient~$\e^0$ using the procedure of Section~\ref{sec:triangulartocanonical}.\\
In the following, we will apply the method of series expansions from differential equations to the basis~$\vec{E}_M$. Although we imposed a minimal number of singular points in order to attain this basis, we emphasize that the method of series expansions does not rely on this criterion. As mentioned in Section~\ref{sec:frobenius}, it can be used to deduce power series from differential equations with any number of singular points, provided that the expansion point is \textit{regular singular}.

\subsection{One-Dimensional Parametrization}
\label{sec:hjdeqell2}

As a next step, we divide the phase space in the exact same way as for the canonical MIs by introducing the parametrizations~$\vec{\gamma}_m$ ($m=1,2,3$) presented in Setion~\ref{sec:hjdeqcan2}. This allows us to calculate the first-order differential equations of the basis~$\vec{E}_M$ with respect to the parameter $\lambda_m$ from the differential equations~\eqref{deqEM} in the variables~$x$, $z$ and $h$ with the help of Eq.~\eqref{deqlambda}. In doing so, we end up with
\begin{align}
\frac{\p}{\p \lambda_m} \vec{E}_M(D;\lambda_m) = &\left( C_M(\lambda_m) + \e \, \tilde{C}_M(\lambda_m) \right) \, \vec{E}_M(D;\lambda_m) \nonumber \\
&+ \e \, \tilde{G}_M(\lambda_m) \, \vec{m}_E(D;\lambda_m) + \mathcal{O}\left(\e^2\right) \,,
\label{deqEMlambda}
\end{align}
where the matrix~$C_M(\lambda_m)$ has the same structure as the matrix~$C_M^{(i)}(\vec{x})$ specified in Eq.~\eqref{matrixCM}. We are now in position to derive the second-order differential equation of~$M_{59}$ with respect to~$\lambda_m$:
\begin{align}
\frac{\p^2 M_{59,m}(\lambda_m)}{\p \lambda_m} + &p_m(\lambda_m) \, \frac{\p M_{59,m}(\lambda_m)}{\p \lambda_m} + q_m(\lambda_m) \, M_{59,m}(\lambda_m) \nonumber \\
&= \e \, \vec{\tilde{g}}_M(\lambda_m) \, \vec{m}_E(D;\lambda_m) + \mathcal{O}\left(\e^2\right) \,.
\label{deqM59lambda}
\end{align}
The coefficient functions $p_1(\lambda_1)$ and $q_1(\lambda_1)$ take a particularly simple form in case of the parametrization~$\vec{\gamma}_1(\vec{x},\lambda_1)$, which is given in Eq.~\eqref{param1} and designed to facilitate a series expansion around the origin~$\vec{x}=0$:
\begin{align}
p_1(\lambda_1) &= \frac{2\,x\,\left(x\,\lambda_1\,(h-z)^2-4\,\left(h\,(x-z)+z\,(x+z)\right)\right)}{d_1(\lambda_1)} \,, \nonumber \\
q_1(\lambda_1) &= \frac{x^2\,(h-z)^2}{4\,d_1(\lambda_1)} \,, \nonumber \\
d_1(\lambda_1) &\equiv x^2\,\lambda_1^2\,(h-z)^2-8\,x\,\lambda_1\,\left(h\,(x-z)+z\,(x+z)\right)+16\,(x+z)^2 \,.
\end{align}
Remarkably, the second-order differential equation of~$M_{59}$ with respect to~$\lambda_1$ has only three singular points including $\lambda_1=\infty$, which correspond to the roots of~$d_1(\lambda_1)$, i.e. two singular points vanish compared to Eq.~\eqref{deqM59} as a result of the parametrization~$\vec{\gamma}_1(\vec{x},\lambda_1)$. Note that the precise expressions of the coefficient functions $p_m(\lambda_m)$ and $q_m(\lambda_m)$ in the parametrizations~$\vec{\gamma}_2(\vec{x},\lambda_2)$ and $\vec{\gamma}_3(\vec{x},\lambda_3)$ are substantially larger and cannot be reproduced here. In what follows, we will however need to know their structure upon expansion around~$\lambda_m=0$, which reads
\begin{equation}
\begin{aligned}
p_1(\lambda_1) &= p_1^{(0)} + \mathcal{O}\left(\lambda_1\right) \,, & \quad q_1(\lambda_1) &= q_1^{(0)} + \mathcal{O}\left(\lambda_1\right) \,, \\
p_2(\lambda_2) &= \frac{p_2^{(-1)}}{\lambda_2} + p_2^{(0)} + \mathcal{O}\left(\lambda_2\right) \,, & \quad q_2(\lambda_2) &= \frac{q_2^{(-2)}}{\lambda_2^2} + \frac{q_2^{(-1)}}{\lambda_2} + q_2^{(0)} + \mathcal{O}\left(\lambda_2\right) \,, \\
p_3(\lambda_3) &= \frac{p_3^{(-1)}}{\lambda_3} + p_3^{(0)} + \mathcal{O}\left(\lambda_3\right) \,, & \quad q_3(\lambda_3) &= \frac{q_3^{(-1)}}{\lambda_3} + q_3^{(0)} + \mathcal{O}\left(\lambda_3\right) \,.
\end{aligned}
\label{pq}
\end{equation}
We verified that the Laurent expansions of $M_{59}$, $M_{60}$ and $M_{62}$ start at order~$\e^4$, whereas $\e^3$ is the first non-trivial order of the Laurent series of $M_{61}$. This can be done either through the method of sector decomposition implemented in \textsc{SecDec} or by using \textsc{Reduze} to express the basis~$\vec{E}_M$ in terms of a purely finite basis and subsequently analyzing the $\e$-coefficients occuring therein.\\
If we combine these findings with the structure of the matrix~$C_M(\lambda_m)$ in Eq.~\eqref{deqEMlambda}, it becomes clear how to proceed using the differential equations with respect to~$\lambda_m$:
\begin{enumerate}
\item Use the first-order differential equation~\eqref{deqEMlambda} of $M_{61}$ to compute its result at order~$\e^3$. To that order of the Laurent expansion, only canonical subsectors contribute.
\item Derive the homogeneous weight-four solutions of the second-order differential equations~\eqref{deqM59lambda} of~$M_{59}$.
\item Derive the inhomogeneous weight-four solutions of the second-order differential equations~\eqref{deqM59lambda} of~$M_{59}$, where the result of step~1 is required.
\item Compute the first derivative of the weight-four result of $M_{59}$ and insert it into the first-order differential equation~\eqref{deqEMlambda} of~$M_{59}$ together with the solution of $M_{59}$ at order~$\e^4$ and all subsector results of lower weights. Next, solve the resulting algebraic equation for $M_{60}$ to determine the result of $M_{60}$ at weight four.
\item Substitute the series expansions of $M_{59}$ and $M_{60}$ at order~$\e^4$ as well as subsector contributions of lower weights into the first-order differential equations~\eqref{deqEMlambda} of $M_{61}$ and $M_{62}$, which allows computing the weight-four results of $M_{61}$ and $M_{62}$.
\end{enumerate}
Step~4 is a simple rearrangement of an algebraic equation, whereas steps~1 and 5 can be achieved in the same way as when determining series expansions from canonical differential equations described in Section~\ref{sec:hjdeq}. In the next two sections, we will elaborate on steps~2 and~3.

\subsection{Homogeneous Solutions of Second-Order Differential Equations}
\label{sec:hjdeqell3}

In this section, we aim to solve the homogeneous part\footnote{For the sake of readability, we omit the superscript~$(k)$ within the notation~$M_{59,m}^{(h,k)}$ and $M_{59,m}^{(p,k)}$ in the following, since we clarified that the Laurent series of $M_{59}$ starts only at order~$\e^4$.}
\begin{equation}
\frac{\p^2 M^{(h)}_{59,m}(\lambda_m)}{\p \lambda_m} + p_m(\lambda_m) \, \frac{\p M^{(h)}_{59,m}(\lambda_m)}{\p \lambda_m} + q_m(\lambda_m) \, M^{(h)}_{59,m}(\lambda_m) = 0
\label{deqM59lambdahom}
\end{equation}
of the second-order differential equation~\eqref{deqM59lambda} with respect to $\lambda_m$ ($m=1,2,3$). This can be done along the lines of Section~\ref{sec:frobenius}, i.e. by making an ansatz of the form~\eqref{ansatzfrobenius}
\begin{align}
M^{(h)}_{59,m}(\lambda_m) &= \sum_{j=0}^p a_j \, \lambda_m^{s+j} \,, \nonumber \\
\frac{\d M^{(h)}_{59,m}(\lambda_m)}{\d\lambda_m} &= \sum_{j=0}^p a_j \, (s+j) \, \lambda_m^{s+j-1} \,, \nonumber \\
\frac{\d^2 M^{(h)}_{59,m}(\lambda_m)}{\d\lambda_m^2} &= \sum_{j=0}^p a_j \, (s+j) \, (s+j-1) \, \lambda_m^{s+j-2}
\label{ansatzM59}
\end{align}
with $a_0\neq 0$. As a next step, these equations are substituted into the homogeneous differential equation~\eqref{deqM59lambdahom}, resulting in indicial equations that can be solved for the parameter~$s$. A requirement for this method to work is that $\lambda_m=0 $ is a regular singular point of the respective differential equation, which we verified. In the following, we elaborate on the details of this derivation for each of the three parametrizations~$\vec{\gamma}_m$ ($m=1,2,3$).\\

\textbf{1) Linear Series Expansion around the Origin Parametrized by~$\boldsymbol{\vec{\gamma}_1(\vec{x},\lambda_1)}$}

After inserting Eq.~\eqref{ansatzM59} into Eq.~\eqref{deqM59lambdahom}, we are left with
\begin{align}
\sum_{j=0}^p a_j \, (s+j) \, (s+j-1) \, \lambda_1^{s+j-2} &+ \left( p_1^{(0)} + \mathcal{O}\left(\lambda_1\right) \right) \, \sum_{j=0}^p a_j \, (s+j) \lambda_1^{s+j-1} \nonumber \\
&+ \left( q_1^{(0)} + \mathcal{O}\left(\lambda_1\right) \right) \, \sum_{j=0}^p a_j \, \lambda_1^{s+j} = 0 \,,
\label{indicial1a}
\end{align}
where we made use of the expansion of the coefficient functions $p_1(\lambda_1)$ and $q_1(\lambda_1)$ given in Eq.~\eqref{pq}. Clearly, the lowest power of $\lambda_1$ occuring in this equation is $\lambda_1^{s-2}$, and requiring that its coefficient must vanish yields the indicial equation:
\begin{equation}
a_0 \, s \, (s-1) = 0 \,.
\end{equation}
Due to the precondition $a_0\neq 0$, the values of~$s$ are straightforwardly obtained as
\begin{equation}
s_1=0 \,, \quad s_2=1 \,.
\end{equation}
The next-to-lowest power in Eq.~\eqref{indicial1a} is given by~$\lambda_1^{s-1}$, whose coefficient reads
\begin{equation}
a_1 \, (s+1) \, s + p_1^{(0)} \, a_0 \, s = 0 \,.
\label{indicial1b}
\end{equation}
For $s_2=1$, the coefficient~$a_1$ of the series expansion in~$\lambda_1$ is obtained to be
\begin{equation}
a_1=-p_1^{(0)}\,a_0/2 \,.
\end{equation}
If $s_1=0$, Eq.~\eqref{indicial1b} is trivially satisfied, implying that $a_1$ can be chosen arbitrarily, so that we take the freedom to set $a_1\equiv 0$. One could continue in this way, thereby determining the coefficients $a_j$ of the two homogeneous solutions with parameters $s_1$ and $s_2$ recursively. As pointed out in Section~\ref{sec:frobenius}, however, we prefer to achieve this up to the desired degree~$p$ of the power series by inserting the ansatz~\eqref{ansatzM59} with fixed $s=s_i$ ($i=1,2$) into the differential equation~\eqref{deqM59lambdahom} and requiring that the coefficients of identical powers of~$\lambda_1$ vanish. Note that this is similar to what is done to derive series expansions for canonical differential equations in Section~\eqref{sec:hjdeq}.\\
Remarkably, we do not need to resort to Eq.~\eqref{liformula2}, although the two solutions~$s_{1,2}$ of the indicial equation differ by an integer. We assume that this is due to the fact that the coefficient~$a_1$ vanishes within the power series involving~$s_1$, so that it becomes linearly independent of the series expansion that contains~$s_2$. By computing the non-vanishing Wronskian of the two homogeneous solutions with the help of Eq.~\eqref{wronskian}, we confirmed that this is the case. Through the combination of both solutions, we finally obtain the homogeneous solution of the second-order differential equation in~$M_{59}$ around the point~$\lambda_1$ at weight four with fully determined coefficients~$a_j$ and $b_j$:
\begin{equation}
M^{(h)}_{59,1}(\lambda_1) = c_{59,1}^{(4)} \left(1+\sum_{j=2}^p a_j \, \lambda_1^j \right) + c_{60,1}^{(4)} \, \lambda_1 \, \left(1+\sum_{j=1}^p b_j \, \lambda_1^j \right) \,.
\label{M59homsol1}
\end{equation}
Therein, we introduced boundary constants $c_{59,1}^{(4)}$ and $c_{60,1}^{(4)}$ in the notation of Section~\ref{sec:hjdeqcan4} in favor of~$a_0$ and $b_0$, recalling that the solution of a second-order differential equation requires two of them. The second one can be understood as the weight-four contribution of the integral~$M_{60}$ coupled to $M_{59}$, whose result involves these two boundary constants as well, since it is determined algebraically from the solution of $M_{59}$ as explained in the previous section. As a check, we used the exact homogeneous solution in the Euclidean region, which is given in terms of complete elliptic integrals of first kind in Ref.~\cite{Bonciani:2016}, to reproduce the series expansion in~$\lambda_1$, thereby finding full agreement.
\newpage

\textbf{2) Linear Series Expansion around the Threshold Parametrized by~$\boldsymbol{\vec{\gamma}_2(\vec{x},\lambda_2)}$}

The equivalent of Eq.~\eqref{indicial1a} for the parametrization~$\vec{\gamma}_2(\vec{x},\lambda_2)$ is obtained by adjusting the structure of the functions~$p_2(\lambda_2)$ and $q_2(\lambda_2)$ according to Eq.~\eqref{pq}:
\begin{align}
\sum_{j=0}^p a_j \, (s+j) \, &(s+j-1) \, \lambda_2^{s+j-2} + \left( \frac{p_2^{(-1)}}{\lambda_2} + p_2^{(0)} + \mathcal{O}\left(\lambda_2\right) \right) \, \sum_{j=0}^p a_j \, (s+j) \lambda_2^{s+j-1} \nonumber \\
&+ \left( \frac{q_2^{(-2)}}{\lambda_2^2} + \frac{q_2^{(-1)}}{\lambda_2} + q_2^{(0)} + \mathcal{O}\left(\lambda_2\right) \right) \, \sum_{j=0}^p a_j \, \lambda_2^{s+j} = 0 \,.
\label{indicial2a}
\end{align}
As before, the deepest pole is given by $\lambda_2^{s-2}$, but this time its coefficient receives contributions from all three sums appearing therein:
\begin{equation}
a_0 \, s \, (s-1) + a_0 \, p_2^{(-1)} \, s + a_0 \, q_2^{(-2)} = 0 \,.
\label{indicial2ab}
\end{equation}
On the condition that $a_0\neq 0$ and with the expansion coefficients
\begin{equation}
p_2^{(-1)}=1 \,, \quad q_2^{(-2)}=-\frac{1}{4} \,,
\label{indicial2b}
\end{equation}
the solution of the indicial equation~\eqref{indicial2ab} yields
\begin{equation}
s_1 = -\frac{1}{2} \,, \quad s_2 = \frac{1}{2} \,.
\end{equation}
As before, it is interesting to have a look at the coefficient of the next-to-leading term~$\lambda_2^{s-1}$ within Eq.~\eqref{indicial2a}:
\begin{equation}
a_1 \, s \, (s+1) + a_1 \, (s+1) \, p_2^{(-1)} + a_0 \, s \, p_2^{(0)} + a_1 \, q_2^{(-2)} + a_0 \, q_2{(-1)} = 0 \,.
\label{indicial2c}
\end{equation}
For $s_2=\nicefrac{1}{2}$, this results in a recursion relation for $a_1$,
\begin{equation}
a_1 = -\frac{a_0 \, p_2^{(0)}}{2} \,,
\end{equation}
where we used the values of $p_2^{(-1)}$ and $q_2^{(-2)}$ indicated in Eq.~\eqref{indicial2b} and the identity
\begin{equation}
q_2^{(-1)} = \frac{1}{2} \, p_2^{(0)} \,.
\label{indicial2d}
\end{equation}
If $s_1=-\nicefrac{1}{2}$, we encounter the same behavior as for the parametrization~$\vec{\gamma}_1(\vec{x},\lambda_1)$ due to the particular relations between the expansion coefficients in Eqs.~\eqref{indicial2b} and \eqref{indicial2d}, finding that Eq.~\eqref{indicial2c} is trivially satisfied. Consequently, we are free to choose $a_1\equiv 0$ in this case, so that the full homogeneous solution of $M_{59}$ in~$\lambda_2$ reads
\begin{equation}
M^{(h)}_{59,2}(\lambda_2) = \frac{c_{59,2}^{(4)}}{\sqrt{\lambda_2}} \, \left(1+\sum_{j=2}^p a_j \, \lambda_2^j \right) + c_{60,2}^{(4)} \, \sqrt{\lambda_2} \, \, \left(1+\sum_{j=1}^p b_j \, \lambda_2^j \right) \,.
\label{M59homsol2}
\end{equation}
As in the previous case, the Wronskian turns out to be different from zero, although $s_1$ and $s_2$ are separated by an integer.\\
\newpage

\textbf{3) Exponential Series Expansion beyond Threshold Parametrized by~$\boldsymbol{\vec{\gamma}_3(\vec{x},\lambda_3)}$}

Let us repeat the procedure for the series expansion of~$M_{59}$ in~$\lambda_3$ for the sake of completeness. Substituting the ansatz~\eqref{ansatzM59} into the homogeneous second-order differential equation~\eqref{deqM59lambdahom} yields
\begin{align}
\sum_{j=0}^p a_j \, (s+j) \, &(s+j-1) \, \lambda_3^{s+j-2} + \left( \frac{p_3^{(-1)}}{\lambda_3} + p_3^{(0)} + \mathcal{O}\left(\lambda_3\right) \right) \, \sum_{j=0}^p a_j \, (s+j) \lambda_3^{s+j-1} \nonumber \\
&+ \left( \frac{q_3^{(-1)}}{\lambda_3} + q_3^{(0)} + \mathcal{O}\left(\lambda_3\right) \right) \, \sum_{j=0}^p a_j \, \lambda_3^{s+j} = 0 \,.
\end{align}
The coefficient of the deepest pole~$\lambda_3^{s-2}$ only receives contributions from the function~$p_3(\lambda_3)$ in this case:
\begin{equation}
a_0 \, s \, (s-1) + a_0 \, s \, p_3^{(-1)} = 0 \,.
\end{equation}
With $a_0\neq 0$ as usual and $p_3^{(-1)}=1$, this leads to the double root
\begin{equation}
s_{1,2} = 0 \,.
\end{equation}
We do not need to go further at this point, since it is obvious that for any power of $\lambda_3$, the recursively determined higher-order coefficients~$a_j$ ($j\geq 1$) will coincide for the two series expansions with~$s_{1,2}$. Deriving the power series representations of the homogeneous solutions in~$\lambda_3$ is therefore exceptional in the sense that it requires the use of Eq.~\eqref{liformula2} as opposed to the parametrizations~$\vec{\gamma}_1(\vec{x},\lambda_1)$ and $\vec{\gamma}_2(\vec{x},\lambda_2)$. As a consequence, the complete homogeneous solution of $M_{59}$ at weight four around the expansion point~$\lambda_3$ reads as follows:
\begin{align}
M^{(h)}_{59,3}(\lambda_3) = \, &c_{59,3}^{(4)} \, \left(1+\sum_{j=1}^p a_j \, \lambda_3^j \right) \nonumber \\
&+ c_{60,3}^{(4)} \, \left[ \log\left(\lambda_3\right) \, \left(1+\sum_{j=1}^p a_j \, \lambda_3^j \right) + 1+\sum_{j=1}^p b_j \, \lambda_3^j \right] \,.
\label{M59homsol3}
\end{align}
Therein, the unknown coefficients~$a_j$ and $b_j$ are fully determined by substituting one of the two homogeneous solutions into the differential equation and solving the resulting algebraic system of equations.

\subsection{Inhomogeneous Solutions of Second-Order Differential Equations}
\label{sec:hjdeqell4}

For the full solution of the elliptic corner integral of sector~$A_{6,215}$ at weight four in terms of series expansions in~$\lambda_m$, it remains to find particular solutions~$M^{(p)}_{59,m}(\lambda_m)$. As explained in great detail in Section~\ref{sec:frobenius2simp}, this can be done by using the exact same ansatz~\eqref{partsol10} as for solving canonical differential equations in Section~\ref{sec:hjdeqcan3}.\\
In fact, this most general ansatz is only required for the power series around the threshold parametrized by~$\vec{\gamma}_2(\vec{x},\lambda_2)$, since the series expansions in~$\lambda_1$ and $\lambda_3$ are carried out around regular points. Moreover, as pointed out previously, the subsector integrals~$\vec{m}_E$ in the second-order differential equation~\eqref{deqM59lambda} are regular in these two expansion points as well, so that we can replace Eq.~\eqref{partsol10} by the simplified form~\eqref{partsol12} in both cases.\\
With the appropriate ansatz at hand, we are in position to go through the usual procedure to determine the coefficients of the particular solution of $M_{59}$ at weight four: First, we substitute it into the inhomogeneous second-order differential equation~\eqref{deqM59lambda} together with the subsector expressions of lower weights and the weight-three result of~$M_{61}$, which was computed previously. In doing so, we observe that the subsector contributions to the $\e^3$-order of the Laurent expansion cancel analytically, thereby serving as a check. The result of~$M_{59}$ at order~$\e^4$ of the Laurent series is then obtained by requiring the coefficients of the powers in~$\lambda_m$ to vanish independently, leading to an algebraic system of equations that can be solved. Alternatively, the inhomogeneous solution can be derived using the variation of constants presented in Section~\ref{sec:frobenius2}. As stated in Eq.~\eqref{partsol2}, we have to integrate analytically over the homogeneous solution in order to accomplish this. This integration was feasible in case of the parametrization~$\vec{\gamma}_1(\vec{x},\lambda_1)$ and led to the same result as using the power series ansatz described before, thus confirming the validity of the inhomogeneous solution.\\
Eventually, we write down the full weight-four result of~$M_{59}$ by combining the homogeneous and inhomogeneous solutions around a given expansion point~$\lambda_m$:
\begin{align}
M_{59,m}(\lambda_m) &= M^{(h)}_{59,m}(\lambda_m) + M^{(p)}_{59,m}(\lambda_m) \nonumber \\
&= c_{59,m}^{(4)} \, M^{(h_1)}_{59,m}(\lambda_m) + c_{60,m}^{(4)} \, M^{(h_2)}_{59,m}(\lambda_m) + M^{(p)}_{59,m}(\lambda_m) \,.
\label{M59sol}
\end{align}
In the next section, we elaborate on how to determine the boundary constants~$c_{59,m}^{(4)}$ and $c_{60,m}^{(4)}$, which appear as coefficients of the homogeneous solutions $M^{(h_1)}_{59,m}(\lambda_m)$ and $M^{(h_2)}_{59,m}(\lambda_m)$ presented in Eqs.~\eqref{M59homsol1}, \eqref{M59homsol2} and \eqref{M59homsol3}.

\subsection{Boundary Conditions and Matching Procedure}
\label{sec:hjdeqell5}

In principle, we fix the boundary constants in the same way as in Section~\ref{sec:hjdeqcan4}, where we determined the boundary conditions of the canonical MIs. This holds in particular for the constants $c_{61,m}^{(3)}$, $c_{61,m}^{(4)}$ and $c_{62,m}^{(4)}$ of the MIs $M_{61}$ and $M_{62}$, which decouple from the other basis integrals $M_{59}$ and $M_{60}$. The two coupled MIs $M_{59}$ and $M_{60}$ share a set of two integration constants $c_{59,m}^{(4)}$ and $c_{60,m}^{(4)}$ per expansion point~$\lambda_m$, that can be determined simultaneously. For this purpose, we impose regularity of the full result in Eq.~\eqref{M59sol} at the point~$\lambda_1=0$, corresponding to the origin~$\vec{x}=0$ and thus to the limit of infinite quark mass. Let us recall that the Laporta basis~$\vec{E}_I$ is regular in this limit and that the integrating factors of the basis~$\vec{E}_M$ provided in Appendix~\ref{sec:hjcan} vanish. As stated previously in Eq.~\eqref{hjboundary}, this leads to the boundary condition
\begin{equation}
\lim_{\vec{x}\to\vec{0}} M_{59,1}(\lambda_1) = \lim_{\lambda_1\to 0} M_{59,1}(\lambda_1) = 0 \,,
\end{equation}
which implies the relation
\begin{equation}
c_{59,1}^{(4)} = -c_{60,1}^{(4)}
\end{equation}
if applied to Eq.~\eqref{M59sol}. At this point, we could retain one of the two undetermined boundary constants, e.g.~$c_{60,1}^{(4)}$, derive the result for~$M_{60}$ in terms of~$M_{59}$ as described in Section~\ref{sec:hjdeqell2} and repeat the procedure. Alternatively, given that the homogeneous solution~\eqref{M59homsol1} is analytic at the point~$\lambda_1=0$, we can take care of the second boundary constant already at the level of the result for~$M_{59}$ by requiring that its first derivative vanishes on top of the function itself: \begin{equation}
\lim_{\vec{x}\to\vec{0}} \frac{\p M_{59,1}(\lambda_1)}{\p\lambda_1} = \lim_{\lambda_1\to 0} \frac{\p M_{59,1}(\lambda_1)}{\p\lambda_1} = 0 \,.
\end{equation}
From this requirement, we obtain the remarkably simple result
\begin{equation}
c_{59,1}^{(4)} = c_{60,1}^{(4)} = 0 \,,
\end{equation}
implying that the full result of the series expansion in~$\lambda_1$ is solely given by the inhomogeneous solution:
\begin{equation}
M_{59,1}(\lambda_1) = M^{(p)}_{59,1}(\lambda_1) \,.
\end{equation}
For all basis integrals $M_{59}$--$M_{62}$ of the elliptic sector~$A_{6,215}$, we have so far fixed the boundary constants of the power series in~$\lambda_1$. The boundary constants of the series expansions in $\lambda_2$ and $\lambda_3$ remain to be determined, but we expect that they can be derived by applying the matching procedure explained in Section~\ref{sec:hjdeqcan4}.

\section[{Differential Equations and Master Integrals: The Planar Elliptic Sector~$A_{7,247}$}]{Differential Equations and Master Integrals:\\The Planar Elliptic Sector~$\boldsymbol{A_{7,247}}$}
\label{sec:hjdeqell247}

The four MIs $I_{67}$--$I_{70}$ belonging to the elliptic sector~$A_{7,247}$ are defined in terms of integral family~$A$ in Appendix~\ref{sec:hjlaporta} and shown in Fig.~\ref{fig:hjmaster}. As we will see shortly, their elliptic nature does not become manifest through the homogeneous solution, but through the inhomogeneous one upon including the results of the elliptic subsector~$A_{6,215}$. In conventional approaches, this would require integrating over elliptic integrals, which is unclear how to accomplish at the moment. In the following, we will resort to the method of series expansions applied previously, thereby substantially simplifying the task of finding solutions for integrals with elliptic subtopologies.\\
We arrive at the basis
\begin{equation}
\vec{F} \equiv \begin{pmatrix} M_{67}\\ M_{68}\\ M_{69}\\ M_{70}\end{pmatrix}
\end{equation}
specified in Appendix~\ref{sec:hjcan} with the help of the method explained in Section~\ref{sec:triangulartocanonical} designed to cast triangular differential equations into canonical ones. This removes all terms proportional to~$\e^0$ at the level of the first-order differential equations in the kinematic invariants~$\vec{x}$,
\begin{align}
\frac{\p}{\p x_i} \vec{F}(D;\vec{x}) = &\,\e \, \left( \tilde{\mathcal{C}}^{(i)}(\vec{x}) \, \vec{F}(D;\vec{x})  + \tilde{\mathcal{G}}^{(i)}(\vec{x}) \, \vec{m}_F(D;\vec{x}) \right) \nonumber \\
&+ \mathcal{C}^{(i)}(\vec{x}) \, \vec{E}_M(D;\vec{x}) + \mathcal{O}\left(\e^2\right) \,,
\end{align}
except for the ones containing the basis~$\vec{E}_M$ of elliptic subsectors provided in Eq.~\eqref{basisEM}. Note that the contributions of the elliptic subsectors~$\vec{E}_M$ with prefactor~$\e^1$ are included in the term involving the subtopology vector~$\vec{m}_F$. The coefficient matrix of the elliptic subsector in $D=4$ takes the form
\begin{equation}
\mathcal{C}^{(i)}(\vec{x}) =
\begin{pmatrix}
0 & 0 & 0 & 0 \\
c_{21} & c_{22} & 0 & 0 \\
0 & 0 & 0 & 0 \\
c_{41} & c_{42} & 0 & 0
\end{pmatrix}
\,,
\label{matrixCF}
\end{equation}
thus making explicit the coupling of the first two components of~$\vec{E}_M$.\\
As a next step, we derive the linear first-order differential equations with respect to the parameters~$\lambda_m$ by exploring Eq.~\eqref{deqlambda} in the usual way, so that we obtain the generic form
\begin{align}
\frac{\p}{\p \lambda_m} \vec{F}(D;\lambda_m) = &\,\e \, \left( \tilde{\mathcal{C}}(\lambda_m) \, \vec{F}(D;\lambda_m)  + \tilde{\mathcal{G}}(\lambda_m) \, \vec{m}_F(D;\lambda_m) \right) \nonumber \\
&+ \mathcal{C}(\lambda_m) \, \vec{E}_M(D;\lambda_m) + \mathcal{O}\left(\e^2\right) \,,
\label{deqellF}
\end{align}
where the matrix~$\mathcal{C}(\lambda_m)$ has the same structure as the one in Eq.~\eqref{matrixCF}. Remarkably, we observe that this matrix takes the even simpler form
\begin{equation}
\mathcal{C}(\lambda_1) =
\begin{pmatrix}
0 & 0 & 0 & 0 \\
0 & c_{22} & 0 & 0 \\
0 & 0 & 0 & 0 \\
0 & c_{42} & 0 & 0
\end{pmatrix}
\label{matrixCF2}
\end{equation}
for the parametrization~$\gamma_1(\vec{x},\lambda_1)$ around the origin, i.e. the elliptic subsector integral~$M_{59}$ decouples in $D=4$ space-time dimensions in this case.\\
Subsequently, we use sector decomposition to determine the first non-trivial $\e$-order of the Laurent expansion of the basis~$\vec{F}$ and find that we are dealing with a basis of only finite MIs. The power series representations of sector~$A_{7,247}$ in~$\lambda_m$ at weight four are then determined as described for canonical differential equations in Section~\ref{sec:hjdeqcan3}, and the determination of the boundary constants of the power series in $\lambda_1$ proceeds along the lines of Section~\ref{sec:hjdeqcan4}. The only difference is that subsector results of the same weight have to be substituted into the system~\eqref{deqellF} of differential equations. According to the structure of the matrix~$\mathcal{C}(\lambda_m)$ occuring therein, they are given by the elliptic integrals~$M_{59}$ and $M_{60}$ in case of the parametrizations~$\gamma_2(\vec{x},\lambda_2)$ and $\gamma_3(\vec{x},\lambda_3)$. As shown in Eq.~\eqref{matrixCF2}, the contribution of~$M_{59}$ does not need to be considered for the parametrization~$\gamma_1(\vec{x},\lambda_1)$ around the origin~$\vec{x}=0$. The only remaining contribution to the $\e^4$-coefficient of the Laurent series of~$\vec{F}$, that is due to the elliptic subsector~$\vec{E}_M$, is given by the non-vanishing weight-three result of the integral~$M_{61}$, which emerges from the multiplication of the subtopology vector~$\vec{m}_F$ with the prefactor~$\e$. Finally, the remaining boundary constants of the series expansions in $\lambda_2$ and $\lambda_3$ can be derived as soon as the corresponding boundary constants of the elliptic subsector~$A_{6,215}$ are determined.

\section[Differential Equations and Master Integrals: Outlook on the Non-Planar Sectors]{Differential Equations and Master Integrals:\\Outlook on the Non-Planar Sectors}
\label{sec:hjdeqellC}

This section is designed to report on the status of decoupling the homogeneous differential equations of the non-planar two-loop MIs for Higgs-plus-jet production with full quark mass dependence, which were analyzed with the help of a collaborator~\cite{Primo}. Beyond that, we give an outlook on whether these differential equations are suited for the derivation of series expansions.\\
Figure~\ref{fig:hjnonplanar} shows the corner integrals of all non-planar sectors with at least one MI, where the number of MIs indicated therein was determined by means of the IBP relations described in Section~\ref{sec:ibp}. In terms of integral family~$C$ defined in Table~\ref{tab:hjtopo}, these corner integrals read
\begin{align}
&C_{6,238,6,0}\,(0,1,1,1,0,1,1,1,0)\,,\nonumber\\
&C_{6,246,6,0}\,(0,1,1,0,1,1,1,1,0)\,,\nonumber\\
&C_{6,303,6,0}\,(1,1,1,1,0,1,0,0,1)\,,\nonumber\\
&C_{6,399,6,0}\,(1,1,1,1,0,0,0,1,1)\,,\nonumber\\
&C_{6,492,6,0}\,(0,0,1,1,0,1,1,1,1)\,,\nonumber\\
&C_{7,254,7,0}\,(0,1,1,1,1,1,1,1,0)\,,\nonumber\\
&C_{7,431,7,0}\,(1,1,1,1,0,1,0,1,1)\,,\nonumber\\
&C_{7,494,7,0}\,(0,1,1,1,0,1,1,1,1)
\label{hjlaportanp}
\end{align}
in the notation of Appendix~\ref{sec:hjlaporta}. Moreover, Fig.~\ref{fig:hjnonplanar} tells us that non-planar sectors occur only for $t=6$ or $t=7$, since any sectors with lower number~$t\leq 5$ of distinct propagators within integral family~$C$ are planar and can be mapped onto the MIs presented in Section~\ref{sec:hjdeq}. In order to analyze whether the MIs of a given sector decouple, we derive the differential equations
\begin{equation}
\frac{\p}{\p x_i} \vec{\mathcal{I}}(D;\vec{x}) = \mathcal{A}^{(i)}(D,\vec{x}) \, \vec{\mathcal{I}}(D;\vec{x}) + \mathcal{B}^{(i)}(D,\vec{x}) \, \vec{m}(D;\vec{x})
\label{deqnonplanar}
\end{equation}
of the Laporta basis~$\vec{\mathcal{I}}$ with respect to the kinematic invariants~$x_i=\{x,z,h\}$ using Eqs.~\eqref{deqmandelstam} and \eqref{deqratios2}. Note that the length of the vector~$\vec{\mathcal{I}}$ is determined by the number of MIs indicated in Fig.~\ref{fig:hjnonplanar} for the sector under consideration and that we do not consider subsector contributions to the differential equations, i.e. $\vec{m}(D;\vec{x})=0$ within Eq.~\eqref{deqnonplanar} in the following. Statements about the decoupling behavior can then be made by studying the entries of the matrix $\mathcal{A}^{(i)}(4,\vec{x})$ in $D=4$ dimensions. More precisely, we try to find a basis $\vec{\mathcal{I}}(D;\vec{x})$ such that the matrix $\mathcal{A}^{(i)}(4,\vec{x})$ is as much as possible in triangular form by applying the guidelines from Section~\ref{sec:basischoice}.\\
\begin{figure}[tb]
\begin{center}
\includegraphics[width=\textwidth]{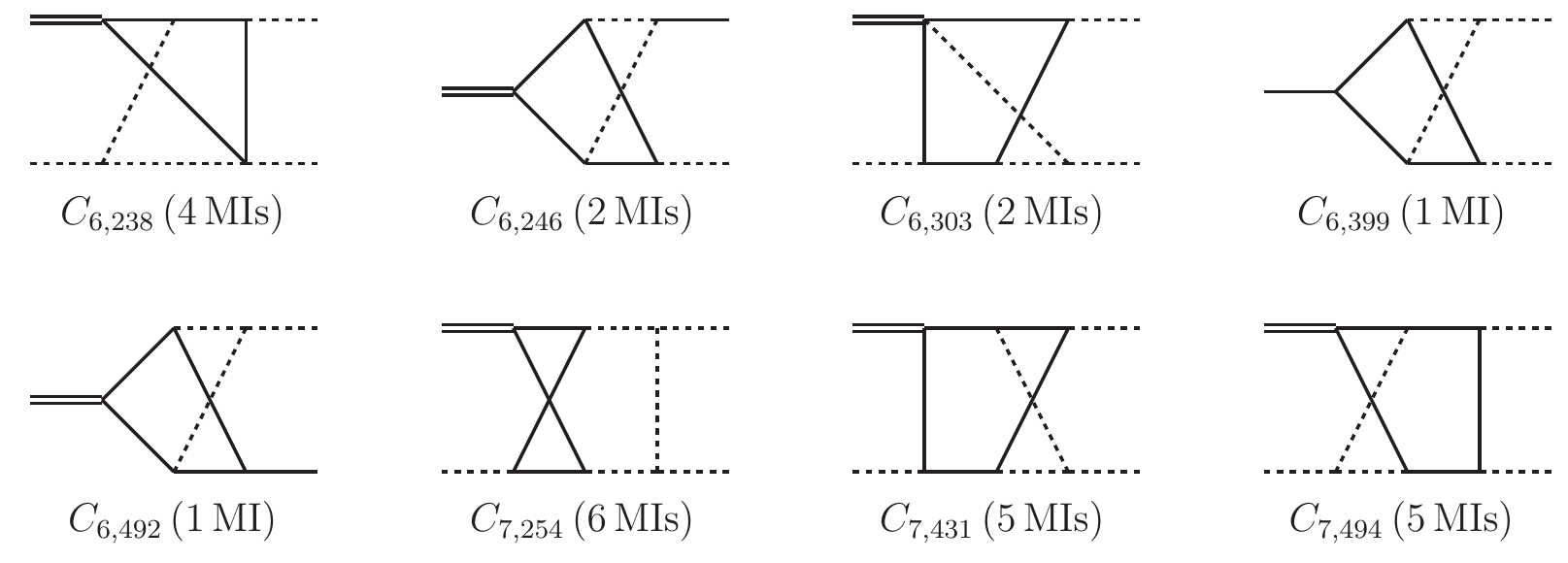}
\caption[Two-loop non-planar corner integrals for the calculation of the amplitude for Higgs-plus-jet production with full quark mass dependence]{\textbf{Two-loop non-planar corner integrals for the calculation of the amplitude for Higgs-plus-jet production with full quark mass dependence.} Dashed lines are massless, whereas internal solid lines denote propagators with mass $m_q$. Double external lines correspond to mass~$m_H$ and solid external lines of up to three-point functions denote virtualities of either~$t$ or $u$, depending on the definition of the Laporta integrals in Eq.~\eqref{hjlaportanp}.}
\label{fig:hjnonplanar}
\end{center}
\end{figure}\noindent
Let us start with sectors~$C_{6,303}$, $C_{6,399}$ and $C_{7,431}$, which form an independent topology tree and contain two, one and five MIs, respectively. We managed to fully decouple the differential equations of these three sectors and conclude that they can be expressed in terms of polylogarithms. However, their contribution to the Feynman amplitude was computed in Section~\ref{sec:hjamp} and turns out to vanish, so that the integration of their differential equations is not required.\\
As a next step, we consider sector~$C_{6,246}$, whose homogeneous solution was determined in the framework of the maximal cut and is known to be elliptic~\cite{Primo:2016,Harley:2017}. Given that this sector includes two MIs, their first-order differential equations can be rewritten as a second-order differential equation in terms of one of them. Since all subsectors must be planar and have at most $t\leq 5$ distinct propagators, we know that they cannot be elliptic. As a consequence, we expect that the approach of solving the second-order differential equation of the corner integral~$M_{59}$, which belongs to the planar elliptic sector~$A_{6,215}$ and was elaborated on in Section~\ref{sec:hjdeqell215}, can be reused here.\\
The only remaining sectors with $t=6$ distinct propagators are~$C_{6,492}$ and $C_{6,238}$. The former has only one MI and trivially evaluates to polylogarithmic expressions, whereas the number of MIs of the latter sector is four. It is at the moment unclear to which extent the differential equations of sector~$C_{6,238}$ decouple, however first observations hint at an elliptic behavior.\\
This leaves us with the two top-level sectors~$C_{7,254}$ and $C_{7,494}$, which possess six and five MIs, respectively. The differential equations of the sector $C_{7,494}$ could be fully decoupled and thus do not evaluate to elliptic functions. However, $C_{6,238}$ appears as a subsector, which becomes clear by pinching the lowermost massive internal line within the diagram of~$C_{7,494}$. Should $C_{6,238}$ turn out to be elliptic, then the differential equations of $C_{7,494}$ could be solved in the same way as for the top-level sector~$C_{7,247}$ of the planar integral family~$A$, which was described in Section~\ref{sec:hjdeqell247}.\\
Finally, we are capable of decoupling four out of six MIs in case of the sector $C_{7,254}$ and recall that the corresponding differential equations receive contributions from the elliptic subsector~$C_{6,246}$ and possibly also from $C_{6,238}$. We expect to be able to solve the second-order differential equation of one of the two coupled MIs of the sector $C_{7,254}$ through a trivial extension of the method of series expansions presented in the previous sections to the case of elliptic homogeneous solutions with elliptic subsectors.\\
To sum up, we anticipate that all non-planar MIs in the context of Higgs-plus-jet production with full quark mass dependence can be solved by using the approach of deriving series expansions from differential equations, which was explained in detail in Sections~\ref{sec:seriesexp}, \ref{sec:onedim} and \ref{sec:matching} and applied to the planar MIs in Sections~\ref{sec:hjdeq}, \ref{sec:hjdeqell215} and \ref{sec:hjdeqell247}. In particular, we can recycle the partioning of the phase space in Section~\ref{sec:hjdeqcan2} as well as the choice of boundary conditions and matching points outlined in Section~\ref{sec:hjdeqcan4}. The only unclear case is given by sector~$C_{6,238}$, which remains to be analyzed.

\section{Numerical Evaluation and Checks}
\label{sec:hjnum}

In this section, we analyze the exceptional feature of series expansions derived from differential equations, which are subject to fast and reliable numerical evaluations, provided that the evaluation point is within the radius of convergence. We point out that the numerical evaluation of the power series in~$\lambda_3$ remains to be studied, thus the following statements are limited to the series expansions in $\lambda_1$ and~$\lambda_2$.

\subsection{Degree of the Series Expansions}

So far, we have neglected discussions on the degree~$p$ of the series expansions required to achieve a result with satisfying accuracy. The desire of highest possible precision with respect to the numerical evaluation of the MIs is confronted by the feasibility to derive series expansions to very high degree~$p$ as well as by the speed required for the evaluation. Let us address the question of how to find the balance between these conflicting interests in case of Higgs-plus-jet production: We have observed that deriving series expansions up to degree~$p=5$ is sufficient in order to make statements about the correctness of the result. In fact, for many integrals the accuracy of the result with $p=5$ is adequate, however we visibly gain accuracy by extending the power series representations to the degree~$p=10$, whose computation is still feasible. This is particularly important for the numerical results obtained from the series expansions in~$\lambda_2$ and $\lambda_3$, which rely on high numerical precision throughout the matching procedure. For the canonical MIs, we were even capable of deducing series expansions up to~$p=15$, for which the benefit in accuracy is however negligible compared to the result of order~$p=10$. Therefore, we report on timings and accuracy of results of degree~$p=10$ in the following, bearing in mind that the speed of the evaluation can be increased substantially by reducing~$p$ if necessary.

\subsection{Timings}

Let us start by considering the timings: We are able to evaluate the series expansions of all planar MIs in~$\lambda_1$ up to degree~$p=10$ and order~$\e^4$ within $0.3$~s for $30$-digit precision within~\textsc{Mathematica}. This number is valid for consecutive numerical evaluation of the MIs, however the procedure is suitable for high parallelization. Hence, the required evaluation time of a single MI might be more instructive and is given by at most $6\cdot 10^{-2}$~s if the $\e$-orders of the given MI are evaluated consecutively. Since the size of the expansion coefficients is considerably larger for the power series in~$\lambda_2$, the two numbers increase in this case to $6.1$~s and $0.6$~s, respectively. Provided that the evaluation time of the amplitude is of similar magnitude, these timings are suitable for the implementation of direct phase space integration routines, particularly if we take into account that the results could be further optimized by implementing them in numerical routines suited for numerical evaluations.

\subsection{Relative Deviation}

Before discussing the accuracy of our numerical results, we have to establish its definition. If the numerical result of a single MI is denoted by~$M_n$ and a result we compare to by~$N_n$, we refer to the quantity
\begin{equation}
\sigma_n = \left|1-\frac{N_n}{M_n}\right|
\end{equation}
as relative deviation of~$N_n$ with respect to~$M_n$. We determine~$\sigma_n$ by comparing to results~$N_n$ that are obtained in two possible ways: If the integral~$M_n$ under consideration has been derived in the context of the NLO corrections to the $H\to Z\,\gamma$ decay rate in Chapter~\ref{chap:hza}, then $N_n$ is given by results in terms of MPLs, which can be evaluated numerically using~\textsc{Ginac}. If the exact expression is not available, which is true of all four-point functions, we obtain numerical results from~\textsc{SecDec3}~\cite{Borowka:2015}. We point out that we were not able to evaluate the MIs belonging to the top-level topologies $B_{7,367}$ and $B_{7,431}$ of integral family~$B$ by means of~\textsc{SecDec3}. In these cases, we switch to the numerical routines implemented in the recently published version of~\textsc{pySecDec}~\cite{Borowka:2017}.\\
Since the amplitude of Higgs-plus-jet production with full quark mass dependence is expected to receive mass corrections of around~$20\,\%$, we require a conservative relative deviation of $\sigma_n<10\,\%$ for the numerical evaluation of the integrals~$M_n$ in one phase space point. In fact, the relative deviation~$\sigma_n$ is much smaller for most of our results, which we calculated for the power series representations of all planar MIs up to degree $p=10$. More precisely, results in the Euclidean region with
\begin{equation}
-1<x<0 \,, \quad -1<z<0 \,, \quad -1<h<0
\end{equation}
are computed using the series expansion in~$\lambda_1$, for which we obtain $10^{-21}<\sigma_n<10^{-6}$. If we evaluate the same series expansion in the physical region below threshold, where we set
\begin{equation}
0<x<2 \,, \quad -1<z<0 \,, \quad h=\frac{1}{2} \,,
\end{equation}
this changes to $10^{-9}<\sigma_n<0.3$. Finally, the comparison of the series expansion in~$\lambda_2$ was carried out in the range
\begin{equation}
2<x<6 \,, \quad -0.1<z<0 \,, \quad h=\frac{1}{2} \,,
\end{equation}
and yields $-10^{-6}<\sigma_n<0.5$. We note that there are a few exceptions with $\sigma_n\approx 15\,\%$, which can be traced back to the top-level sectors of integral family~$B$, whose~\textsc{SecDec} errors are of the same order of magnitude. This is also the reason for which the upper bound of the relative deviation of the series expansion below threshold is given by~$0.3$, although $\sigma_n$ lies below the permille level for most MIs in this case. This applies especially to the elliptic integrals, where we find $\sigma_n\lesssim 10^{-5}$ both in the Euclidean and in the physical region.\\
As a last comment, let us state that we compared the power series representations in~$\lambda_m$ with the exact results in terms of MPLs analytically whenever possible. This can be achieved by converting the MPLs to classical polylogarithms with the help of~\textsc{Gtolrules} and using Eq.~\eqref{landau1} to change from the Landau variables~$\tilde{x}$, $\tilde{h}$ to the kinematic invariants~$x$, $h$. Subsequently, the expression has to be parametrized by~$\vec{\gamma}_m(\vec{x},\lambda_m)$ and expanded around~$\lambda_m=0$.

\subsection{Truncation Error}

We emphasize that we have not yet analyzed the quality of our numerical results towards the border of the physical region with high absolute values of the variable~$z$. This range coincides with the border of the region encircled by the radii of convergence of the series expansions, where we expect the convergence to decline noticeably. Provided that a power series can schematically be written as
\begin{equation}
M_{n,m}^{(p)} = \lambda_m^s \, \sum_{j=0}^p a_j \, \lambda_m^j
\end{equation}
by factorizing appropriate non-integer powers~$\lambda_m^s$ and shifting logarithmic terms into the coefficients~$a_j$, a measure for the convergence behavior can be inferred from the truncation error of the series expansion. The truncation error is assessed by studying the difference of the power series coefficients including powers up to $\lambda_m^p$ and $\lambda_m^{p-1}$:
\begin{align}
\Delta_p M_{n,m} &\equiv \frac{M_{n,m}^{(p)}-M_{n,m}^{(p-1)}}{\lambda_m^s} \nonumber \\
&= \sum_{j=0}^p a_j \, \lambda_m^j - \sum_{j=0}^{p-1} a_j \, \lambda_m^j \nonumber \\
&= a_p \, \lambda_m^p \,.
\label{truncation}
\end{align}
If the absolute values of the real and imaginary parts of $\Delta_p M_{n,m}$ decrease in a given phase space point for increasing~$p$, then the power series~$M_{n,m}$ associated with $\Delta_p M_{n,m}$ converges. An example for such an analysis is provided in Table~\ref{tab:truncation} in case of the elliptic integral~$M_{59}$ in the physical region below threshold.\\
\begin{SCtable}[3][tb]
\caption[Convergence of the power series~$M_{59,1}$ of the elliptic integral~$M_{59}$ in the parameter~$\lambda_1$]{\textbf{Convergence of the power series~$\boldsymbol{M_{59,1}}$ of the elliptic integral~$\boldsymbol{M_{59}}$ in the parameter~$\boldsymbol{\lambda_1}$} expressed through $\Delta_p M_{n,m}$ as defined in Eq.~\eqref{truncation} with $s=\nicefrac{1}{2}$. The weight-four result for $M_{59,1}$ was computed in Section~\ref{sec:hjdeqell215} and $\lambda_1$ is part of the parametrization~$\vec{\gamma}_1(\vec{x},\lambda_1)$ indicated in Eq.~\eqref{param1}. The series expansion is evaluated at the phase space point $x=1.048$, $z=-0.473$, $h=0.500$, where the relative deviation with respect to the \textsc{SecDec} result is $\sigma_{59}=3.5\cdot 10^{-5}$ and the real part vanishes.}
\begin{tabular}{cl}
\toprule
$\Delta_1 M_{59,1}$ & $-0.27\,i$ \\
$\Delta_2 M_{59,1}$ & $-0.046\,i$ \\
$\Delta_3 M_{59,1}$ & $-8.52 \cdot 10^{-3}\,i$ \\
$\Delta_4 M_{59,1}$ & $-1.62 \cdot 10^{-3}\,i$ \\
$\Delta_5 M_{59,1}$ & $-3.24 \cdot 10^{-4}\,i$ \\
$\Delta_6 M_{59,1}$ & $-6.70 \cdot 10^{-5}\,i$ \\
$\Delta_7 M_{59,1}$ & $-1.43 \cdot 10^{-5}\,i$ \\
$\Delta_8 M_{59,1}$ & $-3.12 \cdot 10^{-6}\,i$ \\
$\Delta_9 M_{59,1}$ & $-6.93 \cdot 10^{-7}\,i$ \\
$\Delta_{10} M_{59,1}$ & $-1.57 \cdot 10^{-7}\,i$ \\
\midrule
$M_{59,1}$ & \multicolumn{1}{c}{$-0.33\,i$} \\
\bottomrule
\label{tab:truncation}
\end{tabular}
\end{SCtable}\noindent
At the border of the physical region with high absolute values of~$z$, the convergence behavior could be improved by computing series expansions to higher degree~$p$ in~$\lambda_m$, which is however unfeasible bearing in mind the huge coefficients arising from the IBP reduction and thus in the differential equations. Instead, it is more promising to spread so-called \textit{supporting points} in those domains of the physical region, in which the convergence is unsatisfactory. This could be done by deriving additional power series representations using the generic linear parametrization procedure suggested in Eq.~\eqref{paramalt2}.

\section{Conclusions}
\label{sec:hjconclusions}

In this chapter, we have described the calculation of the planar MIs required for the two-loop corrections to Higgs-plus-jet production with full quark mass dependence in the physical region. Our approach is based on the derivation of series expansions from differential equations with respect to a parameter~$\lambda_m$, which was explained in great detail in Sections~\ref{sec:seriesexp} and Sections~\ref{sec:onedim} and effectively reduces any multi-scale problem to a single-variable one. Subsequently, we have connected multiple series expansions through the matching procedure outlined in Section~\ref{sec:matching} in order to establish results that are valid over the whole physical phase space. Apart from MIs whose differential equations can be cast into canonical form, the computation of the planar MIs required the evaluation of multi-scale elliptic integrals, which were calculated for the first time analytically in the physical region. Moreover, we have demonstrated the strength of this method with respect to the speed and stability of the numerical evaluation.\\
Beyond that, we have given an outlook on the calculation of the non-planar MIs in the physical region, whose differential equations are expected to be solvable using the same methods. The expressions of both the planar and the non-planar MIs are a crucial ingredient of the first fully analytical calculation of the two-loop corrections to the scattering amplitude for Higgs-plus-jet production with full quark mass dependence, which were derived together with the IBP relations necessary to express this amplitude in terms of MIs. This amplitude could help to understand the behavior of the transverse momentum spectrum of the Higgs boson at high energies, which was determined recently~\cite{Jones:2018,Lindert:2018}. Furthermore, it allows to treat pure top quark contributions and top-bottom interference effects to the cross section simultaneously for the first time. Finally, we hope that the promising method of series expansions from differential equations used in this chapter will prove beneficial for the computation of many more multi-scale scattering amplitudes that involve elliptic integrals.


\chapter{Summary and Outlook}
\label{chap:summary}

We achieved the goal of this thesis by computing physical quantities related to the multi-loop multi-scale processes shown in Fig.~\ref{fig:blob}. They include the three-loop corrections to the $Hb\bar{b}$ form factor, the two-loop corrections relevant to the $H\to Z\,\gamma$ decay rate with full quark mass dependence and contributions of the two-loop planar Master Integrals to the amplitude of Higgs-plus-jet production by retaining the full dependence on the internal quark mass. In the following, we recapitulate how this has been accomplished.\\
Initially, we followed the conventional workflow of multi-loop computations, which spans from the application of Feynman rules to the tensor decomposition of scattering amplitudes and finally ends in the reduction of this amplitude in terms of Master Integrals with the help of IBP relations. This enabled us to calculate the third-order QCD corrections to the form factor describing the Yukawa coupling of a Higgs boson to a pair of bottom quarks, which can be used to derive the third-order QCD corrections to Higgs boson production from bottom quark fusion and the fully differential description of Higgs boson decays into bottom quarks.\\
The situation is much more complicated if the Master Integrals, that are required for computing the amplitude of a given process, are unknown. In this case, a suitable approach is to derive differential equations of the Master Integrals with respect to the kinematic invariants, which is particularly useful when it comes to processes with multiple scales. Through the introduction of Multiple Polylogarithms, these differential equations can be solved in terms of iterated integrals in the non-physical region. From the expressions in the non-physical region, results in the physical region are derived by identifying the analytic continuation of the underlying class of functions, whose subsequent numerical evaluation is straightforward. We made use of this powerful tool to calculate the two-loop corrections to the $H\to Z\,\gamma$ decay width by retaining the full dependence on the internal quark mass.\\
The method of differential equations is close to being fully automated for integrals that can be expressed in terms of polylogarithms, or equivalently whose differential equations can be cast into canonical form. In more complicated cases, however, the first-order differential equations of a set of Master Integrals might turn out to be coupled in $D=4$ space-time dimensions, which prevents us from integrating them as described above. In such a case, they can be rephrased as one higher-order differential equation of one of the coupled integrals, whose systematic solution is yet unclear and has to be analyzed case by case. We encountered this behavior in the context of calculating the planar two-loop Master Integrals relevant to Higgs-plus-jet production with full quark mass dependence, so that we developed an approach to derive series expansions from differential equations in a systematic way. By following this procedure, we managed to evaluate up to second-order differential equations of elliptic integrals directly in the physical region. Thanks to the possibility of matching multiple series expansions over the physical range of kinematic invariants, the boundary conditions need to be computed solely in a single point of the phase space. Beyond that, we have given an outlook on the applicability of this method to the non-planar Master Integrals required for the amplitude of Higgs-plus-jet production, which remain to be computed.\\
In conclusion, we are confident that this approach might be helpful in future cases of phenomenological interest, where an exact analytical evaluation of the Master Integrals is not feasible. This applies particularly to theoretical predictions for Standard Model processes with massive particles, that are challenged by increasingly precise measurements thanks to the the Large Hadron Collider experiment. Let us therefore return to the bigger picture, which has served as the framework of this thesis:
\begin{quote}
\textit{``There is a theory in physics that explains, at the deepest level, nearly all of the phenomena that rule our daily lives \ldots It surpasses in precision, in universality, in its range of applicability from the very small to the astronomically large, every scientific theory that has ever existed. This theory bears the unassuming name `The Standard Model of Elementary Particles' \ldots It deserves to be better known, and it deserves a better name. I call it `The Theory of Almost Everything'."} \cite{Oerter:2006}
\end{quote}
We are happy to have contributed to this `Almost Everything', even if just a tiny bit.

\cleardoublepage
\addtocontents{toc}{\protect\newpage\protect}


\phantomsection
\addcontentsline{toc}{chapter}{Appendices}
\appendix


\makeatletter
\addtocontents{toc}{\let\protect\l@paragraph\protect\l@chapter}
\addtocontents{toc}{\let\protect\l@chapter\protect\l@section}
\addtocontents{toc}{\let\protect\l@section\protect\l@subsection}
\addtocontents{toc}{\let\protect\l@subsection\protect\l@subsubsection}
\def\toclevel@chapter{1}
\def\toclevel@section{2}
\def\toclevel@subsection{3}
\makeatother

\fancyhead[LO]{\headfont\nouppercase{\leftmark}}	
\fancyhead[RE]{\headfont\nouppercase{Appendices}}	

\renewcommand{\chaptername}{Appendix}

\setcounter{figure}{0}

\chapter[{Master Integrals for Two-Loop Corrections to the Decay $H\to Z\,\gamma$}]{Master Integrals for Two-Loop Corrections to the Decay $\boldsymbol{H\to Z\,\gamma}$}
\label{chap:hzaMIs}

\section{Laporta Master Integrals}
\label{sec:hzalaporta}

The definition of the Laporta MIs depicted in Fig.~\ref{fig:hzamaster} coincides with the one in the context of Higgs-plus-jet production and is specified in Appendix~\ref{sec:hjlaporta}.

\section{Canonical Master Integrals}
\label{sec:hzacan}

In this appendix, we provide the relations between the two-loop canonical MIs that appear in Eq.~\eqref{canon} and the Laporta MIs in Fig.~\ref{fig:hzamaster} in terms of the Landau variables defined in Eq.~\eqref{landau2}. In order to represent the $H\to Z\,\gamma$ MIs in the notation of Higgs-plus-jet production, we need to extend the set of variables defined therein by
\begin{equation}
z=-\frac{(1-\tilde{z})^2}{\tilde{z}} \,.
\end{equation}
It is important to note that the additional variable $\tilde{z}$ is not independent of the set $\tilde{x},\tilde{h}$ and can be considered spurious in some sense. It is however required to remain within the notation of Higgs-plus-jet production introduced in Eq.~\eqref{ratios}. We stress that we are still dealing with a three-scale problem depending on two independent ratios, which can be either $(\tilde{x},\tilde{h})$ or $(\tilde{z},\tilde{h})$, i.e. the variables $\tilde{x}$ and $\tilde{z}$ never appear in the expression of one MI simultaneously. Therefore, the set of variables $(\tilde{z},\tilde{h})$ can be understood as a copy of the set $(\tilde{x},\tilde{h})$, for which all considerations made in Chapter~\ref{chap:hza} hold as well. The residual independence of the process on the kinematic invariant $z=u/m_q^2$ becomes manifest through the observation that the variable $\tilde{z}$ drops out of the final expression of the two-loop amplitude, which serves as another check of the result.\\
In the following, we denote the crossed MIs mentioned in Section~\ref{sec:hzadeq} and defined in Appendix~\ref{sec:hjlaporta} with indices starting from $201$. In addition, all canonical integrals are normalized such that their Laurent expansion starts at order $\e^0$ and their sign is chosen so that the results are in agreement with the definitions in Appendix~\ref{sec:hjcan} in the Euclidean region. Finally, we extract the mass dimension of the integrals and obtain
\begin{align}
M_1 &= \e^2 \, I_1 \,, \nonumber \\
M_2 &= -m_q^2 \, \e^2 \, \frac{(\tilde{x}+1) (\tilde{x}-1)}{\tilde{x}} \, I_2 \,, \nonumber \\
M_3 &= -m_q^2 \, \e^2 \, \frac{\tilde{x}-1}{\tilde{x}} \, \left[(\tilde{x}+1) \, I_3 + I_4 \right] \,, \nonumber \\
M_4 &= -m_q^2 \, \e^2 \, \frac{(\tilde{x}-1)^2}{\tilde{x}} \, I_4 \,, \nonumber \\
M_5 &= -m_q^2 \, \e^2 \, \frac{(\tilde{h}+1) (\tilde{h}-1)}{\tilde{h}} \, I_5 \,, \nonumber \\
M_6 &= -m_q^2 \, \e^2 \, \frac{\tilde{h}-1}{\tilde{h}} \, \left[(\tilde{h}+1) \, I_6 + I_7 \right] \,, \nonumber \\
M_7 &= -m_q^2 \, \e^2 \, \frac{(\tilde{h}-1)^2}{\tilde{h}} \, I_7 \,, \nonumber \\
M_{13} &= m_q^4 \, \e^2 \, \frac{(\tilde{x}+1)^2 (\tilde{x}-1)^2}{\tilde{x}^2} \, I_{13} \,, \nonumber \\
M_{14} &= - m_q^2 \, \e^3 \, \frac{(\tilde{h}-\tilde{z}) (\tilde{h} \tilde{z}-1)}{\tilde{h} \tilde{z}} \, I_{14} \,, \nonumber \\
M_{15} &= - m_q^2 \, \e^3 \, \frac{(\tilde{h}-\tilde{z}) (\tilde{h} \tilde{z}-1)}{\tilde{h} \tilde{z}} \, I_{15} \,, \nonumber \\
M_{16} &= - m_q^2 \, \e^3 \, \frac{(\tilde{h}-\tilde{z}) (\tilde{h} \tilde{z}-1)}{\tilde{h} \tilde{z}} \, I_{16} \,, \nonumber \\
M_{17} &= - m_q^2 \, \e^2 \, \frac{\tilde{z}\,(\tilde{h}^2+1) - \tilde{h}\,(\tilde{z}+1)}{2\,\tilde{z}\,(\tilde{h}^2+1) - \tilde{h}\,(\tilde{z}+1)^2} \, \left[ -\frac{3}{2} \, \frac{(\tilde{h}-1)^2}{\tilde{h}} I_{7} + \e \, \frac{(\tilde{h}-\tilde{z}) (\tilde{h} \tilde{z}-1)}{\tilde{h} \tilde{z}} \left(2\,I_{15} + I_{16}\right) \right. \nonumber \\
	&\left.\qquad\qquad\qquad\qquad\qquad\qquad\qquad\quad\;\;\;+ m_q^2 \, \frac{(\tilde{z}^2-1) (\tilde{h}^2+1-\tilde{h}\,(\tilde{z}+1))}{\tilde{h} \tilde{z}} \, I_{17} \right] \,, \nonumber \\
M_{18} &= - m_q^2 \, \e^3 \, \frac{(\tilde{h}-\tilde{z}) (\tilde{h} \tilde{z}-1)}{\tilde{h} \tilde{z}} \, I_{18} \,, \nonumber \\
M_{19} &= - m_q^2 \, \e^3 \, \frac{(\tilde{h}-\tilde{z}) (\tilde{h} \tilde{z}-1)}{\tilde{h} \tilde{z}} \, I_{19} \,, \nonumber \\
M_{20} &= - m_q^2 \, \e^2 \, \frac{\tilde{h}\,(\tilde{z}^2+1) - \tilde{z}\,(\tilde{h}+1)}{2\,\tilde{h}\,(\tilde{z}^2+1) - \tilde{z}\,(\tilde{h}+1)^2} \, \left[ -\frac{3}{2} \, \frac{(\tilde{z}-1)^2}{\tilde{z}} I_{4} - \e \, \frac{(\tilde{h}-\tilde{z}) (\tilde{h} \tilde{z}-1)}{\tilde{h} \tilde{z}} \left(2\,I_{18} + I_{19}\right) \right. \nonumber \\
	&\left.\qquad\qquad\qquad\qquad\qquad\qquad\qquad\quad\;\;\;+ m_q^2 \, \frac{(\tilde{h}^2-1) (\tilde{z}^2+1-\tilde{z}\,(\tilde{h}+1))}{\tilde{h} \tilde{z}} \, I_{20} \right] \,, \nonumber \\
M_{21} &= - m_q^2 \, \e^3 \, \frac{(\tilde{h}-\tilde{z}) (\tilde{h} \tilde{z}-1)}{\tilde{h} \tilde{z}} \, I_{21} \,, \nonumber \\
M_{22} &= m_q^4 \, \e^2 \, \frac{(\tilde{h}^2-1) (\tilde{x}^2-1)}{\tilde{h} \tilde{x}} \, I_{22} \,, \nonumber \\
M_{23} &= m_q^4 \, \e^2 \, \frac{(\tilde{h}+1)^2 (\tilde{h}-1)^2}{\tilde{h}^2} \, I_{23} \,, \nonumber \\
M_{28} &= - m_q^2 \, \e^4 \, \frac{(\tilde{h}-\tilde{z}) (\tilde{h} \tilde{z}-1)}{\tilde{h} \tilde{z}} \, I_{28} \,, \nonumber \\
M_{29} &= m_q^4 \, \e^3 \, \frac{(\tilde{h}-\tilde{z}) (\tilde{h} \tilde{z}-1) (\tilde{h}^2-1)}{\tilde{h}^2 \tilde{z}} \, I_{29} \,, \nonumber \\
M_{30} &= - m_q^2 \, \e^4 \, \frac{(\tilde{h}-\tilde{z}) (\tilde{h} \tilde{z}-1)}{\tilde{h} \tilde{z}} \, I_{30} \,, \nonumber \\
M_{31} &= m_q^4 \, \e^3 \, \frac{(\tilde{h}-\tilde{z}) (\tilde{h} \tilde{z}-1) (\tilde{z}^2-1)}{\tilde{h} \tilde{z}^2} \, I_{31} \,, \nonumber \\
M_{43} &= -m_q^2 \, \e^4 \, \frac{(\tilde{h}-\tilde{x}) (\tilde{h} \tilde{x}-1)}{\tilde{h} \tilde{x}} \, I_{43} \,, \nonumber \\
M_{44} &= m_q^4 \, \e^3 \, \frac{(\tilde{h}-\tilde{x}) (\tilde{h} \tilde{x}-1) (\tilde{x}^2-1)}{\tilde{h} \tilde{x}^2} \, I_{44} \,, \nonumber \\
M_{45} &= m_q^4 \, \e^3 \, \frac{(\tilde{h}-\tilde{x}) (\tilde{h} \tilde{x}-1) (\tilde{h}^2-1)}{\tilde{h}^2 \tilde{x}} \, I_{45} \,, \nonumber \\
M_{46} &= m_q^4 \, \e^2 \, \frac{\tilde{h} \tilde{x}-1}{\tilde{h} \tilde{x}} \, \left[ -4 \, (\tilde{h} \tilde{x}-1) \, I_{22} + 2 \, \e \, (\tilde{h}-\tilde{x}) \, \left(\frac{\tilde{x}-1}{\tilde{x}} \, I_{44} - \frac{\tilde{h}-1}{\tilde{h}} \, I_{45} \right) \right. \nonumber \\
	&\qquad\qquad\qquad\quad\;\;\left.+ m_q^2 \, \frac{(\tilde{h}-\tilde{x})^2 (\tilde{h} \tilde{x}-1)}{\tilde{h} \tilde{x}} \, I_{46} \right] \,, \nonumber \\
M_{53} &= m_q^4 \, \e^3 \, \frac{(\tilde{h}-\tilde{x}) (\tilde{h} \tilde{x}-1) (\tilde{x}^2-1)}{\tilde{h} \tilde{x}^2} \, I_{53} \,, \nonumber \\
M_{54} &= - m_q^4 \, \e^3 \, \frac{(\tilde{h}-\tilde{z}) (\tilde{h} \tilde{z}-1) (\tilde{h}^2-1)}{\tilde{h}^2 \tilde{z}} \, I_{54} \,, \nonumber \\
M_{201} &= m_q^2 \, \e^2 \, \frac{(\tilde{z}+1) (\tilde{z}-1)}{\tilde{z}} \, I_{201} \,, \nonumber \\
M_{202} &= m_q^2 \, \e^2 \, \frac{\tilde{z}-1}{\tilde{z}} \, \left[(\tilde{z}+1) \, I_{202} + I_{203} \right] \,, \nonumber \\
M_{203} &= -m_q^2 \, \e^2 \, \frac{(\tilde{z}-1)^2}{\tilde{z}} \, I_{203} \,, \nonumber \\
M_{209} &= - m_q^2 \, \e^3 \, \frac{(\tilde{h}-\tilde{x}) (\tilde{h} \tilde{x}-1)}{\tilde{h} \tilde{x}} \, I_{209} \,, \nonumber \\
M_{210} &= - m_q^2 \, \e^3 \, \frac{(\tilde{h}-\tilde{x}) (\tilde{h} \tilde{x}-1)}{\tilde{h} \tilde{x}} \, I_{210} \,, \nonumber \\
M_{211} &= - m_q^2 \, \e^3 \, \frac{(\tilde{h}-\tilde{x}) (\tilde{h} \tilde{x}-1)}{\tilde{h} \tilde{x}} \, I_{211} \,, \nonumber \\
M_{212} &= - m_q^2 \, \e^2 \, \frac{\tilde{x}\,(\tilde{h}^2+1) - \tilde{h}\,(\tilde{x}+1)}{2\,\tilde{x}\,(\tilde{h}^2+1) - \tilde{h}\,(\tilde{x}+1)^2} \, \left[ -\frac{3}{2} \, \frac{(\tilde{h}-1)^2}{\tilde{h}} I_{7} + \e \, \frac{(\tilde{h}-\tilde{x}) (\tilde{h} \tilde{x}-1)}{\tilde{h} \tilde{x}} \left(2\,I_{210} + I_{211}\right) \right. \nonumber \\
	&\left.\qquad\qquad\qquad\qquad\qquad\qquad\qquad\quad\;\;\;+ m_q^2 \, \frac{(\tilde{x}^2-1) (\tilde{h}^2+1-\tilde{h}\,(\tilde{x}+1))}{\tilde{h} \tilde{x}} \, I_{212} \right] \,, \nonumber \\
M_{213} &= - m_q^2 \, \e^3 \, \frac{(\tilde{h}-\tilde{x}) (\tilde{h} \tilde{x}-1)}{\tilde{h} \tilde{x}} \, I_{213} \,, \nonumber \\
M_{214} &= - m_q^2 \, \e^3 \, \frac{(\tilde{h}-\tilde{x}) (\tilde{h} \tilde{x}-1)}{\tilde{h} \tilde{x}} \, I_{214} \,, \nonumber \\
M_{215} &= - m_q^2 \, \e^2 \, \frac{\tilde{h}\,(\tilde{x}^2+1) - \tilde{x}\,(\tilde{h}+1)}{2\,\tilde{h}\,(\tilde{x}^2+1) - \tilde{x}\,(\tilde{h}+1)^2} \, \left[ -\frac{3}{2} \, \frac{(\tilde{x}-1)^2}{\tilde{x}} I_{4} - \e \, \frac{(\tilde{h}-\tilde{x}) (\tilde{h} \tilde{x}-1)}{\tilde{h} \tilde{x}} \left(2\,I_{213} + I_{214}\right) \right. \nonumber \\
	&\left.\qquad\qquad\qquad\qquad\qquad\qquad\qquad\;\;\;+ m_q^2 \, \frac{(\tilde{h}^2-1) (\tilde{x}^2+1-\tilde{x}\,(\tilde{h}+1))}{\tilde{h} \tilde{x}} \, I_{215} \right] \,.
\label{appendix}
\end{align}
$I_1$ is the two-loop tadpole integral with both propagators taken to be squared. It is given by Eq.~\eqref{tadpole}, so that
\begin{equation}
M_1 = S_\e^2 \,,
\end{equation}
where $S_\e^2$ is the common normalization factor of all two-loop MIs defined in Eq.~\eqref{norm}.

\chapter{Limiting Identities of Multiple Polylogarithms for Small Quark Masses}
\label{chap:hzaexpand}

In this appendix, we specify those MPLs from within the set~\eqref{alphabet}, whose leading-order results do not vanish in the limit $m_q\to 0$ and thus are required for the small quark mass expansion of the $H\to Z\,\gamma$ two-loop amplitude, as outlined in Section~\ref{sec:hzaexp}. It is worth mentioning that all MPLs with at least one of the indices $\{1/\tilde{x},\tilde{c}\}$ with any rational number $\tilde{c}\neq 0$ vanish in that limit. Due to the following first-order approximations of the indices defined in Eq.~\eqref{hzamplindices},
\begin{align}
\lim_{m_q\to 0} J_x = \tilde{x} \,, \nonumber \\
\lim_{m_q\to 0} \frac{1}{J_x} = \frac{1}{\tilde{x}} \,, \nonumber \\
\lim_{m_q\to 0} K_x^+ = c \,, \nonumber \\
\lim_{m_q\to 0} K_x^- = \bar{c} \,, \nonumber \\
\lim_{m_q\to 0} L_x^+ = \frac{1}{\tilde{x}} \,, \nonumber \\
\lim_{m_q\to 0} L_x^- = \tilde{x} \,,
\end{align}
we only need to consider MPLs of the types
\begin{align}
G\left(a_1,\ldots,a_n;\tilde{h}\right) \quad &\text{with} \quad a_i \in \{ 0,\tilde{x} \} \,, \nonumber \\
G\left(b_1,\ldots,b_n;\tilde{x}\right) \quad &\text{with} \quad b_i \in \{ 0,\tilde{x},L_x^- \} \,,
\label{mplexp}
\end{align}
where the latter emerge from the results for $M_{43}\text{--}M_{46}$ and have not been transformed according to Eq.~\eqref{trafo1} for reasons outlined in Section~\ref{sec:hzadeq}. The set~\eqref{mplexp} is reduced by two further constraints: First, not all possible combinations of indices therein occur in the two-loop amplitude of Eq.~\eqref{NLO}. Second, some of the MPLs from that set may occur in the two-loop amplitude, but do not contribute to the leading-order term in the limit $m_q\to 0$. Consequently, we indicate all remaining MPLs from within the set~\eqref{mplexp} at least up to the required order in the expansion.\\
As a final remark, we point out that the only diverging MPLs are of the form
\begin{equation}
\lim_{x\to w_1} G\left(\vec{w};x\right) \,, \qquad \lim_{w_n\to 0} G\left(\vec{w};x\right) \,.
\end{equation}
This agrees with our observation that all limiting identities, which are indicated in the following and contain logarithmic singularities, correspond to one of these two cases.

\section*{Limiting Identities at Weight One}

\begin{align}
\lim_{m_q\to 0} G\left(0;\tilde{h}\right) &= \log\left(\frac{m_q^2}{m_H^2}\right) \,, \nonumber \\
\lim_{m_q\to 0} G\left(\tilde{x};\tilde{h}\right) &= \log\left(1-\frac{m_Z^2}{m_H^2}\right) \,, \nonumber \\
\lim_{m_q\to 0} G\left(0;\tilde{x}\right) &= \log\left(\frac{m_q^2}{m_Z^2}\right) \,, \nonumber \\
\lim_{m_q\to 0} G\left(L_x^-;\tilde{x}\right) &= \log\left(\frac{m_q^2}{m_Z^2}\right) \,.
\end{align}

\section*{Limiting Identities at Weight Two}

\begin{align}
\lim_{m_q\to 0} G\left(0,0;\tilde{h}\right) &= \frac{1}{2} \, \log^2\left(\frac{m_q^2}{m_H^2}\right) \,, \nonumber \\
\lim_{m_q\to 0} G\left(0,\tilde{x};\tilde{h}\right) &= -\mathrm{Li}_2\left(\frac{m_Z^2}{m_H^2}\right) \,, \nonumber \\
\lim_{m_q\to 0} G\left(\tilde{x},0;\tilde{h}\right) &= \log\left(\frac{m_q^2}{m_H^2}\right) \, \log\left(1-\frac{m_Z^2}{m_H^2}\right) + \mathrm{Li}_2\left(\frac{m_Z^2}{m_H^2}\right) \,, \nonumber \\
\lim_{m_q\to 0} G\left(\tilde{x},\tilde{x};\tilde{h}\right) &= \frac{1}{2} \, \log^2\left(1-\frac{m_Z^2}{m_H^2}\right) \,, \nonumber \\
\lim_{m_q\to 0} G\left(0,0;\tilde{x}\right) &= \frac{1}{2} \, \log^2\left(\frac{m_q^2}{m_Z^2}\right) \,, \nonumber \\
\lim_{m_q\to 0} G\left(0,L_x^-;\tilde{x}\right) &= -\zeta_2 \,, \nonumber \\
\lim_{m_q\to 0} G\left(L_x^-,0;\tilde{x}\right) &= \log^2\left(\frac{m_q^2}{m_Z^2}\right) + \zeta_2 \,, \nonumber \\
\lim_{m_q\to 0} G\left(L_x^-,L_x^-;\tilde{x}\right) &= \frac{1}{2} \, \log^2\left(\frac{m_q^2}{m_Z^2}\right) \,.
\end{align}

\section*{Limiting Identities at Weight Three}

\begin{align}
\lim_{m_q\to 0} G\left(0,0,0;\tilde{h}\right) &= \frac{1}{6} \, \log^3\left(\frac{m_q^2}{m_H^2}\right) \,, \nonumber \\
\lim_{m_q\to 0} G\left(0,0,\tilde{x};\tilde{h}\right) &= -\mathrm{Li}_3\left(\frac{m_Z^2}{m_H^2}\right) \,, \nonumber \\
\lim_{m_q\to 0} G\left(0,\tilde{x},0;\tilde{h}\right) &= 2 \, \mathrm{Li}_3\left(\frac{m_Z^2}{m_H^2}\right) - \log\left(\frac{m_q^2}{m_H^2}\right) \, \mathrm{Li}_2\left(\frac{m_Z^2}{m_H^2}\right) \,, \nonumber \\
\lim_{m_q\to 0} G\left(\tilde{x},0,0;\tilde{h}\right) &= \frac{1}{2} \, \log^2\left(\frac{m_q^2}{m_H^2}\right) \, \log\left(1-\frac{m_Z^2}{m_H^2}\right) + \log\left(\frac{m_q^2}{m_H^2}\right) \, \mathrm{Li}_2\left(\frac{m_Z^2}{m_H^2}\right) \nonumber \\
&\quad\, - \mathrm{Li}_3\left(\frac{m_Z^2}{m_H^2}\right) \,, \nonumber \\
\lim_{m_q\to 0} G\left(0,\tilde{x},\tilde{x};\tilde{h}\right) &= \frac{1}{2} \, \log\left(\frac{m_Z^2}{m_H^2}\right) \, \log^2\left(1-\frac{m_Z^2}{m_H^2}\right) + \log\left(1-\frac{m_Z^2}{m_H^2}\right) \, \mathrm{Li}_2\left(1-\frac{m_Z^2}{m_H^2}\right) \nonumber \\
&\quad\, - \mathrm{Li}_3\left(1-\frac{m_Z^2}{m_H^2}\right)  + \zeta_3 \,, \nonumber \\
\lim_{m_q\to 0} G\left(\tilde{x},\tilde{x},0;\tilde{h}\right) &= \frac{1}{2} \, \log\left(\frac{m_q^2}{m_H^2}\right) \, \log^2\left(1-\frac{m_Z^2}{m_H^2}\right) + \frac{1}{2} \, \log\left(\frac{m_Z^2}{m_H^2}\right) \, \log^2\left(1-\frac{m_Z^2}{m_H^2}\right) \nonumber \\
&\quad\, + \log\left(1-\frac{m_Z^2}{m_H^2}\right) \, \mathrm{Li}_2\left(\frac{m_Z^2}{m_H^2}\right) + \log\left(1-\frac{m_Z^2}{m_H^2}\right) \, \mathrm{Li}_2\left(1-\frac{m_Z^2}{m_H^2}\right) \nonumber \\
&\quad\, - \mathrm{Li}_3\left(1-\frac{m_Z^2}{m_H^2}\right)  + \zeta_3 \,, \nonumber \\
\lim_{m_q\to 0} G\left(\tilde{x},\tilde{x},\tilde{x};\tilde{h}\right) &= \frac{1}{6} \, \log^3\left(1-\frac{m_Z^2}{m_H^2}\right) \,, \nonumber \\
\lim_{m_q\to 0} G\left(0,0,0;\tilde{x}\right) &= \frac{1}{6} \, \log^3\left(\frac{m_q^2}{m_Z^2}\right) \,, \nonumber \\
\lim_{m_q\to 0} G\left(0,0,L_x^-;\tilde{x}\right) &= -\zeta_3 \,, \nonumber \\
\lim_{m_q\to 0} G\left(0,L_x^-,0;\tilde{x}\right) &= -\zeta_2 \, \log\left(\frac{m_q^2}{m_Z^2}\right) + 2 \, \zeta_3 \,, \nonumber \\
\lim_{m_q\to 0} G\left(L_x^-,0,0;\tilde{x}\right) &= \frac{1}{6} \, \log^3\left(\frac{m_q^2}{m_Z^2}\right) + \zeta_2 \, \log\left(\frac{m_q^2}{m_Z^2}\right) - \zeta_3 \,, \nonumber \\
\lim_{m_q\to 0} G\left(L_x^-,0,\tilde{x};\tilde{x}\right) &= -\zeta_2 \, \log\left(\frac{m_q^2}{m_Z^2}\right) - 2 \, \zeta_3 \,, \nonumber \\
\lim_{m_q\to 0} G\left(L_x^-,L_x^-,L_x^-;\tilde{x}\right) &= \frac{1}{6} \, \log^3\left(\frac{m_q^2}{m_Z^2}\right) \,.
\end{align}

\section*{Limiting Identities at Weight Four}

\begin{align}
\lim_{m_q\to 0} G\left(0,0,0,0;\tilde{h}\right) &= \frac{1}{24} \, \log^4\left(\frac{m_q^2}{m_H^2}\right) \,, \nonumber \\
\lim_{m_q\to 0} G\left(0,\tilde{x},0,0;\tilde{h}\right) &= -\frac{1}{2} \, \log^2\left(\frac{m_q^2}{m_H^2}\right) \, \mathrm{Li}_2\left(\frac{m_Z^2}{m_H^2}\right) + 2 \, \log\left(\frac{m_q^2}{m_H^2}\right) \, \mathrm{Li}_3\left(\frac{m_Z^2}{m_H^2}\right) \nonumber \\
&\quad\, - 3 \, \mathrm{Li}_4\left(\frac{m_Z^2}{m_H^2}\right) \,, \nonumber \\
\lim_{m_q\to 0} G\left(\tilde{x},0,0,0;\tilde{h}\right) &= \frac{1}{6} \, \log^3\left(\frac{m_q^2}{m_H^2}\right) \, \log\left(1-\frac{m_Z^2}{m_H^2}\right) + \frac{1}{2} \log^2\left(\frac{m_q^2}{m_H^2}\right) \, \mathrm{Li}_2\left(\frac{m_Z^2}{m_H^2}\right) \nonumber \\
&\quad\, - \log\left(\frac{m_q^2}{m_H^2}\right) \, \mathrm{Li}_3\left(\frac{m_Z^2}{m_H^2}\right) + \mathrm{Li}_4\left(\frac{m_Z^2}{m_H^2}\right) \,, \nonumber \\
\lim_{m_q\to 0} G\left(\tilde{x},\tilde{x},0,0;\tilde{h}\right) &= \frac{1}{4} \, \log^2\left(\frac{m_q^2}{m_H^2}\right) \, \log^2\left(1-\frac{m_Z^2}{m_H^2}\right) + \mathcal{O}\left(\log\left(\frac{m_q^2}{m_H^2}\right)\right) \,, \nonumber \\
\lim_{m_q\to 0} G\left(\tilde{x},\tilde{x},\tilde{x},\tilde{x};\tilde{h}\right) &= \frac{1}{24} \, \log^4\left(1-\frac{m_Z^2}{m_H^2}\right) \,, \nonumber \\
\lim_{m_q\to 0} G\left(0,0,0,0;\tilde{x}\right) &= \frac{1}{24} \, \log^4\left(\frac{m_q^2}{m_Z^2}\right) \,, \nonumber \\
\lim_{m_q\to 0} G\left(0,L_x^-,0,0;\tilde{h}\right) &= -\frac{1}{2} \, \zeta_2 \, \log^2\left(\frac{m_q^2}{m_Z^2}\right) + 2 \, \zeta_3 \, \log\left(\frac{m_q^2}{m_Z^2}\right) - 3 \, \zeta_4 \,, \nonumber \\
\lim_{m_q\to 0} G\left(L_x^-,0,0,0;\tilde{h}\right) &= \frac{1}{6} \, \log^4\left(\frac{m_q^2}{m_Z^2}\right) + \frac{1}{2} \, \zeta_2 \, \log^2\left(\frac{m_q^2}{m_Z^2}\right) - \zeta_3 \, \log\left(\frac{m_q^2}{m_Z^2}\right) + \zeta_4 \,, \nonumber \\
\lim_{m_q\to 0} G\left(L_x^-,0,\tilde{x},0;\tilde{h}\right) &= -\zeta_2 \, \log^2\left(\frac{m_q^2}{m_Z^2}\right) + \frac{7}{4} \, \zeta_4 \,, \nonumber \\
\lim_{m_q\to 0} G\left(L_x^-,L_x^-,L_x^-,L_x^-;\tilde{x}\right) &= \frac{1}{24} \, \log^4\left(\frac{m_q^2}{m_Z^2}\right) \,.
\end{align}

\chapter{Master Integrals for Two-Loop Corrections to Higgs-plus-Jet Production}
\label{chap:hjlaportacan}

\section{Laporta Master Integrals}
\label{sec:hjlaporta}

In this appendix, we give the definition of the Laporta integrals depicted in Fig.~\ref{fig:hjmaster}, which leads to a triangular basis in $D=4$ in the sense of Eq.~\eqref{matrixdeq2}. The definitions indicated here have also been used in the context of the two-loop decay rate of $H\to Z\,\gamma$ in Chapter~\ref{chap:hza}. The notation follows
\begin{equation}
I_j = \textit{family}_{t,ID,r,s}\,(b_1,\dots,b_9) \,,
\end{equation}
where $t$ is the number of distinct propagators and the definition of $r$ and $s$ as the sum of powers of all propagators and scalar products, respectively, is given in Eq.~\eqref{rs2}. The indices $b_1,\dots,b_9\in\mathbb{Z}$ describe the denominator powers of the integral representation~\eqref{scalarint3}, where the denominators are given through the definitions of the corresponding \textit{family} in Table~\ref{tab:hjtopo} in the respective order. The crossing $\mathrm{xij}$ of a family indicates a permutation of the external momenta $q_i\leftrightarrow q_j$, and in our case the only occuring permutation is $\mathrm{x13}$, corresponding to the exchange $s\leftrightarrow u$ or equivalently $x\leftrightarrow z$. Finally, the \textit{sector identification number} can be computed from the set of indices by converting it from a binary to a decimal number,
\begin{equation}
ID = \sum_{j=1}^9 2^{b_j-1} \, \Theta(b_j-1) \,,
\end{equation}
with the Heaviside step function
\begin{equation}
\Theta(x) = 
\begin{cases}
0 \quad (x<0)\,,\\
1 \quad (x\geq 0) \,.
\end{cases}
\end{equation}
With these definitions, a sector is fully determined by indicating $\textit{family}_{t,ID}$. We are now ready to give the Laporta basis of the full set of planar MIs for Higgs-plus-jet production at two loops. They are sorted first by increasing $t$ and second by increasing $ID$, and the numbering of the crossed MIs begins with $201$:
\begin{align}
I_{1} &= A_{2,3,4,0}\,(2,2,0,0,0,0,0,0,0) \nonumber \\
&= \int \frac{\d^Dk}{(2\pi)^D} \int \frac{\d^Dl}{(2\pi)^D} \frac{1}{(k^2-m_q^2)^2 \, (l^2-m_q^2)^2} = \frac{S_\e^2}{\e^2} \,, \label{tadpole} \\
I_{2} &= A_{3,35,5,0}\,(2,2,0,0,0,1,0,0,0) \,, \nonumber \\
I_{3} &= A_{3,38,5,0}\,(0,2,2,0,0,1,0,0,0) \,, \nonumber \\
I_{4} &= A_{3,38,5,0}\,(0,2,1,0,0,2,0,0,0) \,, \nonumber \\
I_{5} &= A_{3,131,5,0}\,(2,2,0,0,0,0,0,1,0) \,, \nonumber \\
I_{6} &= A_{3,134,5,0}\,(0,2,2,0,0,0,0,1,0) \,, \nonumber \\
I_{7} &= A_{3,134,5,0}\,(0,2,1,0,0,0,0,2,0) \,, \nonumber \\
I_{8} &= A_{4,43,5,0}\,(1,2,0,1,0,1,0,0,0) \,, \nonumber \\
I_{9} &= A_{4,46,5,0}\,(0,2,1,1,0,1,0,0,0) \,, \nonumber \\
I_{10} &= A_{4,53,5,0}\,(1,0,1,0,2,1,0,0,0) \,, \nonumber \\
I_{11} &= A_{4,53,5,0}\,(1,0,2,0,1,1,0,0,0) \,, \nonumber \\
I_{12} &= A_{4,53,6,0}\,(1,0,1,0,2,2,0,0,0) \,, \nonumber \\
I_{13} &= A_{4,99,6,0}\,(2,2,0,0,0,1,1,0,0) \,, \nonumber \\
I_{14} &= A_{4,139,5,0}\,(1,2,0,1,0,0,0,1,0) \,, \nonumber \\
I_{15} &= A_{4,142,5,0}\,(0,2,1,1,0,0,0,1,0) \,, \nonumber \\
I_{16} &= A_{4,142,5,0}\,(0,1,2,1,0,0,0,1,0) \,, \nonumber \\
I_{17} &= A_{4,142,6,0}\,(0,2,1,1,0,0,0,2,0) \,, \nonumber \\
I_{18} &= A_{4,149,5,0}\,(1,0,1,0,2,0,0,1,0) \,, \nonumber \\
I_{19} &= A_{4,149,5,0}\,(1,0,2,0,1,0,0,1,0) \,, \nonumber \\
I_{20} &= A_{4,149,6,0}\,(1,0,1,0,2,0,0,2,0) \,, \nonumber \\
I_{21} &= A_{4,150,5,0}\,(0,1,1,0,1,0,0,2,0) \,, \nonumber \\
I_{22} &= A_{4,195,6,0}\,(2,2,0,0,0,0,1,1,0) \,, \nonumber \\
I_{23} &= A_{4,387,6,0}\,(2,2,0,0,0,0,0,1,1) \,, \nonumber \\
I_{24} &= A_{5,55,5,0}\,(1,1,1,0,1,1,0,0,0) \,, \nonumber \\
I_{25} &= A_{5,55,6,0}\,(1,1,1,0,1,2,0,0,0) \,, \nonumber \\
I_{26} &= A_{5,62,5,0}\,(0,1,1,1,1,1,0,0,0) \,, \nonumber \\
I_{27} &= A_{5,107,6,0}\,(1,2,0,1,0,1,1,0,0) \,, \nonumber \\
I_{28} &= A_{5,151,5,0}\,(1,1,1,0,1,0,0,1,0) \,, \nonumber \\
I_{29} &= A_{5,151,6,0}\,(1,1,1,0,1,0,0,2,0) \,, \nonumber \\
I_{30} &= A_{5,158,5,0}\,(0,1,1,1,1,0,0,1,0) \,, \nonumber \\
I_{31} &= A_{5,158,6,0}\,(0,1,1,1,1,0,0,2,0) \,, \nonumber \\
I_{32} &= A_{5,171,6,0}\,(1,2,0,1,0,1,0,1,0) \,, \nonumber \\
I_{33} &= A_{5,174,5,0}\,(0,1,1,1,0,1,0,1,0) \,, \nonumber \\
I_{34} &= A_{5,174,6,0}\,(0,1,2,1,0,1,0,1,0) \,, \nonumber \\
I_{35} &= A_{5,174,6,0}\,(0,2,1,1,0,1,0,1,0) \,, \nonumber \\
I_{36} &= A_{5,181,5,0}\,(1,0,1,0,1,1,0,1,0) \,, \nonumber \\
I_{37} &= A_{5,181,6,0}\,(1,0,2,0,1,1,0,1,0) \,, \nonumber \\
I_{38} &= A_{5,181,6,0}\,(1,0,1,0,2,1,0,1,0) \,, \nonumber \\
I_{39} &= A_{5,182,5,0}\,(0,1,1,0,1,1,0,1,0) \,, \nonumber \\
I_{40} &= A_{5,182,6,0}\,(0,1,2,0,1,1,0,1,0) \,, \nonumber \\
I_{41} &= A_{5,182,6,0}\,(0,1,1,0,2,1,0,1,0) \,, \nonumber \\
I_{42} &= A_{5,182,6,0}\,(0,1,1,0,1,1,0,2,0) \,, \nonumber \\
I_{43} &= A_{5,199,5,0}\,(1,1,1,0,0,0,1,1,0) \,, \nonumber \\
I_{44} &= A_{5,199,6,0}\,(1,1,1,0,0,0,2,1,0) \,, \nonumber \\
I_{45} &= A_{5,199,6,0}\,(1,1,1,0,0,0,1,2,0) \,, \nonumber \\
I_{46} &= A_{5,199,7,0}\,(1,1,1,0,0,0,2,2,0) \,, \nonumber \\
I_{47} &= A_{5,211,6,0}\,(2,1,0,0,1,0,1,1,0) \,, \nonumber \\
I_{48} &= A_{5,213,5,0}\,(1,0,1,0,1,0,1,1,0) \,, \nonumber \\
I_{49} &= A_{5,213,6,0}\,(2,0,1,0,1,0,1,1,0) \,, \nonumber \\
I_{50} &= A_{5,213,6,0}\,(1,0,1,0,1,0,1,2,0) \,, \nonumber \\
I_{51} &= A_{5,213,6,0}\,(1,0,2,0,1,0,1,1,0) \,, \nonumber \\
I_{52} &= A_{5,213,6,1}\,(1,0,2,-1,1,0,1,1,0) \,, \nonumber \\
I_{53} &= A_{5,227,6,0}\,(1,2,0,0,0,1,1,1,0) \,, \nonumber \\
I_{54} &= A_{5,395,6,0}\,(1,2,0,1,0,0,0,1,1) \,, \nonumber \\
I_{55} &= A_{6,183,6,0}\,(1,1,1,0,1,1,0,1,0) \,, \nonumber \\
I_{56} &= A_{6,183,6,1}\,(1,1,1,-1,1,1,0,1,0) \,, \nonumber \\
I_{57} &= A_{6,190,6,0}\,(0,1,1,1,1,1,0,1,0) \,, \nonumber \\
I_{58} &= A_{6,190,6,1}\,(-1,1,1,1,1,1,0,1,0) \,, \nonumber \\
I_{59} &= A_{6,215,6,0}\,(1,1,1,0,1,0,1,1,0) \,, \nonumber \\
I_{60} &= A_{6,215,7,0}\,(1,1,1,0,1,0,2,1,0) \,, \nonumber \\
I_{61} &= A_{6,215,7,0}\,(1,1,1,0,1,0,1,2,0) \,, \nonumber \\
I_{62} &= A_{6,215,6,1}\,(1,1,1,0,1,0,1,1,-1) \,, \nonumber \\
I_{63} &= A_{6,221,6,0}\,(1,0,1,1,1,0,1,1,0) \,, \nonumber \\
I_{64} &= A_{6,221,6,1}\,(1,0,1,1,1,-1,1,1,0) \,, \nonumber \\
I_{65} &= A_{6,243,6,0}\,(1,1,0,0,1,1,1,1,0) \,, \nonumber \\
I_{66} &= A_{6,427,7,0}\,(1,2,0,1,0,1,0,1,1) \,, \nonumber \\
I_{67} &= A_{7,247,7,0}\,(1,1,1,0,1,1,1,1,0) \,, \nonumber \\
I_{68} &= A_{7,247,7,1}\,(1,1,1,-1,1,1,1,1,0) \,, \nonumber \\
I_{69} &= A_{7,247,7,1}\,(1,1,1,0,1,1,1,1,-1) \,, \nonumber \\
I_{70} &= A_{7,247,7,2}\,(1,1,1,-1,1,1,1,1,-1) \,, \nonumber \\
I_{71} &= B_{3,35,5,0}\,(2,2,0,0,0,1,0,0,0) \,, \nonumber \\
I_{72} &= B_{3,131,5,0}\,(2,2,0,0,0,0,0,1,0) \,, \nonumber \\
I_{73} &= B_{4,99,6,0}\,(2,2,0,0,0,1,1,0,0) \,, \nonumber \\
I_{74} &= B_{4,142,5,0}\,(0,2,1,1,0,0,0,1,0) \,, \nonumber \\
I_{75} &= B_{4,142,6,0}\,(0,2,1,1,0,0,0,2,0) \,, \nonumber \\
I_{76} &= B_{4,291,6,0}\,(2,2,0,0,0,1,0,0,1) \,, \nonumber \\
I_{77} &= B_{4,387,6,0}\,(2,2,0,0,0,0,0,1,1) \,, \nonumber \\
I_{78} &= B_{5,103,5,0}\,(1,1,1,0,0,1,1,0,0) \,, \nonumber \\
I_{79} &= B_{5,158,5,0}\,(0,1,1,1,1,0,0,1,0) \,, \nonumber \\
I_{80} &= B_{5,171,5,0}\,(1,1,0,1,0,1,0,1,0) \,, \nonumber \\
I_{81} &= B_{5,174,5,0}\,(0,1,1,1,0,1,0,1,0) \,, \nonumber \\
I_{82} &= B_{5,174,6,0}\,(0,2,1,1,0,1,0,1,0) \,, \nonumber \\
I_{83} &= B_{5,182,5,0}\,(0,1,1,0,1,1,0,1,0) \,, \nonumber \\
I_{84} &= B_{5,182,6,0}\,(0,1,2,0,1,1,0,1,0) \,, \nonumber \\
I_{85} &= B_{5,295,5,0}\,(1,1,1,0,0,1,0,0,1) \,, \nonumber \\
I_{86} &= B_{5,295,6,0}\,(1,1,2,0,0,1,0,0,1) \,, \nonumber \\
I_{87} &= B_{5,295,6,0}\,(1,1,1,0,0,1,0,0,2) \,, \nonumber \\
I_{88} &= B_{5,302,5,0}\,(0,1,1,1,0,1,0,0,1) \,, \nonumber \\
I_{89} &= B_{5,302,6,0}\,(0,1,2,1,0,1,0,0,1) \,, \nonumber \\
I_{90} &= B_{5,302,6,0}\,(0,1,1,1,0,1,0,0,2) \,, \nonumber \\
I_{91} &= B_{5,355,6,0}\,(1,1,0,0,0,2,1,0,1) \,, \nonumber \\
I_{92} &= B_{5,391,5,0}\,(1,1,1,0,0,0,0,1,1) \,, \nonumber \\
I_{93} &= B_{6,190,6,0}\,(0,1,1,1,1,1,0,1,0) \,, \nonumber \\
I_{94} &= B_{6,303,6,0}\,(1,1,1,1,0,1,0,0,1) \,, \nonumber \\
I_{95} &= B_{6,303,7,0}\,(1,1,1,1,0,1,0,0,2) \,, \nonumber \\
I_{96} &= B_{6,359,6,0}\,(1,1,1,0,0,1,1,0,1) \,, \nonumber \\
I_{97} &= B_{6,427,7,0}\,(1,2,0,1,0,1,0,1,1) \,, \nonumber \\
I_{98} &= B_{7,367,7,0}\,(1,1,1,1,0,1,1,0,1) \,, \nonumber \\
I_{99} &= B_{7,367,7,1}\,(1,1,1,1,-1,1,1,0,1) \,, \nonumber \\
I_{100} &= B_{7,367,7,1}\,(1,1,1,1,0,1,1,-1,1) \,, \nonumber \\
I_{101} &= B_{7,367,7,2}\,(1,1,1,1,-1,1,1,-1,1) \,, \nonumber \\
I_{102} &= B_{7,431,7,0}\,(1,1,1,1,0,1,0,1,1) \,, \nonumber \\
I_{201} &= A\mathrm{x}13_{3,35,5,0}\,(2,2,0,0,0,1,0,0,0) \,, \nonumber \\
I_{202} &= A\mathrm{x}13_{3,38,5,0}\,(0,2,2,0,0,1,0,0,0) \,, \nonumber \\
I_{203} &= A\mathrm{x}13_{3,38,5,0}\,(0,2,1,0,0,2,0,0,0) \,, \nonumber \\
I_{204} &= A\mathrm{x}13_{4,43,5,0}\,(1,2,0,1,0,1,0,0,0) \,, \nonumber \\
I_{205} &= A\mathrm{x}13_{4,46,5,0}\,(0,2,1,1,0,1,0,0,0) \,, \nonumber \\
I_{206} &= A\mathrm{x}13_{4,53,5,0}\,(1,0,1,0,2,1,0,0,0) \,, \nonumber \\
I_{207} &= A\mathrm{x}13_{4,53,5,0}\,(1,0,2,0,1,1,0,0,0) \,, \nonumber \\
I_{208} &= A\mathrm{x}13_{4,53,6,0}\,(1,0,1,0,2,2,0,0,0) \,, \nonumber \\
I_{209} &= A\mathrm{x}13_{4,139,5,0}\,(1,2,0,1,0,0,0,1,0) \,, \nonumber \\
I_{210} &= A\mathrm{x}13_{4,142,5,0}\,(0,2,1,1,0,0,0,1,0) \,, \nonumber \\
I_{211} &= A\mathrm{x}13_{4,142,5,0}\,(0,1,2,1,0,0,0,1,0) \,, \nonumber \\
I_{212} &= A\mathrm{x}13_{4,142,6,0}\,(0,2,1,1,0,0,0,2,0) \,, \nonumber \\
I_{213} &= A\mathrm{x}13_{4,149,5,0}\,(1,0,1,0,2,0,0,1,0) \,, \nonumber \\
I_{214} &= A\mathrm{x}13_{4,149,5,0}\,(1,0,2,0,1,0,0,1,0) \,, \nonumber \\
I_{215} &= A\mathrm{x}13_{4,149,6,0}\,(1,0,1,0,2,0,0,2,0) \,, \nonumber \\
I_{216} &= A\mathrm{x}13_{4,150,5,0}\,(0,1,1,0,1,0,0,2,0) \,, \nonumber \\
I_{217} &= A\mathrm{x}13_{4,195,6,0}\,(2,2,0,0,0,0,1,1,0) \,, \nonumber \\
I_{218} &= A\mathrm{x}13_{5,55,5,0}\,(1,1,1,0,1,1,0,0,0) \,, \nonumber \\
I_{219} &= A\mathrm{x}13_{5,55,6,0}\,(1,1,1,0,1,2,0,0,0) \,, \nonumber \\
I_{220} &= A\mathrm{x}13_{5,62,5,0}\,(0,1,1,1,1,1,0,0,0) \,, \nonumber \\
I_{221} &= A\mathrm{x}13_{5,158,5,0}\,(0,1,1,1,1,0,0,1,0) \,, \nonumber \\
I_{222} &= A\mathrm{x}13_{5,158,6,0}\,(0,1,1,1,1,0,0,2,0) \,, \nonumber \\
I_{223} &= A\mathrm{x}13_{5,174,5,0}\,(0,1,1,1,0,1,0,1,0) \,, \nonumber \\
I_{224} &= A\mathrm{x}13_{5,174,6,0}\,(0,1,2,1,0,1,0,1,0) \,, \nonumber \\
I_{225} &= A\mathrm{x}13_{5,174,6,0}\,(0,2,1,1,0,1,0,1,0) \,, \nonumber \\
I_{226} &= A\mathrm{x}13_{5,181,5,0}\,(1,0,1,0,1,1,0,1,0) \,, \nonumber \\
I_{227} &= A\mathrm{x}13_{5,181,6,0}\,(1,0,2,0,1,1,0,1,0) \,, \nonumber \\
I_{228} &= A\mathrm{x}13_{5,181,6,0}\,(1,0,1,0,2,1,0,1,0) \,, \nonumber \\
I_{229} &= A\mathrm{x}13_{5,211,6,0}\,(2,1,0,0,1,0,1,1,0) \,, \nonumber \\
I_{230} &= A\mathrm{x}13_{5,395,6,0}\,(1,2,0,1,0,0,0,1,1) \,, \nonumber \\
I_{231} &= A\mathrm{x}13_{6,190,6,0}\,(0,1,1,1,1,1,0,1,0) \,, \nonumber \\
I_{232} &= A\mathrm{x}13_{6,190,6,1}\,(-1,1,1,1,1,1,0,1,0) \,, \nonumber \\
I_{233} &= A\mathrm{x}13_{6,221,6,0}\,(1,0,1,1,1,0,1,1,0) \,, \nonumber \\
I_{234} &= A\mathrm{x}13_{6,221,6,1}\,(1,0,1,1,1,-1,1,1,0) \,, \nonumber \\
I_{235} &= B\mathrm{x}13_{3,35,5,0}\,(2,2,0,0,0,1,0,0,0) \,, \nonumber \\
I_{236} &= B\mathrm{x}13_{4,142,5,0}\,(0,2,1,1,0,0,0,1,0) \,, \nonumber \\
I_{237} &= B\mathrm{x}13_{4,142,6,0}\,(0,2,1,1,0,0,0,2,0) \,, \nonumber \\
I_{238} &= B\mathrm{x}13_{4,291,6,0}\,(2,2,0,0,0,1,0,0,1) \,, \nonumber \\
I_{239} &= B\mathrm{x}13_{5,158,5,0}\,(0,1,1,1,1,0,0,1,0) \,, \nonumber \\
I_{240} &= B\mathrm{x}13_{5,174,5,0}\,(0,1,1,1,0,1,0,1,0) \,, \nonumber \\
I_{241} &= B\mathrm{x}13_{5,174,6,0}\,(0,2,1,1,0,1,0,1,0) \,, \nonumber \\
I_{242} &= B\mathrm{x}13_{5,182,5,0}\,(0,1,1,0,1,1,0,1,0) \,, \nonumber \\
I_{243} &= B\mathrm{x}13_{5,182,6,0}\,(0,1,2,0,1,1,0,1,0) \,, \nonumber \\
I_{244} &= B\mathrm{x}13_{5,295,5,0}\,(1,1,1,0,0,1,0,0,1) \,, \nonumber \\
I_{245} &= B\mathrm{x}13_{5,295,6,0}\,(1,1,2,0,0,1,0,0,1) \,, \nonumber \\
I_{246} &= B\mathrm{x}13_{5,295,6,0}\,(1,1,1,0,0,1,0,0,2) \,, \nonumber \\
I_{247} &= B\mathrm{x}13_{6,190,6,0}\,(0,1,1,1,1,1,0,1,0) \,, \nonumber \\
I_{248} &= B\mathrm{x}13_{6,303,6,0}\,(1,1,1,1,0,1,0,0,1) \,, \nonumber \\
I_{249} &= B\mathrm{x}13_{6,303,7,0}\,(1,1,1,1,0,1,0,0,2) \,.
\end{align}
Note that $S_\e^2$ is the common normalization factor of all two-loop MIs and its definition is given in Eq.~\eqref{norm}.

\section{Canonical Master Integrals}
\label{sec:hjcan}

This appendix is designed to provide the relations between the two-loop canonical basis occuring in Eq.~\eqref{hjdeq2} and the Laporta basis of Fig.~\ref{fig:hjmaster} in terms of the variables defined in Eq.~\eqref{ratios}. The only non-canonical differential equations are given by the integrals $M_{59}$--$M_{62}$ and $M_{67}$--$M_{70}$ corresponding to the sectors $A_{6,215}$ and $A_{7,247}$, respectively. We indicate MIs of crossings, which are described in Appendix~\ref{sec:hjlaporta}, with indices starting from~$201$. Beyond that, all integrals~$M_j$ are normalized such that their Laurent expansion starts at order $\e^0$ and the mass dimension of the integrals is extracted, so that we end up with
\begin{align}
M_{1} &= \e^2 \, I_{1} = S_\e^2 \,, \nonumber \\
M_{2} &= m_q^2 \, \e^2 \, \sqrt{x\left(x-4\right)} \, I_{2} \,, \nonumber \\
M_{3} &= m_q^2 \, \e^2 \, \sqrt{x\left(x-4\right)} \left(I_{3} + \frac{1}{2} \, I_{4}\right) \,, \nonumber \\
M_{4} &= m_q^2 \, \e^2 \, x \, I_{4} \,, \nonumber \\
M_{5} &= m_q^2 \, \e^2 \, \sqrt{\left(-4 + h\right)h} I_{5} \,, \nonumber \\
M_{6} &= \left(\left(2I_{6} + I_{7}\right)m_q^2\e^2\sqrt{\left(-4 + h\right)h}\right)/2 \,, \nonumber \\
M_{7} &= m_q^2 \, \e^2 \, h \, I_{7} \,, \nonumber \\
M_{8} &= m_q^2 \, \e^3 \, x \, I_{8} \,, \nonumber \\
M_{9} &= m_q^2 \, \e^3 \, x \, I_{9} \,, \nonumber \\
M_{10} &= m_q^2 \, \e^3 \, x \, I_{10} \,, \nonumber \\
M_{11} &= m_q^2 \, \e^3 \, x \, I_{11} \,, \nonumber \\
M_{12} &= m_q^2 \, \e^2 \, \sqrt{x\,\left(x-4\right)} \, \left( \e \, \left( I_{10} + \frac{1}{2} I_{11} \right) + \frac{m_q^2}{2} \, I_{12} \right)\,, \nonumber \\
M_{13} &= m_q^4 \, \e^2 \, \sqrt{x\,\left(x-4\right)} \, I_{13} \,, \nonumber \\
M_{14} &= m_q^2 \, \e^3 \, \left(h - z\right) \, I_{14} \,, \nonumber \\
M_{15} &= m_q^2 \, \e^3 \, \left(h - z\right) \, I_{15} \,, \nonumber \\
M_{16} &= m_q^2 \, \e^3 \, \left(h - z\right) \, I_{16} \,, \nonumber \\
M_{17} &= m_q^2 \, \e^2 \, \frac{\sqrt{z\,\left(z-4\right)}}{2\,h-z} \, \left(\frac{3}{4} \, h \, I_{7} - \e \, (h-z) \, \left( I_{15} + I_{16} \right) + m_q^2 \, \left(h\,(h-z)+z\right) \,  I_{17} \right) \,, \nonumber \\
M_{18} &= m_q^2 \, \e^3 \, \left(h - z\right) \, I_{18} \,, \nonumber \\
M_{19} &= m_q^2 \, \e^3 \, \left(h - z\right) \, I_{19} \,, \nonumber \\
M_{20} &= m_q^2 \, \e^2 \, \frac{\sqrt{h\,\left(h-4\right)}}{2\,z-h} \, \left( \e \, (h-z) \, \left( I_{18} + I_{19} \right) + m_q^2 \, \left(h\,(h-z)+z\right) \,  I_{17} + \frac{3}{4} \, h \, I_{203} \right) \,, \nonumber \\
M_{21} &= m_q^2 \, \e^3 \, \left(h - z\right) \, I_{21} \,, \nonumber \\
M_{22} &= m_q^4 \, \e^2 \, \sqrt{x\,\left(x-4\right) \, h\,\left(h-4\right)} \, I_{22} \,, \nonumber \\
M_{23} &= m_q^4 \, \e^2 \, h \, \left(h-4\right) \, I_{23} \,, \nonumber \\
M_{24} &= m_q^2 \, \e^4 \, x \, I_{24} \,, \nonumber \\
M_{25} &= m_q^4 \, \e^3 \, x \, \sqrt{x\left(x-4\right)} \, I_{25} \,, \nonumber \\
M_{26} &= m_q^2 \, \e^4 \, x \, I_{26} \,, \nonumber \\
M_{27} &= m_q^4 \, \e^3 \, x \, \sqrt{x\,\left(x-4\right)} \, I_{27} \,, \nonumber \\
M_{28} &= m_q^2 \, \e^4 \, \left(h - z\right) \, I_{28} \,, \nonumber \\
M_{29} &= m_q^4 \, \e^3 \, \left(h - z\right) \, \sqrt{h\,\left(h-4\right)} \, I_{29} \,, \nonumber \\
M_{30} &= m_q^2 \, \e^4 \, \left(h - z\right) \, I_{30} \,, \nonumber \\
M_{31} &= m_q^4 \, \e^3 \, \left(h - z\right) \, \sqrt{z\,\left(z-4\right)} \, I_{31} \,, \nonumber \\
M_{32} &= m_q^4 \, \e^3 \, \sqrt{x\,z\,(4\,\left(h-x-z)+x\,z\right)} \, I_{32} \,, \nonumber \\
M_{33} &= m_q^2 \, \e^3 \, z \, \left(\left(2\,\e-1\right)\,I_{33} + m_q^2 \, I_{35}\right) \,, \nonumber \\
M_{34} &= m_q^4 \, \e^3 \, \sqrt{x\,z\,(4\,\left(h-x-z)+x\,z\right)} \, \left(I_{34} + I_{35}\right) \,, \nonumber \\
M_{35} &= m_q^4 \, \e^3 \, \sqrt{z\,\left(z\,\left(x-1\right)^2+4\,x\,\left(h-x\right)\right)} \, I_{35} \,, \nonumber \\
M_{36} &= m_q^2 \, \e^3 \, \left(h - x\right) \, \left(\left(2\,\e-1\right) \, I_{36} + m_q^2 \, I_{38} \right) \,, \nonumber \\
M_{37} &= m_q^4 \, \e^3 \, \sqrt{x\,z\,(4\,\left(h-x-z)+x\,z\right)} \, \left(I_{37} + I_{38}\right) \,, \nonumber \\
M_{38} &= m_q^4 \, \e^3 \, \sqrt{(h-x)^2+2\,x\,z\,(h-x)+x\,z^2\,(x-4)} \, I_{38} \,, \nonumber \\
M_{39} &= m_q^2 \, \e^4 \, \left(h - x - z\right) \, I_{39} \,, \nonumber \\
M_{40} &= m_q^4 \, \e^3 \, \sqrt{x\,z\,(4\,\left(h-x-z)+x\,z\right)} \, I_{40} \,, \nonumber \\
M_{41} &= m_q^4 \, \e^3 \, \left(h - x - z\right) \, \left(I_{41} + A_{5,182,6,0}\,(0,2,1,0,1,1,0,1,0)\right) \,, \nonumber \\
M_{42} &= m_q^4 \, \e^3 \, \left(h - x - z\right) \, \left(I_{42} + A_{5,182,6,0}\,(0,1,1,0,1,2,0,1,0)\right) \,, \nonumber \\
M_{43} &= m_q^2 \, \e^4 \, \left(h - x\right) \, I_{43} \,, \nonumber \\
M_{44} &= m_q^4 \, \e^3 \, \left(h - x\right) \, \sqrt{x\,\left(x-4\right)} \, I_{44} \,, \nonumber \\
M_{45} &= m_q^4 \, \e^3 \, \left(h - x\right) \, \sqrt{h\,\left(h-4\right)} \, I_{45} \,, \nonumber \\
M_{46} &= m_q^4 \, \e^2 \, \left(2\,\left(2\,(h+x)-h\,x\right) \, I_{22} - \e \, x \, \left(h-x\right) \, I_{44} + \e \, h \, \left(h-x\right) \, I_{45} \right. \nonumber \\
& \qquad \qquad \left.
+ m_q^2 \, \left(h - x\right)^2 \, I_{46} \right) \,, \nonumber \\
M_{47} &= m_q^4 \, \e^3 \, x \, \sqrt{h\,\left(h-4\right)} \, I_{47} \,, \nonumber \\
M_{48} &= m_q^2 \, \e^4 \, \left(x + z\right) \, I_{48} \,, \nonumber \\
M_{49} &= -\frac{1}{2} \, m_q^4 \, \e^3 \, \sqrt{h\,\left(h-4\right)} \, \left(x \, I_{49} - z \, I_{50}\right) \,, \nonumber \\
M_{50} &= m_q^4 \, \e^3 \, \left(\left(x\,(h-2)-2\,z\right) \, I_{49} + \left(z\,(h-2)-2\,x\right) \, I_{50}\right) \,, \nonumber \\
M_{51} &= m_q^4 \, \e^3 \, \sqrt{x\,z\,(4\,\left(h-x-z)+x\,z\right)} \, I_{51} \,, \nonumber \\
M_{52} &= m_q^2 \, \e^3 \, \left(x + z\right) \, \left(I_{52} -I_{213} - I_{214} \right) \,, \nonumber \\
M_{53} &= m_q^4 \, \e^3 \, \left(h - x\right) \, \sqrt{x\,\left(x-4\right)} \, I_{53} \,, \nonumber \\
M_{54} &= m_q^4 \, \e^3 \, \left(h - z\right) \, \sqrt{h\,\left(h-4\right)} \, I_{54} \,, \nonumber \\
M_{55} &= m_q^4 \, \e^4 \, \sqrt{x\,z\,(4\,\left(h-x-z)+x\,z\right)} \, I_{55}\,, \nonumber \\
M_{56} &= -m_q^2 \, \e^4 \, \left(h - x\right) \, \left(I_{39} - I_{56}\right) \,, \nonumber \\
M_{57} &= m_q^4 \, \e^4 \, \sqrt{x\,z\,(4\,\left(h-x-z)+x\,z\right)} \, I_{57} \,, \nonumber \\
M_{58} &= -m_q^2 \, \e^4 \, z \, \left(I_{39} - I_{58}\right) \,, \nonumber \\
M_{59} &= m_q^4 \, \e^4 \, \left(-x\right)^{3/2} \, I_{59} \,, \nonumber \\
M_{60} &= m_q^6 \, \e^4 \, \left(-x\right)^{3/2} \, I_{60} \,, \nonumber \\
M_{61} &= m_q^6 \, \e^3 \, \sqrt{x\,z\,h\,(h-4)\,\left(4\,(h-x-z)+x\,z\right)} \, I_{61} \,, \nonumber \\
M_{62} &= -m_q^2 \, \e^4 \, x \, \left( I_{48} - I_{62} \right) \,, \nonumber \\
M_{63} &= m_q^4 \, \e^4 \, \sqrt{x\,z\,(4\,\left(h-x-z)+x\,z\right)} \, I_{63} \,, \nonumber \\
M_{64} &= -m_q^2 \, \e^4 \, \left(h - z\right) \, \left(I_{48} - I_{64}\right) \,, \nonumber \\
M_{65} &= m_q^4 \, \e^4 \, x \, \left(h - x\right) \, I_{65} \,, \nonumber \\
M_{66} &= m_q^6 \, \e^3 \, \sqrt{x\,z\,h\,(h-4)\,\left(4\,(h-x-z)+x\,z\right)} \, I_{66} \,, \nonumber \\
M_{67} &= m_q^6 \, \e^4 \, \sqrt{x\,\left(x-4\right) } \, \sqrt{x\,z\,(4\,\left(h-x-z)+x\,z\right)} \, I_{67} \,, \nonumber \\
M_{68} &= m_q^4 \, \e^4 \, \sqrt{x\,\left(x-4\right)} \, \left(\left(h-x\right) \, I_{68} + z \, I_{231}\right) \,, \nonumber \\
M_{69} &= -m_q^4 \, \e^4 \, \sqrt{x\,\left(x-4\right)} \, \left(\left(h-z\right) \, I_{55} - x \, I_{69} - z \, I_{233}\right) \,, \nonumber \\
M_{70} &= -m_q^2 \, \e^3 \, x \left( \frac{1}{2} \, m_q^2 \, x \, \left(I_{25} - I_{27}\right) - \e \, I_{43} - \frac{1}{4} \, m_q^2 \, (h-x) \, I_{44} + \e \, I_{48} + \frac{1}{2} \, m_q^2 \, (h-x) \, I_{53} \right. \nonumber \\
& \qquad \left. + \frac{1}{2} \, m_q^2 \, \e \, (h-z) \, I_{55} + \e \, I_{56} + m_q^2 \, \e \, z \, I_{65} + \frac{1}{2} \, m_q^2 \, \e \, (h-x) \, I_{68} - \frac{1}{2} \, m_q^2 \, \e \, x \, I_{69} - \e \, I_{70} \right. \nonumber \\
& \qquad \left. - \frac{1}{4} \, m_q^2 \, (h-x) \, I_{222} - \frac{1}{2} \, m_q^2 \, \e \, z \, \left(I_{231} + I_{233}\right) \right) \,, \nonumber \\
M_{71} &= m_q^2 \, \e^2 \, x \, I_{71} \,, \nonumber \\
M_{72} &= m_q^2 \, \e^2 \, h \, I_{72} \,, \nonumber \\
M_{73} &= m_q^4 \, \e^2 \, x \, \sqrt{x\,\left(x-4\right)} \, I_{73} \,, \nonumber \\
M_{74} &= m_q^2 \, \e^3 \, \left(h - z\right) \, I_{74} \,, \nonumber \\
M_{75} &= -m_q^2 \, \e^2 \, \sqrt{\left(h - z\right)\,\left(h-z-4\right)}\left(I_{6} + \frac{1}{2} \, I_{7} - m_q^2 \, z \, I_{75}\right) \,, \nonumber \\
M_{76} &= m_q^4 \, \e^2 \, x \, \sqrt{h\,\left(h-4\right)} \, I_{76} \,, \nonumber \\
M_{77} &= m_q^4 \, \e^2 \, h \, \sqrt{h\,\left(h-4\right)} \, I_{77} \,, \nonumber \\
M_{78} &= m_q^2 \, \e^3 \, \left(2\e-1\right) \, x \, I_{78} \,, \nonumber \\
M_{79} &= m_q^2 \, \e^4 \, \left(h - z\right) \, I_{79} \,, \nonumber \\
M_{80} &= m_q^2 \, \e^3 \, \left(\e-1\right) \, x \, z \, I_{80} \,, \nonumber \\
M_{81} &= m_q^2 \, \e^3 \, z \, \left(\left(2\e-1\right) \, I_{81} - m_q^2 \, \left(x-4\right) \, I_{82} \right) \,, \nonumber \\
M_{82} &= m_q^4 \, \e^3 \, z \, \sqrt{x\,\left(x-4\right)} \, I_{82} \,, \nonumber \\
M_{83} &= m_q^2 \, \e^4 \, \left(h - x - z\right) \, I_{83} \,, \nonumber \\
M_{84} &= m_q^4 \, \e^3 \, \sqrt{x\,z\,(4\,\left(h-x-z)+x\,z\right)} \, I_{84} \,, \nonumber \\
M_{85} &= m_q^2 \, \e^4 \, \left(h - x\right) \, I_{85} \,, \nonumber \\
M_{86} &= -m_q^2 \, \e^2 \, \left(
\frac{3}{4} \, \frac{x\,\left(h-4\right)}{h-2\,x} \, I_4
+ \frac{1}{2} \, \left(h-4\right) \, I_5
+ \frac{1}{2} \, \left(h-2\,\left(x-2\right)\right) \, \left(I_6 + \frac{1}{2} \, I_7 \right)
\right. \nonumber \\
& \qquad \left.
+ m_q^2 \, x \, \left(h-4\right) \, I_{76}
- 2 \, \e^2 \, h \, I_{85}
- m_q^2 \, x \, I_{86}
+ \frac{1}{2} \, m_q^2 \, \e \, \left(h-4\right) \, \left(h+x\right) \, I_{87}
\right. \nonumber \\
& \qquad \left.
+ \e \, \frac{h\,\left(2\,h-3\,x-4\right)+4\,x}{h-2\,x} \, I_{213}
+ \frac{1}{2} \, \e \, \frac{\left(h-4\right)\,\left(h-x\right)}{h-2\,x} \, I_{214}
\right. \nonumber \\
& \qquad \left.
- m_q^2 \, \frac{\left(h-4\right)\,\left(h\,\left(x-1\right)-x^2\right)}{h-2\,x} \, I_{215}
- 2 \, \e \, h \, I_{236}
- m_q^2 \, x \, \left(h-x-4\right) \, I_{237}
\right) \,, \nonumber \\
M_{87} &= m_q^4 \, \e^3 \, \left(h - x\right) \, \sqrt{h\,\left(h-4\right)} \, I_{87} \,, \nonumber \\
M_{88} &= m_q^2 \, \e^4 \, \left(x + z\right) \, I_{88} \,, \nonumber \\
M_{89} &= m_q^4 \, \e^3 \, \sqrt{x\,z\,(4\,\left(h-x-z)+x\,z\right)} \, I_{89} \,, \nonumber \\
M_{90} &= -m_q^2 \, \e^2 \left(
\frac{3}{4} \, \frac{x\,(h-2)-2\,z}{h-2\,x} \, I_4
- \e \, \frac{(h-z)\,(x\,(h-2)-2\,z)}{x\,(h-2\,z)} \, \left( I_{18} + \frac{1}{2} \, I_{19} \right)
\right. \nonumber \\
& \qquad \left.
+ m_q^2 \, \frac{(x\,(h-2)-2\,z)\,(h\,(z-1)-z^2)}{x\,(h-2\,z)} \, I_{20}
+ \frac{1}{2} \, m_q^2 \, \e \, \left((x\,(z-2)-2\,z)\right) \, I_{89}
\right. \nonumber \\
& \qquad \left.
- m_q^2 \, \e \, \frac{x^2+z^2-x\,z\,(h-2)}{x} \, I_{90}
- \frac{3}{4} \, \frac{z\,(x\,(h-2)-2\,z)}{x\,(h-2\,z)} \, I_{203}
\right. \nonumber \\
& \qquad \left.
+ \e \, \frac{(h-x)\,(x\,(h-2)-2\,z)}{x\,(h-2\,x)} \, \left( I_{213} + \frac{1}{2} \, I_{214} \right)
\right. \nonumber \\
& \qquad \left.
- m_q^2 \, \frac{(x\,(h-2)-2\,z)\,(h\,(x-1)-x^2)}{x\,(h-2\,x)} \, I_{215}
\right) \,, \nonumber \\
M_{91} &= m_q^4 \, \e^3 \, x \, \left(h - x\right) \, I_{91} \,, \nonumber \\
M_{92} &= m_q^2 \, \e^3 \, \left(2\e-1\right) \, h \, I_{92} \,, \nonumber \\
M_{93} &= m_q^4 \, \e^4 \, x\, z \, I_{93} \,, \nonumber \\
M_{94} &= m_q^4 \, \e^4 \, x \, \left(h - z\right) \, I_{94} \,, \nonumber \\
M_{95} &= m_q^4 \, \e^3 \, x \, \sqrt{h\,\left(h-4\right)} \, \left(I_{89} + m_q^2 \, \left(h - z\right) \, I_{95} - 2 \, I_{241} \right) \,, \nonumber \\
M_{96} &= m_q^4 \, \e^4 \, \left(h - x\right) \, \sqrt{x\,\left(x-4\right)} \, I_{96} \,, \nonumber \\
M_{97} &= m_q^6 \, \e^3 \, x \, z \, \sqrt{h\,\left(h-4\right)} \, I_{97} \,, \nonumber \\
M_{98} &= m_q^6 \, \e^4 \, x \, \sqrt{x\,z\,(4\,\left(h-x-z)+x\,z\right)} \, I_{98} \,, \nonumber \\
M_{99} &= m_q^4 \, \e^4 \, x \, \left(h - x\right) \, I_{99} \,, \nonumber \\
M_{100} &= m_q^4 \, \e^3 \, \sqrt{x\,\left(x-4\right)} \, \left( \frac{1}{2} \, \left(h-2\,z\right) \, \left(I_{37} + I_{38}\right) + \frac{1}{2} \, z \, I_{89} - m_q^2 \, \e \, x \, z \, I_{98} + \e \, x \, I_{100}
\right. \nonumber \\
& \qquad \qquad \qquad \qquad \quad\, \left.
- \frac{1}{2} \left(h-z\right) \, I_{243} \right) \,, \nonumber \\
M_{101} &= -m_q^2 \, \e^2 \, \left(
\frac{1}{4} \, \frac{x^2}{h-x} \, I_4
- \frac{1}{4} \, \frac{h\,x}{h-x} \, I_7
- \frac{1}{2} \, \e \, \frac{x^2}{h-x} \, I_9
+ \frac{1}{2} \, \e \, \frac{x\,(h-z)}{h-x} \, I_{21}
\right. \nonumber \\
& \qquad \left.
- \e \, \left(2\,\e-1\right) \, x \, I_{36}
- \frac{1}{4} \, m_q^2 \, \e \, \left(x\,\left(h-2\,z\right) + \sqrt{x\,z\,(4\,\left(h-x-z)+x\,z\right)}\right) \, I_{37}
\right. \nonumber \\
& \qquad \left.
- \frac{1}{4} \, m_q^2 \, \e \, \left(x\,\left(h-2\,z+4\right) + \sqrt{x\,z\,(4\,\left(h-x-z)+x\,z\right)}\right) \, I_{38}
\right. \nonumber \\
& \qquad \left.
+ \e^2 \, x \, \left( I_{78} - I_{85} + I_{88} \right)
- \frac{1}{4} \, m_q^2 \, \e \, \left(x\,z + \sqrt{x\,z\,(4\,\left(h-x-z)+x\,z\right)}\right) \, I_{89}
\right. \nonumber \\
& \qquad \left.
- m_q^2 \, \e \, x \, z \, I_{91}
- \frac{1}{2} \, m_q^2 \, \e^2 \, x \, \left(h-x\right) \, I_{96}
+ \frac{1}{2} \, m_q^4 \, \e^2 \, x^2 \, z \, I_{98}
- \frac{1}{2} \, m_q^2 \, \e^2 \, x^2 \, I_{100}
\right. \nonumber \\
& \qquad \left.
- \e^2 \, x \, I_{101}
+ \frac{1}{2} \, \e \, \frac{x\,z}{h-x} \, I_{205}
- \frac{1}{2} \, \e \, x \, I_{209}
+ \e^2 \, x \, I_{242}
\right. \nonumber \\
& \qquad \left.
+ \frac{1}{4} \, m_q^2 \, \e \, \left(x\,\left(h-z\right) + \sqrt{x\,z\,(4\,\left(h-x-z)+x\,z\right)}\right) \, I_{243}
\right) \,, \nonumber \\
M_{102} &= -m_q^4 \, \e^4 \, h \, \left( x \, I_{94} - m_q^2 \, x \, z \, I_{102} + z \, I_{248} \right) \,, \nonumber \\
M_{201} &= m_q^2 \, \e^2 \, \sqrt{z\,\left(z-4\right)} \, I_{201} \,, \nonumber \\
M_{202} &= m_q^2 \, \e^2 \, \sqrt{z\,\left(z-4\right)} \, \left(I_{202} + \frac{1}{2} \, I_{203}\right) \,, \nonumber \\
M_{203} &= m_q^2 \, \e^2 \, z \, I_{203} \,, \nonumber \\
M_{204} &= m_q^2 \, \e^3 \, z \, I_{204} \,, \nonumber \\
M_{205} &= m_q^2 \, \e^3 \, z \, I_{205} \,, \nonumber \\
M_{206} &= m_q^2 \, \e^3 \, z \, I_{206} \,, \nonumber \\
M_{207} &= m_q^2 \, \e^3 \, z \, I_{207} \,, \nonumber \\
M_{208} &= m_q^2 \, \e^2 \, \sqrt{z\,\left(z-4\right)} \, \left(\e \, \left(I_{206} + \frac{1}{2} \, I_{207}\right) + m_q^2 \, I_{208}\right) \,, \nonumber \\
M_{209} &= m_q^2 \, \e^3 \, \left(h - x\right) \, I_{209}\,, \nonumber \\
M_{210} &= m_q^2 \, \e^3 \, \left(h - x\right) \, I_{210} \,, \nonumber \\
M_{211} &= m_q^2 \, \e^3 \, \left(h - x\right) \, I_{211} \,, \nonumber \\
M_{212} &= m_q^2 \, \e^2 \, \frac{\sqrt{x\,\left(x-4\right)}}{2\,h-x} \, \left( \frac{3}{4} \, h \, I_7 - \e \, \left(h-x\right) \, \left( I_{210} + \frac{1}{2} \, I_{211} \right) \right. \nonumber \\
& \qquad \qquad \qquad \qquad \quad \;\;\; \left.
+ m_q^2 \, \left(h\,\left(h-x\right)+x\right) \, I_{212} \right) \,, \nonumber \\
M_{213} &= m_q^2 \, \e^3 \, \left(h - x\right) \, I_{213} \,, \nonumber \\
M_{214} &= m_q^2 \, \e^3 \, \left(h - x\right) \, I_{214} \,, \nonumber \\
M_{215} &= -m_q^2 \, \e^2 \, \frac{\sqrt{h\,\left(h-4\right)}}{h-2\,x} \, \left( \frac{3}{4} \, x \, I_4 + \e \, \left(h-x\right) \, \left( I_{213} + \frac{1}{2} \, I_{214} \right) \right. \nonumber \\
& \qquad \qquad \qquad \qquad \qquad \;\; \left.
- m_q^2 \, \left(x\,\left(h-x\right)-h\right) \, I_{215} \right) \,, \nonumber \\
M_{216} &= m_q^2 \, \e^3 \, \left(h - x\right) \, I_{216} \,, \nonumber \\
M_{217} &=m_q^4 \, \e^2 \, \sqrt{h\,\left(h-4\right)\,z\,\left(z-4\right)} \,  I_{217} \,, \nonumber \\
M_{218} &= m_q^2 \, \e^4 \, z \, I_{218} \,, \nonumber \\
M_{219} &= m_q^4 \, \e^3 \, z \, \sqrt{z\,\left(z-4\right)} \, I_{219} \,, \nonumber \\
M_{220} &= m_q^2 \, \e^4 \, z \, I_{220} \,, \nonumber \\
M_{221} &= m_q^2 \, \e^4 \, \left(h - x\right) \, I_{221} \,, \nonumber \\
M_{222} &= m_q^4 \, \e^3 \, \left(h - x\right) \, \sqrt{x\,\left(x-4\right)} \, I_{222} \,, \nonumber \\
M_{223} &= m_q^2 \, \e^3 \, x \, \left(\left(2\,\e-1\right) \, I_{223} + m_q^2 \, I_{225} \right) \,, \nonumber \\
M_{224} &= m_q^4 \, \e^3 \, \sqrt{x\,z\,(4\,\left(h-x-z)+x\,z\right)} \, \left(I_{224} + I_{225}\right) \,, \nonumber \\
M_{225} &= m_q^4 \, \e^3 \, \sqrt{x\,\left(x\,\left(z-1\right)^2+4\,z\,\left(h-z\right)\right)} \, I_{225} \,, \nonumber \\
M_{226} &= m_q^2 \, \e^3 \, \left(h - z\right) \, \left(\left(2\,\e-1\right) \, I_{226} + m_q^2 \, I_{228}\right) \,, \nonumber \\
M_{227} &= m_q^4 \, \e^3 \, \sqrt{x\,z\,(4\,\left(h-x-z)+x\,z\right)} \, \left(I_{227} + I_{228}\right) \,, \nonumber \\
M_{228} &= m_q^4 \, \e^3 \, \sqrt{(h-z)^2+2\,x\,z\,(h-z)+x^2\,z\,(z-4)} \, I_{228} \,, \nonumber \\
M_{229} &= m_q^4 \, \e^3 \, z \, \sqrt{h\,\left(h-4\right)} \, I_{229} \,, \nonumber \\
M_{230} &= m_q^4 \, \e^3 \, \left(h - x\right) \, \sqrt{h\,\left(h-4\right)} \, I_{230} \,, \nonumber \\
M_{231} &= m_q^4 \, \e^4 \, \sqrt{x\,z\,(4\,\left(h-x-z)+x\,z\right)} \, I_{231} \,, \nonumber \\
M_{232} &= -m_q^2 \, \e^4 \, x \, \left(I_{39} - I_{232} \right) \,, \nonumber \\
M_{233} &= m_q^4 \, \e^4 \, \sqrt{x\,z\,(4\,\left(h-x-z)+x\,z\right)} \, I_{233} \,, \nonumber \\
M_{234} &= -m_q^2 \, \e^4 \, \left(h - x\right) \, \left(I_{48} - I_{234} \right) \,, \nonumber \\
M_{235} &= m_q^2 \, \e^2 \, z \, I_{235} \,, \nonumber \\
M_{236} &= m_q^2 \, \e^3 \, \left(h - x\right) \, I_{236} \,, \nonumber \\
M_{237} &= -m_q^2 \, \e^2 \, \sqrt{\left(h-x\right)\,\left(h-x-4\right)} \, \left( I_{6} + \frac{1}{2} \, I_{7} - m_q^2 \, x \, I_{237} \right) \,, \nonumber \\
M_{238} &= m_q^4 \, \e^2 \, z \, \sqrt{h\,\left(h-4\right)} \, I_{238} \,, \nonumber \\
M_{239} &= m_q^2 \, \e^4 \, \left(h - x\right) \, I_{239} \,, \nonumber \\
M_{240} &= m_q^2 \, \e^3 \, x \, \left(\left(2\,\e-1\right) \, I_{240} - m_q^2 \, \left(z-4\right) \, I_{241} \right) \,, \nonumber \\
M_{241} &= m_q^4 \, \e^3 \, x \, \sqrt{z\,\left(z-4\right)} \, I_{241} \,, \nonumber \\
M_{242} &= m_q^2 \, \e^4 \, \left(h - x - z\right) \, I_{242} \,, \nonumber \\
M_{243} &= m_q^4 \, \e^3 \, \sqrt{x\,z\,(4\,\left(h-x-z)+x\,z\right)} \, I_{243} \,, \nonumber \\
M_{244} &= m_q^2 \, \e^4 \, \left(h - z\right) \, I_{244} \,, \nonumber \\
M_{245} &= -m_q^2 \, \e^2 \, \left(
\frac{1}{2} \, \left(h-4\right) \, I_5
+ \frac{1}{2} \, \left(h-2\,\left(z-2\right)\right) \, \left(I_6 + \frac{1}{2} \, I_7 \right)
\right. \nonumber \\
& \qquad \left.
+ \e \, \frac{h\,\left(2\,h-3\,z-4\right)+4\,z}{h-2\,z} \, I_{18}
+ \frac{1}{2} \, \e \, \frac{\left(h-4\right)\,\left(h-z\right)}{h-2\,z} \, I_{19}
\right. \nonumber \\
& \qquad \left.
- m_q^2 \, \frac{\left(h-4\right)\,\left(h\,\left(z-1\right)-z^2\right)}{h-2\,z} \, I_{20}
- 2 \, \e \, h \, I_{74}
- m_q^2 \, z \, \left(h-z-4\right) \, I_{75}
\right. \nonumber \\
& \qquad \left.
+ \frac{3}{4} \, \frac{z\,\left(h-4\right)}{h-2\,z} \, I_{203}
+ m_q^2 \, z \, \left(h-4\right) \, I_{238}
- 2 \, \e^2 \, h \, I_{244}
- m_q^2 \, z \, I_{245}
\right. \nonumber \\
& \qquad \left.
+ \frac{1}{2} \, m_q^2 \, \e \, \left(h-4\right) \, \left(h+z\right) \, I_{246}
\right) \,, \nonumber \\
M_{246} &=m_q^4 \, \e^3 \, \left(h - z\right) \, \sqrt{h\,\left(h-4\right)} \,  I_{246} \,, \nonumber \\
M_{247} &= m_q^4 \, \e^4 \, x \, z \, I_{247} \,, \nonumber \\
M_{248} &= m_q^4 \, \e^4 \, z \, \left(h - x\right) \, I_{248} \,, \nonumber \\
M_{249} &= -m_q^4 \, \e^3 \, z \, \sqrt{h\,\left(h-4\right)} \, \left( 2 \, I_{82} - I_{89} -m_q^2 \, \left(h - x\right) \, I_{249} \right) \,. 
\end{align}

\makeatletter
\addtocontents{toc}{\let\protect\l@subsubsection\protect\l@subsection}
\addtocontents{toc}{\let\protect\l@subsection\protect\l@section}
\addtocontents{toc}{\let\protect\l@section\protect\l@chapter}
\addtocontents{toc}{\let\protect\l@chapter\protect\l@paragraph}
\def\toclevel@chapter{0}
\def\toclevel@section{1}
\def\toclevel@subsection{2}
\makeatother

\fancyhead[LO]{\headfont\nouppercase{\rightmark}}	
\fancyhead[RE]{\headfont\nouppercase{\leftmark}}	

\cleardoublepage

\chapter*{Acknowledgements}
\addcontentsline{toc}{chapter}{Acknowledgements}

First and foremost, I am deeply grateful to my supervisor Prof. Dr. Thomas Gehrmann for giving me the opportunity to spend my PhD at the Physics Institute of the University of Zurich by providing me with this challenging and exciting topic. Knowing that this is rare in the particle physics community, I very much appreciate the intensive supervision and the fruitful discussions, which were accompanied by a well-balanced stimulating yet stress-free atmosphere. He always took the time for questions or problems that I was confronted with.\\
In addition, it is a pleasure to thank Prof. Dr. Massimo Passera for his willingness to serve as an external referee for this thesis. Beyond that, I would like to express my sincere gratitude to Prof. Dr. Massimiliano Grazzini, Prof. Dr. Stefano Pozzorini and Prof. Dr. Florencia Canelli for being part of my PhD committee. Further current and former members of the institute deserve my special thanks for the pleasant working atmosphere, thereby motivating me to come to the office full of joy every morning within the past four and a half years. Dedicated thanks go to Federico Buccioni, Daniel Hulme, Jan Niehues and Marius Wiesemann for proofreading many parts of this thesis. I would also like to thank my collaborators Andreas von Manteuffel, Peter Marquard, Jens Hoff and Amedeo Primo for many useful discussions. I moreover wish to thank Andreas von Manteuffel and Erich Weihs for their assistance with the use of \textsc{Reduze} and the inhouse \textsc{Mathematica} package, respectively, as well as Gudrun Heinrich and Sophia Borowka for their help with \textsc{SecDec}. Special thanks go to my `predecessor' Lorenzo Tancredi at the Physics Institute, who introduced the multi-loop world to me and had an open ear when it came to physics questions.\\
Although my scientific career is going to end with this thesis, I am convinced that my way of thinking has changed for the good over the many years of studying and doing my PhD in physics, enabling me to benefit from the experiences I have gained in the past outside of the physics world in the future. More importantly, however, physics has provided me with some of the most valuable things in life: Friendship. This thesis would not exist without Benjamin Weber, Christian Gorenflo and Shawn Williamson, who accompanied me throughout my physics studies. The studies ended, the friendship has not. This equally holds for Federico Buccioni, Jan Niehues, Marius Wiesemann and Michel Coban, with whom I share countless memorable moments from our common time in Zurich and beyond.\\
Finally, I would like to dedicate thanks to my family, who supported me in every possible way. My parents were not offered many educational opportunities so that they tirelessly motivated me to benefit from the ones I have. Special thanks go to my sister Sophie and my `brother' Benjamin~K. for being there whenever I need them. The same applies to my~Fussel, my soulmate, who I happened to find during my PhD in Zurich. I can always count on you.

\cleardoublepage
\phantomsection
\addcontentsline{toc}{chapter}{List of Figures}
\listoffigures

\cleardoublepage
\phantomsection
\addcontentsline{toc}{chapter}{List of Tables}
\listoftables

\cleardoublepage
\phantomsection
\addcontentsline{toc}{chapter}{\bibname}

												  
\bibliographystyle{utphys}

\bibliography{thesis}

\end{document}